\documentstyle[12pt,epsf]{article}
\def\journal{\topmargin .3in	\oddsidemargin .5in
	\headheight 0pt	\headsep 0pt
	\textwidth 5.625in 
	\textheight 8.25in 
	\marginparwidth 1.5in
	\parindent 2em
	\parskip .5ex plus .1ex		\jot = 1.5ex}

 \catcode`\@=11

\@addtoreset{equation}{subsection}
\@addtoreset{footnote}{section}
\def\theequation{\thesubsection.\arabic{equation}}

\newtoks\@stequation

\def\subequations{\refstepcounter{equation}%
  \edef\@savedequation{\the\c@equation}%
  \@stequation=\expandafter{\theequation}
  \edef\@savedtheequation{\the\@stequation}
  \edef\oldtheequation{\theequation}%
  \setcounter{equation}{0}%
  \def\theequation{\oldtheequation\alph{equation}}}

\def\endsubequations{\setcounter{equation}{\@savedequation}%
  \@stequation=\expandafter{\@savedtheequation}%
  \edef\theequation{\the\@stequation}\global\@ignoretrue
  \vspace*{-12pt} \\}

\def\fig#1#2#3#4{
\begin{figure}
\begin{center}
\mbox{\epsfysize #1 \epsffile{#2}}
\end{center}
\caption{#3}
\label{#4}
\end{figure}}

\def\tabl#1#2#3#4{
\begin{table}[tb]
\begin{center}
\mbox{\epsfysize #1 \epsffile{#2}}
\end{center}
\caption{#3}
\label{#4}
\end{table}}

\journal

\catcode`\@=11
\def\marginnote#1{}
\catcode`\@=11
\def\section{\@startsection {section}{1}{0pt}{-3.5ex plus -1ex minus
 -.2ex}{2.3ex plus .2ex}{\raggedright\large\bf}}
\catcode`\@=12
\newskip\humongous \humongous=0pt plus 1000pt minus 1000pt
\def\caja{\mathsurround=0pt}
\def\eqalign#1{\,\vcenter{\openup1\jot \caja
	\ialign{\strut \hfil$\displaystyle{##}$&$
	\displaystyle{{}##}$\hfil\crcr#1\crcr}}\,}
\newif\ifdtup

\def\R{{\rm I\!R}}
\def\N{{\rm I\!N}}
\def\one{{\mathchoice {\rm 1\mskip-4mu l} {\rm 1\mskip-4mu l}
{\rm 1\mskip-4.5mu l} {\rm 1\mskip-5mu l}}}
\def\Q{{\mathchoice
{\setbox0=\hbox{$\displaystyle\rm Q$}\hbox{\raise 0.15\ht0\hbox to0pt
{\kern0.4\wd0\vrule height0.8\ht0\hss}\box0}}
{\setbox0=\hbox{$\textstyle\rm Q$}\hbox{\raise 0.15\ht0\hbox to0pt
{\kern0.4\wd0\vrule height0.8\ht0\hss}\box0}}
{\setbox0=\hbox{$\scriptstyle\rm Q$}\hbox{\raise 0.15\ht0\hbox to0pt
{\kern0.4\wd0\vrule height0.7\ht0\hss}\box0}}
{\setbox0=\hbox{$\scriptscriptstyle\rm Q$}\hbox{\raise 0.15\ht0\hbox to0pt
{\kern0.4\wd0\vrule height0.7\ht0\hss}\box0}}}}
\def\C{{\mathchoice
{\setbox0=\hbox{$\displaystyle\rm C$}\hbox{\hbox to0pt
{\kern0.4\wd0\vrule height0.9\ht0\hss}\box0}}
{\setbox0=\hbox{$\textstyle\rm C$}\hbox{\hbox to0pt
{\kern0.4\wd0\vrule height0.9\ht0\hss}\box0}}
{\setbox0=\hbox{$\scriptstyle\rm C$}\hbox{\hbox to0pt
{\kern0.4\wd0\vrule height0.9\ht0\hss}\box0}}
{\setbox0=\hbox{$\scriptscriptstyle\rm C$}\hbox{\hbox to0pt
{\kern0.4\wd0\vrule height0.9\ht0\hss}\box0}}}}

\font\fivesans=cmss10 at 4.61pt
\font\sevensans=cmss10 at 6.81pt
\font\tensans=cmss10
\newfam\sansfam
\textfont\sansfam=\tensans\scriptfont\sansfam=\sevensans\scriptscriptfont
\sansfam=\fivesans
\def\sans{\fam\sansfam\tensans}
\def\Z{{\mathchoice
{\hbox{$\sans\textstyle Z\kern-0.4em Z$}}
{\hbox{$\sans\textstyle Z\kern-0.4em Z$}}
{\hbox{$\sans\scriptstyle Z\kern-0.3em Z$}}
{\hbox{$\sans\scriptscriptstyle Z\kern-0.2em Z$}}}}

\mathchardef\endbar="375

\def\ceilfill{$\raise3pt\hbox{$\mathsurround=0pt\mathord\endbar$}
  \mkern-2mu \xleaders\hbox{$\mkern-5mu
  \mathord-\mkern-5mu$}\hfill\mkern-7mu
  \raise3pt\hbox{$\mathsurround=0pt\mathord\endbar$}$}

\def\floorfill{$\raise9pt\hbox{$\mathsurround=0pt\mathord\endbar$}
  \mkern-2mu \xleaders\hbox{$\mkern-5mu
  \mathord-\mkern-5mu$}\hfill\mkern-7mu
  \raise9pt\hbox{$\mathsurround=0pt\mathord\endbar$}$}

\def\overcontract#1{\mathop{\vbox{\ialign{##\crcr\noalign{\kern3pt}
  \ceilfill\hskip6pt\crcr\noalign{\kern3pt\nointerlineskip}
  $\hfil\displaystyle{#1}\hfil$\crcr}}}}

\def\undercontract#1{\mathop{\vtop{\ialign{##\crcr
  $\hfil\displaystyle{#1}\hfil$\crcr\noalign{\kern3pt\nointerlineskip}
  \floorfill\hskip6pt\crcr\noalign{\kern3pt}}}}}
\def\a{\alpha}
\def\b{\beta}
\def\g{\gamma}
\def\d{\delta}
\def\e{\epsilon}
\def\m{\mu}
\def\n{\nu}
\def\t{\theta}
\def\p{\pi}
\def\ps{\psi}
\def\r{\rho}
\def\th{\theta}
\def\s{\sigma}
\def\l{\lambda}
\def\o{\omega}

\def\f{\phi}
\def\x{\xi}
\def\et{\eta}
\def\L{\Lambda}
\def\S{\Sigma}

\def\Fi{\Phi}
\def\O{\Omega}
\def\D{\Delta}

\def\bq{\vec{q}}
\def\bp{\vec{p}}
\def\br{\vec{r}}

\def\bs{\bar{s}}

\def\bA{\bar{A}}
\def\bB{\bar{B}}

\def\bE{\bar{E}}

\def\tL{\tilde{L}}
\def\tT{\tilde{T}}
\def\tD{\tilde{\Delta}}

\def\sL{{\cal L}}
\def\J{{\cal J}}
\def\F{{\cal F}}

\def\G{{\cal G}}
\def\K{{\cal K}}
\def\cD{{\cal D}}

\def\ctD{\tilde{{\cal D}}}
\def\T{{\cal T}}
\def\C{{\cal C}}
\def\cO{{\cal O}}

\def\hg{\hat{g}}
\def\op{\oplus}
\def\au{{\rm Aut}\,g}
\def\pa{\partial}

\def\df{\delta \phi}
\def\tdf{\tilde{\delta}\phi}

\def\ra{\rightarrow}
\def\lra{\leftrightarrow}
\def\ti{\times}

\def\xx{\hbox{ }^*_*}
\def\oo{\hbox{ }^o_o}
\def\bon{{\bf 1}. }
\def\btw{{\bf 2}. }
\def\bth{{\bf 3}. }
\def\bfo{{\bf 4}. }
\def\bfi{{\bf 5}. }
\def\bsi{{\bf 6}. }
\def\bse{{\bf 7}. }
\def\bei{{\bf 8}. }
\def\bni{{\bf 9}. }
\def\Ab{{\bf A}. }
\def\Bb{{\bf B}. }
\def\Cb{{\bf C}. }
\def\Db{{\bf D}. }

\def\bul{$\bullet \;\,$}
\def\Ut{SU(n)_{metric}[\scriptstyle{N=1 \atop t=0} \displaystyle]}

\def\Uppt{SU(\Pi_i n_i)_{metric}[\scriptstyle{N=1 \atop t=0} \displaystyle]}
\def\G{{\cal G}}
\def\I{{\cal I}_M({\rm Aut}\,g)}
\def\Is{{\cal I}_M({\rm Aut}(g\times SO({\rm dim}\,g)))}

\def\St{SO(n)_{diag}[\scriptstyle{N=1 \atop t=0} \displaystyle]}
\def\gt{g_{metric}[\scriptstyle{N=1 \atop t=0} \displaystyle]}
\def\Sd{SO(n)_{diag}}
\def\hh{\tilde{h}}\def\ep{\hfill $\Box$}

\def\rd{{\rm d}}
\def\u{\underline}
\def\ni{\noindent}
\def\nl{\newline}
\def\ed{\end{document}}

\def\be{\begin{equation}}
\def\ee{\end{equation}}
\def\bs{\begin{subequations}}
\def\es{\end{subequations}}

\def\bibtem#1{\bibitem{#1} }

\def\le#1{\label{#1} \ee}
\def\ls#1{\label{#1} \es}
\def\lam#1{}

\def\sp{\quad, \quad}
\def\pe{\quad . }
\def\ben{\begin{enumerate}}
\def\een{\end{enumerate}}
\def\ben{\begin{itemize}}
\def\een{\end{itemize}}

\newlength{\boxsize} 
\def\limit#1#2{ \smash{ \mathop{#1} \limits_{#2} } }
\def\limitt#1#2#3{ \settowidth{\boxsize}{\scriptsize $#2$}
         \,
         \makebox{\makebox[\boxsize][c]{$#1$}
         \hspace{-\boxsize}
         \hspace{-12pt}
         \raisebox{-4pt}{\scriptsize $#2$}
         \hspace{-\boxsize}
         \hspace{-12pt}
         \raisebox{-9pt}{\scriptsize $#3$} } }

\def\alphab{\bar{\a}}
\def\bT{\bar{T}}
\def\bz{\bar{z}}
\def\tP{\tilde{P}}
\def\tA{\tilde{A}}
\def\ta{\tilde{a}}
\def\tY{\tilde{Y}}
\def\ty{\tilde{y}}
\def\tz{{\tilde{z}}}
\def\tw{{\tilde{w}}}
\def\tg{\tilde{\g}}
\def\tu{\tilde{u}}
\def\tF{\tilde{F}}
\def\tW{\tilde{W}}
\def\tps{\tilde{\ps}}
\def\~{\tilde}

\def\hga{\hat{\g}}
\def\penclose#1{\left(#1\right)}
\def\sbenclose#1{\left[#1\right]}
\def\ident{\equiv}
\def\half{{1\over2}}
\def\ys{{y^*}}
\def\del{\partial}
\def\Tr{{\rm Tr}}
\def\rank{{\rm rank\,}}
\def\lrp{\stackrel{\leftrightarrow}{\partial}}
\def\lrbp{\stackrel{\leftrightarrow}{\bar\partial}}
\def\tpa{\tilde{\partial}}
\def\ep{\hfill $\Box$}
\def\un{\underline}
\def\>{\rangle}
\def\<{\langle}
\def\lt{\tilde L}
\def\ld{{\dot L}}

\begin{document}
\begin{titlepage}

$\mbox{}$ \hfill UCB-PTH-95/02  \\
hep-th/9501144 \hfill     LBL-36686  \\
$\mbox{}$ \hfill CERN-TH.95/2 \\
$\mbox{}$ \hfill  CPTH-A342.0195
\begin{center}

{\large \bf Irrational Conformal Field Theory}
\footnote{To appear in Physics Reports.}\footnote{The work of MBH
and KC was supported in part by the Director,
Office of
Energy Research, Office of High Energy and Nuclear Physics, Division
of
High Energy Physics of the U.S. Department of Energy under Contract
DE-AC03-76SF00098 and in part by the National Science Foundation
under
grant PHY90-21139.
The work of NO was supported in part by a
S\'ejour Scientifique de Longue Dur\'ee of the Minist\`ere des
Affaires
Etrang\`eres.}

\vskip .3in
M.B. Halpern${}^1$\footnote{e-mail: MBHALPERN@LBL.GOV,
HALPERN@PHYSICS.BERKELEY.EDU}
, E. Kiritsis${}^2$\footnote{e-mail: KIRITSIS@NXTH04.CERN.CH}
, N.A. Obers${}^3$\footnote{e-mail: OBERS@ORPHEE.POLYTECHNIQUE.FR }
, K. Clubok${}^1$\footnote{e-mail: CLUBOK@PHYSICS.BERKELEY.EDU}
\vskip .1in
{\em  1. Department of Physics, University of California and\\
      Theoretical Physics Group, Lawrence Berkeley Laboratory\\
      Berkeley, California 94720, USA}
\vskip .1in
{\em  2. Theory Division,  CERN, CH-1211 \\
      Geneva 23, SWITZERLAND}
\vskip .1in
{\em 3. Centre de Physique Th\'eorique \\
Ecole Polytechnique, F-91128 Palaiseau, FRANCE}

\end{center}
\vskip .1in
\begin{abstract}
This is a review of irrational conformal field theory, which includes
rational conformal field theory as a small subspace.
Central topics of the review include the Virasoro master equation, its
solutions and the dynamics of irrational conformal field theory.
Discussion of the dynamics includes the generalized Knizhnik-Zamolodchikov
equations on the sphere, the corresponding heat-like systems on the torus
and the generic world-sheet action of
irrational conformal field theory.
\end{abstract}
\end{titlepage}
\renewcommand{\thepage}{\roman{page}}
\setcounter{page}{1}
\setcounter{footnote}{1}
\renewcommand{\theequation}{\thesection.\arabic{equation}}
\tableofcontents
\newpage

\renewcommand{\thesubsection}{\thesection.\arabic{subsection}}
\renewcommand{\theequation}{\thesubsection.\arabic{equation}}

\renewcommand{\thepage}{\arabic{page}}
\setcounter{page}{1}
\addcontentsline{toc}{section}{Introduction}
\section*{Introduction}

This is a review of irrational conformal theory, a subject which has grown from
the discovery \cite{hk,rus,nuc} of unitary conformal field theories (CFTs)
with irrational central charge.

More precisely, irrational conformal field theory (ICFT) is defined to include
all conformal field theories, and, in particular, ICFT includes rational
conformal field theory (RCFT)
$$
\mbox{ICFT}\;\;\;\supset\supset\;\;\;\mbox{RCFT}
$$
as a very small subspace (see Fig.\ref{f11}).

The central tool in the study of ICFT is the general affine-Virasoro
construction \cite{hk,rus},
$$
T=L^{ab}\xx J_{a}J_{b}\xx
$$
where $T$ is a conformal stress tensor and $J_{a}$, $a=1\ldots$dim$\,g$
are the currents of a general affine Lie algebra.
The matrix $L^{ab}$ is called the inverse inertia tensor of the ICFT,
in analogy with the spinning top.

The Virasoro condition on $T$ is summarized by the Virasoro master
equation (VME), which is a set of quadratic equations for the
inverse inertia tensor.
The solution space of the VME is called affine-Virasoro space.
This space contains the conventional RCFTs (which include
the affine-Sugawara and
coset constructions) and a vast array of new, generically-irrational conformal
field theories.

Generic irrationality of the central charge (even on positive integer level of
affine compact Lie algebras) is a simple
consequence of the quadratic nature of the VME, and, similarly, it
is believed that the space of all unitary theories
is dominated by irrational central charge.
Many candidates for new unitary
RCFTs, beyond the conventional RCFTs, have also been found.

A coarse-grained picture of affine-Virasoro space is obtained by
thinking in terms of the symmetry of the inverse inertia tensor
in the spinning top analogy:
The conventional RCFTs are very special cases of relatively
high symmetry, while the
generic ICFT is completely asymmetric.

Many exact unitary solutions with irrational central charge
have been found, beginning with the
solutions of Ref.\ \cite{nuc}.  More generally,
the conformal field theories of affine-Virasoro space have been
partially classified, using graph theory \cite{gt,lie}
and generalized graph theory \cite{ggt}.
Remarkably, all the ICFTs so far classified are unitary on positive
integer level of the compact affine algebras.

This review is presented in three parts,

\centerline{I.~~The General Affine-Virasoro Construction}

\centerline{II.~~Affine-Virasoro Space\phantom{uuuuuuuuuuuuuuuu}}

\centerline{III.~~The Dynamics of ICFT\phantom{uuuuuuuuuuuuuuu}}

\ni which reflect the major stages in the
development of ICFT.
\fig{3cm}{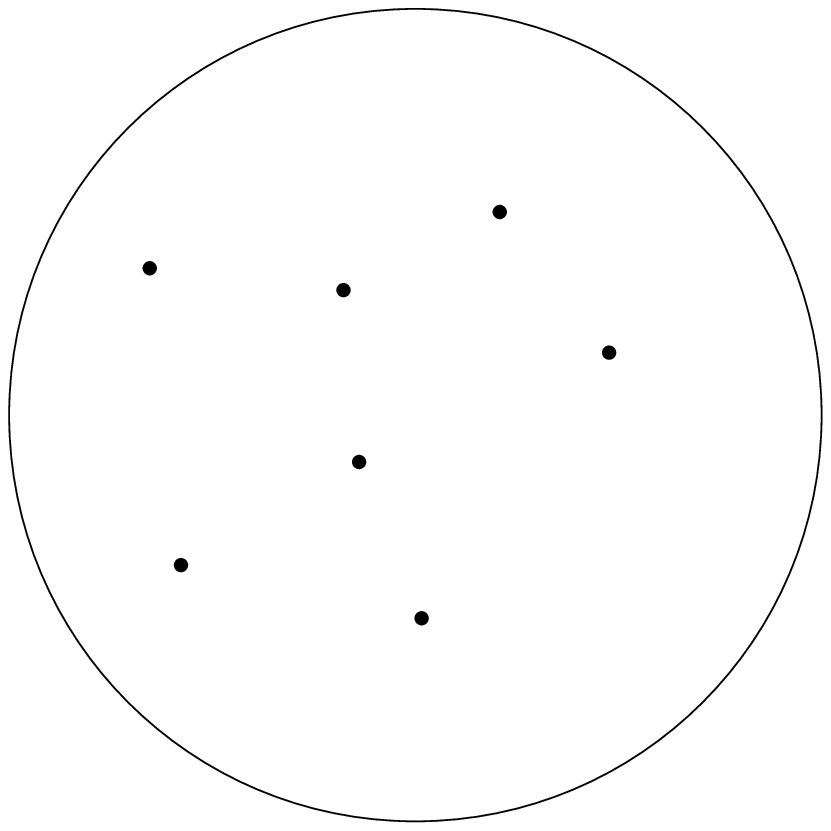}{RCFTs are rare in the space of all ICFTs.}{f11}

\vskip .5cm

$\bullet$ Part I. The first part includes a discussion of the VME,
and the associated N=1 and N=2 superconformal master equations
\cite{sme}.
General constructions satisfying the $W_{3}$ algebra are also reviewed.
This part also includes  the geometric reformulation \cite{gme}
of the VME as an Einstein equation
on the group manifold, the exact C-function \cite{cf}
on affine-Virasoro space
and the possibility \cite{k} that conformal field
theories may exist which are more general
than those of the VME.

$\bullet$ Part II. The second part discusses the solution space
of the master equations in further detail, including
the known exact unitary solutions
with irrational central charge.
The high-level expansion \cite{hl} of the master equations and the
partial classification of affine-Virasoro space by graph
theory are also reviewed here.
The partial classification by graph theory involves a
new connection
between groups and graphs, called generalized graph theory on Lie $g$
\cite{ggt}, which may be of interest in mathematics.

$\bullet$ Part III. In the third part we discuss the dynamics of ICFT, which
includes the generalized Knizhnik-Zamolodchikov
equations [103-105] on the sphere 
and the corresponding heat-like systems \cite{tor} on the torus.
These equations describe the correlators and characters of the general
ICFT, including a new description of the coset constructions on the
sphere and the torus.
As applications, we review the associated new results for
coset constructions and the more general high-level
solutions for the generic ICFT on simple $g$.
This part also contains a review of the generic world-sheet
action \cite{gva,bch} of ICFT, including a speculation on
its relation to $\sigma$-models.

Three short reviews of ICFT [87-89] have also appeared, 
which follow the same stages in the development of the subject.

\vskip .4cm
ICFT is not a finished product. Here is a list of some of the central
outstanding problems.
\vskip .4cm
\noindent
{\bf 1.} Classification. Large as they are, the graph theories classify
only small regions of affine-Virasoro space (see Section \ref{nhl}).
A complete classification is an important open problem.

\noindent
{\bf 2.} Correlators. Although the high-level correlators of ICFT are
known (see Section \ref{hc21}), it is an important open problem to obtain the
finite-level correlators of any of the exactly known unitary theories
 with irrational central charge.

\noindent
{\bf 3.} Other approaches to ICFT. Beyond the approach reviewed here,
complementary approaches to ICFT are an important open problem.
In this connection, we mention the promising directions
through non-compact coset constructions \cite{dpl,barn}
and through subfactors \cite{j,po} in mathematics.

\vskip .4cm
Other open problems are discussed as they arise in the text.

\newpage
\part{The General Affine-Virasoro Construction\label{kk0}}

\section{Affine Lie Algebras and the Conventional Affine-Virasoro
Constructions\label{sec1}}
\subsection{Affine Lie Algebra\label{sec1.1}}
\setcounter{equation}{0}

Mathematically an affine Lie algebra  \cite{km,moo,bh} $\hat g$
is the  loop algebra of a finite dimensional Lie algebra $g$
(maps from $S^1\to g$) with a
central extension. The generators of the affine algebra can
be represented as $J_{a}(\theta)$, $a=1\ldots$dim$\,g$
with $\theta$ the angle parametrizing $S^1$. The generators
$J_{a}(\theta)$ are often called the currents because
they are Noether currents in field-theoretic
realizations of affine algebras. The Fourier
modes $J_{a}(m)$, $m\in\Z$,  of the currents are often
used as a convenient basis.
See Ref.\ \cite{kac} for a detailed discussion of
affine Lie algebras in mathematics.

The most general untwisted affine Lie algebra $\hat g$ is
\be
[J_{a}(m),J_{b}(n)]=i{f_{ab}}^{c}J_{c}(m+n)+mG_{ab}\d_{m+n,
0}\,\,\,,\,\,\,\;\;m,n\in \Z\label{kaffine}
\ee
where $a,b,c=1\ldots $dim $g$ and ${f_{ab}}^{c}$ and $G_{ab}$
are respectively the structure
constants and generalized  Killing metric of $g$, not necessarily compact
or semisimple.
The zero modes $J_{a}(0)$ of the affine algebra satisfy the Lie algebra of~$g$.

The generalized Killing metric $G_{ab}$ satisfies the conditions,

\centerline{$G_{ab}$ is symmetric}

\centerline{${f_{ab}}^{d}G_{dc}$ is totally antisymmetric.}

\noindent
For non-semisimple algebras the generalized metric is not limited to the
Killing metric of the Lie algebra \cite{nw}.
For semisimple algebras $g=\oplus_{I} g_{I}$,
the generalized metric can be written as
\be
G_{ab}=\oplus_{I}k_{I}\eta^{I}_{ab}\label{kmetric}
\ee
where $\eta^{I}_{ab}$ is the Killing metric of $g_{I}$.
Each coefficient $k_{I}$ is related to the (invariant)  level
$x_{I}$ of the affine algebra by
\be
x_{I}=2k_{I}/\psi_{I}^{2}\label{klevel}
\ee
where $x_{I}$ is independent of the scale of the highest root $\psi_{I}$.
Level $x$ of affine Lie $g$ is often denoted by $g_{x}$.

Unitarity of the representations of compact affine algebras requires the
level to be a non-negative integer.
Other important numbers are the
dual Coxeter number $\tilde h_{g}$ of $g$ and the (invariant)
quadratic Casimir operators,
\bs\be
\qquad\qquad
{\tilde h}_{I}=Q_{I}/\psi_{I}^{2}\;\;,\;\;{f_{ac}}^{d}{f_{bd}}^{c}=
-\oplus_{I}Q_{I}\eta^{I}_{ab}\label{kdcn}
\qquad\qquad
\ee\be\label{casimir}
\qquad
{\cal I}({\cal T}^{I})=Q({\cal T}^{I})/\psi_{I}^{2}\;\;\;,\;\;\;
{({\cal T}_{a}^{I}\eta^{ab}_{I}{\cal T}_{b}^{I})_{J}}^{K}=
{\delta_{J}}^{K}Q({\cal T}^{I})
\qquad
\ee\be
\qquad\qquad
{\cal T}=\oplus_{I}{\cal T}^{I}\;\;\;,\;\;\; J,K=1\ldots\dim {\cal T}
\qquad\qquad
\ee\es
where ${\cal T}^{I}$ is a matrix irreducible representation (irrep)
of $g_{I}$.

Affine modules or highest weight representations of $\hat g$
are constructed as Verma modules.
The raising operators are $J_{a}(n\geq 1)$ and the
raising operators $J_{a}(0)$, $a\in \Phi_{+}$
of the positive roots.
A highest weight state $|R_{\cal T}\rangle$ is annihilated by the raising
operators and these states are classified by the highest weights of
irreps ${\cal T}$ of $g$.
The highest weight representation corresponding to ${\cal T}$
is generated by the action of the lowering operators on the highest weight
state.

In physics it is convenient to consider the collection of states
$|R_{\cal T}\rangle^{J}$, $J=1\ldots
$dim$\,{\cal T}$, called the affine primary states, which are
 generated by the action of the zero modes of the algebra
on the highest weight state.
Thus, the affine-primary states satisfy
\be
J_{a}(m\geq 0)|R_{\cal T}\rangle^{J}=\delta_{m,0}
|R_{\cal T}\rangle^{K}{({\cal T}_{a})_{K}}^{J}\label{kafpri}
\ee
where ${\cal T}$ is the corresponding matrix irrep of $g$.
The other states in the affine module are generated by the action
of the negative modes $J_{a}(m\leq 0)$ of the currents
on the affine primary states.

A special representation of $\hat g$ is the one corresponding to
the trivial or 1-dimensional representation of $g$.
We will denote its affine primary state by $|0\rangle$ and
call it the affine vacuum,
since it is the ground state in unitary field-theoretic
realizations of $\hat g$.

Affine Lie algebras are realized in quantum field theory as
algebras of local currents.
The local form of the currents is
\be
J_{a}(z)=\sum_{m\in \Z}{J_{a}(m)\over z^{m+1}}\label{kmodes}
\ee
and the operator product expansion (OPE) of the
currents\footnote{See Ref.\ \cite{bbss1} for a systematic
approach to the OPEs of composite operators.},
\be
J_{a}(z)J_{a}(w)={G_{ab}\over
(z-w)^2}+i{f_{ab}}^{c}{J_{c}(w)\over (z-w)}+{\rm reg.}
\label{kope}\ee
is equivalent to the mode algebra (\ref{kaffine}).
We will also need the affine-primary fields,
\bs\label{kprim}\be
J_{a}(z)R^{J}_{g}({\cal T},w)=
{R^{K}_{g}({\cal T},w)\over (z-w)}{({\cal T}_{a})_
{K}}^{J}+{\rm reg.} \label{kprim1}
\ee
\be
R^{J}_{g}({\cal T},0)|0\rangle=
|R_{\cal T}\rangle^{J}\label{kstate}
\ee\es
which are the interpolating fields of the affine-primary states.

\subsection{Virasoro Algebra\label{sec1.2}}

The Virasoro algebra is \cite{vir},
\bs\label{kvir1}\be
[L(m),L(n)]=(m-n)L(m+n)+{c\over 12}m(m^2-1)\delta_{m+n,0}
\label{kvirasoro}
\ee
\be
T(z)T(w)={c/2\over (z-w)^4}+{2T(w)\over
(z-w)^2}+{\pa_{w}T(w)\over (z-w)}
+{\rm reg.}\label{kvope}
\ee\be
T(z)=\sum_{m\in \Z}{L(m)\over z^{m+2}}
\ee\es
where $T$ is called the stress tensor and $c$ is called the central charge.
The central term was first observed by Weis \cite{weis}.
A Virasoro primary field $\Phi_{\Delta}$ of conformal weight $\Delta$
satisfies
\bs\be
T(z)\Phi_{\Delta}(w)=\Delta{\Phi_{\Delta}(w)\over (z-w)^2}+
{\partial \Phi_{\Delta}(w)\over (z-w)}+{\rm reg.}
\ee\be
\Phi_{\Delta}(0)|0\rangle =|\Phi_{\Delta}\rangle
\ee\be
L(m\geq 0)|\Phi_{\Delta}\rangle = \delta_{m,0}
\Delta|\Phi_{\Delta}\rangle
\ee\es
where $|\Phi_{\Delta}\rangle$ is a Virasoro primary state and
$|0\rangle$ is the $SL(2,\R)$-invariant ground state with
$\Delta=0$.
See Ref.\ \cite{bpz} for further discussion of the foundations of
conformal field theory.

The {\em affine-Virasoro constructions} are Virasoro operators built from the
currents of affine Lie algebra, and $|0\rangle$ is the affine vacuum
in this case.

\subsection{Affine-Sugawara constructions\label{sec1.3}}

The simplest affine-Virasoro constructions generalize the notion of quadratic
Casimir operators in finite Lie algebras.
These operators are the affine-Sugawara constructions \cite{bh,h1,kz,s2}
on semisimple $g$, which read
\bs\label{affinesuga}\be
L_{g}(m)=L^{ab}_{g}\sum_{n\in \Z}
\xx J_{a}(m-n)J_{b}(n)\xx\label{ksuga}
\ee\be
T_{g}(z)=L^{ab}_{g}\xx J_{a}(z)J_{b}(z)\xx
\ee\be
L_{g}^{ab}=\oplus_{I}{\eta_{I}^{ab}\over 2k_{I}+Q_{I}}\;\;\;,\;\;\;
c_{g}=\sum_{I}{x_{I}{\rm dim}\,g_{I}\over x_{I}+{\tilde
h}_{I}}\label{kcentral}
\ee\es
where $\eta_{I}^{ab}$ is the inverse Killing metric of $g_{I}$.
The normal-ordering symbol $\xx\xx$ means that negative modes of the currents
are on the left.

Under the affine-Sugawara construction, the currents and the affine primary
fields are Virasoro primary fields,
\bs\be
T_g(z)J_{a}(w)={J_{a}(w)\over (z-w)^2}+{\pa_{w}J_{a}(w)\over (z-w)}
+{\rm reg.}
\label{kvec}
\ee\be
T_{g}(z)R_{g}^{J}({\cal T},w)=\Delta_{g}({\cal T})
{R_{g}^{J}({\cal T},w)\over (z-w)^2}
+{\pa_{w}R_{g}^{J}({\cal T},w)\over (z-w)}+{\rm reg.}\label{kvir}
\ee\be
\Delta_{g} ({\cal T})=\sum_{I}{{\cal I}({\cal T}^{I})
\over x_{I}+\tilde h_{I}}\label{kdim}
\ee\es
with conformal weights one and $\Delta_{g}({\cal T})$ respectively.
In (\ref{kdim}), ${\cal I}({\cal T})$ is the quadratic Casimir operator
(\ref{casimir}).

The affine-Sugawara constructions are realized field-theoretically in the WZW
models \cite{nov,wit1}, which are reviewed in Section \ref{hc44}.

\subsection{Coset constructions\label{coset}}

The next simplest affine-Virasoro constructions are the $g/h$ coset
constructions \cite{bh,h1,gko}.

When $h\subset g$ is a subalgebra of $g$, the stress tensor of the $g/h$ coset
construction is
\bs\be
T_{g/h}(z)=T_{g}(z)-T_{h}(z)=L^{ab}_{g/h}\xx J_{a}(z)J_{b}(z)\xx
\ee\be
L^{ab}_{g/h}=L_{g}^{ab}-L_{h}^{ab}\;\;\;,\;\;\;c_{g/h}=c_{g}-c_{h}
\ee\es
where $T_{g}$ and $T_{h}$ are the stress tensors of the affine-Sugawara
construction on $g$ and $h$.
The coset stress tensor commutes with the $\hat h$ current algebra and the
affine-Sugawara construction on $h$,
\bs\be
T_{g/h}(z)J_{A}(w)={\rm reg.}\;\;\;\;A=1\ldots{\rm dim}\,h
\ee\be
T_{g/h}(z)T_{h}(w)={\rm reg.}\;\;.
\ee\es
The commutativity of the stress tensors implies that, for each $g/h$, the
affine-Sugawara construction $T_{g}=T_{g/h}+T_{h}$ is a tensor product CFT,
formed by tensoring the coset CFT with the CFT on $h$.

Coset constructions are  realized field-theoretically as gauged WZW
models \cite{brs,gk,gk2,kahs,kas}.

\addcontentsline{toc}{section}{Appendix: History of Affine Lie Algebra
and the Affine-Virasoro \hspace*{1.8cm}Constructions}
\section*{Appendix: History of Affine Lie Algebra and the
\phantom{Appendix: }Affine-Virasoro Constructions.\label{kk-1}}

Early development of affine Lie algebra and the affine-Virasoro constructions
followed independent lines in mathematics and physics.

\noindent
{\bf 1.} \underline{Affine Lie algebra}. Affine Lie algebra,
or current algebra on the circle,
was discovered independently in mathematics \cite{km,moo} and physics
\cite{bh}.

In mathematics, the affine algebras were introduced as natural generalizations
of finite dimensional Lie algebras, while in physics the affine algebras
were introduced by example, in order to describe current-algebraic
spin  and internal symmetry on the string.
The examples in physics included the affine central extension some years before
it was recognized in mathematics.

Affine Lie algebras are also known as centrally-extended loop algebras.
They comprise a special case of the more general
system known as Ka\v c-Moody algebra \cite{km,moo}, which also
includes the hyperbolic algebras.

\noindent
{\bf 2.} \underline{World-sheet fermions}. World-sheet fermions
were given independently in
\cite{bh} and \cite{r}, which describe the (Weyl) half-integer moded
and (Majorana-Weyl) integer moded cases respectively.
Half-integer moded Major\-ana-Weyl fermions were
introduced later in \cite{ns}.
World-sheet fermions played  a central role in the early representation
theory of affine Lie algebras \cite{bh} and superconformal
symmetry \cite{r,ns}.

\noindent
{\bf 3.} \underline{Representations of affine Lie algebras}.
The first concrete realization \cite{bh} of affine Lie algebra was
untwisted $SU(3)_{1}$. This realization followed the quark model
\cite{gen} to construct the level-one currents
from world-sheet fermions in the $3$ and $\bar 3$ of $SU(3)$.
Other fermionic realizations were given on orthogonal
groups in \cite{bh,h1}.

The vertex operator constructions of affine Lie algebra began in
\cite{h2,h2a,bhn}, which gave the construction of untwisted $SU(n)_1$
from compactified spatial dimensions.
This work extended the Coleman-Mandelstam bosonization of a single fermion on
the line \cite{cole,mand}
to the bosonization of many fermions on the circle,
and hence, through the fermionic realizations, to the bosonization
of the affine currents.
Technically, the central ingredients in this construction
were the Klein transformations
\cite{klein} or cocycles which are necessary for many fermions,
and the recognition of the structural analogy between the string vertex
operator \cite{fv} and Mandelstam's bosonized fermion on the line.

Twisted scalar fields  were first studied in \cite{ht,ws} and
twisted vertex operators were
introduced in  \cite{Cf}.

Concrete realizations of affine Lie algebra came later in
mathematics, where the first realization \cite{lw} was
the vertex operator construction of twisted $SU(2)_{1}$.
The untwisted vertex operator construction  of $SU(n)_{1}$ in physics was
also generalized \cite{fk,s1} to level one of simply laced $g$.

The vertex operator  construction plays a
central role in string theory
as the internal symmetry of the heterotic string \cite{gh}.

\noindent
{\bf 4.} \underline{Affine-Sugawara constructions}.
The simplest set of  affine-Virasoro constructions
are  the affine-Sugawara constructions, on the currents
of affine Lie algebra.
(Sugawara's model \cite{sug,som} was in four dimensions on a different
algebra.)
The first examples of affine-Sugawara
constructions were given in \cite{bh,h1},
using the fermionic representations of the affine algebra.
The affine-Sugawara constructions were later generalized in
\cite{kz,s2},
and the corresponding WZW action was given in \cite{nov,wit1}.

\noindent
{\bf 5.} \underline{Coset constructions}. The next set of affine-Virasoro
constructions were the $g/h$ coset constructions.
The first examples of coset constructions were given implicitly in \cite{bh}
and explicitly in \cite{h1}.
They were later generalized in \cite{gko} and the corresponding
gauged WZW action was given in \cite{brs,gk,gk2,kahs,kas}.

\noindent
{\bf 6.} \underline{Beyond the coset constructions}.
Ref.\ \cite{bh} also gave another
affine-Vira\-soro construction, the spin-orbit construction, which was
more general than the affine-Sugawara and coset constructions.
The spin-orbit construction provided a central motivation
for the discovery of the Virasoro master equation \cite{hk,rus}, which
collects all possible affine-Virasoro constructions.
An independent motivation was provided in Ref.\ \cite{k},
which considered general Virasoro constructions using arbitrary
(2,0) operators.

\vskip .4cm
Other historical developments in conformal field theory are
noted, as they arise, in the
introductory sections of the text.

\renewcommand{\thesubsection}{\thesection.\arabic{subsection}}
\renewcommand{\theequation}{\thesubsection.\arabic{equation}}

\section{The Virasoro Master Equation\label{sec2}}
\setcounter{subsection}{0}
\subsection{Derivation of the Virasoro Master Equation\label{sec3}}
\setcounter{equation}{0}

In Section \ref{sec1} we reviewed the simplest affine-Virasoro
constructions, that is,
the affine-Sugawara constructions and coset constructions.
In this section, we discuss the {\em general affine-Virasoro
construction} \cite{hk,rus},
\be
T=L^{ab}\xx J_{a}J_{b}\xx\label{kqua}
\ee
where $J_a, a=1\ldots\dim g$ are the currents of affine $g$ and
$L^{ab}=L^{ba}$ is called the {\em inverse inertia tensor} in analogy with the
spinning top.

{}To set up the construction, one first defines  the
normal-ordered current bilinears via the current-current
OPE,
\def\pw{\partial_w}
$$
J_{a}(z)J_{b}(w)={G_{ab}\over (z-w)^{2}}+i{f_{ab}}^{c}[{1\over
z-w}+{1\over 2}
\pw +{1\over 6}(z-w)\pw^{2}]J_{c}(w)\hspace*{3cm}$$
\be\hspace*{1.5cm}+[1+{1\over 2}(z-w)\pw]T_{ab}(w)+(z-w)X_{ab}(w)+{\cal
O}[(z-w)^{2}]\;\;.\label{kjj}
\ee
This expansion  defines the spin-two composite operators
\be
T_{ab}(z)=T_{ba}(z)=\xx J_{a}J_{b}\xx
\ee
and the
spin-three operators $X_{ab}(z)=-X_{ba}(z)$, both operators being
quasiprimary with respect to the affine-Sugawara construction on $g$.
The two-point functions of $T_{ab}$ and $X_{ab}$,
\be
\langle T_{ab}(z)T_{cd}(w)\rangle ={P_{ab,cd}\over
(z-w)^{4}}\,\,\,,\,\,\,
\langle X_{ab}(z)X_{cd}(w)\rangle ={H_{ab,cd}\over
(z-w)^{6}}\label{k2p}
\ee
can be computed from (\ref{kjj}), where\footnote{We use
$A_{(a}B_{b)}\equiv A_a B_b + A_b B_a$ and
$A_{[a}B_{b]}\equiv A_a B_b - A_b B_a$.}
\be
P_{ab,cd}=G_{a(c}G_{d)
b}-{1\over 2}{f_{a(c}}^{e}{f_{d)b}}^{f}G_{ef}\label{kP}
\ee
and $H_{ab,cd}$ is given in \cite{cf}.
\setcounter{footnote}{0}
For affine compact $g$, the matrix $P_{ab,cd}$ is non-negative when
each level $x_{I}$
of the simple components $g_{I}$ is some positive integer.

The next step is to compute the $T_{ab}J_c$ OPE,
\begin{eqnarray}
T_{ab}(z)J_{c}(w)&=&{M_{ab,c}}^{d}\left[{1\over (z-w)^{2}}+{1\over
(z-w)
}\pw+{1\over 2}\pw^{2}\right] J_{d}(w) \nonumber \\
\label{kTJ}
& & + {N_{ab,c}}^{de}\left[{1\over (z-w)}+{3\over 4}\pw\right]T_{de}(w)+
W_{abc}(w) \nonumber \\
& & +{K_{ab,c}}^{de}X_{de}(w)+{\cal O}(z-w)
\end{eqnarray}
where
\be
{M_{ab,c}}^{d}= {\d_{(a}}^{d}G_{b)c}+{1\over
2}{f_{e(a}}^{d}{f_{b)c}}^{e}\;\;\;,\;\;\;
{N_{ab,c}}^{de}={i\over 2}{\d_{(a}}^{(d}{f_{b)c}}^{e)}\label{kM}
\ee
and (\ref{kTJ}) serves as a definition of the spin-three quasiprimary
composite operators $W_{abc}(z)$.

Finally one  computes the OPE among the current bilinears $T_{ab}$,
$$T_{ab}(z)T_{cd}(w)={P_{ab,cd}\over (z-w)^{4}}+
{R_{ab,cd}}^{ef}\left[{1\over (z-w)^{2}}+{1\over
2(z-w)}\pw\right]T_{ef}(w) \hspace*{3cm}
$$
$$
\hspace*{1.5cm}
+{Q_{ab,cd}}^{e}\left[
{1\over
(z-w)^{3}}+{1\over 2(z-w)^{2}}\pw +{1\over
6(z-w)}\pw^{2}\right]J_{e}(w)
$$
\be+{S_{ab,cd}}^{efg}{W_{efg}(w)\over
(z-w)}+{U_{ab,cd}}^{ef}{X_{ef}(w)\over
(z-w)}+{\rm reg.}\hspace*{.8cm}
\label{kTT} \ee
where
$Q$, $S$ and $U$ are antisymmetric under the interchange
$(ab\leftrightarrow cd)$ while $R$ is symmetric.
The expressions for $Q$, $R$ and $S$ are given in \cite{hk,cf} and
$P_{ab,cd}$ is defined in (\ref{kP}).

We are now ready to answer the question: What are the conditions on the inverse
inertia tensor so that $T(L)$ in (\ref{kqua}) satisfies the Virasoro algebra?

Using (\ref{kTT}), it is not difficult to see that $L^{ab}$ must
satisfy a system of quadratic equations
\be
2L^{ab}=L^{cd}L^{ef}R_{cd,ef}{}^{ab}\;\;\;, \;\;\;
c=2L^{ab}L^{cd}P_{ab,cd}\;\;.\label{kME1}
\ee
Using the form of $R$ in Ref.\ \cite{hk}, eq.(\ref{kME1})
gives the explicit form of the {\em Virasoro master equation} \cite{hk,rus},
\bs\be
L^{ab}=2L^{ac}G_{cd}L^{db}-L^{cd}L^{ef}{f_{ce}}^{a}
{f_{df}}^{b}-L^{cd}
{f_{ce}}^{f}{f_{df}}^{(a}L^{b)e}\label{kME2}
\ee\be
c=2G_{ab}L^{ab}\label{kME3}
\ee\es
which is often abbreviated as VME below.
The linear form of the central charge in (\ref{kME3})
is obtained by using the VME.

In summary, for each solution $L^{ab}$ of the VME,
one obtains a conformal stress tensor $T(L)=L^{ab}\xx J_a J_b\xx$
with central charge (\ref{kME3}).

\vskip .4cm
\noindent
\underline{A Feigin-Fuchs generalization}
\vskip .3cm

A more general Virasoro construction has also been considered,
where the stress tensor contains terms linear in the
first derivatives of the currents
\cite{hk}
\be
T=L^{ab}\xx J_{a}J_{b}\xx+D^{a}\partial J_{a}\;\;.\label{kkk}
\ee
In this case, the Virasoro conditions are the generalized VME \cite{hk,gme},
\bs\label{kkk1}\be
L^{ab}=2L^{ac}G_{cd}L^{db}-L^{cd}L^{ef}{f_{ce}}^{a}
{f_{df}}^{b}-L^{cd}
{f_{ce}}^{f}{f_{df}}^{(a}L^{b)e}+i{f_{cd}}^{(a}L^{b)c}
D^{d}
\ee
\be
D_{a}(2G^{ab}
L_{be}+{f^{ab}}_{d}L_{bc}{f^{cd}}_{e})=D_{e}\;\;\;,\;\;\;
c=2G_{ab}(L^{ab}-6D^{a}D^{b})\label{keigen}
\ee\es
which includes the Feigin-Fuchs constructions \cite{f1,gn,FF} and the
more general c-changing linear deformations in \cite{fh}.

Another generalization with terms linear in the currents \cite{hk},
\bs\be
L(m)=L^{ab}T_{ab}(m)+D^{a}J_{a}(m) +{1\over 2}G_{ab}D^{a}D^{b}\d_{m,0}
\ee\be
c=2G_{ab}L^{ab}
\ee\es
is summarized by the VME and the eigenvalue condition on
$D$ in (\ref{keigen}).
This generalization includes the original example \cite{bh} of these
constructions, and the more general c-fixed linear deformations
(or inner-automorphic twists) in \cite{fh}.

We also note that a class of affine-Virasoro constructions has been found
using higher powers of the currents \cite{gs}, but these constructions
are automorphically equivalent to the quadratic constructions.

\vskip .4cm
\noindent
\underline{A geometric formulation of the VME}
\vskip .3cm

The VME has been  identified  \cite{gme} as an Einstein-like
equation on the group manifold $G$.

The central components in this identification are the left-invariant Einstein
metric $g_{ij}$ on $G$ and the left-right invariant
affine-Sugawara metric $g_{ij}^{(g)}$,
\be
g_{ij}={e_{i}}^{a}{e_{j}}^{b}L_{ab}\;\;,\;\;g_{ij}^{(g)
}={e_{i}}^{a}{e_{j}}^{b}L^{g}_{ab}
\ee
where ${e_{i}}^{a}$ is the left-invariant vielbein on $G$
and $L_{ab}$ is the inertia tensor.
Then the VME may be reexpressed as the Einstein system,
\be
\hat R_{ij}+g_{ij}=g^{(g)}_{ij}\;\;,\;\;\hat R_{ij}\equiv R_{ij}-{1\over
2}\tau_{kl(i}
{\tau_{j)}}^{kl}\;\;,\;\;c={\rm dim}\,g-4R\label{kkk4}
\ee
where $R_{ij}$  and $R$ are the Ricci tensor and curvature scalar of $g_{ij}$
and  $\tau$ is the
natural  contorsion on $G$.
In the case of the generalized VME (\ref{kkk1}), one obtains
 an Einstein-Maxwell system on the group manifold.

\vskip .4cm
\noindent
\underline{VME on non-semisimple algebras}
\vskip .3cm

We emphasize that the VME is valid on non-semisimple
algebras.
In this case, the Killing metric $\eta_{ab}$ of the algebra
(in $G_{ab}=k\eta_{ab}$)
is degenerate so the set of solutions of the VME will
not include an analogue of the affine-Sugawara construction.
However, there are non-degenerate invariant metrics $\hat \eta_{ab}$
 on some non-semisimple Lie algebras \cite{nw}
which allow the alternate choice $G_{ab}=k\hat \eta_{ab}$ in the VME
on the same semisimple algebra.
In this case, one obtains an analogue
\cite{nw,kk,sf3,ors,moh2,moh3,ffs,kem,kkl,ke}
of the affine-Sugawara construction with \cite{moh2,ffs}
\be
[(2k+Q)\eta]^{-1}\to [2k\hat \eta +Q\eta]^{-1}
\ee
in (\ref{affinesuga}).
Moreover, all the structure of the VME described below
is valid in this case, including K-conjugation and the
coset constructions \cite{kk,sf,sf1,sf3,ao}.

\subsection{Affine-Virasoro Space\label{sec44}}
\subsubsection{Simple properties\label{sec444}}
\setcounter{equation}{0}

The solution space of the VME is called {\em affine-Virasoro space}.
Here are some simple properties of this space.

\vskip .4cm
\noindent
{\bf 1}. Affine-Sugawara constructions \cite{bh,h1,wit1,kz}.
The affine-Sugawara construction $L_{g}$,
\be
L_{g}^{ab}=\oplus_{I}{\eta_{I}^{ab}\over 2k_{I}+Q_{I}}
\,\,\,\;\;,\,\,\,\;\;
c_{g}=\sum_{I}{x_{I}{\rm dim}\,g_{I}\over x_{I}+
{\tilde h}_{I}}\label{kkk3}
\ee
is a solution of the VME for arbitrary level of any semisimple  $g$
and similarly for $L_{h}$ when $h\subset g$.

\vskip .4cm
\noindent
{\bf 2}. K-conjugation \cite{bh,h1,gko,k,hk}.
{\em K-conjugation covariance} is  one of the most important features of the
VME, playing a major role in the structure of affine-Virasoro space and
the dynamics of ICFT.

When $L$ is a solution of the master equation on $g$, then so is
the
K-conjugate partner $\lt$ of $L$,
\be\label{kKco}
\lt^{ab}=L^{ab}_{g}-L^{ab}
\;\;\;,\;\;\;
\tilde c=c_{g}-c
\ee
and the corresponding stress tensors $T\equiv T(L)$ and
$\tilde T\equiv T(\lt)$ form
a commuting pair of Virasoro operators,
\bs\be
T(z)+\tilde T(z)=T_{g}(z)
\ee\be
T(z)\tilde T(w)={\rm reg.}
\ee\es
which add to the affine-Sugawara stress tensor $T_g$.

The decomposition $T_{g}=T+\tilde T$ strongly suggests that, for each
K-conjugate pair $T$ and $\tilde T$,
the affine-Sugawara construction is a tensor product CFT, formed
by tensoring the conformal field theories of $T$ and $\tilde T$.
In practice, one faces the inverse problem, namely the definition of the $T$
theory by modding out the $\tilde T$ theory and vice versa.
In the operator approach to the dynamics of
ICFT, this procedure is called factorization
(see Part III).
K-conjugation also plays a central role in the world-sheet action of ICFT
(see Section \ref{hc41}).

\vskip .3cm
\noindent
{\bf 3}. Coset constructions \cite{bh,h1,gko}.
K-conjugation generates new solutions from old.
The simplest examples are the $g/h$ coset constructions,
\be
\tilde L=L_{g/h}=L_{g}-L_{h}\;\;,\;\; \tilde
T=T_{g/h}=T_{g}-T_{h}\;\;,\;\;
\tilde c=c_{g/h}=c_{g}-c_{h}
\ee
which follow by K-conjugation from the subgroup construction
$L_{h}$ on $h\subset g$.

\vskip .3cm
\noindent
{\bf 4}. Affine-Sugawara nests \cite{wit2,nuc}.
Repeated K-conjugation on nested subalgebras
$g\supset h_{1}\supset h_{2}\cdots\supset h_{n}$
gives the {\em affine-Sugawara nests},
\bs\label{kkk5}\be
L_{g/h_1 / \ldots /h_n}  = L_g - L_{h_1/ \ldots /h_n} = L_g +
\sum_{j=1}^n
(-1)^j
L_{h_j}
\ee\be
\qquad c_{g/h_1/ \ldots / h_n}  =c_g - c_{h_1/\ldots /h_n} =c_g +
\sum_{j=1}^n (-1)^j c_{h_j}\qquad
\ee\es
which include the affine-Sugawara and coset constructions
as the simplest cases.  In what follows, we refer to the cases with
$n\ge 2$ as the higher affine-Sugawara nests.

The nest stress tensors in (\ref{kkk5}) may be rearranged as sums of
mutually-commuting\footnote{The component stress tensors in (\ref{kkk6})
commute because the currents of $h$ (and hence all constructions
$L^\#_h$ on $h$) commute
with the $g/h$ coset construction for any $g\supset h$.}
stress tensors of $g/h$ and $h$,
\bs\label{kkk6}\be
 T_{g/h_1/ \ldots  /h_{2m+1} }  = T_{g/h_1} + \sum_{i=1}^{m}
T_{h_{2i}/h_{2i+1}}
\ee
\be
T_{g/h_1/ \ldots  /h_{2m} } = T_{g/h_1} + \sum_{i=1}^{m-1}
T_{h_{2i}/h_{2i+1}} + T_{h_{2m}}
\ee\es
so the conformal field theories of the higher affine-Sugawara nests are
expected to
be tensor-product field  theories, formed by tensoring the indicated
constructions on $h$ and $g/h$.
This was established at the level of conformal blocks in Ref.\ \cite{wi2}.

\vskip .4cm
\noindent
{\bf 5.} Affine-Virasoro nests \cite{nuc}.
Repeated K-conjugation on nested subalgebras
$g_{n}\supset\cdots\supset g_{1}\supset g$ also generates the
{\em affine-Virasoro nests},
\bs\label{avnests}\be
T^{\#}_{g_{2m}/.../g_{1}/g}=\sum_{i=1}^{m}T_{g_{2i}
/g_{2i-1}}+T^{\#}_{g}
\ee\be
T^{\#}_{g_{2m+1}/.../g_{1}/g}=\sum_{i=1}^{m}T_{g_{2i+1}
/g_{2i}}+T^{\#}_{g_{1}/g}\label{avb}
\ee\es
with an arbitrary construction $T^{\#}_{g}$ on $g$ at the
bottom of the nest.
In (\ref{avb}), $T^{\#}_{g_{1}/g}=T_{g_{1}}-T^{\#}_{g}$ is the K-conjugate
partner of $T^{\#}_{g}$ on $g_{1}$.
In parallel with the form (\ref{kkk6}) of the affine-Sugawara nests,
we have written the affine-Virasoro nest stress tensors
as sums of mutually-commuting stress tensors.
As above, the commutativity of the component
stress tensors strongly suggests that the
affine-Virasoro nests are tensor-product theories of the indicated coset
constructions with  the constructions $T^{\#}_{g}$ and $T^{\#}_{g_{1}/g}$.

\vskip .4cm
\noindent
{\bf 6.} Irreducible constructions \cite{nuc}.
Affine-Virasoro space exhibits a two-dimension\-al structure, with
affine-Virasoro nesting as the vertical
direction ($g$ above $h$ when $g\supset h$).
The horizontal direction is the set of all constructions at fixed
$g$, which
contains in particular the subset of $irreducible$ $constructions$ on $g$
\be
\{L^{\rm irr}_{g}\}\,\,:\,\,(L^{\rm irr}_{g})_{(0)}=L_{g}\,\,
,\,\,(L^{\rm irr}_{g})_{(1)}\,\,,
\,\,(L^{\rm irr}_{g})_{(2)}\,\,,\,\,....\label{kirr}
\ee
which are not obtainable by nesting from below.
Among the irreducible constructions on $g$, only one, the affine-Sugawara
construction $L_{g}$, is a conventional RCFT.
All the rest of the irreducible constructions are new CFTs,
called the $new$  $irreducible$ $constructions.$

The number of new irreducible constructions is very large on large
group manifolds and the new irreducible constructions have been enumerated
\cite{gt} in the graph theory ansatz on $SO(n)$ (see Section \ref{sod}).
The graph theory also indicates that, on large manifolds, the generic
ICFT is a new irreducible construction.

\vskip .4cm
\noindent
{\bf 7}. Unitarity \cite{fqs,gko,nuc}. Unitary constructions
$L(m)^\dagger=L(-m)$ on positive
integer level of affine
compact $g$ are easily recognizable.
If in a specific basis the currents satisfy
\be
J^{\dagger}_{a}(n)={\rho_{a}}^{b}J_{b}(-n)
\ee
then a unitary construction $L$ satisfies
\be
(L^{ab})^{*}=L^{cd}{(\rho^{-1})_{c}}^{a}{(\rho^{-1})_{d}}^{b}\;\;.
\ee
In particular  $L^{ab}=$real  is sufficient for unitarity in any Cartesian
basis.

Unitarity
guarantees that $c(L)\geq 0$, and K-conjugate partners of unitary
constructions are also unitary with $c(\tilde{L}) \geq 0$.
The double inequality
\be
 0 \leq c(L) \leq c_g \label{kineq}
\ee
follows for all unitary Virasoro constructions on affine compact $g$.
Moreover, all unitary high-level central charges \cite{nuc,hl} on simple
compact $g$ are integer valued from
0 to dim$\,g$ (see Section \ref{hlv}).

It is also known \cite{gko} that all
unitary Virasoro
constructions satisfying $0 \leq c(L) < 1$ can be realized as
$g/h$ coset
constructions. It follows that all unitary affine-Virasoro
constructions
satisfying $c_g-1<c(L) \leq c_g$
are realizable as the K-conjugate partners $T_{h}$ of the coset
constructions with central charge
between 0 and 1. Moreover, all unitary solutions of the VME with
high-level central charge 0\,,\,1\,,\,dim$\,g-1$ and dim$\,g$
are known \cite{hl}
on simple compact $g$ (see Section \ref{hlv}).

\vskip .4cm
\noindent
{\bf 8}. $SL(2,\R)$-invariance \cite{hk,nuc,cf}. The affine vacuum $|0\rangle$
satisfies
\bs\label{ksl2}\be
J_a(m\geq 0)|0\rangle=T_{ab}(m\geq -1)|0\rangle=0
\ee\be
X_{ab}(m\geq
-2)|0\rangle=W_{abc}(m\geq -2)|0\rangle=0
\ee\es
so that all
the conformal field theories of affine-Virasoro
space are $SL(2,\R)$-invariant with
$L(m\geq -1)|0\rangle =0$.

\vskip .4cm
\noindent
{\bf 9.} Spectrum \cite{nuc}. The operator $L(0)$ can be diagonalized
level by level in the affine modules and all the eigenvalues
(conformal weights) are real and positive semidefinite for unitary
constructions.

As an example, consider the $L^{ab}$-broken affine primary states
\bs\be
|R_{\cal T}\rangle^{\alpha} = |R_{\cal T}\rangle ^{J}{\xi_{J}}^{\alpha}
\;\;\;,\;\;\;\alpha=1\ldots{\rm dim}\, {\cal T}
\ee\be\qquad
L^{ab}{({\cal T}_a{\cal T}_b)_{J}}^{K}{\xi_{K}}^{\alpha} =
\Delta_{\alpha}({\cal T}){\xi_{J}}^{\alpha}\qquad
\label{kspec}
\ee\es
which diagonalize the conformal weight matrix
$H=L^{ab}{\cal T}_a{\cal T}_b$.
These states are Virasoro primary states of $T=L^{ab}\xx J_{a}J_{b}\xx$
with conformal weight $\Delta_{\alpha}({\cal T})$. See Section
\ref{hc3} for discussion of these states in a broader context.

There is an easy explicit demonstration of the consequences of unitarity
in this case.
When $L^{ab}$ is real on compact $g$, we have  hermitian $H$,
real $\Delta_{\alpha}({\cal T})$ and unitary $\xi$.
Positivity of the space $\{|R_{\cal T}\rangle^{\alpha}\}$, and hence
$\Delta_{\alpha}({\cal T})\geq 0$,
follows from positivity of the unitary affine highest weight
states.

\vskip .4cm
\noindent
{\bf 10.}  Automorphically equivalent CFTs \cite{gme,hl,gt}.
The elements $\o\in {\rm Aut}\,g$
of the automorphism group of $g$ satisfy
\bs
\be
{f_{ab}}^c= {\o_a}^d {\o_b}^e {(\o^{-1})_g}^c {f_{de}}^g \label{kf}
\ee
\be
 G_{ab}={\o_a}^c {\o_b}^d G_{cd} \label{kmet}
\ee
\es
and (\ref{kmet}) may be written as
\be
{(\o^{T})_{a}}^{b}\equiv G^{bc}{\o_{c}}^{d}G_{da}={(\o^{-1})_{a}}^{b}
\label{kmet1}
\ee
so that $\o$ is a (pseudo) orthogonal matrix which is an element of
$SO(p,q)$, $p+q=$dim$\,g$ with metric $G_{ab}$. The automorphism
group includes the outer automorphisms of $g$ and the inner
automorphisms
\be
g(\o) {\cal T}_a g^{-1}(\o) =
{\o_a}^b {\cal T}_b \;\,,\;\;\;g(\o)\, \in\, G
\label{kinn}
\ee
where ${\cal T}$ is any matrix irrep of $g$.

The automorphisms of $g$ induce automorphically-equivalent CFTs in the VME
as follows. The transformation
\be
{J'_a}(m) ={\o_a}^b
J_b(m)\,\,\,,\,\,\,\o\,\in\,{\rm Aut}\,g\label{ktran}
\ee
is an automorphism of the affine algebra (\ref{kaffine}), and
$(L')^{ab}$ defined by
\be\label{k1}
(L')^{ab}=L^{cd} {(\o^{-1})_c}^a  {(\o^{-1})_d}^b
\ee
is an automorphically-equivalent solution of the VME when $L^{ab}$ is
a solution.

Complete gauge-fixing or modding out by the automorphisms of $g$ \cite{nuc,gt}
is an important physical problem which has been completely solved only
for the known graph theory units of ICFT (see Sections \ref{nhl} and
\ref{gth}).

\vskip .4cm
\noindent
{\bf 11}. Generalized K-conjugation \cite{lie}.
K-conjugation, which applies on all affine-Virasoro space,
was described in item
2 of this list. A number of generalized K-conjugations also hold on
the space of Lie $h$-invariant CFTs, which are those theories with a
Lie symmetry $h\subset g$ (see Sections \ref{ca}, \ref{lhg}, and \ref{hc39}).
We mention in particular the case of $K_{g/h}$-conjugation,
\be
(\tilde L)_{g/h}=L_{g/h}-L\;\;,\;\;(\tilde c)_{g/h}=c_{g/h}-c
\label{kgh}
\ee
whose action is conjugation through the coset stress tensor $T_{g/h}$ instead
of $T_{g}$. This covariance holds in the subspace of the affine
Lie-$h$ invariant CFTs, which is the set of all ICFTs with a
local Lie symmetry.
Similarly, generalized K-conjugations through the higher affine-Sugawara nests
$T_{g/h_{1}/.../h_{n}}$ were discussed in \cite{lie}.

\subsubsection{A broader view\label{sec4445}}

We collect here a number of more general features of affine-Virasoro space.

\vskip .4cm
\noindent
{\bf 1.} Level-families.
Except for solutions at sporadic levels, the basic units of
affine-Virasoro space are the conformal {\em level-families} $L^{ab}(k)$,
which are essentially smooth functions of the affine level.

\vskip .4cm
\noindent
{\bf 2.} Counting. Because the VME has an equal number of equations and
unknowns, one may count the generically-expected number $N(g)$ of
conformal level-families on each $g$ \cite{nuc},
\be
N(g)=2^{{\rm dim}\,g({\rm dim}\, g-1)/2}\label{knuml}\;\; .
\ee
These numbers are very large on large groups and, in particular, one
finds that
\be
N(SU(3))\simeq {1\over 4} {\rm billion}
\ee
so most conformal constructions await discovery,
\be
\mbox{ICFT}\;\;\;\supset\supset\;\;\;\mbox{RCFT}\;\;.
\ee
The number of physically-distinct level-families on $g$ is less than
(\ref{knuml}) due to residual automorphisms \cite{gt}, but the corrections are
apparently negligible on large groups (see Section \ref{fcg}).

\vskip .4cm
\noindent
{\bf 3.} Generalized graph theories \cite{gt,sme,mb,nsc,sin,ggt,lie}.
In the (partial) classification of
the CFTs of affine-Virasoro space,  one finds that
the Virasoro master equation
and the superconformal master equation generate
generalized graph theories on Lie $g$,
including conventional graph theory as a special case on the  orthogonal
groups. In the generalized graph theories, each generalized
 graph labels a conformal level-family.

This development began with conventional  graph theory in \cite{gt}, and
the general case is axiomatized in \cite{ggt}.
We will return to this subject in Sections \ref{nhl} and~\ref{gth}.

\vskip .4cm
\noindent
{\bf 4.} Unitary irrational central charge
[94, 95, 160, 96-98, 101, 102, 93]. 
It is clear from the form of the master equation that the generic
level-family  has generically-irrational central charge,
and, similarly, it is believed that
the space of all unitary theories
is dominated by irrational central charge.  Indeed (on positive
integer level of affine compact Lie algebras)
unitary irrational central charge is  generic for
 all the known exact level-families
(see Section \ref{ucicc}) and for all the level-families so far classified
by the graph theories.

As an example, the value at level 5 of $SU(3)$ \cite{hl},
\be
c\left((SU(3)_{5})^{\#}_{D(1)}\right)=2\left(1-{1\over \sqrt{61}}\right)
\simeq 1.7439
\label{neednom}
\ee
is the lowest unitary irrational central charge yet observed.
The nomenclature in (\ref{neednom}) is discussed in Section \ref{ucicc}.

\subsubsection{Special categories\label{sec445}}

Beyond the generic level-families, we comment on a number of smaller categories
seen so far in the VME.

\vskip .4cm
\noindent
{\bf 1.} Conventional RCFTs.
In the development of ICFT,
{\em rational conformal field theory} (RCFT) is defined as
the small subspace of theories with
rational central charge and rational conformal weights.
The {\em conventional RCFTs} are defined  as the set of affine-Sugawara
nests, which includes the affine-Sugawara constructions, the coset
constructions, and the higher affine-Sugawara nests.

\vskip .4cm
\noindent
{\bf 2.} New RCFTs \cite{ks1,gep,nsc,ssc}.
Beyond the conventional RCFTs, many candidates
for new RCFTs have been found (see Sections \ref{spc} and \ref{bks}).

\vskip .4cm
\noindent
{\bf 3.} CFTs with a symmetry $H\subset$ Aut\,$g$ \cite{lie}.
Although most solutions $L^{ab}$ of the VME on $g$ have no symmetry
(corresponding to a completely asymmetric spinning top), one sees
a hierarchy of symmetry-breaking in the VME on $g$,
\be
{\rm Aut}\,g\to {\rm Lie}\,h\to H\to 1
\ee
where Aut $g$ is the maximal symmetry on $g$ and 1 is complete
asymmetry.
The $H$-invariant CFTs and the Lie $h$-invariant CFTs, which include the
conventional RCFTs, are discussed in Sections \ref{ca}, \ref{lhg}
and \ref{hc39}.

\vskip .4cm
\noindent
{\bf 4.} Quadratic deformations
\cite{rus,nuc,rus2,gt,bl,rs,sme,cf,mb,nsc,sta2}.
These are continuous manifolds of solutions in the VME, which
occur at sporadic levels. The quadratic deformations are
examples of quasi-rational CFTs, which have fixed rational central
charge but generically-continuous conformal weights.
There are no manifestly unitary constructions with
continuously-varying central charge in the VME \cite{cf}.

A set of sufficient conditions (see Section \ref{sec5}) has been found
for the occurence of quadratic deformations \cite{cf}, and other
systematics are discussed in \cite{nuc}.  An explicit example of these
constructions is reviewed in Section \ref{su2}, and the names of all
known exact quadratic deformations are listed in Section \ref{spc}.

\vskip .4cm
\noindent
{\bf 5.} Self K-conjugate constructions \cite{gt,mb,cf}.
In these constructions, the K-conjugate partners $\tilde L$ and $L$
are automorphically equivalent,
\be
{\tilde L}=\o L\o^{-1}\,\,\,\,\,\,\hbox{for
some}\;\o\,\in\,{\rm Aut}\,g\label{kself1}
\ee
so that each construction in the pair has
half affine-Sugawara central charge,
\be
c={c_{g}\over 2}\;\;.
\ee
The self K-conjugate constructions are found on
Lie group manifolds of even dimension.
The first set of these constructions (see Section  \ref{sod})
was found on $SO(4n)$ and $SO(4n+1)$, where they are in one-to-one
correspondence with the self-complementary graphs of graph theory.
The names of all known exact self K-conjugate
 level-families are listed in Section
\ref{spc}, including examples
on $SU(3)$ and $SU(5)$.

\vskip .4cm
\noindent
{\bf 6}. Self K$_{g/h}$-conjugate constructions \cite{lie}.
These constructions are generalizations of the self K-conjugate
constructions, in which K$_{g/h}$-conjugate pairs
(see eq. (\ref{kgh})) are automorphically equivalent.
In analogy with the self K-conjugate constructions, the self
K$_{g/h}$-conjugate constructions have half coset central charge,
\be
c={c_{g/h}\over 2}
\ee
and the names of the known exact level-families of this type are given
in Section \ref{spc}.
It has been conjectured \cite{lie} that self-conjugate constructions exist
for all the generalized K-conjugations described above.

\subsection{More General Virasoro Constructions\label{sec4}}
\setcounter{equation}{0}

The VME constructs Virasoro operators out
of bilinears in affine currents. A presumably more  general construction
along these lines employs arbitrary (2,0) operators \cite{k,dh}.

Consider an initial  conformal field theory which, in
addition to its stress
tensor $T_{\bullet}(z)$ with central charge $c_{\bullet}$,
contains also a number of (2,0) operators
$\Phi_{i}(z)$. These (2,0) operators can be either Virasoro
primary under $T_{\bullet}$ or
derivatives of (1,0) operators. We will not consider  the second
possibility, which leads to generalizations of the Feigin-Fuchs
type \cite{f1,gn,FF}.
The most general OPE among the operators $\Phi_i$, compatible with
conformal invariance, is
\begin{eqnarray}
\Phi_i(z)\Phi_j (w) &=& {\delta_{ij}\over (z-w)^4} +
{4\over c_{\bullet} }\delta_{ij} {T_{\bullet}(w)\over (z-w)^2} +
C_{ijk} {\Phi_k(w)\over (z-w)^2 } \nonumber \\
& & + {2\over c_{\bullet} }\delta_{ij} {\partial_wT_{\bullet}(w)
\over (z-w)}+ {1\over 2} C_{ijk} {\partial_w\Phi_k(w)\over (z-w)}+
{\rm reg.} \label{ss3}
\end{eqnarray}
where possible spin-one and spin-three terms have been
suppressed because
they do not contribute in the construction below.
Eq.(\ref{ss3}) also assumes the choice of
 an orthonormal  basis for the $\Phi_{i}$.
Associativity constrains the OPE
coefficients $C_{ijk}$ to be completely symmetric in $i,j,k$.

One now considers a general linear combination of $T_{\bullet}$ and
$\Phi_{i}$
\be
T(z) = \l_{\bullet} T_{\bullet}(z) + \sum_{i=1}^N \l_i \Phi_i (z)\label{ss5}
\ee
and demands that it satisfies the Virasoro algebra.
Using the OPE (\ref{ss3}), one  obtains the following system of quadratic
equations,
\bs\label{ss6}\be
\sum_{i=1}^N (\l_i)^2 = {1\over 2 } \l_{\bullet} (1-\l_{\bullet})
c_{\bullet}\ee\be
\sum_{i,j = 1 }^N C_{ijk} \lambda_i \lambda_j = 2 (1-2\lambda_{\bullet}
	)\lambda_k \quad k = 1,2,3\ldots  N
\ee\es
and the  central charge $c $ of $ T(z)$ is  given by
$c = \l_{\bullet}c_{\bullet}$.
The K-conjugation covariance (\ref{kKco}) seen in the VME is also a
feature of the  more general Virasoro construction:
If $(\l_{\bullet} , \l_i)$ is a solution of
(\ref{ss6}), then one can verify that the K-conjugate partner
$([1-\l_{\bullet}], -\l_i)$ is
also  a solution.
Moreover, the corresponding K-conjugate
pair of stress tensors commute and sum
to $T_{\bullet}$, as in the VME.

The more general Virasoro construction (MGVC) in (\ref{ss5}) contains at
least two representations,
\be
T_{\bullet}=\left\{\matrix{T_{g}\cr T_{\rm free\,\,bosons}\cr}\right.
\;\;\;\;\;,\;\;\;\;\;\;
\left\{\Phi_{i}\right\}=\left\{\matrix{\xx J_{a}J_{b}\xx\cr
\partial \phi^{i}\partial\phi^{j}\;\;\;,\;\;\;
e^{i\vec \alpha \cdot \vec \phi}\;\;\;,\;\;\; \partial \phi^{i} e^{i\vec
\beta\cdot\vec\phi}\cr}\right.
\ee
\vskip .3cm
\noindent
where the top line leads to the VME and the bottom line
describes the
Interacting Boson Models (IBMs) studied in \cite{k,dhs,dhs2,ks,gep}.
Any affine-Virasoro construction can be bosonized into
an IBM, so one sees the hierarchy of conformal constructions
\be
{\rm MGVC}\;\;\;\supset\;\;\; {\rm IBMs}\;\;\;\supset\;\;\; {\rm VME}\;\;.
\ee
We believe that the sets in this hierarchy are progressively larger towards
the left, but this has not yet been demonstrated: The
only known CFTs which may go  beyond the VME
are the N=2 IBMs discussed in Section \ref{bks}.

\renewcommand{\theequation}{\thesection.\arabic{equation}}
\section{The C-Function and a C-Theorem\label{sec5}}
\setcounter{equation}{0}

In this section, the notion of affine-Virasoro space is extended
to include all inverse inertia tensors $L^{ab}$, whether
the operator $T(L)$ is Virasoro or not.
On this space, there is an exact C-function and an associated flow
which satisfies a C-theorem \cite{cf}.
The notions of C-function and C-theorem in conformal field
theory were introduced in
Ref.\ \cite{z3}.

This development begins with the following cubic function on
affine-Virasoro space,
\bs\label{cfunction}\be
A(L)\equiv 2L^{ab}P_{ab,cd}(L^{cd}-2\beta^{ab}(L))
\ee\be
P_{ab,cd}\equiv z^4\langle T_{ab}(z)T_{cd}(0)\rangle=
G_{a(c}G_{d)b}-{1\over 2}{f_{a(c}}^{e}{f_{d)b}}^{f}G_{ef}
\ee\be
\beta^{ab}(L)\equiv -L^{ab}+2L^{ac}G_{cd}L^{db}-L^{cd}L^{ef}
{f_{ce}}^{a}{f_{df}}^{b}-L^{cd}
{f_{ce}}^{f}{f_{df}}^{(a}L^{b)e}
\ee\es
where $T_{ab}=\xx J_{a}J_{b}\xx$ and $P_{ab,cd}$ is the natural
metric on the space. The function $A(L)$ is the correct action on
affine-Virasoro space because the equation for its extrema
\be
{\partial A(L)\over \partial L^{ab}}=-12P_{ab,cd}
\beta^{cd}(L)=0\label{cfun}
\ee
is the Virasoro master equation.
Moreover, the action is a C-function on affine-Virasoro
space because
\be
A(L)|_{\beta=0}=c(L)
\ee
is the central charge (\ref{kME3}) of the conformal stress tensor $T(L)$.

All the conformal field theories of the VME are fixed points of the associated
flow,
\be
\ld^{ab}=\b^{ab}(L)\label{kflow}
\ee
where overdot is derivative with respect to an auxiliary variable.  This
flow
satisfies the C-theorem,
\be
{\dot A}(L)\leq 0\label{kc16}
\ee
on positive integer level of affine compact $g$.

Ref.\ \cite{cf} discusses the flow equation (\ref{kflow})
in further detail, including the
flow near a fixed point and closed subflows. Closed subflows are associated
to each consistent ansatz (see Section \ref{ca}) of the VME.
Of particular interest is the subflow in the ansatz $SO(n)_{diag}$,
which is a flow on the space of graphs.
Two other applications are discussed below.

\vskip .4cm
\noindent
{\bf 1}. Morse theory. The C-function is a
Morse function \cite{morse} on affine-Virasoro space. Morse polynomials
and the Witten index can be  defined in the standard way for
any closed subflow. In the case of the flow on the space of
graphs, the Morse polynomials turn out to be well-known
functions in graph enumeration.

\vskip .4cm
\noindent
{\bf 2.} Prediction of sporadic deformations.
Using the flow equation,  a set of sufficient  conditions
was derived
for the occurence of a quadratic deformation (see Section \ref{sec445})
which, in this context, is a
continuous family of fixed points:

\noindent
If there are two level-families of the VME,

a) which are flow-connected at high level, and

b) whose central charges cross as the level is lowered,

\noindent
then there is a continuous family of fixed points at the level where
the central charges cross for the first time.

Concerning condition a), what must be checked in high-level
perturbation theory  is that there is a flow which starts from
a small neighborhood of one of the level-families
and ends at the other level-family.
Condition b) was noted independently in \cite{str}.
This method was applied successfully in \cite{cf,mb,sta2}, leading to the
exact forms of  several new quadratic deformations in the VME.

Further discussion of quadratic deformations is found in Sections
\ref{su2}, \ref{spc} and Ref.\  \cite{nuc}

\vskip .4cm
\noindent
\underline{Speculation on the flow equation}
\vskip .3cm

The flow equation (\ref{kflow}) bears a strong resemblance to the expected form
of an exact renormalization group equation, but the flow equation cannot be the
conventional renormalization group equation.
The reason is that the conventional renormalization group analysis
is restricted to Lorentz-invariant paths between CFTs,
while the generic path of the flow equation is apparently
non-relativistic.

More precisely, the flow equation, if indeed it corresponds to a
renormalization group equation, is expected to run over the set of
quantum field theories whose Hamiltonian,
\be
H=L^{ab}\sum_{m\in \Z}\xx [J_{a}(-m)J_{b}(m)+
\bar J_{a}(-m)\bar J_{b}(m)]\xx\label{ham}
\ee
is the natural extension \cite{rd} of the generic affine-Virasoro Hamiltonian
(see Section \ref{hc43}) to arbitrary $L^{ab}$.

As noted in \cite{rd}, the theories in this set
are scale invariant (due to normal ordering) and the generic theory is
non-relativistic.
At a fixed point, where $L^{ab}$ satisfies  the VME,
Lorentz and conformal invariance are restored ($H=L(0)+\bar L(0)$)
and the theory develops a spin-two gauge symmetry \cite{gva}
due to K-conjugation.
See Section \ref{hc31} for further discussion of this Hamiltonian at the
fixed
points, including the corresponding world-sheet action of the generic ICFT.

The overall picture here may be formulated as a  conjecture
that the renormalization group equation of the large
set of theories (\ref{ham}) is the flow equation (\ref{kflow}).
Moreover, all these theories (including the CFTs) may be integrable models
because their spectrum is solvable level-by-level in the affine modules.

\renewcommand{\thesubsection}{\thesection.\arabic{subsection}}
\renewcommand{\theequation}{\thesubsection.\arabic{equation}}

\section{General  Superconformal Constructions\label{sec6}}

Superconformal algebras play a special role in CFT and string theory. The
simplest example is the N=1 superconformal algebra \cite{r,ns}
which contains in addition to the stress tensor an extra fermionic
field (the supercurrent) with spin 3/2.
This algebra  is the gauge algebra of the superstring (see e.g.\ \cite{gsw}).
Consequently, the
representations of the algebra are crucial in finding
superstring ground states.
Extended superconformal algebras also play an important role in
superstring theory: N=2 superconformal symmetry \cite{ad,ad2}
is needed \cite{bdfm} to construct
ground states of the heterotic string \cite{gh} with
at least N=1 spacetime supersymmetry in four dimensions.
If N=4 superconformal symmetry \cite{ad2}
is present then one obtains at least N=2 spacetime supersymmetry
\cite{bd}.

The N=1 and N=2 superconformal master equations, which collect
the N=1 and N=2 superconformal solutions of
the VME, were obtained by Giveon and three of the authors in Ref.\ \cite{sme}.
In what follows we review both of these systems.

\subsection{The N=1 Superconformal Master Equation\label{sec7}}
\setcounter{equation}{0}

The N=1 superconformal algebra is \cite{r,ns},
\bs\be
[L(m),L(n)]=(m-n)L(m+n)+\frac{c}{12}m(m^2-1)
\d_{m+n,0}\label{s1}
\ee
\be
\{G(r),G(s)\}=2L(r+s)+\frac{c}{3}(r^2-\frac{1}{4})\d_{r+s,0}
\label{s2}
\ee\be
[L(m),G(r)]=(\frac{m}{2}-r)G(m+r)\label{s3}
\ee\es
where $G(r)$ is the supercurrent, with $r\in \Z+{1\over 2}$ (
Neveu-Schwarz) or $\Z$ (Ramond).
A property of this algebra is that Jacobi identities and
the relations (\ref{s2},c) imply
the Virasoro algebra (\ref{s1})
so that only the $GG$ and $LG$ algebra is required for superconformal
symmetry.

Standard representations of this algebra involve
world-sheet fermions \cite{bh,r,ns}
\bs
\be
S_I(z)S_J(w) = \eta_{IJ} \Delta(z,w) + \oo S_I(z)S_J(w) \oo
\,\,\,,\,\,I,J=1\ldots F  \label{s8}
\ee\be
\Delta(z,w)=\left\{ \matrix{ {1 \over z-w} & (\mbox{BH-NS}) \cr
                        {z+w \over 2\sqrt{zw}(z-w)} & (\mbox{R})}
\right. \label{s9}
\ee
\es
where $S_{I}(z)=\sum_{r}S_{I}(r)z^{-r-1/2}$, $r\in \Z+{1\over 2}$
(Bardakci-Halpern-Neveu-Schwarz) or $r\in \Z$ (Ramond).
The fermionic metric $\eta_{IJ}$ is the metric on the carrier space
of some antisymmetric representation ${\cal T}$,  and,
without loss of generality, one may consider the fermions
to be in the vector representation  of $SO(p,q)_{\tau}$, $p+q=F$
whose currents
\be
J_{IJ}=i\sqrt{{\tau \psi^2 \over 2}}\oo S_I S_J \oo,\;\;\;\;I<J
\label{s10}
\ee
have level $\tau =1$ for $F\neq 3$ and $\tau= 2$ for $F=3$.

The general superconformal construction is carried out  on the manifold
\be
g_{x}\times {\rm fermionic}\;\, SO(p,q)_{\tau}\;\;,\;\;\;
g_x \equiv \oplus_I g_{x_I}\label{s11}
\ee
with a general set $J_{A}$, $A=1\ldots \dim g$ of ``bosonic" currents on
$g_{x}$, which commute with the fermions and satisfy
the current algebra (\ref{kjj}).
Except for terms linear in the fermions,
the most general N=1 supercurrent is
\be
G(z)=e^{AI}J_A(z)S_I(z)+\frac{i}{6}t^{IJK}\oo S_I(z)S_J(z) S_K(z) \oo
\label{s14}
\ee
where $e^{AI}$ and $t^{IJK}$ are  called the {\em vielbein}
and the {\em three-form} respectively.
The addition of linear terms $\partial S_{I}$
has also been considered \cite{sme}.

N=1 superconformal symmetry is summarized by the {\em N=1
superconformal master equation} \cite{sme},
\bs\label{s27}\be
\label{s28}
\eqalign{2E^{AI}(e,t)\equiv&-e^{AI}+e^{BJ}e^{CK}e^{DI}(\d_B^AG_{CD}+
{f_{EB}}^A{f_{CD}}^E)\eta_{JK}\cr
&+t^{IJK}(\frac{1}{2}t^{MNL}e^{AR}\eta_{LR}+e^{BM}e^{CN}
{f_{BC}}^A)\eta_{JM}\eta_{KN}=0\cr}
\ee
\be
\label{s29}
\eqalign{2T^{IJK}(e,t)\equiv&-t^{IJK}+e^{AL}e^{B[K}
t^{IJ]M}G_{AB}\eta_{LM}+
2e^{AI}e^{BJ}e^{CK}{f_{AB}}^D G_{DC} \cr
&+(\frac{1}{2}t^{P[IJ}t^{K]MN}t^{RLQ}+2t^{MPI}t^{NQJ}
t^{LRK})\eta_{PQ}\eta_{MR}
\eta_{NL}=0\;\;\;\;\;\;  \cr}\ee\be
c=\frac{3}{2}e^{AI}e^{BJ}G_{AB}
\eta_{IJ}+\frac{1}{4}t^{IJK}t^{LMN}\eta_{IL}
\eta_{JM}\eta_{KN}
\label{s299}\ee\es
which is often abbreviated as SME in what follows.
In the SME,
$A^{[IJ}B^{K]}\equiv A^{IJ}B^K + A^{JK}B^I + A^{KI} B^J$
is totally antisymmetric in $I,J,K$ when $A^{IJ}$
is antisymmetric.

The SME collects all the constructions of the VME which have at least N=1
superconformal symmetry.
Historically, a special case of the SME was considered earlier by Mohammedi
\cite{moh} and another special case was considered independently
by Ragoucy and Sorba \cite{rs}.

The stress tensor of the general construction, whose coefficients
satisfy the VME, is
\bs
\be
T(z)={\cal L}^{AB} T_{AB}+{\cal F}^{IJ}(\oo S_I(z)
{\mathop\partial^{\leftrightarrow}}_z S_J(z) \oo -{\epsilon
\eta_{IJ}\over 4z^2}) \hspace*{3cm}\label{s15}
\ee
$$\hspace*{1cm} +i{\cal M}^{IJA}J_A(z) \oo S_I(z)S_J(z)\oo +
{\cal R}^{IJKL}\oo S_I(z)S_J(z) S_K(z)S_L(z)\oo
$$
\label{s17}\be
{\cal L}^{AB}=\frac{1}{2}e^{AI}e^{BJ}\eta_{IJ}
\ee\be
{\cal F}^{IJ}=-\frac{1}{4}e^{AI}e^{BJ}G_{AB}-
\frac{1}{8}t^{IKL}t^{JMN}\eta_{KM}\eta_{LN}
\ee\be
{\cal M}^{IJA}=\frac{1}{2}e^{BI}e^{CJ}{f_{BC}}^A+
\frac{1}{2}t^{IJK}e^{AL}\eta_{KL}
\ee\be
{\cal R}^{IJKL}=-\frac{1}{24}t^{MI[J}t^{KL]N}\eta_{MN}
\ee
\be
A{\mathop\partial^{\leftrightarrow}}B\equiv A(\partial B)-(\partial
B)A,\;\;\;
\;\;\epsilon \equiv \left\{\matrix{0 & (\mbox{BH-NS}) \cr 1 & (\mbox{R})}
\right.\;\;.
\label{s16}
\ee
\es
We note that the K-conjugate partner $\tilde T=T_{g}-T$
of a superconformal construction $T$, although a commuting Virasoro
operator, is not generally superconformal.

Remarkably, the SME (\ref{s27}) is a
``consistent ansatz'' \cite{nuc} of the Virasoro master equation, in
that the
superconformal system contains the same number of (cubic) equations
as unknowns.  As a consequence,
one may also count the generically-expected number of
superconformal constructions \cite{sme}.
The generic level-family of the SME has irrational central charge,
and unitary solutions of the SME are recognized when the vielbein
and three-form are real in any Cartesian basis of compact $g_x \times
SO(F)_{\tau}$, $x_I \in \N$.
As in the VME, irrational central  charge is expected to dominate  the
space of all unitary solutions of the SME.

The SME contains all the standard N=1 superconformal constructions, including
the GKO N=1 coset constructions \cite{gko2}, the non-linear realizations
\cite{w,abkw} and the N=1 Kazama-Suzuki coset constructions
\cite{ks1,ks2}.
Known unitary
 irrational solutions of the SME are discussed in Sections \ref{sus},
\ref{sma} and \ref{gsm}.
The super C-function of the SME is given in Ref.\ \cite{sme}.

\subsection{The N=2 Superconformal Master Equation\label{sec8}}
\setcounter{equation}{0}

The N=2 superconformal algebra is \cite{ad,ad2},

\bs\label{n2}\be
G_{i}(z)G_j(w)={2c/3 \d_{ij} \over (z-w)^3}
+\left(\frac{2}{(z-w)^2}+\frac{1}{z-w}
\partial_{w}\right)i\e_{ij}J(w)
+\frac{2\d_{ij}}{z-w}T(w)+{\rm reg.}\label{s30}
\ee\be
T(z)G_i(w)=\left(\frac{3/2}{(z-w)^2}+\frac{1}{z-w}
\partial_{w}\right)G_i(w)+{\rm reg.}\label{s31}
\ee\be
J(z)G_i(w)={i \e_{ij} \over z-w}G_j(w)+{\rm reg.}\label{s32}
\ee
\be
T(z)T(w)={c/2 \over (z-w)^4} +\left(\frac{2}{(z-w)^2}+\frac{1}{z-w}
\partial_{w}\right)T(w)+{\rm reg.}\label{s33}
\ee\be
T(z)J(w)=\left(\frac{1}{(z-w)^2}+\frac{1}{z-w}
\partial_{w}\right)J(w)+{\rm reg.}\label{s34}
\ee\be
J(z)J(w)={c/3 \over (z-w)^2}+{\rm reg.}\label{s35}
\ee\es
where $i,j=1,2$ and $\epsilon_{ij}$ is antisymmetric with
$\epsilon_{12}=1$.

Except for linear terms in the fermions, the most general N=2 supercurrents are
\be
G_i(z)=e_i^{AI}J_A(z)S_I(z)+\frac{i}{6}t_i^{IJK}\oo S_I(z)S_J(z)
S_K(z)\oo\;\;,
\;\;i=1,2\label{s36}\ee
which are $(e_1,t_1)$ and $(e_2,t_2)$ copies of (\ref{s14}), and the
$U(1)$ current
\bs
\be
J(z)=A^AJ_A(z)+iB^{IJ}\oo S_I(z) S_J(z) \oo\label{s37}
\ee\be
A^A=\frac{1}{2}e_1^{BI}e_2^{CJ}{f_{BC}}^A\eta_{IJ}\label{s38}
\ee\be
B^{IJ}=-\frac{1}{4}(e_1^{AI}e_2^{BJ}-e_2^{AI}e_1^{BJ})G_{AB}
-\frac{1}{8}(t_1^{IKL}t_2^{JMN}-t_2^{IKL}t_1^{JMN})
\eta_{KM}\eta_{LN}\label{s39}
\ee
\es
follows from the OPE (\ref{s30}).
The forms of the N=2 stress tensor and the N=2 central
charge are those given in
(\ref{s17}) and (\ref{s299}), now written in terms of either
($e_{1},t_{1}$) or ($e_{2},t_{2}$).

In Ref.\ \cite{sme}, the N=2 superconformal construction is
summarized in two parts. First, one obtains two copies
of the N=1 SME
\be
E^{AI}(e_i,t_i)=T^{IJK}(e_i,t_i)=0,\;\;\;i=1,2\label{s399}
\ee
which follow from the OPE (\ref{s31}).
Second, one obtains the N=2 SME \cite{sme},
\bs\label{s40}\be
(e_i^{AI}e_1^{BJ}-\e_{ij}e_j^{AI}e_2^{BJ})\eta_{IJ}=0
\ee\be
(e_i^{AI}e_1^{BJ}-\e_{ij}e_j^{AI}e_2^{BJ})G_{AB}+
\frac{1}{2}(t_i^{IKL}t_1^{JMN}-\e_{ij}t_j^{IKL}
t_2^{JMN})\eta_{KM}\eta_{LN}=0
\ee\be
(e_i^{BI}e_1^{CJ}-\e_{ij}e_j^{BI}e_2^{CJ}){f_{BC}}^A+
(t_i^{IJK}e_1^{AL}-\e_{ij}t_j^{IJK}e_2^{AL})
\eta_{KL}=0
\ee\be
(t_i^{IM[J}t_1^{KL]N}-\e_{ij}t_j^{IM[J}t_2^{KL]N})
\eta_{MN}=0
\ee
$$
\e_{ij}e_j^{AI}=\frac{1}{2}e_1^{DJ}e_2^{EK}e_i^{CI}
{f_{DE}}^B{f_{BC}}^A\eta_{JK}+
\frac{1}{2}e_i^{AK}(e_1^{BJ}e_2^{CI}-e_2^{BJ}
e_1^{CI})G_{BC}\eta_{JK}
$$
\be
+\frac{1}{4}e_i^{AK}(t_1^{JMN}t_2^{IPQ}-t_2^{JMN}
t_1^{IPQ})\eta_{MP}\eta_{NQ}
\eta_{JK}
\ee\be
\e_{ij}t_j^{IJK}=\left(\frac{1}{2}e_2^{B[K}t_i^{IJ]M}
e_1^{AL}G_{AB}\eta_{LM}
+\frac{1}{4}t_i^{M[IJ}t_2^{K]QR}t_1^{LNP}\eta_{LM}\eta_{NQ}
\eta_{PR}\right)\ee\es
$$-(1\leftrightarrow 2)$$
which follows from the OPEs (\ref{s30},c).
All the other OPEs in (\ref{n2}) are then satisfied.
Moreover, it was shown in \cite{ff1,getz}
that the N=1 SMEs in  eq.(\ref{s399}) are
redundant, so that the N=2 SME (\ref{s40}) is the complete
description of the general N=2 construction.
The inclusion of linear terms in the N=2 supercurrents (\ref{s36}) was
considered by Figueroa-O'Farrill in \cite{ff}.

The N=2 SME contains the standard N=2 superconformal constructions,
including the N=2 Kazama-Suzuki coset constructions \cite{ks1,ks2}.
It is an important open problem to find unitary solutions of the
N=2 SME with irrational central charge.

\section{Related Constructions\label{sec9}}
\setcounter{equation}{0}

\boldmath
\subsection{Master Equation for the W$_{3}$ Algebra\label{sec10}}
\unboldmath

W-algebras \cite{z1,fz,fl} are extended conformal algebras,
including a Virasoro
subalgebra, with extra spin $\geq 3$ bosonic generators.
These algebras play an important  role in CFT,
2-d gravity and integrable models.
For a review of W-algebras, see Ref.\ \cite{bs}.

The simplest of these extended algebras is the non-linear $W_{3}$ algebra,
which involves the Virasoro subalgebra and the OPEs \cite{z1},
\bs\label{s46}\be
T(z)W(w)=3{W(w)\over (z-w)^2}+{\partial_w W(w)\over (z-w)}+
{\rm reg.}\label{s466}\ee
$$
W(z)W(w)={c/3\over (z-w)^6}+
{16\over 22+5c}\left[{2\over
(z-w)^2}+{\partial_w\over(z-w)}\right]\Lambda(w)\hspace*{3cm}$$
\be
\hspace*{.5cm}
+\left[{2\over (z-w)^4}+{\partial_w\over
(z-w)^3}
+{3\over 10}{\partial_w^2\over (z-w)^2}+{1\over 15}{\partial_w^3\over
(z-w)}\right]T(w)+{\rm reg.}
\label{s47}\ee\es
where $\Lambda(z)\sim \xx T^{2}(z)\xx$ and
(\ref{s466}) says that $W(z)$ is a spin-three
Virasoro primary field.

The $W_{3}$ master equation must collect all solutions of the VME
with $W_{3}$ symmetry.
The most general
operators in this construction are
\bs\be
T=L^{ab}\xx J_{a}J_{b}\xx+D^{a}\partial J_{a}
\ee
\be
W=K^{abc}W_{abc}+F^{ab}X_{ab}+
N^{ab}\partial\xx J_{a}J_{b}\xx +M^{a}\partial^2 J_{a}
\ee\es
where
$X_{ab}\sim \xx J_{a}\lrp
J_{b}\xx$ and $W_{abc}\sim \xx J_{a}J_{b}J_{c}\xx$
are spin-three operators (with respect to $T_{g}$)
whose precise definition is given
in (\ref{kjj}) and (\ref{kTJ}).

The $W_{3}$ master equation follows from the OPEs (\ref{s46}), and can be
written as a set of algebraic relations
among the coefficients $K,F,N$ and $M$.
The exact form of this master equation is not known, but three
special cases have been considered.

The case $F=N=M=D=0$ was discussed by Belov and Lozovik \cite{bl5},
and Romans \cite{ro} has studied the full system on abelian $g$.

The $W_{3}$-master equation for $N=M=D=0$ was also considered \cite{hkou}
at high level on
simple compact $g$, assuming $L^{ab}={\cal O}(k^{-1})$.
This asymptotic behavior (see Section \ref{hla})
is believed to include all unitary level
families and implies that $F^{ab}={\cal O}(k^{-1})$ and
$K^{abc}={\cal O}(k^{-3/2})$. On simple $g$,
the leading term of the high-level expansion is equivalent to an effective
abelianization (see Section \ref{hc22}), so the high-level form
of the $W_{3}$-master equation,
\bs\label{highkw}\be
L^{ab}=9k^2K^{acd}{K^{b}}_{cd}-12kF^{ac}{F^{b}}_{c}\label{wa}
\ee\be
2kL^{d(a}{K^{bc)}}_{d}=3K^{abc}\;\;\;,\;\;\;kL^{c[a}{F^{b]}}_{c}=-F^{ab}
\label{wb}
\ee\be
L^{bc}{K_{bc}}^{a}=L^{c(a}{F^{b)}}_{c}=K^{abd}{F^{c}}_{d}=0
\label{wc}
\ee\be
9k^2K^{acd}{K^{b}}_{cd}-32kF^{ac}{F^{b}}_{d}={32\over 22+5c}L^{ab}
\label{wd}
\ee\be
9kK^{ea(b}{K^{cd)}}_{e}={32\over 22+5c}L^{a(b}L^{cd)}
\label{we}\ee\es
can be read from Ref.\ \cite{ro}.
In these relations, indices are raised and lowered with the Killing metric and
$A^{(a}B^{bc)}\equiv A^{a}B^{bc}+A^{b}B^{ca}+A^{c}B^{ab}$.

Substitution of (\ref{wa}) in (\ref{wd}) gives the intermediate result,
\be
F^{ac}{F^{b}}_{c}=\;{9k\over 32}\,{c-2\over c+2}K^{acd}{K^{b}}_{cd}
\label{ww}
\ee
and further analysis of the system shows that $F^{ab}=0$.
Using equations (\ref{wa},b) one can show that
$K^{acd}{K^{b}}_{cd}=0$ implies $K^{abc}=0$, which gives the surprising
result that the high-level central charge of all solutions is equal
to two!

This result should be considered in parallel with the fact that all unitary
$W_{3}$-symmetric theories with $c\leq 2$ are known RCFTs \cite{fz}
and the conjecture that the high-level central charge is an upper bound
on the central charge of any level-family (see Section \ref{hlv}).
{}Together, the result, the fact and the conjecture imply the negative
conclusion that there are no new unitary level-families of the $N=M=D=0$ $W_3$
master equation, all level-families being equivalent to
the $(SU(3)_{x}\times SU(3)_{1})/SU(3)_{x+1}$ construction
\cite{bbss2} of the standard $W_{3}$ minimal models.

Including the linear terms, a class of $W_{3}$ constructions was given in
\cite{ro,ds}.

There are other W-algebras whose general construction
should be analyzed, including  $W_{N}$ algebras with $N>3$ and their infinite
spin limits, for example $W_{\infty}$ \cite{bakas,pope}.

\subsection{Virasoro Constructions on Affine Superalgebras\label{sec11}}

The general Virasoro construction on affine Lie superalgebras was given
independently in Refs.\ \cite{dt} and \cite{frt}.

The general affine Lie superalgebra \cite{kac2},
\bs\label{supera}\be
J_{a}(z)J_{b}(w)={k\eta_{ab}\over (z-w)^2}+i{f_{ab}}^{c}{J_{c}(w)\over (z-w)
}+{\rm reg.}
\ee\be
J_{a}(z)F_{I}(w)=-{({\cal T}_{a})_{I}}^{J}{F_{J}(w)\over (z-w)}+{\rm reg.}
\ee\be
F_{I}(z)F_{J}(w)={k\hat\eta_{IJ}\over (z-w)^{2}}+{R_{IJ}}^{a}
{J_{a}(w)\over (z-w)}+{\rm reg.}
\ee\es
contains an affine subalgebra $\hat g$ and a set of dim$\;{\cal T}$ spin-one
fermionic currents $F$ in matrix representation ${\cal T}$ of $g$.
The quantities $\eta_{ab}$ and $\hat \eta_{IJ}$ are the Killing metric of $g$
and a symplectic form respectively.
The affine superalgebras do not have unitary representations\footnote{This
is easily understood in
standard representations, which have the form
$F\sim \bar B \psi+\bar \psi B$, where $\psi$ and $B$ are respectively
spin-1/2 world-sheet fermions \cite{bh,r} and spin-1/2 world-sheet bosons
\cite{bh}. The bosonic fields are now called spin-1/2 ghosts \cite{fms}
or symplectic bosons \cite{gow}.}.

In this case, the most general Virasoro operator has the form
\be
T=L^{ab}\xx J_{a}J_{b}\xx +iD^{IJ}\xx F_{I}F_{J}\xx +D^{a}\partial J_{a}
\ee
where $D^{IJ}$ is antisymmetric.
We give the generalization of the VME in the case $D^{a}=0$
\cite{dt,frt},
\bs\be
L^{ab}=2k\eta_{cd}L^{ac}L^{bd}-L^{cd}L^{eg}
{f_{ce}}^{a}{f_{dg}}^{b}-L^{cd}{f_{ce}}^{g}{f_{dg}}^{(a}
L^{b)e}
\ee
$$
+D^{IJ}D^{KL}{R_{IK}}^{a}{R_{JL}}^{b}
-iL^{c(a}{R_{JK}}^{b)}D^{IJ}{({\cal T}_c)_{I}}^{K}
$$
\be
D^{IJ}=-2ik\hat \eta_{KL}D^{IK}D^{JL}+iD^{KL}D^{M[I}{({\cal T}_{a})
_{L}}^{J]}{R_{KM}}^{a}\hspace*{1cm}
\ee
$$\hspace*{1cm}
-L^{ab}D^{K[I}{({\cal T}_{a})_{L}}^{J]}{({\cal T}_{b})_{K}}^{L}
+L^{ab}D^{KL}{({\cal T}_{a})_{K}}^{[I}{({\cal T}_{b})_{L}}^{J]}\;\;.
$$
\es
This system contains the VME itself when $D^{IJ}=0$.

The general structure of the VME is also seen here, including
affine-Sugawara constructions, K-conjugation and coset constructions.
Moreover, the general system was analyzed on several small superalgebras,
and a number of exact solutions were found.

\newpage

\renewcommand{\thepage}{\arabic{page}}
\part{Affine-Virasoro Space}
\section{Exact Solutions of the Master Equations}

In this section, we review the known new
exact solutions of both the VME and N=1
SME given in eqs.(\ref{kME2}) and (\ref{s27}). We begin with a discussion of
consistent ans\"atze and other relevant concepts that are useful for obtaining
solutions. Then we give the explicit form of the following four examples of
new solutions\footnote{The nomenclature for new solutions is discussed
in Section \ref{ucicc}.},

\bul generalized spin-orbit constructions \cite{hk}

\bul $SU(2)_4^\#$ \cite{rus2,nuc}

\bul $\mbox{simply-laced}\,g^\# $ \cite{nuc}

\bul $ SU(n)^{\#}[m(N=1),rs]$ \cite{sin}.

\ni
Here our choice is partly motivated by historical reasons, though each of these
solutions also serves as an illustration of more general features of
affine-Virasoro space. Finally, we give the complete list of all known exact
unitary constructions with irrational central charge \cite{nuc}
 and lists of the known
quadratic deformations \cite{rus2,nuc}, the
self K-conjugate constructions \cite{gt,mb},
the self K${}_{g/h}$-conjugate constructions \cite{lie},
and the candidates for new RCFTs \cite{ks,gep,nsc,ssc}
so far found in the master
equations.

\subsection{Solving the Master Equations}

\subsubsection{Consistent ans\"atze and group symmetry \label{ca} }
\lam{ca}

The numerical equality of equations and unknowns in the master equation
reflects  the solvability of the system, and, in
particular, one  can estimate the
number of solutions of such a system, given that the coefficients are generic:
For $n$ quadratic equations, the number of solutions is $2^n$. Moreover,
the numerical equality  reflects closure under OPE of the operator subset
$\{ L^{ab}\xx J_aJ_b\xx,\,\,\,\forall L^{ab}\}$ as seen in  Section \ref{sec3}.

Similarly, a {\em consistent ansatz} $\{ L^{ab}({\rm ansatz})\}\subset
\{ L^{ab}\}$ for the master equation is a set of restrictions on $L^{ab}$
which, when imposed on the VME, gives a reduced VME which maintains
numerical equality of equations and unknowns. The consistency of an
ansatz implies
closure of the operator subset $\{L^{ab}({\rm ansatz})\xx J_aJ_b\xx\}$.
We recall
that the VME has been identified (see Section \ref{sec3})
as an Einstein-like system on the
group manifold, and, as in general relativity, consistent ans\"atze and
subans\"atze have played a central role in obtaining exact solutions of the
master equations.

\vskip .4cm
\ni \u{The $H$-invariant CFTs}
\vskip .3cm

As an example, a broad class of consistent ans\"atze is associated to a
symmetry of the inverse inertia tensor under subgroups of $\au$.
In the VME on affine $g$, the {\em $H$-invariant ansatz}
$A_{g}(H)$ is \cite{lie},
\be
A_{g}(H)\,\,\,:\,\,\,\,L=\o L\o^{-1}\,\,\,,\,\,\,\forall\,\,\o\,\in\,H\,
\subset\, \au
\label{hia} \ee
\lam{hia}
where $H$ can be taken as any finite subgroup or Lie subgroup of $\au$, and
may involve inner or outer automorphisms of $g$.
The ansatz $A_{g}(H)$ is a set of linear relations on $L^{ab}$ which requires
that all the conformal field theories (CFTs) of the ansatz are invariant under
$H$. Collectively, all such CFTs with a group symmetry are called the
{\em $H$-invariant CFTs}.

As a set, the $H$-invariant CFTs follow the subgroup embeddings in
$\au$, so that
\bs
\be
  A_{g}(\au) \subset A_{g}(H_1) \subset A_{g}(H_2) \cdots \subset A_{g}(H_n)
 \subset A_{g}(1)
\ee
\be
\au \supset H_1 \supset H_2 \cdots \supset H_n \supset 1
\ee
\label{sue} \es
\lam{sue}
where $ 1 $ is the trivial subgroup. Subgroup embedding in $\au$ is a
formidable problem, but many examples can be obtained.

The largest $H$-invariant ansatz $A_{g}(1)$, associated to the smallest
symmetry group $H=1$,  is the VME itself on affine $g$.
The smallest $H$-invariant ansatz $A_{g}(\au)$, $g=\op_I g_I$ contains only
$L=0$ and the affine-Sugawara constructions on all subsets of
$g_{I}$, and hence no new solutions. Some less trivial examples include
the graph-symmetry subans\"atze \cite{gt,lie} (see Section \ref{sod}),
outer automorphic ans\"atze \cite{lie}, ans\"atze following from grade
automorphisms \cite{lie}, and the metric ansatz on $SU(n)$ \cite{mb,lie}
(see Section \ref{mbm}).

\vskip .4cm
\ni \u{The Lie $h$-invariant CFTs}
\vskip .3cm

The {\em Lie $h$-invariant CFTs}
with $h\subset g$ are the subset of $H$-invariant
CFTs when  $H$ is a  Lie group. These CFTs are described by the {\em Lie
$h$-invariant ans\"atze} $A_g({\rm Lie}\,h) \subset A_g(H)$.

When $H\subset G$ is a connected  Lie subgroup, the ansatz $A_{g}(H)$ in
(\ref{hia}) is equivalently described by its infinitesimal form
\be
A_{g}({\rm Lie}\,h)\,\,\,:\,\,\,\, \d L^{ab}(\psi)=L^{c(a}{f_{cd}}^{b)}
\psi^d =0
\label{lhs} \ee
\lam{lhs}
where $\psi^a$, $a=1\ldots {\rm dim}\,h$ parametrizes $H$ in the neighborhood
of the origin, so that $\d L$ is an infinitesimal transformation of $L$ in $H$.
The symmetry (\ref{lhs}) is also a consistent ansatz when $H$ is disconnected.
The reduced VME of the
generic Lie $h$-invariant ansatz is, like the VME itself, a large set of
coupled quadratic equations, so the generic Lie $h$-invariant
CFT has irrational central charge.

Since the affine-Sugawara construction $L_{g}$ in (\ref{kkk3}) is
Lie $h$-invariant
for all $h\subset g$, it follows from (\ref{lhs}) that the K-conjugate partner
${\tilde L}=L_{g}-L$ of a Lie $h$-invariant CFT is also Lie $h$-invariant.
In particular, it follows that the $h$ and $g/h$ constructions of RCFT are
Lie $h$-invariant CFTs.

\vskip .4cm
\ni \u{Symmetry hierarchy in ICFT}
\vskip .3cm

It follows from the discussion above that there is a symmetry hierarchy
in ICFT,
\be
\mbox{ICFT} \supset \supset H\mbox{-invariant CFTs} \supset \supset
{\rm Lie}\,h\mbox{-invariant CFTs} \supset \supset \mbox{RCFT}
\pe
\le{shi}
This hierarchy shows that  RCFT is a very small subspace, of relatively
high symmetry, in the much larger space of ICFTs with a symmetry. The generic
ICFT is completely asymmetric \cite{gt}, and the generic theory with a symmetry
has irrational central charge.

As an example, the number of $SO(m)$-invariant level-families on $SO(n)$ is
\cite{lie},
\be
|A_{SO(n)}(SO(m))|=2^{1+{(n-m)(n-m+1)((n-m)(n-m-1)+6)\over 8}}={\cal O}(
2^{n^{4}/8})\,\,{\rm at}\,\,{\rm fixed}\,\,m
\ee
This  number  has not been corrected for residual continuous or discrete
automorphisms of the ansatz, but this estimate and the corresponding
Lie $h$-invariant fraction
\be
{|A_{SO(n)}(SO(m))|\over N(SO(n))}={\cal O}(2^{-n^{3}m/2})\,\,{\rm at}\,\,
{\rm fixed}\,\,m
\ee
shows that Lie $h$-invariant CFTs, although not generic, are copious in the
space of CFTs. Here, $N(SO(n))$ is the generic number (\ref{knuml}) of
level-families in the VME on $SO(n)$. A more precise statement of these
conclusions is obtained with graph theory in Ref.\ \cite{gt}
(see Section \ref{lhg}).

\subsubsection{Other features of the equations}

Solving coupled quadratic equations is a non-trivial task.
Some concepts which help are as follows.

\ni \bon Basis choice.
The choice of Lie algebra basis is important because a given ansatz generally
has its simplest form in a particular basis. (See, for example, the metric
ansatz discussed in Section \ref{ggth}.)

\ni \btw Factorization.
{}To solve the  coupled quadratic equations, one typically tries to eliminate
variables, thereby increasing the order of the equations until some
higher-order polynomial equation in one variable is obtained. In this process,
simplifications occur when one finds that a quadratic equation can be
factorized  into two linear equations, thus reducing the system into
smaller subsystems called sectors. An example of factorization into sectors
is given in Section \ref{soc}.

\ni \bth K-conjugation covariance.
The K-conjugation property $L+\tL=L_g$ reduces a $(2n)$th-order algebraic
equation to $n$th order \cite{nuc}.
As a result, subsystems of up to three coupled
quadratic equations can always be solved analytically\footnote{A central
role of the symmetry ans\"atze discussed above is to reduce the master
equation to such a manageable form.  As a corollary, no completely
asymmetric solution has so far been obtained in closed form.}.

The role of K-conjugation from submanifolds of $G$ is even more important.
The high order of the equations on a given manifold $G$ is partially explained
by the presence of affine-Virasoro nests $\tL = L_g - L_h^\#$, where $L_h^\#$
is a construction on a smaller manifold $h \subset g$. (This includes the $g/h$
coset constructions on $g$). The new solutions $ L_{g}^{\#}$ on $g$ (more
precisely, the new irreducible constructions on $g$ -- see Section
\ref{sec444}) are thus
obscured by the nesting from below. Conversely, the nesting of known
solutions is partially responsible for the fact that ans\"atze can be further
factored down to simpler subsystems.

\ni \bfo High-level analysis  \cite{hl}.
The high-level analysis of consistent ans\"atze (see Section \ref{hla})
can provide useful clues in solving the equations, as illustrated explicitly
in Ref.\ \cite{hl}. For example, high-level analysis enables one to predict
which solutions will appear in which sectors of the equations, and indicates
which of the solutions are already known and which are new. This knowledge is
useful, since if one knows that a particular high order equation contains a
known solution, one can reduce the order of the equation by factoring out this
solution.

\subsection{Four Examples of New Constructions}

\subsubsection{Generalized spin-orbit constructions \label{soc}}
\lam{soc}

The original spin-orbit constructions \cite{bh,ma} were the first examples of
affine-Vira\-soro constructions beyond the affine-Sugawara and coset
constructions.
These stress tensors involve a spin-orbit term $\pi \cdot J$,
\bs
\be
\D T=\pi^{\m} J_{\m} \quad , \quad \m =0,1\ldots D-1
\ee
\be
\pi^\m=i\pa\phi^{\m} \quad,\quad J_{\m}=\bar{\ps} \g_{\m} \ps
\ee
\es
where $\pi^\m$ are the abelian currents
(orbital operators) of the string, and
$J_{\m}$ are $SO(D-1,1)$ currents which carry  spin in
the spacetime spinor $\ps$. The motivation for the spin-orbit model was the
ghost-free introduction of current-algebraic spin on the string, using
the K-conjugate pair of Virasoro operators to eliminate the spin ghosts along
with the orbital ghosts.  The model was overshadowed by the later development
of the NS model \cite{ns}, but the original motivation was realized in
\cite{ma}, which argued that, in fact, the spin-orbit model was equivalent
to the NS model in 10 dimensions.  The spacetime spinor $\ps$ of the
spin-orbit model
now plays a central role in the Green-Schwarz \cite{grs} formulation of the
superstring.

The spin-orbit constructions also provided a central motivation for the
Virasoro master equation in Ref.\ \cite{hk}, where the  generalization of the
original constructions was also given.

{}To understand the generalized spin-orbit constructions, consider a Lie
algebra $g$ (not necessarily compact), and a coset $g/h$ with semisimple
$h \subset g$. One also needs a set of abelian currents $\p_I = i \pa \phi_I$,
$I=1 \ldots {\rm dim}\,g/h$. In the notation of the master equation,
the generalized
spin-orbit constructions are on the manifold $g_x \times U(1)^{\dim g/h}$, and
the affine currents are ordered as
$J_a =(\pi_I,J_I ,J_A)  $  where $A=1\ldots {\rm dim}\,h$
and $J_I$ are the coset currents.

The generalized Killing metric $G_{ab}$ of the master equation is
\be
G_{ab} = k \left( \matrix{ \e \et_{IJ} & 0 & 0 \cr 0 & \et_{IJ} & 0 \cr
0 & 0 & \et_{AB} \cr} \right)
\ee
where $\et_{AB}$, $\et_{IJ}= \pm 1$ is the Killing metric in a Cartesian
basis of $g$ and $\e=\pm 1$.
The invariant level of affine $g$ is $x=2k/\ps_g^{2}$, where $\ps_g$ is
the highest root of $g$.  The spin-orbit
ansatz for the inverse inertia tensor is
\bs
\be
L^{ab} = \ps_g^{-2} \left( \matrix{ \et^{IJ} \left( \matrix{ \e \l_\p &
\l_{so} \cr \l_{so}
& \l_{g/h} \cr}
\right) &  \;\, \matrix{0 \cr 0 \cr}  \cr  \;\;\;\;
\matrix{ 0 & \;\;\;\; 0 \cr}
 &  \et^{AB} \l_h  \cr} \right)
\ee
\be
T(L) = \ps_g^{-2} \xx [\e \l_\p \p^2 + 2 \l_{so} \p \cdot J_{g/h} +
\l_{g/h} (J_{g/h})^2 +
\l_h (J_h)^2 ] \xx
\ee
\label{soa} \es
\lam{soa}
where $\l_{so}$ is the spin-orbit coupling.

Consistency of the ansatz in the VME requires that $G/H$ is a symmetric space,
which includes
\bs
\be
{ SU(n)_x \over SO(n)_{2x}}\;\;
(n \geq 4 ) \sp {  SU(3)_x \over SU(2)_{4x} } \sp
{SU(2)_x \over U(1)_x}
\ee
\be
{(E_{6})_{x}\over
Sp(4)_{x}} \sp {(E_{7})_{x}\over
SU(8)_{x}}\sp {(E_{8})_{x}\over
SO(16)_{x}}
\ee
\be
 {SU(2n)_x \over Sp(n)_x} \sp {SO(n+1)_x \over SO(n)_x}
\sp {(F_4)_x \over Spin(9)_x} \sp { (E_6)_x \over (F_4)_x}
\ee
\be
{SO(2n)_{x}\over SO(n)_{x}\ti SO(n)_{x} } \sp
 {Sp(2n)_x \over  Sp(n)_x \ti Sp(n)_x }  \sp
{ (G_2)_x \over SU(2)_x \ti SU(2)_x }
\ee
\es
and their non-compact generalizations. The original spin-orbit construction
\cite{bh,ma} employed a noncompact generalization of $SO(n+1)/SO(n)$.

Substituting the ansatz into the VME, one obtains the consistent ansatz or
reduced VME,
\bs
\be
\l_{\pi} =x(\l_{\pi}^{2}+\epsilon \l_{so}^{2})
\ee
\be
\l_{so}=\l_{so} [x(\l_{\pi} +\l_{g/h})+
{1 \over 2}(\l_{g/h} +\l_h)\hh_{g}]
\le{feq}
\be
\l_{g/h}=(x+\hh_{g})\l_{g/h}^{2}+x\epsilon \l_{so}^{2}
\ee
\be
 \l_h=(x+r^{-1}\hh_{h})\l_h^{2}+\l_{g/h}(2\l_h-\l_{g/h})(\hh_{g}-r^{-1}\hh_{h})
\ee
\be
c=x[(\l_{\pi} +\l_{g/h})\dim {g/h} +\l_h \dim {h}]
\ee
\es
where $\hh_g$ and $\hh_h$ are the dual Coxeter numbers of $g$ and $h$, and
$r$ is the embedding index of $h\subset g$.
Eq.(\ref{feq}) factorizes into two sectors. For the sector
$\l_{so} =0$, one finds only affine-Sugawara and coset constructions. For
the sector $\l_{so} \neq 0$, one obtains the
generalized spin-orbit constructions \cite{hk},
\bs
\be
\l_\p ={1 \over 2x }\left( 1+\eta
F^{-1}(4x +4r^{-1} \hh_{h}-3\hh_{g})\right)
\ee
\be
\l_{so}  =\eta \s F^{-1}\sqrt{(-\epsilon /k)(2 x+2r^{-1}\hh_{h}-\hh_{g})}
\ee
\be
\l_{g/h} ={{1-\eta F^{-1}(4x +4r^{-1}\hh_{h}-\hh_{g})} \over
{2(x+\hh_{g})}}
\ee
\be
\l_h  = {{1+\eta F^{-1}(5\hh_{g}-4r^{-1}\hh_{h})} \over
{2(x+\hh_{g})}}
\ee
\be
\eqalign{
c&={1 \over 2}\left[{x{\rm dim}\,g \over {x+\hh_{g}}}+{\rm dim}\,g/h\right]
\cr
&+{{\eta\left[
x(5\hh_{g}-4r^{-1}\hh_{h}){\rm dim}\,h+\hh_{g}(2x+4r^{-1}\hh_{h}-
3\hh_{g}){\rm dim}\,g/h
\right]} \over
{2 F(x+\hh_{g})}}
\cr}
\ee
\be
F=\sqrt{(3\hh_{g}-4r^{-1}\hh_{h})^{2}-16 x (\hh_{g}-r^{-1}\hh_{h})}
\ee
\es
where $ \et = \pm 1 $ and $\s = \pm 1$.
The two values of $\eta$ correspond to K-conjugation, while the
sign change $\s$ in the spin-orbit coupling labels automorphic copies under
the $U(1)$ outer automorphism $\p_I \ra  - \p_I,\; \forall I  $.

The central charges of the generalized spin-orbit constructions are
generically irrational but the constructions are not manifestly unitary.
It may be possible, however,  to find unitary subspaces
\cite{br,godt,df} of the constructions, and, indeed,
Mandelstam \cite{ma} used the K-conjugate pair of Virasoro operators to
argue that the spin-orbit construction for level one of
$SO(9,2)/SO(9,1)$ is equivalent to the NS model
\cite{ns} in 10 dimensions.

\subsubsection{${\bf SU(2)_4^\# }$ \label{su2}}
\lam{su2}

The only simple algebra for which the VME has been completely
solved\footnote{See also \cite{dt}, where the VME is completely solved
on the superalgebra $SU(2|1)\supset SU(2)\times U(1)$.}
 is $g=SU(2)$. Beyond the affine-Sugawara and coset constructions,
one unexpected construction, called
$SU(2)^\#_4$, is found at level four \cite{rus,nuc,cf}.
This construction was the first example of a {\em  quadratic deformation}
(see Section \ref{sec445}),
which is any solution of the master equations with continuous
parameters.
The quadratic deformations are also called sporadic because they occur only
rarely, at sporadic levels, as seen in this case. Although these constructions
have fixed rational central charge, the continuous parameters imply
generically-continuous conformal weights, so these deformations are sometimes
called quasi-rational.

A complete solution on $SU(2)$ is possible because one can use
the inner automorphisms of $SU(2)$ to gauge-fix the inverse inertia tensor
to the diagonal form,
\be
L^{ab} = {\l_a \over \ps_g^2} \d_{ab}
\sp T(L) = \ps_g^{-2} \sum_{a=1}^3 \l_a \xx J_a J_a \xx
\label{sua} \ee
\lam{sua}
in the standard Cartesian basis of $SU(2)$.
Then the master equation reduces to three coupled equations,
\bs
\be
\l_a (1-x \l_a) \d_{ab} = \sum_{c,d=1}^3
\l_c (\l_a + \l_b - \l_d) \e_{cda} \e_{cdb}
\ee
\be
c = x  \sum_{a=1}^3 \l_a
\ee
\es
where $\e_{abc}$ is the Levi-Civita density. The solutions of this system
 for generic level $x$ are the expected affine-Sugawara and
$SU(2)/U(1)$ coset constructions.

For the particular level $x=4$, one also finds the
quadratic deformation $SU(2)^\#_4$
\cite{rus,nuc,cf}\footnote{The connection with the
form of $SU(2)_{4}^{\#}$ in \cite{nuc} is:
$\l_{3}=\alpha^{2}\l$, $\l_{1}=(L^{\alpha,-\alpha}+L^{\alpha,
\alpha})$, $\l_{2}=(L^{\alpha,-\alpha}-L^{\alpha,\alpha})$,
$-1/12\leq \l \alpha^{4} \leq 1/4$.}
\bs
\be
\l_{1}(\phi)={1\over 12  }(1+\sqrt{3}\cos \phi+\sin \phi)
\ee
\be
\l_{2}(\phi)={1\over 12  }(1-2\sin \phi)
\ee
\be
\l_{3}(\phi)={1\over 12  }(1-\sqrt{3}\cos \phi+\sin \phi)
\ee
\be
c=1
\ee
\es
which is unitary for  $0\leq \phi <2\pi$.
The solution exists for complex  $\phi$, but unitarity is lost.

The circle described by $\phi$ is shown in Fig.\ref{f1},
where we have also indicated the location
of six rational points. In fact,
 the existence of a quadratic deformation at the particular level $x=4$
can be predicted by the method described in \cite{cf} and reviewed in
Section \ref{sec5}: At this level,
the central charges of the
$U(1)$ and $SU(2)/U(1)$ level-families coincide, while these two
level-families are
flow-connected at high-level. The six rational points on the circle
are also predicted by this method.
\fig{5cm}{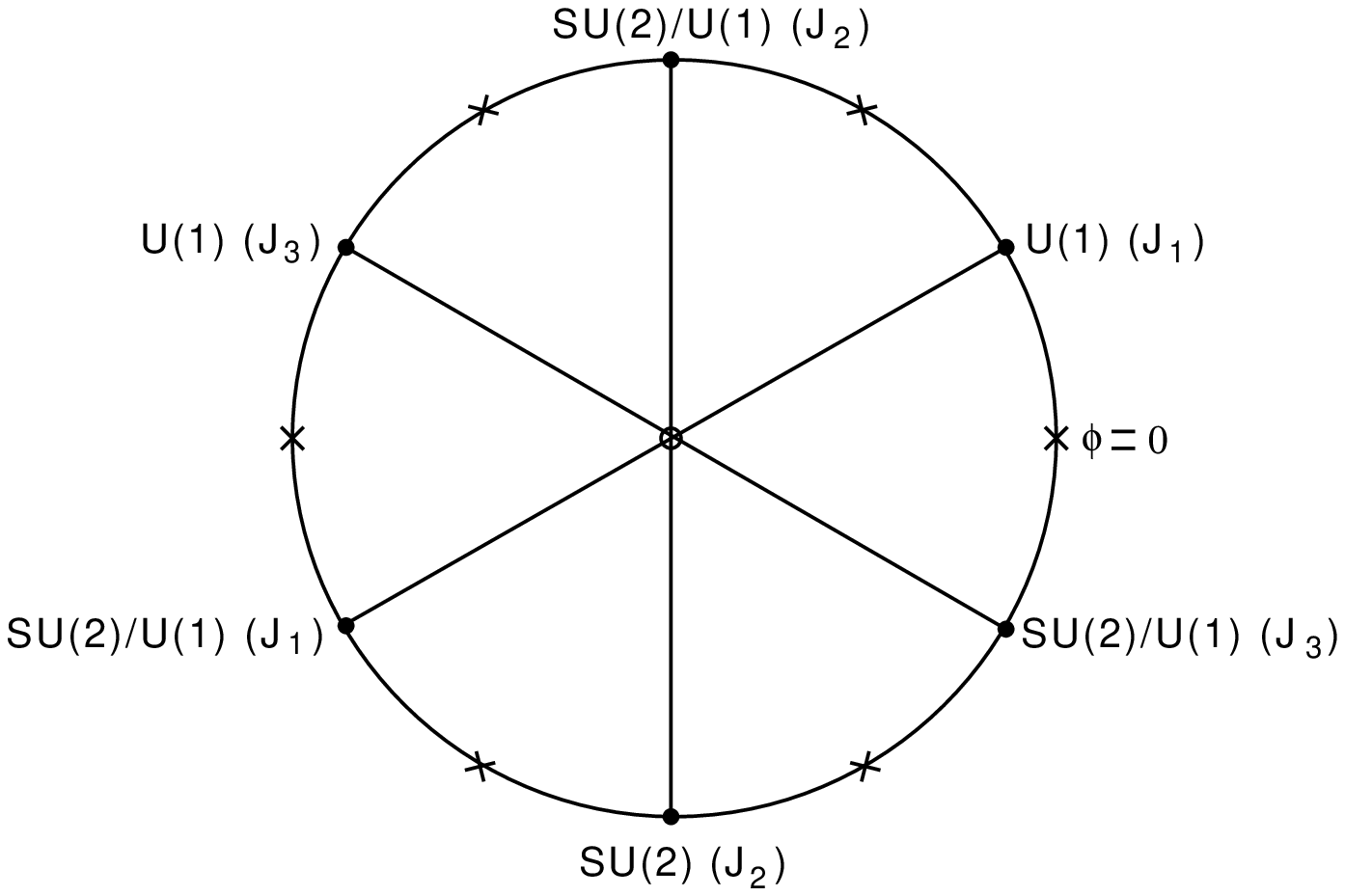}{$SU(2)_4^\#$ is a circle. The dots are the six rational
constructions $h$ and $g/h$.}{f1}

$SU(2)_4^\#$ is closed under K-conjugation on the $SU(2)$ manifold,
which acts on the circle as reflection through the origin.
The action of the residual automorphisms ($S_3 \sim$ permutations of $\l_a$)
of the stress tensor (\ref{sua})
is reflection about each of the three axes shown
in the Figure. It follows that the points marked by crosses in Fig.\ref{f1}
are automorphically-equivalent copies of a single self-K-conjugate
construction in $SU(2)_4^\#$. Modding by $S_3$, each $\p/3$ arc from a
$g/h$ to an $h'$ is a fundamental region of the solution \cite{nuc}.

Generically-continuous conformal weights  have been verified for
$SU(2)_4^\#$. The eigenvalue problem (\ref{kspec})
for spin $j$ of $SU(2)$ is that
of the
completely asymmetric spinning top, and the conformal weights are generically
non-degene\-rate. For example
\be
\Delta(j=1)={1\over 12}\left(
2(1 + \cos \phi ), 2 - \sin \phi \pm \sqrt{3} \cos \phi \right)
\ee
is obtained for the splitting of spin one, while
$\Delta(j=\frac{1}{2})=\frac{1}{16}$ is independent of the deformation.

It turns out that $SU(2)^\#_4$ is a new formulation of an old theory:
The spectral data of the deformation strongly
 suggest \cite{nuc} that $SU(2)^{\#}_{4}$ is a chiral version
of the line of
$\Z_{2}$ orbifold models at $c=1$, where the two primary fields with
fixed
dimension $1/16$ are the twist fields of the orbifold line.
This has been verified in some detail in Ref.\ \cite{rus2}.

General discussion of quadratic deformations is found in Refs.\
\cite{nuc,cf}, and the names of
all known exact quadratic deformations are listed in
Section \ref{spc}.

\subsubsection{Simply-laced $g^{\#}$ \label{sl}}
\lam{sl}

The first unitary constructions with irrational central charge were obtained
by Porrati, Yamron and three of the authors in Ref.\ \cite{nuc}. These
constructions, which live on $SU(2)\ti \cdots \ti SU(2) $ and
simply-laced$\,g$, are called
$(SU(2)_x^q)^\#$ and simply-laced$\,g_x^\#$ respectively.

{}To obtain simply-laced$\,g_x^\#$,
 it is convenient to use the Cartan-Weyl basis $(J_A,E_\r)$, $\r \in \Phi$
 of simple compact $g$. Then one restricts the discussion to
 simply-laced $g$ in the maximally-symmetric ansatz,
\bs
\be
L^{AB}=\l\psi_g^{-4}\sum_{\r}\r^{A}\r^{B}=\l\psi_g^{-2}\hh_g\delta^{AB}
\sp A,B =1 \ldots  {\rm rank}\, g \ee
\be
L^{\r
,\pm\r}=\psi_g^{-2}L_{\pm}
\ee
\label{msa} \es
\lam{msa}
where $\r \in \Phi$ are the roots of $g$, and all other components of the
inverse inertia tensor are zero.
The name ``maximally-symmetric'' derives from the operator form of the ansatz
\bs
\be
T(L) = \psi_g^{-2}( \l \hh_g \sum_A T_{AA} + L_-
\sum_\r T_{\r,-\r} + L_+ \sum_\r
T_{\r \r})
\ee
\be
= \psi_g^{-2}\left(\l\hh_g\sum_{A}\xx J_A J_A \xx +{1\over
2}(L_{+}-L_{-})\sum_{\r>0}\xx
(E_{\r}-E_{-\r})^{2}\xx+\right.
\ee
$$
+\left. {1\over 2}(L_{+}+L_{-})\sum_{\r>0}\xx(E_{\r}+E_{-\r})
^{2}\xx\right)
$$
\label{mso} \es
\lam{mso}
which
shows complete symmetry under permutations of the positive (or negative) roots
of $g$. Unitary solutions on $x\in \N$ require that
$\l$ and $ L_{\pm}$ are  real.

{}To understand the consistency of
this ansatz, note that for any Lie $g$  the operators
$\{i(E_{\r}-E_{-\r})\,,\,\forall\,\r>0\}$
generate a subgroup $h\subset
g$ such
that $g/h$ is a symmetric space.
This is the symmetric space with maximal ${\rm dim}\,g/h$ at fixed $g$.
The ansatz (\ref{msa}) corresponds to the three-subset decomposition
\be
g = h + \mbox{Cartan}\,g + R
\le{gdc}
where an asymmetry is allowed between the components
$\mbox{Cartan}\,g $ and $R$ in  $ g/h$. Substituting the ansatz into the
VME, one finds that the
maximally-symmetric ansatz is  consistent because, in this basis,
 the commutator of any two currents in $g$ contains
the currents of at most one of the three subsets in (\ref{gdc}).

\newpage
The ansatz contains a number of affine-Sugawara and coset
constructions,
 and one new pair of K-conjugate
constructions, called
simply-laced $g_x^{\#}$ \cite{nuc},
\bs
\be
\l={1\over 2\hh_g(\hh_g+x)}[1-\eta
B^{-1}(2\hh_g^{2}+\hh_g(4-x)-2x^{2}+10x-16)]
\ee
\be
L_{-}={1\over 2(\hh_g+x)}[1+\eta
B^{-1}(x\hh_g-6x+16)]\,\,,\,\,L_{+}=-\eta
B^{-1}(x-4)
\ee
\be
c={x\, {\rm rank}\,g\over 2(\hh_g+x)}[\hh_g+1+\eta
B^{-1}(\hh_g^{2}(x-2)+\hh_g(12-5x)+2x^{2}-10x+16)]
\label{csl} \ee
\lam{csl}
\be
B\equiv
\sqrt{\hh_g^{2}x^{2}+4\hh_g(x^{3}-13x^{2}+40x-32)+4(x^{4}-10x^{3}+41x^{2}
-80x+64)}
\ee
\es
where the values $\eta =\pm 1 $
correspond to K-conjugation. Simply-laced$\,g_x^\#$ is
 completely unitary with generically-irrational central charge
across all levels $x \in \N$ of simply-laced $g$.

Simply-laced $g_x^{\#}$ is rational for $SU(2)_x$, $SO(4)_x$ and also
for
$x=1,2,4$ and for $\hh_g=2n$, $x=n+3$. All
these points may be identified with known constructions $h$ and $g/h$.
The lowest irrational central charges of the construction are found
at level
three, and, in particular, the value on $SU(3)$
\be
c(SU(3)_{3}^{\#})=2\left(1-{1\over \sqrt{73}}\right)\simeq 1.7659
\ee
is the lowest
irrational central charge
in the ansatz.
More generally, the central charges in
(\ref{csl})
increase
 with $x$ and ${\rm rank}\, g$ at fixed $\eta$, giving the high-level
central charge,
\be
\lim_{k\to\infty}c={\rm dim}\,\Phi_{+}(g)+{1+\eta\over 2}{\rm
rank}\,g
\pe
\le{suc}
This is an example of the general result (see Section \ref{hla}) that
 the high-level central charges of the VME on simple $g$
are always integers between 0 and ${\rm dim}\,g$.

We also remark on the irrational conformal weights $\Delta =c/6x$ of
$SU(3)_{x}
^{\#}$, computed with (\ref{kspec}) for the $3$ or $\bar 3$ of $SU(3)_{x}$;
these degenerate weights apparently lie in a general family of one fermion
conformal
weights $\Delta (\mbox{one fermion})=c/2c_{(free\;\,fermions)}$ noted
for
fermionic affine-Sugawara constructions in \cite{h3}.

Generalization of simply-laced $g^\#$ to semisimple algebras was given in
Ref.\
\cite{slg}, and related $SU(3)$ level-families with less than maximal symmetry
(the basic ansatz on $SU(3)$) were found independently by Schrans and Troost
\cite{str} and by two of the authors \cite{hl}.
The complete list of all known exact unitary irrational level-families is
given in Section \ref{ucicc}.

\subsubsection{${\bf  SU(n)^{\#}[m(N=1),rs]}$ \label{sus} }
\lam{sus}

The first set of unitary irrational N=1 superconformal constructions
\cite{sin}, called
$ SU(n)_x^{\#}$ $[m(N=1),rs]$,
was found by solving the superconformal master equation (SME).

This solution is obtained in the Pauli-like basis of $SU(n)$ (see Section
\ref{kgg}), whose trigonometric structure constants are irrational.
In this basis the adjoint index $a=\bp=(p_1,p_2)$ is an integer-valued
two-vector, and the
ansatz for the supercurrent on $SU(n) \ti SO(n^2-1)$ is
\bs
\be
 G( n ;rs) =- \sqrt{\l \over k}[ J_{(r,0)}S_{(n-r,0)}+ J_{(n-r,0)}S_{(r,0)}
+ J_{(0,s)}S_{(0,n-s)} + J_{(0,n-s)}S_{(0,s)}]
\ee
\be
1 \leq r,s \leq [ (n-1) /2]
\ee
\label{sca} \es
where $n$ is the $n$ of $SU(n)$, and $r,s$ label the ansatz.
Unitarity on $x\in \N$ requires $\l \geq 0$.

Remarkably, the SME reduces to a single linear equation for each
$n,r,s$, and the
corresponding solutions, called
$ SU(n)_x^{\#}[m(N=1),rs]$, have
the irrational central charges,
\bs
\be
 \l(SU(n)_x^{\#}[m(N=1),rs])={nx \over nx + 8\sin^2(rs \pi/n) }
\ee
\be
 c(SU(n)^{\#}_x[m(N=1),rs])={6nx \over nx + 8\sin^2(rs \pi/n) }
\pe
\ee
\es
Irrationality of the central charge arises in this case
from the trigonometric structure constants of the
basis.

The superconformal constructions
$ SU(n)_x^{\#}[m(N=1),rs]$
 are the simplest unitary irrational
level-families found so far.
In fact, these constructions are only  a small subset of
a much larger graph-theoretic set of superconformal level-families,
which is also described by linear equations (see
Section \ref{sma}).

\subsection{Lists of New Constructions}

\subsubsection{Unitary constructions with irrational central charge
\label{ucicc}}

We list here the names of all known exact unitary solutions of the
master equations with irrational
central charge.
\bs
\be
 ((\hbox{simply-laced}\,g_x)^q)^{\#}_M
\;\;\;\;\;\;\;\;  \cite{nuc,slg}
\label{slg} \ee
\lam{slg}
\be
 SU(3)^{\#}_{BASIC}=\left\{ \matrix{SU(3)^{\#}_M \cr
SU(3)_{D(1)}^{\#},\;\;SU(3)_{D(2)}^{\#},\;\;SU(3)_{D(3)}^{\#} \cr
SU(3)_{A(1)}^{\#},\;\;SU(3)_{A(2)}^{\#}} \right.
\;\;\;\;\;\;\;\;  \cite{str,hl}
\label{su3} \ee
\lam{su3}
\be
 SO(n)^{\#}_{diag}=\left\{ \matrix{
\left. \matrix{ SO(2n)^{\#}[d,4],\;\,n\geq 3 \cr
SO(5)^{\#}[d,6]_2 \cr
SO(2n+1)^{\#}[d,6]_{1,2},\;\,n \geq 3 \cr } \right\}
 & \;\;\;\;\;\;\;\;  \cite{gt} \cr
\left. \matrix{ SO(6)^\#[d(SO(2) \ti SO(2)),5 ]_{1,2} \cr
SO(6)^\#[d(SO(2) \ti SO(2)),8]_1 \cr
SO(6)^\#[d(SO(2)), 7']_{1,2} \cr } \right\}
& \;\;\;\;\;\;\;\;  \cite{lie} \cr
}\right.
\label{sond} \ee
\lam{sond}
\be
 SU(5)^{\#}[m,2]
\;\;\;\;\;\;\;\;  \cite{mb}
\label{su5} \ee
\lam{su5}
\be
 SU(n)^{\#}[m(N=1),rs]
\;\;\;\;\;\;\;\;  \cite{sin}
\label{sc1} \ee
\lam{sc1} \be
 SU(\Pi^s_i n_i)^{\#}[m(N=1);\{r\}\{t\}]
\;\;\;\;\;\;\;\;  \cite{ggt}
\label{sc2} \ee
\lam{sc2}
\label{lis} \es
\lam{lis}
Each of these entries is a collection of conformal level-families, defined on
all levels of affine $g$.
On $x\in\N$, each of these level-families is generically unitary with
irrational central charge, as seen in the example of Section \ref{sl}.
In the level-families on simple $g$, the lowest
level with unitary irrational central charge is $x=3$ (simply-laced$\,g^\#$ and
$SO(2n)^\#[d,4]$). The list also contains
 the references in which the explicit forms
of these constructions can be found.

The nomenclature in the list is as follows. \nl
{\bf a)} The symbol $\#$ indicates that the construction is new, i.e. not
an affine-Sugawara nest. \nl
{\bf b)} The Lie algebraic part of the
name (e.g. $SU(3)$, $SO(n)$ etc.)
 denotes the affine Lie algebra $g$ on which the
construction is found. In the case of the superconformal constructions in
(\ref{sc1})
and (\ref{sc2}),
labeled by N=1, it is understood that the construction is on
$ g \ti SO({\rm dim}\,g)_1 $.\nl
{\bf c)} Each name includes
a label for the ansatz (and subansatz) of the VME (or SME) in which the
construction occurs. For example, the label $M$ in (\ref{slg}) stands for
$M$=Maximal symmetric ansatz, which includes the level-families
simply-laced$\,g_x^\#$ given explicitly in Section \ref{sl}. In (\ref{su3}),
we see the nested subsans\"atze
Basic $\supset$ D(=Dynkin) $ \supset$ M(=Maximal symmetric), which show
increasing symmetry toward the right.
The level-family $SU(3)_M^\#$ is included in simply-laced $g^\#$.
Eqs.(\ref{su5}), (\ref{sc1}) and (\ref{sc2}) are
constructions in the $m$=metric
ansatz on $SU(n)$ and its superconformal analogue.
The explicit form of $SU(n)^{\#}[m(N=1),rs]$ was given in Section \ref{sus}.

In (\ref{sond}), the notation $SO(n)[d,R]$ indicates that the construction
is found in an $R$-dimensional subansatz of the $d=$diagonal ansatz on
$SO(n)$.
The last three sets of level-families on $SO(6)$ have a Lie $h$ symmetry
denoted by $d(h)$. The constructions
 $SU(3)^\#_{D(1,2)}$ and $SU(3)^\#_{A(1,2)}$ also have
a Lie $U(1)$ symmetry. \nl
{\bf d)} Extra numbers and symbols distinguish between
various constructions in a given ansatz.

The diagonal $(d)$ and metric $(m)$ ans\"atze are discussed in
Sections \ref{gth} and~\ref{ggth}.

\subsubsection{Special categories \label{spc}}
\lam{spc}

\ni \u{Quasi-rational solutions}
\vskip .3cm

Beyond the unitary irrational constructions, the master equations have
also generated a large number of unitary quasi-rational solutions, which are
constructions  with rational central charge and continuous or
generically-irrational conformal weights.

These constructions come in various categories
mentioned in Section \ref{sec445}, and we list below all known
examples for which the exact forms are known.

\ni \bon Quadratic deformations (continuous).
\bs
\be
SU(2)_4^\#
\;\;\;\;\;\;\;\; \cite{rus,nuc}
\ee
\be
{\rm Cartan}\,g^\#
 \sp (SU(2)_x \ti SU(2)_x )^\# \;\;\; (x \neq 4)
\;\;\;\;\;\;\;\; \cite{nuc}
\ee
\be
SO(2n+1)_2^\# [d,6] \;\;, \;\; n \geq 2
\;\;\;\;\;\;\;\; \cite{gt}
\ee
\be
SO(n)_2^\# [d]
\;\;\;\;\;\;\;\; \cite{bl}
\ee
\be
SO(4)^\#_2 [d,6]_1  \sp SO(4)_2^\# [d,6]_2
\;\;\;\;\;\;\;\; \cite{sta2}
\ee
\be
SU(3)_3^\#
\;\;\;\;\;\;\;\; \cite{mb}
\ee
\be
({\rm Cartan}\,g)^\#_{N=1}
\;\;\;\;\;\;\;\; \cite{sme,nsc}
\ee
\be
  (SU(2)_2 \ti SU(2)_2 )^\#_{N=1}
\;\;\;\;\;\;\;\; \cite{rs}
\ee
\es
The explicit form of $SU(2)_4^\#$ is given in Section \ref{su2}.
The deformation $(SU(2)_x\times SU(2)_x)^\#$ is supersymmetric at level
two, where it is included in the supersymmetric deformation
$(SU(2)_2  \times SU(2)_2)^\#_{N=1}$.  All possible deformations on level
one of simply-laced $g$ are included \cite{nuc} in the deformation
Cartan $g^\#$.

\ni \btw  Self K-conjugate constructions
 ($c=c_g/2$).
\bs
\be
SO(4)^\#[d,4] \sp SO(5)^\#[d,6]_1 \sp SO(5)^\#[d,2]
\;\;\;\;\;\;\;\; \cite{gt}
\le{sks}
\be
SU(3)^\#[m,2]
\;\;\;\;\;\;\;\; \cite{mb}
\ee
\es

\ni \bth Self K${}_{g/h}$-conjugate constructions
($c=c_{g/h}/2$).
\be
SO(6)^\# [ d(SO(2),7']_{3,4}
\;\;\;\;\;\;\;\; \cite{lie}
\le{sck}

\ni \bfo Other quasi-rational CFTs.
The superconformal constructions on triangle-free graphs \cite{nsc}
and on simplicial complexes \cite{ssc} form two other
large classes  of new quasi-rational constructions which will be discussed
in Section \ref{gsm}.

\vskip .4cm
\ni \u{New RCFTs} \nl
\vskip .3cm

Another category of new constructions are the candidates for new RCFTs,
which are non-standard RCFTs beyond the affine-Sugawara nests. \nl
\ni \bul superconformal constructions on edge-regular triangle-free
graphs \cite{nsc}. \nl
\ni \bul superconformal constructions on regular triplet-free
2-complexes \cite{ssc}. \nl
\ni \bul bosonic N=2 superconformal constructions \cite{ks,gep}.

\section{The Semi-Classical Limit and Generalized
Graph Theories \label{nhl}}
\lam{nhl}

\subsection{Overview}
In Section \ref{hla}, we review the method of {\em high-level expansion},
in powers of the inverse level $k^{-1}$.  The leading term of this expansion
was given in \cite{nuc} and higher-order corrections were studied
systematically in \cite{hl}.  A special case of the expansion was
also studied in \cite{rai}.

High-level expansion is a basic tool for the systematic study of the
master equations, enabling one to see the structure
of affine-Virasoro space at high level, including
unitarity, symmetries, counting and other properties, without
knowing the exact form of the constructions.
The expansion of the VME is unique on simple $g$, but we will
also remark on the various high-level expansions of the SME \cite{sme},
which is formulated on semisimple $g$.

Using the high-level expansion, the first connection between ICFT
and graph theory was found by Halpern and Obers in Ref.\
\cite{gt}. Following this observation,
the high-level expansion of the  VME and SME has yielded
a {\em partial classification of affine-Virasoro space} by generalized graph
theory on Lie $g$ \cite{gt,sme,mb,nsc,sin,ggt,lie},
including conventional graph theory on the orthogonal groups
\cite{gt,nsc,lie}. In each generalized graph theory, the graphs are in
one-to-one correspondence with conformal or superconformal level-families.

In the course of this work,
 an unsuspected Lie group- and conformal field-theoretic structure
was seen in
(generalized) graph theory \cite{gt,sme,mb,nsc,sin,ggt,lie}.
We believe that this development, called {\em generalized graph theory on
Lie $g$}, is a
fundamental connection between Lie groups and (generalized) graph theory,
which will be important in mathematics.
The subject has been axiomatized \cite{ggt} without reference to its origin
and applications in ICFT, and this development is summarized in Section
\ref{agt}.
 Then we return to conformal field theory in Section \ref{ggth}
and show how generalized
graph theory arises as a natural structure in the master equations.

So far, six {\em graph theory units}  of conformal level-families,
\be
\eqalign{
SO(n)_{diag} &:  \mbox{graphs of}\ SO(n) \;\;\cite{gt,lie} \cr
SU(n)_{metric} &:\mbox{sine-area graphs of} \ SU(n) \;\;\cite{mb} \cr
SU(\Pi^s_1 n_i)_{metric} &: \mbox{sine} (\oplus \mbox{area) graphs of}
\ SU(\Pi_i^s n_i) \;\;\cite{ggt} \cr
 SO(n)_{diag}\hbox{$ [ {{N=1}\atop{t= 0}}]$} &:\mbox{signed graphs of}\ SO(n)
\times SO(n(n-1))/2) \;\; \cite{sme,nsc} \cr
SU(n)_{metric}\hbox{$[ {{N=1}\atop{t=0}}]$} &:\mbox{signed sine-area graphs
of}\ SU(n) \times SO(n^2 -1)\;\; \cite{sin} \cr
SU(\Pi_i^s n_i)_{metric}\hbox{$ [{{N=1}\atop{t=0}} ]$}&: \mbox{signed sine}
(\oplus\mbox{area)  graphs of }\  SU(\Pi^s_i n_i) \times SO(\Pi^s_i n^2_i -1)
\cr
&\hspace*{3.9in} \cite{ggt}\cr
}
\ee
have been found in the master equations, and it is expected that there exist
many others. In particular, the bosonic N=2 superconformal constructions
of Kazama and Suzuki \cite{ks,gep} may be considered as a seventh
graph theory unit (See Section \ref{bks}).

The master equations are also expected to generate other
geometric categories on Lie $g$, beyond generalized graph theory.
In particular, Ref.\ \cite{ssc} discusses
an eighth geometric unit,
\be
SO({\rm dim}\,SO(n))\hbox{ $[{{N=1}\atop{t_f \ }}]$ }: \quad
\mbox{2-complexes of }
SO({\rm dim}\,SO(n))
\ee
in which the level-families are classified by the
two-dimensional simplicial complexes. This structure was found in the SME
on orthogonal groups and generalizations of this category are expected on
other groups.

Large as they are, the known graph theory units cover only very small
regions of affine-Virasoro space, a situation which is depicted in
Fig.\ref{fgt}.
At present, the most promising direction for classifying larger regions
lies in finding additional magic bases of Lie $g$ (see Section
\ref{mbg}) and their corresponding graph theory units.
\fig{4cm}{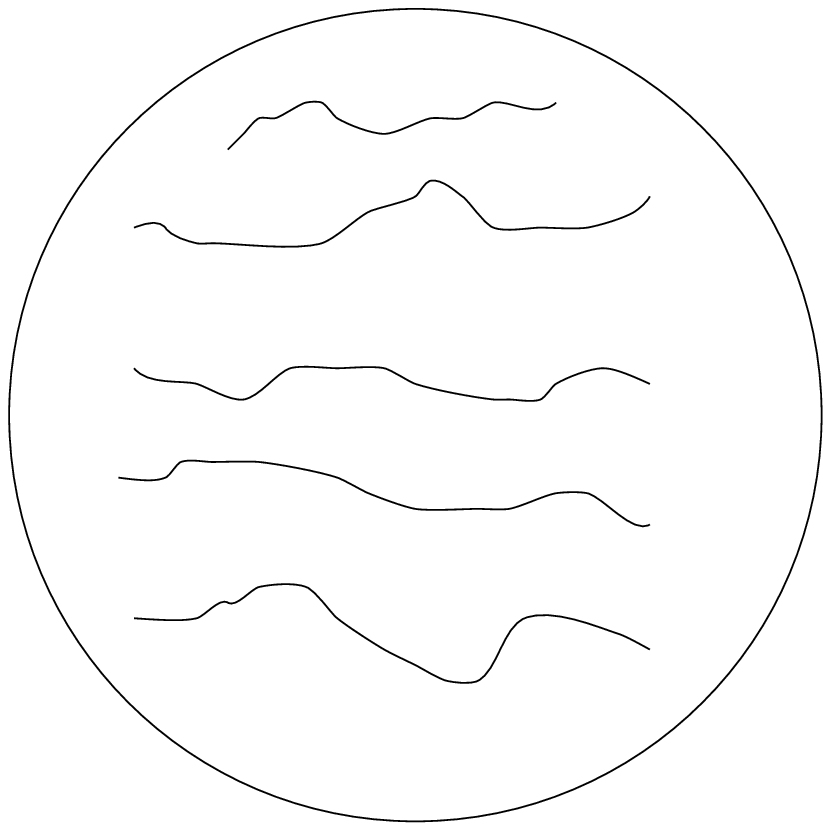}{Known graph theory units in ICFT.}{fgt}

The high-level expansion has also been used to study the correlators of
ICFT on the sphere and the torus (see Sections \ref{hc21} and \ref{hc39}),
and the leading term of the expansion plays a central role in
the generic world-sheet of ICFT (see Section \ref{hc41}).

\subsection{High-Level Analysis of the Master Equations \label{hla}}
\lam{hla}

\subsubsection{The semi-classical limit \label{slc}}
\lam{slc}

The class of
 {\em high-level smooth} CFTs \cite{nuc,hl,gva}
on semisimple $g=\oplus_I g_I$ is defined as the set of all CFTs
 whose inverse inertia
tensor is $\cO(k^{-1})$ at high level, where all levels $k_I$ are taken
large. It is believed \cite{gva,gt} that  the
high-level smooth CFTs, \nl
1) are precisely the generic level-families on simple $g$, and \nl
2) include all unitary level-families on $x\in\N$ of simple compact $g$ \nl
where (2) follows from (1). The evidence for (1) is discussed in  the remarks
around eq.(\ref{hls}). An independent argument for (2) was also given in
\cite{nuc}.
In what follows we restrict the discussion to simple compact $g$.

For this class of theories, each level-family $L^{ab}(k)$ has the
high-level expansion \cite{nuc,hl},
\bs
\be
L^{ab} = \frac{1}{k} \sum_{p=0}^{\infty} L^{ab}_{(p)} k^{-p}
= {P^{ab} \over 2k } + \cO (k^{-2}) \sp P^{ab} \equiv {L^{ab}_{(0)} \over 2}
\ee
\be
c =\sum_{p=0}^{\infty} c_{p} k^{-p} = c_{0} + \cO(k^{-1})
\ee
\label{hl1}
\lam{hl1}
\be
P_a{}^b\equiv \eta_{ac}P^{cb}=P_a{}^c P_c{}^b \sp c_0 = {\rm rank}\,P
\le{hln}
\es
whose leading term $L^{ab}=P^{ab}/2k,\; P^2=P$ was first given in
\cite{nuc}.  Note in particular
that $P$ is a projector, called the projector of the $L$ theory,
 and that the high-level central charge $c_0$
is an integer between 0 and ${\rm dim}\,g$.

Similarly, one may use K-conjugation $\tL=L_g-L$ and the high-level form
$L^{ab}_g=\et^{ab}/2k+\cO(k^{-2})$ of the affine-Sugawara construction
to obtain the leading term of the K-conjugate theory \cite{nuc,hl},
\bs
\be
\tL^{ab} = {\tilde{P}^{ab} \over 2k} + \cO (k^{-2})
\ee
\be
\tilde{P}+ P = \one  \sp \tilde{P} P =0 \sp
\tilde{c}_0 = {\rm rank}\,\tilde{P} = {\rm dim}\,g - c_0
\label{hln2}
\ee
\es
where $\tilde{P}$ is the projector of the $\tL$ theory.

The results in (\ref{hln}) and (\ref{hln2}) are
necessary conditions for
$L^{ab} \simeq P^{ab}/2k, \tL^{ab}\simeq \tP^{ab}/2k $ to be
leading-order solutions of the VME, but, as we will discuss below, higher
orders in the expansion can give  further restrictions on the projectors.
The development in the next two subsections follows the general analysis
of Ref.\ \cite{hl}.

\subsubsection{Radial and angular variables}
In general high-$k$ analysis, it is useful to introduce a new
set of variables for the master equation.
Unitarity on positive integer level of affine compact
$g$ requires \cite{gko,nuc}
\be
L^{ab}=L^{ba}={\rm real}
\ee
in any Cartesian basis, so all unitary solutions are included in the
eigenbasis
\be
L^{ab}=\sum_c \Omega^{ac}\Omega^{bc}\l_c
\label{eib} \ee
\lam{eib}
where $\l_a \in \R$
 and $\O \in SO({\rm dim}\,g)$ are called the radial variables and the
angular variables respectively.
In this eigenbasis,
the master equation  takes the form
\bs
\be
\l_a(1-2k\l_a)=\sum_{cd}\l_c(2\l_a-\l_d)\hat{f}_{cda}^2
\ee
\be
0=\sum_{cd}\l_c(\l_a+\l_b-\l_d)\hat{f}_{cda}\hat{f}_{cdb} \sp a<b
\label{ode} \ee
\lam{ode}
\be
\hat{f}_{abc} \equiv f_{a'b'c'}\O^{a'a}\O^{b'b}\O^{c'c}
\ee
\be
c=2k\sum_a\l_a
\ee
\label{me2} \es
\lam{me2}
where all angular dependence has been absorbed into the
$SO({\rm dim}\,g)$-twisted structure constants $\hat{f}_{abc}$ of $g$.

A natural gauge-fixing for the system is the coset decomposition of the angular
variables
$
\O (SO({\rm dim}\,g))=\O(g)\O(SO({\rm dim}\,g) /g )
$
since the twisted structure constants do not depend on the inner automorphisms
of $g$.
The gauge-fixed form of (\ref{me2}) shows $N(g)$ (in (\ref{knuml}))
equations on the same number of variables $\l$ and $\O(SO({\rm dim}\,g)/g)$,
in agreement with the counting in the general basis.

\subsubsection{High-level analysis of the VME \label{hlv}}
\lam{hlv}
We discuss  the high-level expansion of the inverse inertia tensor
in the radial-angular form (\ref{eib}),
\be
\l_a=\frac{1}{k}\sum_{p=0}^{\infty}\l_a^{(p)}k^{-p},\;\;\;
\O^{ab}=\sum_{p=0}^{\infty}\O^{ab}_{(p)}k^{-p}
\sp \hat{f}_{abc}=\sum_{p=0}^{\infty}\hat{f}_{abc}^{(p)}k^{-p}
\label{hle}\ee
\lam{hle}
so that in particular
\bs
\be
L_{(0)}^{ab}= \sum_c \O_{(0)}^{ac}\O^{bc}_{(0)}\l_c^{(0)}
\ee
\be
\O_{(0)}\O_{(0)}^T=1,\;\;\;
\O_{(0)}\O_{(1)}^T+\O_{(1)}\O_{(0)}^T=0
\ee
\be
\hat{f}_{abc}^{(0)}=f_{a'b'c'}\O^{a'a}_{(0)}\O^{b'b}_{(0)}\O^{c'c}_{(0)}
\pe
\ee
\es
The all-order expansion preserves the total antisymmetry of
the $p$th-order twisted structure constants $\hat{f}_{abc}^{(p)}$.

Substitution of the expansion (\ref{hle}) into the coupled system
(\ref{me2}) gives the
zeroth-order solution for the radial variables,
\be
\l_a^{(0)}=\frac{\th_a}{2} \sp \th_a=0\;{\rm or}\,1
\sp a=1\ldots{\rm dim}\,g
\ee
 and then, for each choice of
$\{ \th_a \}$, the zeroth-order quantization condition
\be
0=\sum_{cd}\th_c(\th_a+\th_b-\th_d)\hat{f}_{cda}^{(0)}\hat{f}_{cdb}^{(0)}
\sp a<b
\label{qco} \ee
\lam{qco}
constrains the zeroth-order angular variables $\O_{(0)}^{ab}$, whose
solutions may be discrete or continuous. Evaluation of the projection
operators $P^{ab}= 2 L_{(0)}^{ab}=\sum_c \O_{(0)}^{ac}\O_{(0)}^{bc}\th_c$,
completes order $p=0$ of the expansion, and the result
through order $p=1$,
\bs
\be
L^{ab}=\frac{1}{2k}\sum_c\O_{(0)}^{ac}\th_c\O_{(0)}^{bc}+
\frac{1}{k^2}\sum_c(\O_{(0)}^{ac}\l_c^{(1)}\O_{(0)}^{bc}+
\O_{(0)}^{ac}\frac{\th_c}{2}\O_{(1)}^{bc}+\O_{(1)}^{ac}\frac{\th_c}{2}
\O_{(0)}^{bc})+{\cal O}(k^{-3})
\ee
\be
\l_a^{(1)}=\frac{1}{4(1-2\th_a)}
\sum_{cd}\th_c(2\th_a-\th_d)(\hat{f}_{cda}^{(0)})^2
\label{o1r} \ee
\lam{o1r}
\be
c=\sum_a\th_a+\frac{2}{k}\sum_a\l_a^{(1)}+{\cal O}(k^{-2})
\ee
\es
was given in Ref.\ \cite{hl}.

More generally, the order $p \geq 1$  expansion shows that the radial variables
$\l_a^{(p)}$  are unambiguously determined in terms
of the lower-order data, but the angular variables are more subtle.
In particular, eq.(\ref{ode}) at order $p$ may contain other constraints like
(\ref{qco})  which quantize lower-order angular
variables that were originally continuous.
(This behavior
is familiar in higher-order degenerate perturbation theory in quantum
mechanics.) In other words,  some of
the zeroth-order continuous solutions may be high-$k$ artifacts, which
quantize at higher order. See Ref.\ \cite{hl} for further discussion of the
higher-order expansion.

\vskip .4cm
\ni \u{General properties of the high-level expansion}
\vskip .3cm

\ni \bon  The asymptotic central charge \cite{nuc,hl} of each
level-family,
\bs
\be
c_0 \equiv  \lim_{k \rightarrow \infty}c =\sum_a\th_a={\rm rank}\, P
\ee
\be
c_0 \in \{ 0,1\ldots{\rm dim}\,g \}
\ee
\ls{acc}
is the number of non-zero $\th$'s  in $\{\l_a^{(0)}=\th_a/2\}$, and
 $c_0$ is called the sector number of the level-family.
Each sector $c_0$
exhibits a large  degeneracy
at high level, shown
schematically in Fig.\ref{f2},
which is generally lifted at higher order.\nl
\fig{7cm}{figr2.ps}{Affine-Virasoro space: high-level central charges
 on simple compact $g$.}{f2}
\ni \btw Finite-order irrational central charge can be seen in the high-level
 expansion
when the structure constants of $g$ and hence $\O_{(0)}^{ab}$ are irrational
(see Section \ref{ggth}). \nl
\ni \bth The approach to each $c_0$ is from below, since
\be
\l_a^{(1)} \leq 0\;\;\;,\;\forall \,a
\ee
is easily verified from (\ref{o1r}). On the basis of this result and
the $k$-dependence of all known exact level-families, the exact universal
behavior
\be
 \rd c(k)/ \rd k < 0
\ee
was conjectured for all level-families in Ref.\ \cite{hkou}. \nl
\ni \bfo The low and high sectors
\be
c_0=\left\{ \matrix{ 0,\,1 \cr {\rm dim}\,g-1,\,{\rm dim}\,g \cr}
\right.
\ee
of the expansion have been solved to all orders \cite{hl}.
The low sectors contain
only $L=0$ and the $c=1$ sector of the quadratic deformation
 Cartan$\,g^{\#}$ \cite{nuc}, while the high sectors contain only their
 K-conjugate partners $L_g$ and the $c=c_g-1$ sector of
$g/{\rm Cartan}\,g^{\#}$. \nl
\ni \bfi A conjecture, based on the collection of all known exact
solutions,
is that all unitary irrational level-families satisfy the inequalities
\cite{gt,mb},
\be
{\rm rank}\,g \leq  c_0 \leq  {\rm dim}\,g-{\rm rank}\,g
\le{ncj}
so that the new constructions are centrally located in Fig.\ref{f2}.

\vskip .4cm
\ni \u{Applications of the high-level expansion}
\vskip .3cm

The high-level expansion was first employed \cite{hl}
to see all the level-families in
the ansatz $SU(3)_{BASIC}$, where it provided sufficient
structural clues to find the exact
form of all unitary irrational solutions $(SU(3)_x)^\#_{BASIC}$ in the basic
ansatz \cite{nuc,slg,str,hl}. The first isomorphism with graph theory
\cite{gt}  was
seen in the high-level expansion of the diagonal ansatz
on $SO(n)$, and high-level expansion  on other
groups revealed the structure known as generalized graph theory
on Lie $g$ \cite{mb,sin,ggt} (see Sections \ref{agt} and \ref{ggth}).

The high-level
expansion is also an important approximation technique in the dynamics
of ICFT [103-106] 
(see Sections \ref{hc21} and \ref{hc39}) and
the leading term of the expansion plays a central role in
the generic world-sheet action of ICFT \cite{gva} (see Section \ref{hc41}).

\subsubsection{High-level analysis of the SME \label{hlsm}}
\lam{hlsm}

The relative simplicity of the high-level expansion in the VME
on simple $g$ is due
to the fact that there is
an effective abelianization of the dynamics at high
level (see Section \ref{hc22}).
On semisimple $g$, other high-level expansions, with some levels held fixed,
are also possible. These expansions are generally more complicated
because some of the dynamics remain non-abelian.

In the SME, for example, the level of the fermionic currents (\ref{s10})
is fixed at $\tau =1$ or 2, and two
distinct high-level expansions, which select two classes of solutions,
 have been considered \cite{sme},
\bs
\be
e^{AI} = \frac{1}{\sqrt{k}}\sum_{p=0}^{\infty} e^{AI}_{(p)}k^{-p} \sp
t^{IJK} = \frac{1}{\sqrt{k}}\sum_{p=0}^{\infty} t^{IJK}_{(p)}k^{-p}
\le{shl1}
\be
e^{AI} = \frac{1}{k}\sum_{p=0}^{\infty} e^{AI}_{(p)}k^{-p} \sp
t^{IJK} = \sum_{p=0}^{\infty} t^{IJK}_{(p)}k^{-p} \pe
\le{shl2}
\ls{shl}
Both these classes correspond to integer powers of
$k^{-1}$ for $L^{ab}$.

 The class of
solutions (\ref{shl1}) is said to be {\em vielbein-dominated} because
the leading equation
\be
 e^{AI}_{(0)}=ke^{AJ}_{(0)}e^{CK}_{(0)}e^{DI}_{(0)}\et_{CD}\et_{JK}
\ee
involves only the vielbein, while the equation for the leading term
$t^{IJK}_{(0)}$ of the 3-form depends on $e^{AI}_{(0)}$. This class contains
the N=1 affine-Sugawara and Kazama-Suzuki coset constructions
\cite{ks1,ks2} and the large
sets  of superconformal constructions classified by signed generalized graph
theory \cite{nsc,sin,ggt} (see Sections \ref{sma} and \ref{gsm}).

Similarly, the class of solutions (\ref{shl2}) is said to be
{\em 3-form-dominated}
because the leading equation,
\be
t_{(0)}^{IJK} =
(\frac{1}{2} t_{(0)}^{P[IJ}t_{(0)}^{K]MN}t_{(0)}^{RLQ}+ 2 t_{(0)}^{MPI}
t_{(0)}^{NQJ}t_{(0)}^{LRK})\et_{PQ}\et_{MR}
\et_{NL}
\ee
involves only the 3-form. This class contains the
GKO N=1 coset constructions \cite{gko2}, the nonlinear realizations
\cite{w,abkw} and the large class of superconformal
constructions on two-dimensional simplicial complexes \cite{ssc}
(see Section \ref{gts}).

\subsection{Groups and Graphs:
The Axioms of Generalized Graph Theory \label{agt}}
\lam{agt}
In this subsection, we review the
axiomatic formulation of generalized graph theory on Lie $g$
\cite{ggt}. Because this
development may be of future interest in mathematics, we present the subject
first without discussion of its
 origin and application in conformal field theory, which is postponed to
Section \ref{ggth}. The less mathematically-oriented reader may wish to begin
with Section \ref{sod}, where the original classification of conformal
level-families by conventional graphs is reviewed as a special case of the
more general development here.

\boldmath
\subsubsection{The magic bases of Lie $g$ \label{mbg}}
\unboldmath
\lam{mbg}

The development of generalized graph theory depends on a special class of
bases, called the magic bases of Lie $g$. In this section, $g$ may refer to
either the Lie group or its Lie algebra. Historically, this set of bases was
defined as a sufficient condition for the consistency of a certain class of
ans\"atze in the VME (see Section \ref{ggth}).

We begin with the  algebra of simple Lie $g$
\be
 [J_a,J_b]=i {f_{ab}}^c J_c \sp a,b,c=
1\ldots{\rm dim}\,g
\ee
and its  Killing metric $\et_{ab}$. The Killing metric is used to raise and
lower
the adjoint indices $a,b,c$, viz. $f_{abc}= {f_{ab}}^d \et_{dc}$.
\newline
\ni \Ab A {\em magic basis} $g_M$ of Lie $g$ satisfies
\bs
\be
 \et_{ab}=\et^{ab} \sp f_{abc}=f^{abc}
\ee
\be
 {f_{ab}}^c \neq 0 \;\;\mbox{for at most one}\;\, c
\ee
\label{mb1} \es
\lam{mb1}
so that the Killing metric is an involutive automorphism of $g$, and no two
generators commute to more than a single generator.
A magic basis of $g=\oplus_I g_I$ is obtained iff the basis of
each $g_I$ is magic. For simplicity, the discussion below is limited to the
magic bases of simple $g$.\nl
\ni \Bb The subclass of {\em real magic bases} satisfies
the additional restrictions,
\be
 \et_{ab}=\mbox{real}\sp {f_{ab}}^c={\rm real}
 \sp \T_a^{\dagger}=\sum_{b}\et_{ab}\T_b
\le{mb2}
where $\{\T_a\}$ is any matrix irrep of $g$ and dagger is matrix
adjoint.\nl
\ni \Cb  The set of all magic bases is not known, and the
 known examples include
\be
\label{kmb}
 \eqalign{
&\mbox{\bul the standard Cartesian basis of}\;\,SO(n) \;\;
\cite{gt,sme,nsc} \cr
&\mbox{\bul the Pauli-like basis of}\;\,SU(n) \;\;\cite{pat,mb,sin} \cr
&\mbox{\bul the tensor-product bases of}\;\,SU(\Pi{}_i n_i)
\;\; \cite{ggt} \pe \cr}
\ee
\lam{kmb}
The explicit forms of the first two bases are given in Section \ref{kgg}, and
each of the three bases  is in fact a real magic basis. \nl
\ni \Db
Let $h \subset g$ be a Lie subgroup of Lie $g$ and $g_M$
a magic basis of $g$ with adjoint indices $\{a\}$. An {\em $M$-subgroup}
$h^M(g_M)$ of $g_M$ is any subgroup $h$ whose adjoint indices $\{A \in h\}$
are a subset of $\{a\}$ and
\be
 \et_{AI}=0\sp I \in g/h\;\,.
\label{msu} \ee
\lam{msu}
The induced basis $h_M$ of an $M$-subgroup
$h^M(g_M)$  of $g_M$
\be
h_M:\;\;\;\et_{AB},\;{f_{AB}}^C \sp A,B,C \in h^M(g_M)
\ee
is a magic basis of Lie $h$.

\boldmath
\subsubsection{Generalized graph theory on Lie $g$ \label{ggg}}
\lam{ggg}
\unboldmath

Each magic basis $g_M$  of Lie $g$ supports a generalized
graph theory on Lie $g$ \cite{gt,mb,sin,ggt}
\be g_M \rightarrow \mbox{generalized graph theory of}\;\,g_M \;\, ,
\le{mbp}
where the notion of adjacency in the generalized graphs derives from the
structure constants of Lie $g$ in the magic basis.

The central definitions in this development are: \nl
\ni \Ab  The {\em edge-function} of a  magic basis $g_M$ of Lie $g$ is
\be
 \th_a \in \{0,1\}\;\,,\;\;\;\;\;a=1\ldots{\rm dim}\,g
\label{efn}\ee
\lam{efn}
and the $2^{{\rm dim}\,g}$ choices  $\{\th_a\}$ define the same number of
generalized graphs $\G$ on $g_M$, where
\be
 E(\G)=\{a\;|\;\th_a(\G)=1,\,a\in (1\ldots{\rm dim}\,g)\}
\le{eli}
is the {\em edge-list} of $\G$.
The edge-function in (\ref{efn})
generalizes the adjacency matrix
$\th_{ij}(\G),\;1\leq i <j \leq n$ of the graphs of order $n$ on $SO(n)$. \nl
\ni \Bb {\em Edge-adjacency}  in a
generalized graph is determined by the
structure constants ${f_{ab}}^c$ of the magic basis $g_M$:
\be
 a,b\in E(\G)\;\;  \mbox{are adjacent in}\;\;  \G \; \; \mbox{iff}
\;\; \sum_c ({f_{ac}}^b)^2 \neq 0
\label{ead} \ee
\lam{ead}
The magic basis identity $ \sum_c (f_{bc}{}^a)^2 = \sum_c (f_{ac}{}^b)^2$
guarantees that edge-adjacency is
symmetric.
Although
the edges  $a$ and $b$ in (\ref{ead}) are not necessarily distinct,
self-adjacent edges do not occur in
the generalized graphs of the known magic bases (\ref{kmb}). \nl
\ni \Cb
The {\em generalized edge-adjacency matrix} of a
generalized graph $\G$ of $g_M$ is
\be {\cal A}_{ab}(\G) = \left\{ \matrix{
2\psi^{-2}_g \sum_c ({f_{ac}}^b)^2 & , & a,b \in E(\G)
\;\,\mbox{adjacent in}\;\,\G  \cr
0 & , & a,b \in E(\G)
\;\,\mbox{not adjacent in}\;\,\G\;\,.  } \right.
\label{eam} \ee
\lam{eam}
which is a symmetric matrix in
the space of generalized graph edges of $\G$. \nl
\ni \Db
The {\em generalized edge degree} of edge $a$ in $\G$ is
\be
 {\cal D}_a(\G)= \sum_{b\in E(\G)} {\cal A}_{ab}(\G)=2\psi^{-2}_g
\sum_{b \in E(\G)} \sum_c ({f_{ac}}^b)^2 \;\,,\;\;\;\;
a \in E(\G)
\label{edg} \ee
\lam{edg}
and edge-regular generalized graphs have uniform generalized
edge degree ${\cal D}(\G)$, $\forall a\in E(\G)$.

Some important categories of generalized graphs are: \nl
\ni \bon A {\em symmetry-constrained} generalized graph
satisfies
\be \forall\;\,a,b:\;\;\;\;\;\;
\th_a(\G)=\th_b(\G)\;\;{\rm when}\;\, \et_{ab}\neq 0\;\,.
\label{scg} \ee
\lam{scg}
\ni \btw A {\em generalized graph triplet} is a set of three
generalized graph edges which satisfy
\be \{a,b,c\in E(\G)\;|\;f_{abc}\neq 0\}\;\,,
\label{tri} \ee
\lam{tri}
and in the  known magic bases (\ref{kmb}),
 the generalized edges
of a generalized graph triplet are mutually adjacent.
A {\em triplet-free} generalized graph
\be
\forall\;\,a,b,c:\;\;\;\;\;\;
 \th_a(\G) \th_b(\G) \th_c(\G) =0 \;\;{\rm when}\;\,f_{abc} \neq 0
\label{tfg} \ee
\lam{tfg}
has no generalized graph triplets. \nl
\ni \bth
The {\em complete graph} $\K_g$ on $g$ satisfies
\be  \K_g : \;\;\;\;  \th_a(\K_g)=1 \sp  a=1\ldots{\rm dim}\,g \pe
\label{cgr} \ee
\lam{cgr}
\ni \bfo  The {\em complement} $\tilde{\G}$ of a graph $\G$ is given by
\be \tilde{\G}=\K_g-\G : \;\;\;\;
\th_a(\tilde{\G})=1-\th_a(\G)
\le{ctg}
and the set of symmetry-constrained generalized graphs is closed under
complementarity. \nl
\ni \bfi  The {\em coset graphs} of $g_M$ satisfy
\be \G_{g/h}=\K_g-\K_h
\le{cog}
where $\K_h$ is the complete graph of any $M$-subgroup $h^M(g_M)$ of $g_M$. \nl
\ni \bsi
Let $g_M$ be a magic basis of Lie $g$ and
\be
 g \supset h_N^M(g_M) \supset \ldots \supset h_1^M(g_M)
\le{nsq}
be a nested sequence of $M$-subgroups of $g_M$.
Then,
the {\em affine-Sugawara nested graphs} of the sequence (\ref{nsq}),
\be \{\K_{h_1},\;\K_{h_2}-\K_{h_1},\;\K_{h_3}-(\K_{h_2}-\K_{h_1}),\ldots,
\; \K_g-(\K_{h_N}-(\ldots)) \}
\le{asg}
are obtained by repeated complementarity on the nested $M$-subgroups of the
sequence. The set of all affine-Sugawara nested graphs
of $g_M$ are those obtained in this
way on all possible nested $M$-subgroup sequences of $g_M$. \nl
\ni \bse
The {\em affine-Virasoro nested graphs} of the sequence (\ref{nsq}),
\be \{\G_{h_1},\;\K_{h_2}-\G_{h_1},\;\K_{h_3}-(\K_{h_2}-\G_{h_1}),\ldots,
\; \K_g-(\K_{h_N}-(\ldots)) \}\;\,,\;\;\forall \;\,\G_{h_1}
\le{avg}
are obtained by repeated complementarity on the nested $M$-subgroups of the
sequence, where $\G_{h_1}$ is any generalized graph in the generalized sub
graph theory of $(h_1)_M$. The set of all
 affine-Virasoro nested graphs of $g_M$ are
those obtained in this way on all possible nested $M$-subgroup
sequences of $g_M$. \nl
\ni \bei
The {\em irreducible graphs} of $g_M$ are those generalized
graphs of $g_M$ which are not obtainable by affine-Virasoro nesting from the
generalized graphs of any submagic basis $h_M \neq g_M$.

\subsubsection{Generalized graph isomorphisms \label{ggi} }
\lam{ggi}

In generalized graph theory on $g_M$, the generalized graph isomorphisms
\cite{mb,ggt} live in the isomorphism group,
\be
\I \subset \au
\ee
which is a permutation subgroup
of the automorphism group of Lie $g$. The permutations $a\rightarrow \p(a)$
act on the adjoint index $a=1\ldots{\rm dim}\,g$ of Lie $g$. Although the
permutations act on the edges of the generalized graphs, we will see in
Section \ref{kgg} that $\I$ includes the conventional graph isomorphisms
(permutation of labels on
points) when the special case of  the conventional graphs is considered.

The basic structure of $\I$ includes the following. \nl
\ni \Ab For each magic basis $g_M$ of $g$, an adjoint permutation $\p$
is an element of $\I$ when there exists a set of
non-zero numbers $\{\g_{\p}(a) \}$ such that
\bs
\be
\g_{\p}(a)\g_{\p}(b)\et_{\p(a)\p(b)}=\et_{ab}
\ee
\be
 \g_{\p}(a)\g_{\p}(b) {f_{\p(a)\p(b)}}^{\p(c)}=
{f_{ab}}^{c} \g_{\p}(c)
\ee
\be
 \g_{\p}^2(a)\g_{\p}^2(b) = \g_{\p}^2(c) \;\;\;\;\;\;
{\rm when}\;\, {f_{ab}}^c \neq 0
\ee
\be
\g_{\p}(a)=e^{i\p \n_{\p}(a)} \;\;\mbox{when}\;\,g_M\;\,
\mbox{is a real magic basis} \;\,.
\ee
\ls{iso}
$\I$ is a finite subgroup of $\au$, and its elements are (real) magic-basis
preserving automorphisms $J_a' = \g_\p (a) J_{\p(a)}$ of $g$.\nl
\ni \Bb
A central feature of the isomorphism group is  that the
squared structure constants of the magic basis are
preserved
\be
({f_{\p(a)\p(b)}}^{\p(c)})^2 =({f_{ab}}^c)^2\;\,,\;\;\;\;\;\p\in\I
\ee
by the permutations in $\I$. \newline
\ni \Cb
The isomorphism group
acts as edge permutations on the adjoint index $a=1\ldots{\rm dim}\,g$ of
the edge-function $\th_a$ of $\G$,
\be \th_a'(\G) \equiv \th_{\p(a)}(\G)\;\,,\;\;\;\;\;\p \in \I \pe
\ee
Generalized graphs on Lie $g$ are {\em isomorphic} or equivalent
when their edges are related by an edge permutation in $\I$,
\be
  \G' \mathrel{\mathop\sim_{\p}} \G \;\;\;{\rm when}\;\,
\th_a(\G')= \th_{\pi(a)}(\G) \;\,,\;\;\;\;\;\p \in \I \pe
\ee
The isomorphism class of $\G$ is the set of all generalized graphs isomorphic
to $\G$.

Some simple properties of the isomorphism groups are as follows: \newline
\ni \bon An isomorphism $\p$ is in the {\em symmetry group} auto$\,\G$
of a generalized graph $\G$ when
\be
 \th_{\p(a)}(\G)=\th_a(\G)\;\,,\;\;\;\;\;\p \in {\rm auto}\,\G \subset \I \;\,.
\ee
\ni \btw Both the symmetry-constrained graphs and the triplet-free graphs
are closed under $\I$. \nl
\ni \bth It follows from (\ref{iso}) that
edge-adjacency in generalized graphs is preserved under
generalized graph isomorphisms,
\be
 \eqalign{
a,b\in E(\G)\;\,&\mbox{adjacent in}\;\,\G \rightarrow \cr
&\p(a),\p(b)\in E(\G')\;\,
\mbox{adjacent in}\;\,\G' \mathrel{\mathop\sim_{\p}}\G
  \cr}
\ee
and
generalized edge-adjacency matrices of
isomorphic graphs  are related by edge relabelling,
\be
 {\cal A}_{\p(a)\p(b)}(\G')={\cal A}_{ab}(\G)\;\,\;
{\rm when}\;\;\G'\mathrel{\mathop\sim_{\p}} \G \;\,.
\ee
\ni \bfo
An {\em invariant graph function} $I(\G)$ satisfies
\be I(\G')=I(\G)\;\;{\rm when}\;\,\G' \sim \G\;\,.
\le{igf}
For example,
any graph function
$I_{\S}[\{\th_a(\G)\},\{({f_{ab}}^c)^2\}]$ which is summed over all adjoint
indices $a,b,c$ is an invariant graph function. \newline
\ni \bfi
Complements of generalized
isomorphic graphs are isomorphic
\be
 \widetilde{(\G')} \mathrel{\mathop\sim_{\p}} \tilde{\G} \;\;\;{\rm when}\;\;\;
\G' \mathrel{\mathop\sim_{\p}} \G \;\,.
\ee
\ni \bsi
A {\em self-complementary generalized graph}
 satisfies
\be
 \tilde{\G} \sim \G\;\,.
\ee
The self-complementary generalized graphs live with
${\rm dim}\,E(\G)=\frac{1}{2}{\rm dim}\,g$ on Lie group
manifolds of even dimension.

\subsubsection{Signed generalized graph theory \label{sgt}}
\lam{sgt}

For each magic basis $g_M$ of Lie $g$, one may also define a signed
generalized graph theory
on $ g_M \times SO({\rm dim}\,g)$,
\be
g_M \ti SO({\rm dim}\,g)   \ra \mbox{signed generalized graph theory of}
\;\, g_M \ti SO({\rm dim}\,g)
\le{spr}
in close analogy to eq.(\ref{mbp}).
Historically, this development \cite{sme,nsc,sin,ggt} arose in the study
of metric ans\"atze in the SME (see Section \ref{sma}).

The signed generalized graphs of $g_M \times SO({\rm dim}\,g)$ are the
generalized graphs of $g_M$ with an extra + or $-$ sign on each generalized
graph edge.
Edge-adjacency in signed generalized graphs is the same
as in unsigned generalized
graphs and symmetry-constrained and triplet-free graphs are also
defined as above.

The isomorphism group $\Is$ of
the signed generalized graphs is somewhat more involved.
In particular, this group contains
 the
edge-permutation subgroup  $\I$ and a
 sign-flip subgroup.
Using the sign-flip subgroup it has been  shown that the signs
of the symmetry-constrained signed generalized graphs
are isomorphic, so that a
 symmetry-constrained unsigned generalized graph of
$g_M$ can be taken as the representative of each signed isomorphism
class \cite{nsc,sin,ggt}.

\subsubsection{Known generalized graph theories \label{kgg}}
\lam{kgg}

The generalized graph theories that correspond to the list (\ref{kmb}) of
known magic bases are
\be
\label{upr}
\eqalign{
\mbox{Cartesian basis of}\;\,SO(n)  &\rightarrow \mbox{graphs of}\;\,SO(n) \;\,
\cite{gt,nsc,ggt}
 \cr
\mbox{Pauli-like basis of}\;\,SU(n)  & \rightarrow
\mbox{sine-area graphs of}\;\,SU(n)\;\,\cite{mb,ggt}  \cr
\mbox{product bases of}\;\,SU(\Pi_i n_i) &  \rightarrow
\mbox{sine}(\oplus\mbox{area) graphs of}\;\,
SU(\Pi_i n_i)\;\;\;\;  \cite{ggt} \cr}
\ee
\lam{upr}
so that conventional graph theory \cite{har}
 is a special case, on the orthogonal groups,
of generalized graph theory on Lie $g$. Similarly, the prescription
(\ref{spr}) generates the signed analogues of the generalized graph
theories in (\ref{upr}).
In what follows, we discuss the three examples (\ref{upr})
 in further detail \cite{ggt}.

\vskip .4cm
\ni \u{The graphs of $SO(n)$}
\vskip .3cm

As a first example of generalized graph theory, we work out
 the definitions
of Sections \ref{mbg}--3  for the Cartesian basis
of  $SO(n)$, called the graphs of $SO(n)$.
 As we will see from these definitions, the graphs of $SO(n)$ are the usual
graphs of conventional graph theory.

The standard Cartesian basis of $SO(n)$ is a real magic basis \cite{gt,mb,ggt},
in which the adjoint indices $a=ij$, $1 \leq i <j\leq n$ are ordered pairs of
vector indices of $SO(n)$. The non-zero structure constants of the basis are
\be
 {f_{ij,il}}^{jl}=-\sqrt{{\tau_n\psi_n^2 \over 2}}\;\,,\;\;\;\;\;i<j<l
\sp \tau_n = \left\{ \matrix{ 1 & n \neq 3 \cr 2 & n=3 \cr} \right.
\le{scs}
where $\psi_n$ is the highest root of $SO(n)$.

The generalized graphs of this basis are the $2^{({n \atop 2})}$
conventional graphs of order $n$, with edge-lists
\be
 E(\G_n)=\{(ij)\;|\; \th_{ij}(\G_n)=1,\;1\leq i<j \leq n\}
\ee
so that $\th_{ji}(\G_n)\equiv \th_{ij}(\G_n)$ is the adjacency matrix of a
graph
$\G_n$ of order $n$.
In the present viewpoint, however, edge-adjacency has not yet been specified
in the graphs.

Using the non-zero structure constants (\ref{scs}) of Cartesian $SO(n)$,
the definition of edge-adjacency in (\ref{ead}) becomes the statement,
\be
 {\rm edges}\;\,(ij)\;{\rm and}\;\,(kl)\;\,\mbox{are adjacent when
they share a graph point}
\pe
\ee
This is the usual definition of adjacency in conventional
graph theory, and the usual
edge-adjacency matrix \cite{har},
\be
\label{gea}
 {{\cal A}}_{ij,kl}(\G_n) = A^{(1,1)}_{ij,kl} (\G_n) =
 \left\{ \matrix{
1 & , & (ij),(kl)\in E(\G_n) \;\,\mbox{adjacent in}\;\,\G_n \cr
0 & , & (ij),(kl)\in E(\G_n) \;\,\mbox{not adjacent in}\;\,\G_n \cr } \right.
\ee
\lam{gea}
follows from eq.(\ref{eam}). The generalized edge-degree (\ref{edg})
reduces in this case to the usual edge-degree of graph theory, which counts
the number of edges adjacent to an edge.

Furthermore, the metric is diagonal so all conventional graphs solve the
symmetry-constraint (\ref{scg}), and (\ref{tri}) says that generalized graph
triplets are graph triangles in this case.
The $M$-subgroups of the Cartesian basis of $SO(n)$,
\be
h(\Sd) = \ti_{i=1}^N SO(m_i) \sp \sum_{i=1}^N m_i = n
\le{msg}
were identified in Ref.\ \cite{gt}.

The defining relations
of the isomorphism group ${\cal I}_M({\rm Aut}\,SO(n))$ of the
graphs of order $n$ can be obtained by substituting
the metric and  structure
constants  of Cartesian $SO(n)$ into the general relations (\ref{iso}).
These relations are
unfamiliar in
conventional graph theory, but they have been solved in Ref.\ \cite{ggt} to
find
that
\be
 {\cal I}_M ({\rm Aut}\,SO(n))  =\mbox{permutation group of}\;\,n\;\,
\mbox{graph points}
\pe
\ee
This is the usual isomorphism group of conventional graph theory,
so that each isomorphism class is represented by an unlabeled graph.
\fig{2cm}{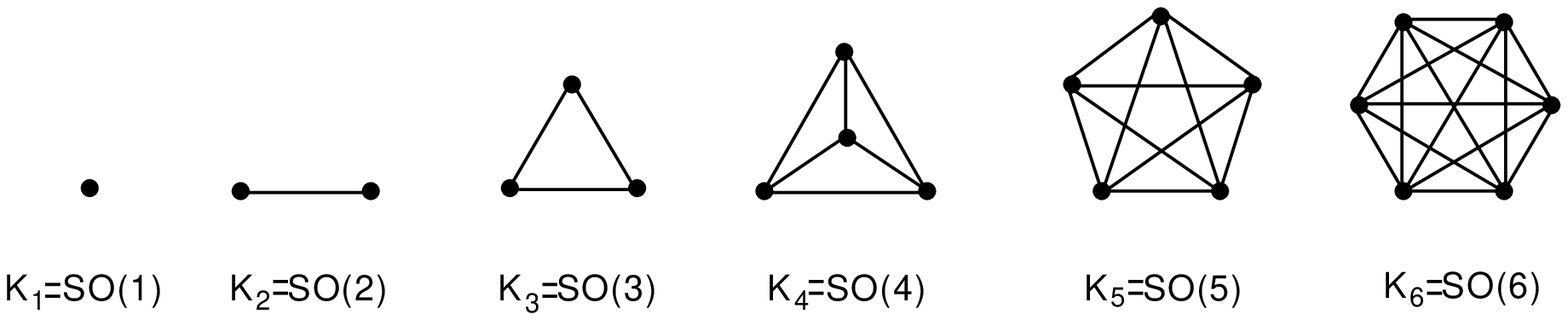}{Complete graphs $\K_n$ = affine-Sugawara construction on
$SO(n)$.}{f3}

Following the definitions of Section \ref{ggg}, one finds further agreement
with the usual notions of conventional
graph theory:  \nl
{\bf a)} The generalized complete graphs in (\ref{cgr}) reduce to the
conventional comple graphs
 $\K_n$  with all possible
edges among $n$ points. The first six complete graphs are shown in Fig.\ref{f3}
 \nl
{\bf b)} The complement of a generalized graph reduces to the conventional
complement
$\tilde{\G}_n$ of a graph $\G_n$, obtained by removing the
lines of $\G_n$ from
the complete graph $\K_n$ (see Fig.\ref{f4}).\nl
\fig{2.5cm}{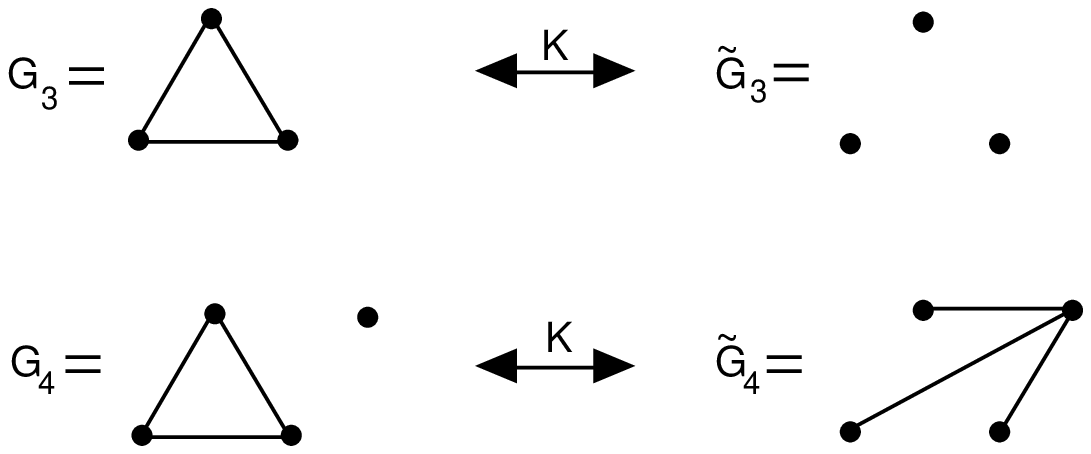}{Complementary graphs on $SO(3)$ and $SO(4)$.}{f4}
{\bf c)} The self-complementary generalized graphs live on
$SO(4n)$ and $SO(4n+1)$, where they  are identified as the usual
self-complementary graphs \cite{har} of conventional
graph theory (see Fig.\ref{f5}).  \nl
\fig{1.5cm}{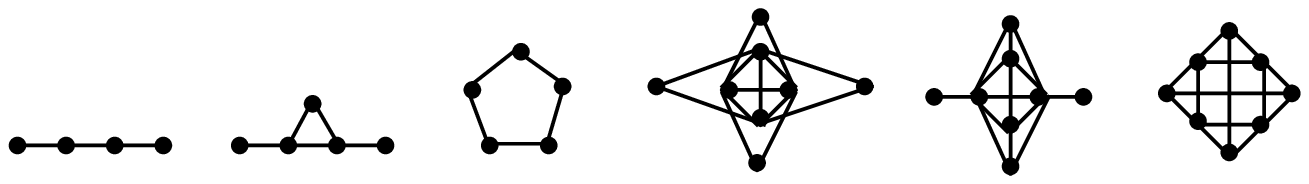}{The first six self-complementary graphs.}{f5}
Further discussion of the Lie group- and conformal field-theoretic
structure of conventional
 graph theory \cite{gt,nsc} is found in Section \ref{gth}.

\vskip .4cm
\ni \u{The sine-area graphs of $SU(n)$}
\vskip .3cm

The Pauli-like basis \cite{pat,ffz1,ffz2,b1,mb} of $SU(n)$
 is a real magic basis, in which the adjoint indices
 $a= \bp= (p_1,p_2)\in F_n'$ are vectors on a
two-dimensional lattice with period
$n$ and origin (0,0). The structure constants of the basis are
trigonometric,
\bs
\be
\et_{\bp,\bq}=\et^{\bp,\bq}=
\s(\bp)\d_{(\bp+\bq)({\rm mod}\,n),\vec{0} }
\ee
\be
{f_{\bp,\bq}}^{\br}=-\sqrt{{2 \psi_n^2 \over n}}\s(\bp,\bq)
\sin(\frac{\pi}{n}(\bp \times \bq))\d_{(\bp+\bq)({\rm mod}\,n),\br}
\ee
\es
where $\psi_n$ is the highest root of $SU(n)$ and $\s(\bp)$, $\s(\bp,\bq)$
are equal to $\pm 1$.

The generalized graphs of this basis are the $2^{n^2-1}$ sine-area graphs
\cite{mb,sin,ggt} of $SU(n)$, with edge-lists
\be  E(\G)=\{\bp\;|\;\th_{\bp}(\G)=1,\;\bp\in F'_n\}
\pe
\ee
 Each  sine-area graph $\G$ displays the vectors $\bp$,
 from the origin, in
the edge-list of $\G$, and, as an example, Fig.\ref{f6} shows the sine-area
graphs of $SU(2)$.
Using the natural periodicity of the Pauli-like bases
\cite{ffz1,ffz2,b1}, it may also be possible to draw the sine-area graphs on a
torus.
\fig{6cm}{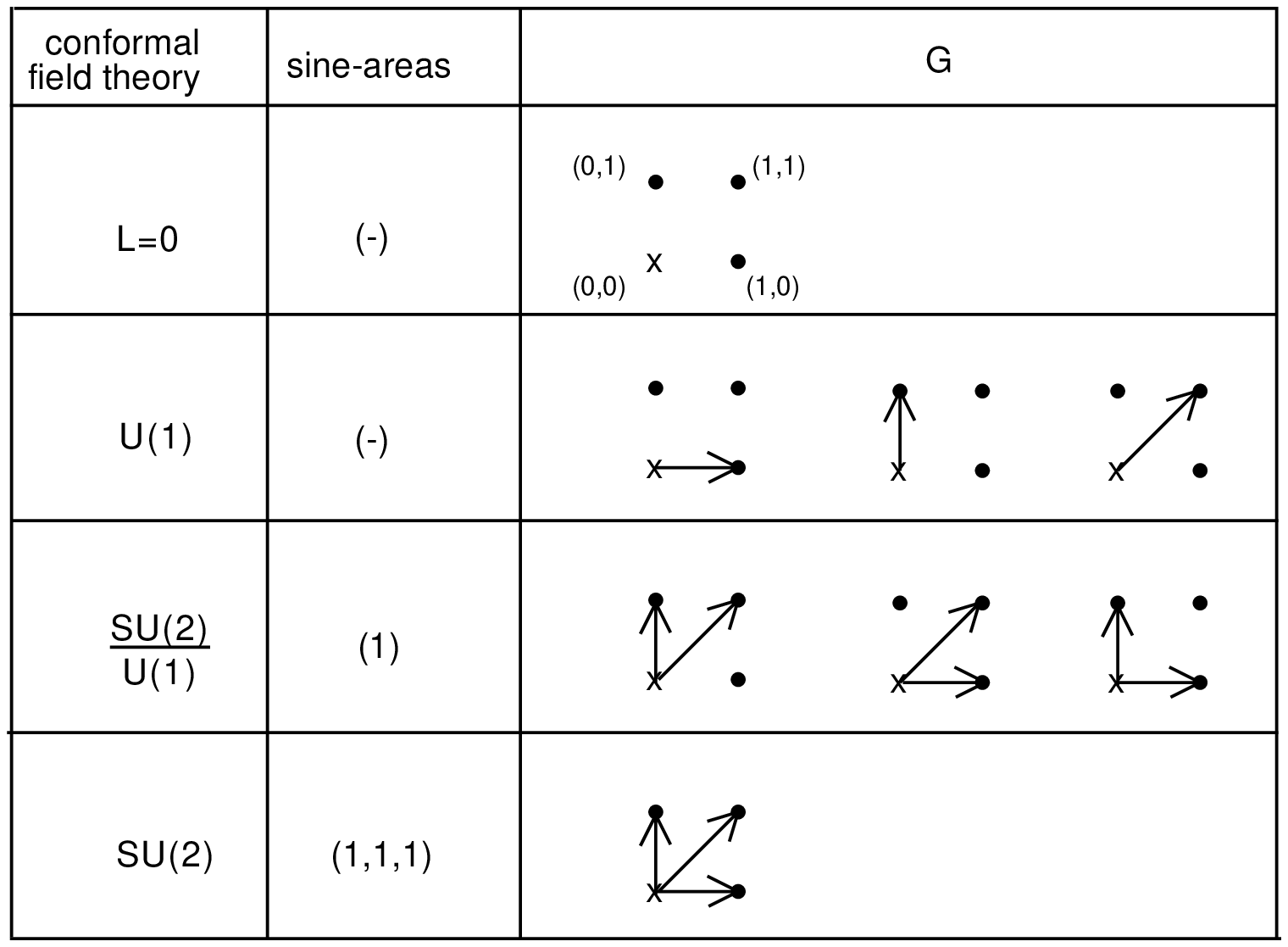}{The sine-area graphs of $SU(2)$.}{f6}

Edge-adjacency in sine-area graphs,
\be
 \bp,\bq\in E(\G)\;\,\mbox{are adjacent iff}\;\,(\mbox{sine-area})_{\bp,\bq}
\neq 0
\ee
follows from (\ref{ead}), where
\be (\mbox{sine-area})_{\bp,\bq} \equiv |\sin(\frac{\pi}{n}(\bp \times \bq))|
\;\,,\;\;\;\;\;\bp,\bq \in E(\G)
\ee
is the sine-area of each edge-pair $(\bp,\bq)$ in $\G$. In the sine-area
graphs of $SU(2)$, all edges are adjacent, but this is not true for larger
unitary groups.
The entries in the generalized edge-adjacency matrix and the generalized
edge degree of  a sine-area graph
are generically irrational, which is the source of generically irrational
central charge in the conformal and superconformal constructions on sine-area
graphs \cite{mb,sin}.

Except for $SU(2)$ and $SU(3)$, the $M$-subgroups of the Pauli-like bases
have not been determined, though
the irrational structure constants suggest that
embedding of Pauli-like bases in each other is sporadic.
The defining relations of the  isomorphism group
${\cal I}_M({\rm Aut}\,SU(n))$ of the
sine-area graphs have not been completely solved beyond $SU(2)$.
Fig.\ref{f6} shows the four isomorphism classes of the sine-area graphs of
$SU(2)$ \cite{mb,ggt}.
For $SU(3)$,
a preliminary coarse-grained set of isomorphism classes was given in
Ref.\ \cite{ggt}.
More generally, the isomorphisms of ${\cal I}_M({\rm  Aut}\,SU(n))$
 preserve all invariant graph functions,
such as the set of sine-areas of the graphs.

\vskip .4cm
\ni \u{The sine$(\op$area) graphs of $SU(\Pi_i n_i)$}
\vskip .3cm

The product bases of
$SU(\Pi_{i=1}^s n_i)$ \cite{b2,ggt} are an infinite set of real
magic bases which are tensor products of the Pauli-like bases on
$U(n_i)$.
 A sine$(\oplus$area) graph $\G$ of $SU(\Pi_{i=1}^s n_i)$  is
 a sequence of $s$ ``sine-area graphs''
$G_i,\;i=1\ldots s$ drawn
according to a set of rules specified in Ref.\ \cite{ggt}.
Using the natural periodicities of the  product bases,
it may also be possible to draw the sine($\op$area) graphs of
$SU(\Pi_{i=1}^s n_i)$ on Riemann surfaces
of genus $s$ (see Ref.\ \cite{b2}).

\subsection{Generalized Graph Theory and the Classification of Conformal
Field Theory \label{ggth}}
\lam{ggth}

We return now to conformal field theory and the application of the magic bases
and generalized graph theories to the master equations of ICFT.
Historically, this development occured in the opposite order from that
presented here: Generalized graph theory on Lie $g$ was first observed
\cite{mb} in the high-level expansion of the metric ansatz
$g_{metric}$ of the Virasoro master equation on affine $g$.
A short review of graph theory in  the classification problem is found in
Ref.\ \cite{rd}.

\newpage
\subsubsection{Magic bases and metric ans\"atze in the VME \label{mbm}}
\lam{mbm}

For every magic basis $g_M$ (see (\ref{mb1})) of simple Lie $g$, the
Virasoro master equation admits a consistent ansatz, called the
metric ansatz
 $g_{metric}$,
\bs
\be
 L^{ab} =\psi_g^{-2} L_a  \et_{ab}
\sp T(L) = \psi_g^{-2} \sum_{a,b} L_a \et_{ab} \xx J_a J_b \xx
\;\;\;\; ,\;\;a=1\ldots{\rm dim}\,g
\label{mana}
\ee
\be
 L_a(1-xL_a)=\sum_{cd}L_c(2L_a-L_d)({f_{cd}}^a)^2 /\psi_g^2
\le{mme}
\be
 (L_a-L_b)\et_{ab}=0
\le{sco}
\be
 c=x\sum_a L_a
\ee
\ls{man}
where $\psi_g$ is the highest root of $g$ and $x=2k/\psi_g^2$ is the
invariant level of affine $g$. The condition (\ref{sco}) enforces the
symmetry $L^{ba}=L^{ab}$ in the ansatz.
The consistency of $g_{metric}$ in the VME
is intimately related to the structure
 constants of the magic basis $g_M$, and, in fact, the definition
(\ref{mb1}) of a magic basis was originally obtained \cite{mb} as the
sufficient condition for the consistency of this ansatz in the VME.
When the magic basis is real (see eq.(\ref{mb2})),
one has $J_a(m){}^{\dagger}
= \sum_b \et_{ab} J_b(-m)$ and  unitary solutions on $x \in \N$ of simple
compact $g$ require $L_a =$real.

The generalized graph theory of $g_M$ is seen in the high-level expansion
of $g_{metric}$,
\bs
\be
 L_a(\G,x) = {\th_a(\G) \over x}+\sum_{p=2}^{\infty}
L_a^{(p)}[\th_a(\G)]x^{-p}
\le{hlf}
\be
 \th_a(\G) \in  \{0,1\}\;\,,\;\;\;\;\;\;a=1\dots{\rm dim}\,g
\ee
\be
 (\th_a(\G)-\th_b(\G))\et_{ab}=0
\le{con}
\ls{hlm}
whose leading term involves the edge-function
$\th_a(\G)$ of the generalized graphs $\G$ (see Section \ref{agt}).
The high-level form (\ref{hlf}) is
 easily obtained from the left side of
(\ref{mme}), which determines that $L_a \simeq 0$ or $x^{-1}$ for each $a$
at this order.
 Moreover, the constraint
(\ref{con}), which follows from (\ref{sco}),
 selects only the symmetry-constrained graphs of $g_M$ (see Section \ref{ggg}).
The higher-order terms in $L_a(\G)$ are  uniquely
determined by $\th_a(\G)$, so each symmetry-constrained generalized graph
$\G$ labels an entire level-family $L_a(\G,x)$ of conformal field theories.

Comparing (\ref{mana}) and (\ref{hlm}) with the high-level form
$L^{ab}=P^{ab}/2k$ in (\ref{hl1}), one obtains the explicit form of the
high-level projector $P$ of the $L$ theory in $g_{metric}$,
\begin{equation}
P_a{}^b=\d_a{}^b \theta_b(\G)
\end{equation}
so the high-level projectors are essentially the edge-functions of the
generalized graphs (that is, the adjacency matrix in conventional graph
theory).

The central charge of the
level-family on $\G$ is
\bs
\be
c(\G,x)= \sum_{p=0}^{\infty} I_{\Sigma}^{(p)}(\G)x^{-p}
={\rm dim}\,E(\G)-\frac{1}{x}(\sum_{a\in E(\G)}{\cal D}_a(\G)+\frac{1}{2}
I_{\S}^B(\G)) +
{\cal O}(x^{-2})
\le{chl}
\be
I_{\S}^B(\G) =  \frac{2}{\psi_g^2}\sum_{abc}\t_b(\G)
\t_c(\G)(1- 2 \t_a(\G) )
({f_{bc}}^a)^2
\ee
\ls{hlc}
where ${\rm dim}\,E(\G)$ is the number of edges in $\G$,
 ${\cal D}_a(\G)$ is the generalized edge degree of $\G$ (see
(\ref{edg})) and
$I_{\S}^{B}(\G)$ is an invariant graph function. In agreement with
eq.(\ref{acc}), the leading-order contribution
\be
c_0 =\sum_a \t_a(\G) =  {\rm dim}\, E(\G)
\ee
is an integer between 0 and ${\rm dim}\,g$. This integer may
 be viewed as the
simplest invariant graph function, and
all the terms  in the high-level expansion of the central charge are
invariant graph functions
because the central charge is invariant under
$\au  \supset \I$.
Similarly, the set $\{\D(\T;\G,x)\}$ of broken conformal weights
of each affine representation $\T$ of $g$
provides a large family of invariant graph functions.

Even at this finite order of the high-level expansion,
the result (\ref{hlc})
shows generic irrationality of the central charge when the
squared structure constants of the magic basis are irrational. This applies
in particular to the conformal constructions on sine-area graphs \cite{mb}
and sine($\op$area) graphs \cite{ggt}. When $g_M$ is a real magic basis
on compact $g$, the high-level expansion  strongly suggests \cite{gt,mb}
that the level-family of each generalized graph is generically unitary
on $x\in  \N$, and this has been verified explicitly for all the known
examples in $\{g_{metric}\}$.

Other generalized graph-theoretic properties of $g_{metric}$ are as follows
\cite{gt,mb}. \newline
\bon The affine-Sugawara construction \cite{bh,h1,kz,s2}
\be
L_a(\K_g)={1\over  x+\tilde{h}_g }\;\,,\;\;\;\;\;
c(\K_g)={x{\rm dim}\,g \over x+\tilde{h}_g }
\ee
lives on the complete graph $\K_g$ with $\t_a=1$, $a=1 \ldots  {\rm dim}\,g$.
\nl
\btw The K-conjugate partner \cite{bh,h1,gko,hk} $\tilde{L}_a(\G)$ of the
construction $L_a(\G)$ on $\G$
\be
 \tilde{L}_a(\G)=L_a(\tilde{\G})=\frac{1}{x+\tilde{h}_g}-L_a(\G)\;\,,\;\;\;\;\;
\tilde{c}(\G)=c(\tilde\G)=c(\K_g)-c(\G)
\ee
lives on the complementary graph $\tilde{\G}$ with
$\t_a(\tilde{\G})=1-\t_a(\G)$.
 \newline
\bth Automorphisms and isomorphisms. We have discussed above that an ansatz
on $g$ may involve a residual automorphism group $ \au({\rm
ansatz})\subset  \au $,
which defines physically-equivalent CFTs. Moreover, each generalized graph
theory involves an isomorphism group $\I$. In fact, these two
groups are identical \cite{gt,mb,ggt}
\be
{\rm Aut}\,g_{metric} =\I \subset \au
\ee
in the ansatz $g_{metric}$. This means that  level-families which live on
isomorphic generalized graphs,
\be
 L_a(\G')=L_{\p(a)}(\G) \;\;{\rm when}\;\;\G' \mathrel{\mathop\sim_{\p}} \G
\ee
are automorphically equivalent as CFTs in $g_{metric}$.\nl
\bfo The level-family $L_a(\G)$ carries the symmetry of its generalized graph,
\be
 L_{\p(a)}(\G) =L_a(\G)\;\;{\rm when}\;\,\t_{\p(a)}(\G)=\t_a(\G)
\ee
which gives rise to  a very large number of consistent graph-symmetry
subans\"atze \cite{gt,lie}. \nl
\bfi The self K-conjugate constructions \cite{gt,mb} of $g_{metric}$
\be
 L_a(\tilde{\G})=\tilde{L}_a(\G)=L_{\p(a)}(\G)\;\;{\rm when}\;\;
\tilde{\G} \mathrel{\mathop\sim_{\p}} \G
\ee
live on the self-complementary symmetry-constrained
graphs\footnote{In particular \cite{gt}, self K-conjugate constructions
live on all the
self-complementary graphs of conventional
graph theory because the symmetry constraint
(\ref{sco})  is
trivial in this case (see Section \ref{sod}).}. The
central charge of a self K-conjugate construction is $c(\K_g)/2$ because
$L_a(\G)$ and $L_a(\tilde{\G})$ are automorphically equivalent (see Section
\ref{sec44}). \newline
\bsi Any Lie subgroup construction
$(L_h)_a,\;h\subset g$ in $g_{metric}$ satisfies
\be
 \t_A=1\;\,,\;\;\;\t_I=0 \sp
A\in h\;,\;I \in g/h
\ee
so that the symmetry constraint (\ref{con})
implies $\et_{AI}=0$, $\forall A,I$. It follows that $h$ is an $M$-subgroup
$h^M(g_M)$ of $g_M$ (see eq.(\ref{msu})) and the Lie subgroup
constructions of $g_{metric}$ live on the complete graphs $\K_h$ of the
$M$-subgroups. \newline
\bse The $g/h$ coset constructions \cite{bh,h1,gko} of $g_{metric}$
\be
L(\G_{g/h})=L(\K_g)-L(\K_h)
\ee
live on the coset graphs $\G_{g/h}=\K_g-\K_h$ (see eq.(\ref{cog})). \newline
\bei The affine-Sugawara nests \cite{wit2,nuc} of $g_{metric}$ contain all the
standard rational conformal field theories of $g_{metric}$ (including the
affine-Sugawara construction on $g$ and the $g/h$ coset constructions).
These constructions are generated by repeated K-conjugation
on nested  $M$-subgroups
$g \supset h^M_N(g_M) \ldots \supset h_1^M(g_M) $ of $g_M$, and they
live on the
affine-Sugawara nested graphs of $g_M$ (see eq.(\ref{asg})), e.g.
\be
 L(\K_{h_1})\;\,,\;\;\;\;\;L(\K_{h_2}-\K_{h_1})=L(\K_{h_2})-L(\K_{h_1})\;\,,
\ee
$$ L(\K_{h_3}-(\K_{h_2}-\K_{h_1}))=L(\K_{h_3})-L(\K_{h_2})+L(\K_{h_1})\;\,,
\ldots \;\,. $$
The affine-Sugawara nested graphs are not generic on $g_M$, so
 the generic construction in $g_{metric}$ is a new conformal field theory
on affine $g$. \newline
\bni Similarly, the affine-Virasoro nests \cite{nuc} of $g_{metric}$
live on the symmetry-con\-strained
affine-Virasoro nested graphs of $g_M$ (see eq.(\ref{avg})), and the
irreducible constructions \cite{nuc} of $g_{metric}$ live on the
symmetry-constrained irreducible graphs of $g_M$ (see Section \ref{sec44}).

\vskip .4cm
\ni \u{Overview: Generalized graph theory in the VME}
\vskip .3cm

The structure described above is a prescription \cite{mb},
\be
\label{pr1}
 g_M\;\,{\nearrow \atop \searrow}
\matrix{ \mbox{conformal level-families of}\;\,g_{metric} \cr
                \uparrow \downarrow \cr
          \mbox{generalized graph theory of}\;\,g_M\;\;\;\;}
 \ee
\lam{pr1}
in which each magic basis $g_M$ of
Lie $g$ provides both a generalized graph theory on
$g_M$ and the generically new conformal field theories of $g_{metric}$, whose
level-families are classified by the symmetry-constrained generalized graphs.
The set of CFTs in each $g_{metric}$ is called a {\em graph theory unit} of
conformal level-families.

The known examples of this prescription are,
\bs
\be
\label{csg}
\qquad \mbox{Cartesian basis of}\;\,SO(n)\;\,{\nearrow \atop \searrow}
\matrix{ \Sd \cr
                \uparrow \downarrow \cr
          \mbox{graphs} } \qquad
 \ee
\lam{csg}
\be
\label{psg}
 \mbox{Pauli-like basis of}\;\,SU(n)\;\,{\nearrow \atop \searrow}
\matrix{ SU(n)_{metric} \cr
                \uparrow \downarrow \cr
          \mbox{sine-area graphs} }
\ee
\lam{psg}
\be
 \mbox{product bases of}\;\,SU(\Pi_i n_i)\;\,{\nearrow \atop \searrow}
\matrix{ SU(\Pi_in_i)_{metric} \cr
                \uparrow \downarrow \cr
          \mbox{sine}(\op \mbox{area) graphs} }
\ee
\es
which were discussed in \cite{gt,lie}, \cite{mb} and \cite{ggt} respectively.
As a detailed example of this prescription,  we will discuss the first case
(the classification by  conventional graphs)
in Section \ref{sod}. In fact, this case provided the first example of
 Lie group-theoretic structure in graph theory and the closely-related
  graph-theoretic
structure of ICFT, leading eventually to the more general structure
in (\ref{pr1}).

\subsubsection{Superconformal metric ans\"atze \label{sma}}
\lam{sma}

In parallel with (\ref{pr1}), each magic basis $g_M$ of $g$ also gives
a superconformal
prescription \cite{sin},
\be
\label{pr2}
 g_M\times SO({\rm dim}\,g) \;\,{\nearrow \atop \searrow}
\matrix{ \mbox{superconformal level-families of}\;\,\gt\;\;\;\;\;\;\;\;\;\; \cr
                \uparrow \downarrow \cr
          \mbox{signed generalized graph theory of}\;\,
                 g_M \times SO({\rm dim}\,g)}
\ee
\lam{pr2}
which provides both a
signed
generalized graph theory \cite{sme,nsc,sin,ggt}
on $g_M \times SO({\rm dim}\,g)$ and the
superconformal field theories of the superconformal  metric ansatz \cite{sin}
$\gt$ in the N=1 SME.

The known examples of this prescription,
\bs
\be
\label{tfg2}
 (\mbox{Cartesian basis of}\;SO(n))\times SO(n(n-1)/2) \;\,
{\nearrow \atop \searrow}
\matrix{ \St \cr
                \uparrow \downarrow \cr
         \mbox{signed graphs} }
 \ee
\lam{tfg2}
\vskip .05cm
\be
\label{plb}  (\mbox{Pauli-like basis of}\;SU(n))\times SO(n^2-1) \;\,
{\nearrow \atop \searrow}
\matrix{ \Ut \cr
                \uparrow \downarrow \cr
          \mbox{signed sine-area graphs} }
\ee
\lam{plb}
\vskip .05cm
\be
\label{ppb}
(\mbox{product bases of}\;SU(\Pi_i n_i))\times SO(\Pi_i n_i^2 -1) \;\,
{\nearrow \atop \searrow}
\matrix{ \Uppt \cr
                \uparrow \downarrow \cr
          \mbox{signed sine(}\op\mbox{area) graphs} }
\ee
\lam{ppb}
\es
were discussed in \cite{sme,nsc}, \cite{sin}  and \cite{ggt} respectively.

The superconformal metric ansatz $\gt$ is
\be
A= I \equiv a=1 \ldots {\rm dim}\,g
: \;\;\;\; e^{ab} = e_a \et_{ab} \sp t^{abc} = 0
\ee
where  $\et_{ab}$ is the Killing metric on $g$, and $e^{ab}$ and
$t^{abc}$ are
the vielbein and three-form in the N=1 SME (see Section \ref{sec7}).
Symmetry of the vielbein
requires the symmetry constraint $(e_a -e_b) \et_{ab} =0$, and an
additional constraint $e_a e_b e_c f_{abc}=0$ is obtained from
eq.(\ref{s29}) at zero three-form.  Consistency  of the ansatz
 in the SME also requires a magic basis of $g$.
Using the
vielbein-dominated high-level expansion (\ref{shl1}), it was shown
\cite{nsc,sin} that
the level-families of this ansatz live on
the
symmetry-constrained triplet-free
signed generalized graphs of $g_M \ti SO({\rm dim}\,g)$. (Signed
generalized graphs were reviewed in Section \ref{sgt}.) The
symmetry-constraint and triplet-free character of the signed graphs
follow from the two constraints mentioned above. The absence of triplet
structures signals a partial abelianization of the system
which  is important in obtaining simple subans\"atze and solutions of the
SME.

It was shown in Refs.\ \cite{nsc,sin}
that the signing of the generalized graphs
is an automorphic equivalence in the ansatz, so that
 the level-families can be gauge-fixed to those that live
on the unsigned symmetry-constrained triplet-free
generalized graphs $\G$ of {\samepage $g_M$,
\bs
\be
 \t_a(\G)=\t_b(\G)\;\;{\rm when}\;\, \et_{ab}\neq 0
\ee
\be
 \t_a(\G) \t_b(\G) \t_c(\G) =0 \;\;{\rm when}\;\,f_{abc} \neq 0 \pe
\ee
\ls{stc} }
In what follows, we restrict the discussion to this gauge. On the generalized
graphs (\ref{stc}), one finds that $e_a$ is proportional to $\t_a(\G)$,
so  $e_a$ vanishes on the
missing edges of the  graphs, and only the {\em edge variables}
\be
\l_a(\G,x)  \equiv  k e_a^2 \sp a \in E(\G)
\ee
on the graph edges $\t_a (\G) =1$ remain to be determined.

The supercurrent and stress tensor of $\gt$ can be expressed in
terms of the edge variables,
\bs
\be
 G(\G,z)=\sum_{a,b \in E(\G)} \sqrt{{\l_a(\G,x) \over k}} \et_{ab}J_a(z) S_b(z)
\;\,,\;\;\;\;\;x=2k/\psi^2_g
\le{scm}
\be
T(\G,z)=\frac{1}{2k}\sum_{a,b\in E(G)}\l_a(\G,x)\et_{ab}(\xx J_a (z)J_b(z)\xx -
\frac{k}{2}\oo S_a(z)
{\mathop\partial^{\leftrightarrow}}S_b (z)\oo)
\le{sct}
$$
\;\;\;\;\;\;\;+\frac{i}{2k}\sum_{a,b,c\in E(G)}
\sqrt{\l_b(\G,x)\l_c(\G,x)}
f^{abc}J_a(z) \oo S_b(z) S_c (z)\oo +\frac{\e c}{24z^2}
$$
\es
where $J_a$ and $S_a$, $a=1\ldots{\rm dim}\,g$ are  respectively
the currents of affine $g$ and a set of fermions in the adjoint of $g$.
The fermions are (BH-NS,R) when $\e=(0,1)$, and the square roots in the
stress tensor, which indicate irrational
conformal weights, tell us that the generic construction in the ansatz
is not an RCFT.
When the magic basis is real, unitarity on $x \in \N$ of simple
compact $g$ requires that $\l_a \geq 0$.

Remarkably, the superconformal metric ansatz $\gt$ reduces the
third-order SME to a set of  linear
equations on the edge-variables \cite{sin},
{\samepage
\bs
\be
\sum_{b\in E(\G)} ( \one +\frac{1}{x} {\cal A}(\G))_{ab} \l_b(\G,x)=1
\le{leq}
\be
{\cal A}_{ab}(\G) = \left\{ \matrix{
2\psi^{-2}_g \sum_c ({f_{ac}}^b)^2 & , & a,b \in E(\G)
\;\,\mbox{adjacent in}\;\,\G  \cr
0 & , & a,b \in E(\G)
\;\,\mbox{not adjacent in}\;\,\G\;\,.  } \right.
 \le{eam2}
\be
 c(\G,x)=\frac{3}{2}\sum_{a \in E(\G)}\l_a(\G,x) \;\,.
\ee
\ls{lsc} }
Here, $x$ is the invariant level of $g$,
$ \one$ is the unit matrix in the space of generalized graph edges and
$ {\cal A}_{ab}(\G)$ is the generalized edge-adjacency matrix of a
generalized graph $\G$ of $g_M$, given also in (\ref{eam}).
Historically, it was noticed \cite{nsc,sin} that the
matrix ${\cal A}$ in (\ref{eam2}) reduced to the edge-adjacency matrix of
conventional graph theory when $g=SO(n)$, and this intuition was used
to define the generalized  edge-adjacency matrix
in generalized graph theory on Lie $g$.

Some important properties of the system (\ref{lsc}) are listed below. \nl
\ni \bon Generalized lattices. In the system (\ref{lsc}), the edge
variable  $\l_a(\G,x)$ couples to the edge variable $\l_b(\G,x)$ only when
$a,b \in E(\G)$ are adjacent in $\G$. In other words, each level-family
$\{\l_a(\G,x)\}$  of $\gt$
lives with ``nearest neighbor'' coupling on the generalized lattice defined by
its generalized graph $\G$. \nl
\ni \btw High-level expansion. The leading terms in the high-level expansion of
$\gt$  are
\bs
\be
 \l_a(\G,x)=1-{ {\cal D}_a(\G) \over x} +{\cal O}(x^{-2})
\ee
\be
 c(\G,x)=\frac{3}{2}[{\rm dim}\,E(\G)- \frac{1}{x}\sum_{a \in E(\G)}
{\cal D}_a(\G) +{\cal O}(x^{-2})]
\ee
\es
where ${\cal D}_a(\G)$ is the generalized edge degree (\ref{edg})
 of edge $a$ in $\G$.
The expansion is convergent at least for $x > - a_{min}(\G) $, where
$a_{min}(\G)$ is the smallest eigenvalue of ${\cal A}(\G)$.

As in $g_{metric}$ (see Section \ref{mbm}), the finite-order
expansion shows generic irrationality of the central charge
when the squared structure constants of the magic basis are irrational,
which includes the graph theory units (\ref{plb}) and (\ref{ppb}).
Conversely, when the squared structure constants are
rational, both the edge-variables and the central charge are rational on
$x \in \N$, but the roots in the stress tensor (\ref{sct}) still
indicate that the generic theory is not an RCFT.

When $g_M$ is a real magic basis on compact $g$, the all-order expansion
indicates \cite{nsc,sin} that the superconformal level-family of each
generalized graph is generically unitary for $x\in  \N$. This has been
verified for all known examples in
$\{\gt\}$. \nl
\ni \bth Level-families and deformations \cite{nsc}. At generic level,
one may invert the operator $(1 + \frac{1}{x}{\cal A} ) $ to obtain the
level-families of the ansatz. This operator is not invertible at the
particular levels $x \in \{ - a_I(\G)\} $, where $\{a_I(\G)\}$ is the set of
eigenvalues of ${\cal A}(\G)$, and $c$-fixed quadratic deformations may
occur at these points. \nl
\ni \bfo Edge-regular generalized graphs. The simplest superconformal
level-families in $\gt$,
\bs
\be
\l_a(\G)=\l(\G) ={x \over x+{\cal D}(\G) }\;\,,\;\;\;\;\;\forall\;\,a\in E(\G)
\ee
\be
c(\G)={3 x{\rm dim}\,E(\G) \over 2(x+{\cal D}(\G))}
\ee
\ls{msc}
live on the edge-regular generalized graphs (see eq.(\ref{edg})),
with uniform generalized edge degree ${\cal D}(\G)$. When the magic basis is
real on compact $g$, which includes all known examples, the superconformal
level-families (\ref{msc}) are completely unitary for all $x\in  \N$.
As an example, the unitary irrational superconformal level-families discussed
in Section \ref{sus} were obtained on the edge-regular triplet-free sine-area
graphs of Fig.\ref{f7}.
\fig{2.5cm}{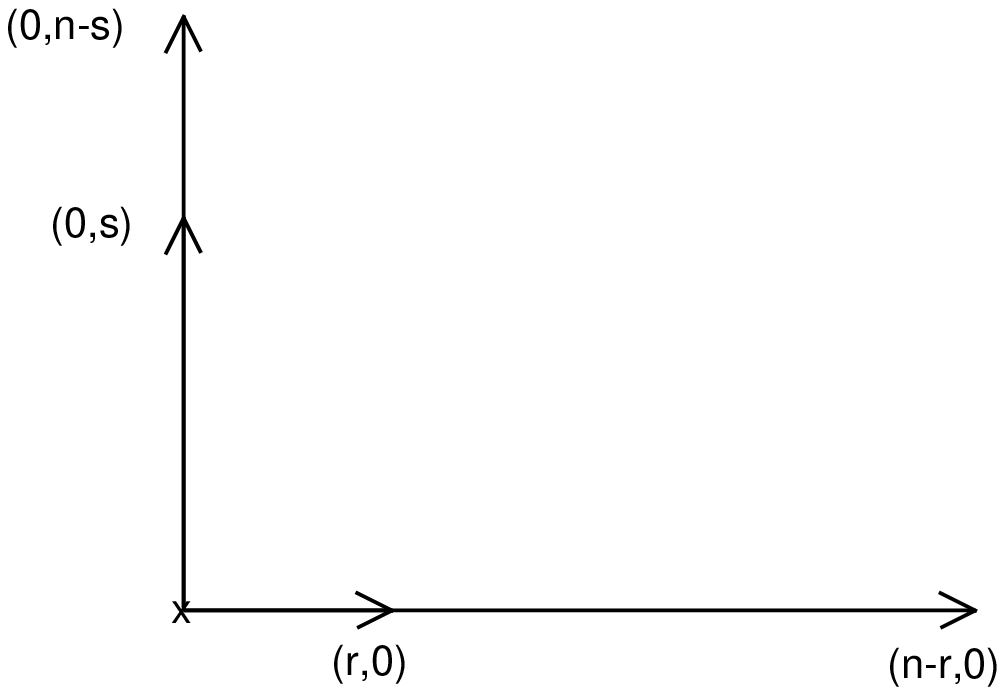}{The superconformal level-families $SU(n)_x^\#[m(N=1),rs]$
live on these edge-regular triplet-free sine-area graphs.}{f7}

\section{Conventional Graph Theory in the Master Equations \label{gth} }
\lam{gth}

In Sections \ref{agt} and \ref{ggth} we have discussed generalized graph
theory and its application to the classification of conformal field theories
in the master equations. In this section, we focus on the simplest
graph theory units in eqs.(\ref{csg}) and (\ref{tfg2}), namely the conformal
and superconformal constructions on $SO(n)$ \cite{gt,lie} and
$SO(n) \ti SO( {\rm dim}\,SO(n)) $ \cite{sme,nsc}, which live on
the graphs of conventional graph theory \cite{har}.
The ansatz on $SO(n)$ also exhibits a closed subflow on affine-Virasoro
space (see Section \ref{sec5}), which, in this case, is a flow on the space of
graphs \cite{cf}.

\subsection{Conventional Graph Theory in the VME \label{sod}}
\lam{sod}

\subsubsection{The ansatz ${\bf SO(n)_{diag}}$ \label{aso}}
\lam{aso}

In the standard Cartesian basis of $SO(n)$ (see (\ref{scs})),
the metric ansatz (\ref{man})
takes the form \cite{gt},
\bs
\be
L^{ab}= L^{ij,kl} = \ps_n^{-2} L_{ij} \d_{ij,kl}
\le{dan}
\be
T(L) = \ps_n^{-2} \sum_{i<j} L_{ij} \xx J_{ij}^2 \xx
\ee
\es
where $i,j,k,l=1\ldots n$ and
$\ps_n$ is the highest root of $SO(n)$. Because the inverse inertia
tensor (\ref{dan}) is diagonal, this ansatz is called $\Sd$ or the diagonal
ansatz on $SO(n)$.
The reduced master equation of $\Sd$ is
\bs
\be
 L_{ij}(1-xL_{ij})+\tau_n \sum_{l\neq i,j}^n
[L_{il}L_{l j} - L_{ij} (L_{il} + L_{jl})]
=0
\le{soe}
\be
L_{ji} =L_{ij} \sp L_{ii}\equiv 0 \sp \tau_n =
\left\{ \matrix{1 & n \neq 3 \cr 2 &  n =3 \cr} \right.
\le{tdf}
 \be
c=x\sum_{i<j}L_{ij}
\ee
\ls{dia}
where $\tau_n$ is the embedding index of Cartesian $SO(n)$ in  $SO(p >n)$
and unitarity requires $L_{ij}= $real.

The reduced master equation is $( {n \atop 2})$ coupled quadratic
equations, so there will be
2$^{ ({{n}\atop{2}})}$ solutions (the level-families) generically.
All these solutions are seen\footnote{For the
graph theory ansatz, this confirms the
belief \cite{gva,gt} that the high-level
smooth CFTs ($L^{ab}= \cO (k^{-1})$)  are precisely the
generic level-families. This identification also holds
for all metric ans\"atze.}
in the high-level expansion (\ref{hlm})
of the inverse inertia tensor,
\bs
\be
 L_{ij}(\G_n,x) = \frac{1}{x}\t_{ij}(\G_n)+ O(x^{-2})
\ee
\be
\t_{ij}(\G_n)
 \in \{0,1\}
\ee
\be
  c_0(\G_n, x) ={\rm dim}\,E(\G_n)+ O(x^{-1})
\ee
\be
T(\G_n, x) = {1\over x\psi^2_n} \sum_{(ij)\in E(\G_n)}\xx J^2_{ij}\xx +
\cO (x^{-2})   \pe
\ee
\ls{hls}
Here $\G_n$ is any conventional
 graph of order $n$ (on the points
$ i=1 \ldots  n$ with adjacency matrix $\th_{ij}$)
so the level-families $L(\G_n,x)$
of $\Sd$ are classified by the graphs of order $n$.
Each level-family is
unitary on $x \in \N$
 down to some finite critical level, which is quite small in all
known exact solutions, and
the central charges of the generic level-family
are generically irrational.

\subsubsection{Features of the classification by graphs \label{fcg}}
\lam{fcg}

\u{Graph Theory $\ra$ CFT}
\vskip .3cm

Here is an overview of the classification of $\Sd$ by graph theory.

\ni \bon  The residual automorphisms of $\Sd$ are the
graph isomorphisms $\G'_n \sim \G_n$, which are the permutations of the
labels on the points of the graphs. It follows that
the automorphically-inequivalent
level-families of $SO(n)_{diag}$ are in one-to-one correspondence
with the unlabelled graphs. \nl
\ni \btw   The level-family of $\G_n$ has the symmetry of its graph, that
is,
\be
L_{\pi (i)\pi(j)} (\G_n, x) = L_{ij}(\G_n, x)
\le{gsy}
when $\pi$
is a permutation in auto$\,\G_n$. For each possible graph symmetry $H \in
S_n = {\rm Aut}\,SO(n)$, the linear relations (\ref{gsy}) define the consistent
graph-symmetry subansatz which collects the $H$-invariant level-families of
$\Sd$ (see Section \ref{ca}).
The $H$-invariant level-families include the Lie $h$-invariant level-families,
discussed in Section \ref{lhg}. \nl
\ni \bth The affine-Sugawara construction on $SO(n)$ is the complete graph
$\K_n,$ with all possible graph edges among $n$ points.\nl
\ni \bfo Given a level-family $L(\G_n, x)$ of a graph $\G_n$, the
$K$-conjugate level-family $\widetilde{L}(\G_n, x) = L(\widetilde{\G}_n, x)$
lives on the complementary graph $\widetilde{\G}_n = \K_n - \G_n$. \nl
\ni \bfi The subgroup constructions $L_{h(\Sd)}$
 live on the disconnected subgroup graphs,
\be
 \G( h(\Sd) = \K_{m_1} \cup \K_{m_2} \cup \ldots \cup \K_{m_N}
\le{sgg}
where $h(\Sd)$ in (\ref{msg}) is the set of
 M-subgroups of the Cartesian basis. \nl
\ni \bsi The coset constructions of $SO(n)_{diag}$ live on
the coset graphs,
\be
\G(SO(n)/h(\Sd))=\tilde{\G}(h(\Sd))=
\tilde{\K}_{m_1} + \tilde{\K}_{m_2} + \ldots  + \tilde{\K}_{m_N}
\le{npg}
where $\tilde{\K}_m$ is the completely disconnected graph on $m$ points
and the join $\G^{(1)} + \G^{(2)} $
is defined by connecting every point in $\G^{(1)}$ to every
point in $\G^{(2)}$. In graph theory \cite{har}, the coset graphs (\ref{npg})
are called the complete
$N$-partite graphs of order $n$. \nl
\ni \bse The self K-conjugate level-families of $SO(n)_{diag}$ live with
$c= c_g/2$ on the self-complementary graphs $\widetilde{\G}_n \sim \G_n$
of order $n$. These graphs are found only on $SO(4n)$ and $SO(4n+1)$, and
the first six self-complementary graphs are shown in Fig.\ref{f5}.
The self-complementary graphs have been enumerated \cite{hp}, and one finds,
for example, that the number $s_n$ of self K-conjugate level-families on
$SO(n)$ is
\be
\matrix{n &  4 & 5 & 8 & 9 & 12 & 13 & 16 & 17 \cr
    s_n & 1 & 2 & 10 & 36 & 720 & 5600 & 703,760 & 11,220,000 \cr}
\pe
\ee
\ni \bei The high-level conformal weights of the vector representation
of $SO(n)$,
\be
\D_i(\G_n,x) =\tau_n{ d_i(\G_n) \over 2x} + \cO (x^{-2})
\ee
are proportional
to the degrees of the points in $\G_n$, where the degree $d_i$ is the number
of edges connected to the $i$th point.

\vskip .4cm
\ni \u{CFT $\ra$ Graph Theory}
\vskip .3cm

In the list above, we have seen that standard
 categories in graph theory
are useful in the classification of conformal field theories. We turn now
to a number of {\em new} graph-theoretic categories, whose definition
in generalized graph theory on Lie $g$ (see Section \ref{agt}) was motivated
by the structure of
conformal field theory. Another new category, the Lie $h$-invariant
graphs, is discussed in Section \ref{lhg}.
\vskip .3cm
\ni \bon The affine-Sugawara nests on $g \supset h_1 \supset \cdots \supset
h_n$
(see Section \ref{sec44}),
live on the {\em affine-Sugawara nested graphs},
which then classify all the conventional RCFTs in $\Sd$.
Schematically, these graphs are  obtained by a nesting
procedure which involves alternate
 subtraction and addition of the lines of the subgroup graphs in (\ref{sgg}).
See Ref.\ \cite{gt} (and eq.(\ref{asg})) for the precise characterization
of these graphs and their enumeration. \nl
\ni \btw The affine-Virasoro nests (see Section \ref{sec44}) live on the
{\em affine-Virasoro nested graphs} \cite{gt}. The precise characterization
of these graphs is given in Ref.\ \cite{gt} and eq.(\ref{avg}). They are
formed in a fashion similar to the
 affine-Sugawara nested graphs, allowing general graphs
at the bottom of
the nest. \nl
\ni \bth The irreducible level-families (see Section \ref{sec44}) of the master
equation  are those which cannot be obtained
by affine-Virasoro nesting from smaller manifolds.
 They include the affine-Sugawara constructions and the new
conformal field theories on a given manifold. In $\Sd$, this class of
constructions lives on the {\em irreducible graphs} of order $n$, which
include the complete graph $\K_n$ and the
{\em new irreducible graphs} \cite{gt},
\be
\eqalign{
\;\;\;\;\;&\G\;\,
\mbox{is a new}\;\,\mbox{irreducible graph iff}\;\,\G\;\,{\rm and}\;\,
\tilde{\G} \cr
&\mbox{are both non-trivial connected graphs} \cr}
\ee
where the trivial graph is the empty graph. These graphs were enumerated
in \cite{gt}, and the number of (unlabelled) new irreducible graphs
\be
ir^{\#}_n=2C_n-g_n
\le{nig}
is the number of physically-distinct
new irreducible level-families in $\Sd$. In (\ref{nig}),
 $g_n$ and $C_n$ are the numbers of unlabelled graphs and connected graphs
respectively at order $n$.

An immediate consequence of this enumeration is that new
 irreducible level-families are generic on large manifolds
(see also Table \ref{t2}).

\begin{table}
\begin{center}
\begin{tabular}{||c|r|r||} \hline \hline
manifold & all$\;\;\;\;\;\;\;\;\;$   & new$\;\;\;\;\;\;\;\;$ \\
 & level-families  & irreducible$\;\;\;$  \\
       &            &  level-families  \\
$SO(n)_{diag}$ & $g_n\;\;\;\;\;\;$ & $ir^\#_n\;\;\;\;\;\;$ \\ \hline \hline
$SO(2)$ & 2 & 0 \\ \hline
$SO(3)$ & 4 & 0 \\ \hline
$SO(4)$ & 11 & 1 \\ \hline
$SO(5)$ & 34 &  8  \\ \hline
$SO(6)$ & 156 & 68 \\ \hline
$SO(7)$ & 1,044 & 662  \\ \hline
$SO(8)$ & 12,346 & 9,888 \\ \hline
$SO(9)$ & 274,668 & 247,492  \\ \hline
$SO(10)$ & 12,005,168 & 11,427,974 \\ \hline \hline
\end{tabular}

\end{center}
\caption{Irreducible level-families on $SO(n)$.}
\label{t2}
\end{table}

In affine-Virasoro space, all CFTs can be uniquely constructed by
affine-Virasoro nesting from the irreducible constructions \cite{nuc,gt}.
In parallel, all
the graphs of graph theory
can be uniquely constructed by affine-Virasoro nesting
from the irreducible graphs.

The known exact level-families on new irreducible graphs \cite{gt,lie}
are listed
in eqs.(\ref{slg}) (the case $SO(2n)_M^\#= SO(2n)^\#[d,3]$), (\ref{sond}),
(\ref{sks}) and
(\ref{sck}).
The solutions $SO(2n)^\#_M$, which are included explicitly in the example
of Section \ref{sl}, are the most symmetric new irreducible
level-families in $\Sd$.
 All these constructions are
obtained by using the symmetry of the corresponding graphs, which
determines the smallest consistent subansatz in which the
constructions are  found.

\ni \bfo The complete classification of all 156 distinct level-families in
$SO(6)_{diag}$ is given in Ref.\ \cite{gt}.

\vskip .4cm
\ni \u{Counting ICFTs}
\vskip .3cm

In graph theory, the generic graph is an identity graph, which
has no symmetry (see Fig.\ref{f8}).
 On the other hand, the affine-Sugawara nested graphs
(the conventional RCFTs) always have at least
a $\Z_2$-symmetry \cite{gt}. It follows that the generic level-family in $\Sd$
is a set of new (unitary, irrational) conformal field theories.
\fig{2cm}{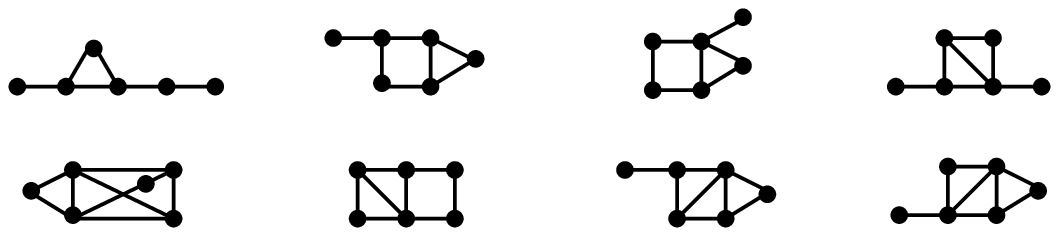}{The first eight identity graphs are new level-families in
$SO(6)_{diag}$.}{f8}

The same conclusion is seen in the
asymptotic forms \cite{gt},
\bs
\be
N_1 (\hbox{total on affine} \;\,  SO(n)) = O(e^{n^4\ln 2/8} )
\le{tns}
\be
N_2 (SO(n)_{diag} = \hbox{unlabelled graphs of order}\ n) = O (e^{n^2\ln
2/2})
\ee
\be
N_3 (\hbox{affine-Sugawara nests in } SO(n)_{diag}) \leq O (e^{2n\ln 2})
\ee
\be
N_1 \gg N_2 \gg N_3
\ee
\es
where $N_1$ is the total number of level-families in the VME on $SO(n)$,
and
$N_2 \gg N_3$ is the conclusion of the previous paragraph.

Note that, in $\Sd$, the number of graphs
($2^{({n \atop 2})}=$ \# of level-families) has the same asymptotic
form as the number of unlabeled graphs ($N_2=$ \# of inequivalent
level-families). Similarly, it has been conjectured \cite{rd} that $N_1$
in (\ref{tns}) is the asymptotic form of the number of inequivalent
level-families in the VME on $SO(n)$. The
 comparison $N_1 \gg N_2$ shows that ICFT is much larger than any
particular graph-theory unit of ICFTs.

\boldmath
\subsubsection{The Lie $h$-invariant graphs \label{lhg}}
\lam{lhg}
\unboldmath

The {\em Lie $h$-invariant graphs} \cite{lie} classify the
Lie $h$-invariant level-families of $\Sd$. These graphs form an
 important new category in graph theory on Lie $g$, because,
in this category, one sees explicitly the action of the group on the
graphs. In particular, the Lie $h$-invariant graphs satisfy
\bs
\be
\Theta (\G_n) = \o \Theta (\G_n) \o^{-1}
\le{hin}
\be
\Theta_{ij,kl} (\G_n) \equiv \d_{ik} \d_{jl} \t_{ik} (\G_n)
\sp \forall \; \o_{ij,kl} \in h(\Sd)
\ee
\be
h (\Sd) = \ti_{i=1}^N SO(m_i) \sp \sum_{i=1}^N m_i = N
\ee
\ls{dhi}
 where $\t_{ij}(\G_n)$ is the adjacency matrix of $\G_n$ and $\o$ is the
adjoint action in the $M$-subgroups of the Cartesian basis of $SO(n)$
(see Section \ref{kgg}).

Before solving (\ref{hin}) to obtain a characterization of the Lie
$h$-invariant graphs, we indicate how this category arises naturally
in the conformal field theories of $\Sd$.

The Lie $h$-invariant ans\"atze were discussed in Section \ref{ca}.
The $h(\Sd)$-invariant
subansatz
of $SO(n)_{diag}$,
\be
A_{SO(n)_{diag}}(h(SO(n)_{diag}))\,\,:\,\,(L_{ij}-L_{kl})f_{ij,kl,rs}=0\,\,\,,
\,\,\,\forall\,rs\,\in\,h(SO(n)_{diag})
\le{lso}
follows from the general form (\ref{lhs}) of a Lie $h$-invariant
ansatz and the form (\ref{dan}) of the stress tensor in $\Sd$.
The subansatz (\ref{lso}) collects all the  $h(SO(n)_{diag})$-invariant
CFTs in $\Sd$.

With  $L_{ij}(\G_n,x) =
\theta_{ij}(\G_{n})/x + \cO (x^{-2}) $, one finds that the high-level
form of this subansatz
\be
(\theta_{ij}(\G_{n})-\theta_{kl}(\G_{n}))f_{ij,kl,rs}=0\,\,\,,\,\,\,\forall\,
rs\,\in\,h(SO(n)_{diag})
\le{lhi}
is the infinitesimal form of the definition (\ref{dhi}). The definition
(\ref{dhi}) is also the high-level limit of the finite form
(\ref{hia}) of the Lie $h$-invariant subansatz. It follows that the
Lie $h$-invariant level-families of $\Sd$ are classified by the Lie
$h$-invariant graphs \cite{lie}.

{}To obtain a visual characterization of the Lie $h$-invariant graphs,
one  solves (\ref{lhi}) for a given $M$-subgroup of Cartesian $SO(n)$. As an
example,
consider the $SO(m)$-invariant graphs of order $n \geq m$, for which
one needs the vector-index decomposition
\be
i=(\mu,I) \sp \m = 1 \ldots m \sp I = m+1 \ldots n
\ee
where Greek letters are vector indices of $SO(m)$. Then, the solution of
 (\ref{lhi}) is the set of $SO(m)$-invariant graphs,
\be
\matrix{\hbox{{\it SO(m)}-invariant}\cr
\hbox{graphs of}\cr \hbox{order}\;n\cr}\;\;\;\;\;\left\{\;\;\;\;\;\matrix{
\theta_{\mu\nu}(\G_{n})\;\hbox{is independent of}\;\mu,\nu\cr
\theta_{\mu I}(\G_{n})\;\hbox{is independent of}\;\mu\cr
\theta_{IJ}(\G_{n})\;{\rm arbitrary}\cr}\right.
\ee
which classify the $SO(m)$-invariant level-families of $\Sd$.

The $SO(m)$-invariant graphs of order $n$ are shown schematically in
Fig.\ref{f9},
which distinguishes two cases,
\bs
\be
\theta_{\mu\nu}=0 \;: \;\;\;\;
\G_{n-m}\oplus_{p}{\tilde \K}_{m}
\ee
\be
\;\;\;\; \theta_{\mu\nu}=1 \;: \;\;\;\; \G_{n-m}\oplus_{p}\K_{m} \pe
\ee
\es
Here, $\G_{n-m}$ is any graph of order $n-m$, and the {\em partial join}
$\oplus_{p}$ is defined to connect $p\leq n-m$ points of $\G_{n-m}$ to all
points in ${\tilde \K}_{m}$ or $\K_{m}$.
\fig{2cm}{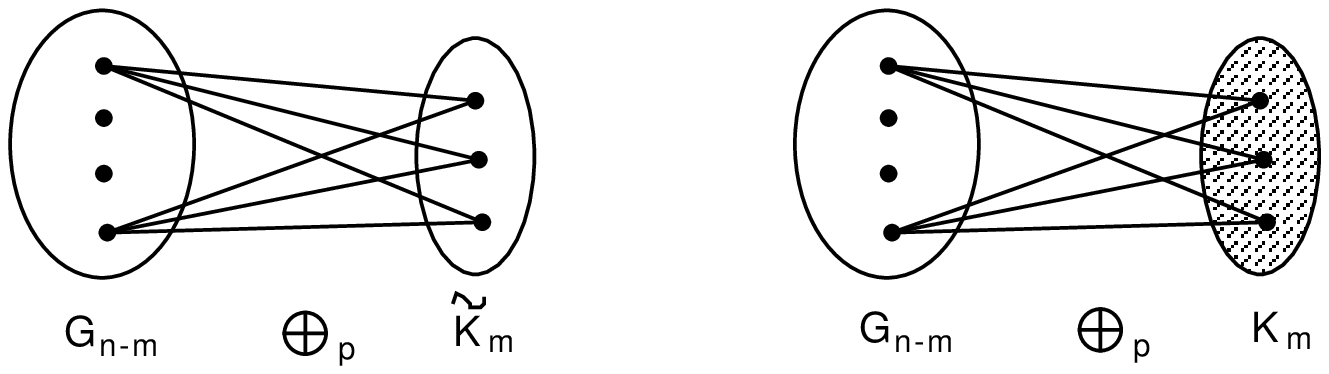}{The $SO(m)$-invariant graphs of order $n$.
The partial join $\oplus_p$ has $p=2$ in these examples.}{f9}

The $SO(m)$-invariant graphs have a
 {\em graph-local} discrete symmetry $S_{m}$ which permutes the labels
of the  points
of $\tilde{\K}_m$ or $\K_m$, and, conversely, any graph with a graph-local
symmetry $S_{m}$ is an $SO(m)$-invariant graph. A prescription to
construct the general $h(\Sd)$-invariant graph is given in
Ref.\ \cite{lie}.

Some further properties of Lie $h$-invariant graphs are as follows. \nl
\ni \Ab Symmetry hierarchy in graph theory. The graph hierarchy,
$${\rm graphs} \supset \supset
 \mbox{graphs  with symmetry} \phantom{WWWWWWWWWWW}
$$
$$ \supset \supset
\;\mbox{Lie}\;h \mbox{-invariant graphs}\phantom{W}
$$
\be
\phantom{WWWWWWWWWWWW}
\supset \supset \;
\mbox{affine-Sugawara  nested graphs}
\ee
is a special case of the symmetry hierarchy (\ref{shi}) in ICFT. In this
hierarchy, the graphs with any symmetry classify the $H$-invariant CFTs
of $\Sd$. The Lie $h$-invariant graphs, which classify the Lie
$h$-invariant CFTs of $\Sd$, are those with the special graph-local
symmetry described above. The affine-Sugawara nested graphs classify the
affine-Sugawara nests in $\Sd$, which are
the conventional RCFTs of the ansatz.
It follows that the generic Lie $h$-invariant
level-family is a set of new (generically unitary and irrational) CFTs.

\ni \Bb Graphical identification of (1,0) and (0,0) operators.
The Lie $h$-invariant graphs also tell us how the Lie $h$ symmetry is
realized in the Lie $h$-invariant CFTs. In particular, it has been shown
\cite{lie} that the $h=SO(m)$ currents $J_{ij}=J_{\m \n}$ are
either (1,0) or (0,0) operators,
\bs
\be
J_{ij}\;{\rm is}\; {\rm a}\; (1,0)\;{\rm operator}\;{\rm of}\;
L(\G_{n-m}\oplus_{p} \K_{m})\;{\rm when}\;i,j\in \K_{m}
\ee
\be
J_{ij}\;{\rm is}\; {\rm a}\; (0,0)\;{\rm operator}\;{\rm of}\;L(\G_{n-m}
\oplus_{p}{\tilde \K}_{m})\;{\rm when}\;i,j\in{\tilde \K}_{m}
\ee
\es
that is, the edges or missing edges in $\K_m$ or $\tilde{\K}_m$ correspond
to (1,0) or (0,0) currents respectively. It follows that the Lie
subgroup symmetry is realized globally for the theories with a $\K_m$
component and locally for the theories with a $\tilde{\K}_m$ component.
(The $g/h$ coset constructions are examples of the latter case.)
Similarly, each subgroup $SO(m_i)$ in the general case
$h(\Sd)= \ti_{i=1}^N SO(m_i)$ is realized either locally or globally
\cite{lie}.

\ni \Cb Generalized complementarity in graph theory. {\em
Generalized K-conjugation}
in ICFT (and in particular K${}_{g/h}$-conjugation) was reviewed in Section
\ref{sec44}. The generalized K-conjugations are realized in graph theory
as new generalized complementarities in the space of Lie
$h$-invariant graphs. In particular, K${}_{g/h}$-conjugation corresponds to
a new K${}_{g/h}$-complementarity through the coset graphs (\ref{npg}), which
is
defined on the subspace of graphs with a local Lie $h$ symmetry.

Moreover,
the self K$_{g/h}$-conjugate level-families of $SO(n)_{diag}$ live with
$c=c_{g/h}/2$ on the {\em self K$_{g/h}$-complementary graphs}, for
which the graph and its K$_{g/h}$-complement are isomorphic.
Examples of these graphs are given in \cite{lie}, which also gives the
exact form of the  self K${}_{SO(6)/SO(2)}$-conjugate level-families
(\ref{sck}).  Although these graphs have been counted on small manifolds
\cite{lie}, enumeration of the self K${}_{g/h}$-complementary graphs is an
open problem in graph theory.

Many other generalized K-conjugations exist \cite{lie},
which correspond to generalized
complementarities through the affine-Sugawara nested graphs.
In parallel with the self K${}_{g/h}$-complementary graphs,
it would be interesting to find examples of self-complementary graphs for
all the generalized complementarities.

\ni \Db The development above describes only the Lie $h$-invariant graphs
of conventional graph theory, which live on $SO(n)$. It would be interesting
to combine the general theory of Lie $h$-invariant CFTs (see Section \ref{ca})
and generalized graph theory on Lie $g$ to characterize the Lie
$h$-invariant graphs on other manifolds.

\subsection{Conventional Graph Theory in the SME \label{gsm}}
\lam{gsm}

\subsubsection{Graph theory of superconformal level-families \label{gts}}
\lam{gts}

The superconformal  ansatz $\Sd [N=1]$ is a consistent ansatz of the SME
on $SO(n)_x \ti SO( {\rm dim}\,SO(n))_{\tau_n}  $ with supercurrent \cite{sme}
\bs
\be
G(z) =\sum_{i < j} \L_{ij} J_{ij}(z) S_{ij}(z) - i
\sqrt{ \tau_n  \ps_n^2 \over 2}
\sum_{i  < j < l} t(ijl) \oo S_{ij}(z) S_{il}(z) S_{jl}(z) \oo
\ee
\be
e^{ij,kl} = \L_{ij} \d_{ik} \d_{jl} \sp t^{ij,il,jl} = t(ijl) f^{ij,il,jl}
\ee
\ls{sss}
where $\L_{ij}$ and $t(ijl)$ are the vielbein and the three-form variables
of the ansatz, and $\tau_n$ is defined in (\ref{tdf}).
The ansatz contains at least two large classes of superconformal
field theories, which are seen respectively in the vielbein-dominated
and the three-form-dominated high-level expansions (\ref{shl}).

In the case of the vielbein-dominated expansion, one finds a (signed)
graph-theoretic classification in which the high-level vielbein
$\L_{ij}^{(0)}$ defines the edges $ij$ of the graphs $\G$ and the
high-level three-form $t(ijl)^{(0)}$ lives on the unordered triplets
$ijl$ of the graphs. The roles of $\L$ and $t$ are reversed in the
three-form-dominated expansion, with the vielbein living on the
1-boundaries of a set of 2-complexes $\C$ defined by the 3-forms.
In both cases, the set of superconformal field theories is much larger
than a graph theory unit because the variables which live on the geometric
structures can assume many values.

Within this ansatz, detailed studies have been made of the following
two consistent subans\"atze \cite{nsc,ssc},
\bs
\be
\St \;\; : \;\;\; t(ijl) =0 \sp \mbox{no graph triangles in}\;\, \G
\le{tga}
\be
SO({\rm dim}\, SO(n))[N=1] \;\; : \L_{ij} =0 \sp
\mbox{no simplex-triplets in}\;\, \C
\le{tca}
\es
which live in the vielbein- and three-form-dominated expansions
respectively.
The first subansatz (\ref{tga}) is the superconformal metric ansatz
$\St \subset \gt$ (see Section \ref{sma}), which, as discussed below,
is classified by the triangle-free conventional graphs. The purely
fermionic ansatz (\ref{tca}) is classified by the triplet-free two-dimensional
simplicial complexes, where a simplex-triplet is three mutually-adjacent
2-simplices.

Both of these subans\"atze reduce the SME to a set of linear equations
(see also Section \ref{sma}), whose solutions are generically-new
superconformal field theories with rational central charge. In both cases,
moreover, one finds many candidates for new RCFTs. In what follows we discuss
the constructions on triangle-free graphs in further detail, referring
the reader to \cite{ssc} for a discussion of the constructions on the
2-complexes.

\subsubsection{Superconformal constructions on triangle-free graphs
\label{stf}}
\lam{stf}

We discuss the superconformal metric ansatz $\St$ (see Ref.\ \cite{nsc} and
eq.(\ref{tfg2})), whose automorphically-inequivalent level-families live
on each tri\-angle-free conventional graph $\G_n$ of order $n$,
\be
\forall i,j, k: \ \ \t_{ij}(\G_n) \t_{jk}(\G_n)\t_{ki}(\G_n) =0
\pe
\ee
The edge-variables $\l_{ij} (\G_n, x) = k\L_{ij}^2 (\G_n,x)$ of the
level-families live on the
edges $(ij)$
of each triangle-free graph. The supercurrent of the ansatz is a special
case of eq.(\ref{scm}),
\be
 G(\G_n,x) =\sum_{(ij)\in E(\G)}\sqrt{{\l_{ij}(\G_n, x) \over k}}J_{ij}S_{ij}
\ee
and unitarity on $x \in \N$ requires $\l_{ij} \geq 0$.

The linear equations of the ansatz
\bs
\be
 \sum_{k <l}({\one}+\frac{\tau_n}{x} A^{(1,1)}(\G_n))_{ij,kl}\l_{kl}(\G_n, x)=1
\ee
\be
 c(\G_n, x)=\frac{3}{2}\sum_{i<j}\l_{ij}(\G_n, x)
\ee
\ls{lea}
are a special case of eq.(\ref{lsc}). Here, $\one$ is the unit matrix in the
space of graph edges of $\G_n$ and
$ A^{(1,1)}_{ij,kl}(\G_n)$
is the edge-adjacency matrix (\ref{gea}) of $\G_n$.
The linear system (\ref{lea}) shows nearest-neighbor coupling
of the edge-variables on the
generalized lattice defined by each conventional graph.

The level-families of $SO(n)_{diag} [{{N=1}\atop{t = 0}}]$ are generically
unitary with rational central charge, but the generic
level-family is a new set of superconformal field theories because: \nl
\bon  the generic stress tensor has irrational coefficients, so that
irrational conformal weights occur in the generic construction, \nl
\btw the conventional N=1 constructions in the
subansatz are a small subset of the level-families on triangle-free graphs, \nl
\bth the central charge of the generic construction cannot be expressed as sums
and differences of affine-Sugawara central charges. \nl
In these broad features, the level-families on triangle-free graphs
parallel the self K-conjugate constructions \cite{gt,mb}, which have
half affine-Sugawara central charge and generically irrational conformal
weights.
\tabl{10cm}{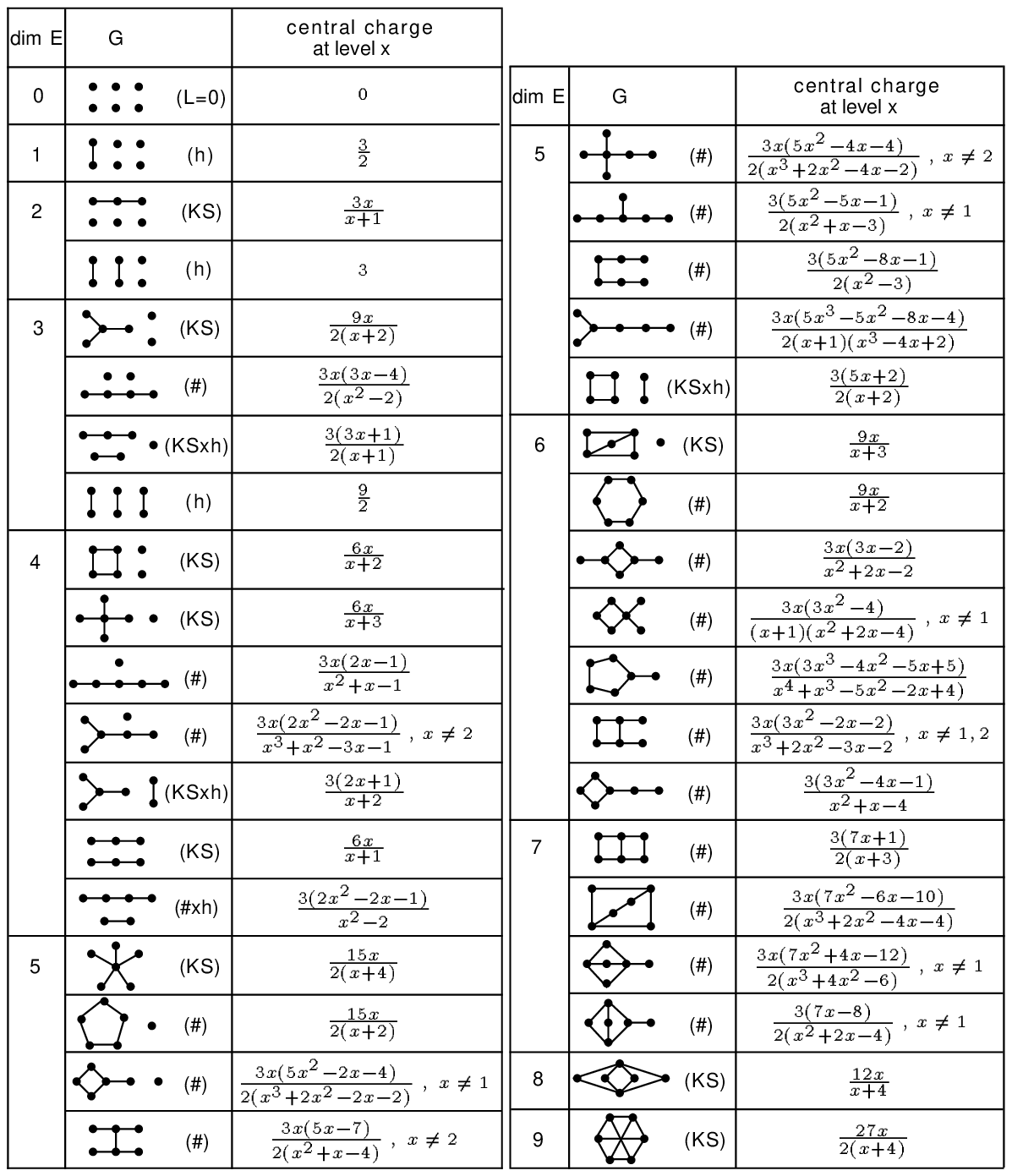}{The triangle-free graphs of
$SO(6)_{diag}[ {N=1 \atop t=0}]$.}{t1}

{}To understand item 3 above, see Table \ref{t1}, which shows the
central charges of the level-families of $SO(6)_{diag}
 [{{ N=1}\atop{t=0}}]$ on
the triangle-free graphs of order 6.
In this table, all constructions are  unitary on  level $x \in \N$
unless indicated explicitly (e.g., $x \neq 1$) with the central charge.
Conventional superconformal constructions are labelled by $h$
(subgroup construction) and KS (Kazama-Suzuki
 coset construction \cite{ks2}), while new superconformal
level-families are marked with the usual \#.

Consider, for example, the first new level-family of the table, which is
 the path graph of order 4.
The denominator of this central charge has roots at irrational level
$x= \pm \sqrt{2}$, so the level-family cannot be obtained by coset
construction. (Sums and differences of affine-Sugawara central charges
can diverge only at rational levels.)
Similarly, the generic level family $\lambda_{ij}(\G_n, x)$ is new because the
roots of the denominator of $c(\G_n, x)$ are among the generically-irrational
eigenvalues of the edge-adjacency matrix (\ref{gea}) of its graph $\G_n$.

\vskip .4cm
\ni \u{Candidates for new RCFTs}
\vskip .3cm

Because of the square roots in the stress tensor, it is expected that
the generic level-family in  $SO(n)_{diag} [{{N=1}\atop{t=0}}]$
has irrational conformal weights. On the other hand, there are many
level-families which appear to be new RCFTs (beyond $h$ and $g/h$) because
the square roots cancel. These constructions
live among the level-families (\ref{msc}) on
the edge-regular triangle-free graphs, which have
\be
 \l(\G)={x \over x+\tau_n D(\G)} \sp
 c(\G)=\frac{3x{\rm dim}\,E(\G)}{2(x+\tau_n D(\G))}
\ee
where $D(\G)$ is the edge-degree of the graph. In this case, the stress
tensor has rational coefficients on $x \in \N$ because
 $\sqrt{\l_{rs} \l_{kl}}=\l$. The conventional superconformal level-families
of the subansatz are a small subset of the level-families on edge-regular
graphs, but many candidates for new RCFTs are also included (e.g. the
cycles $C_5$ and $C_6$ in Table \ref{t1}). Low-lying conformal weights in
these theories are rational, but it is an open question whether these
theories are truly new RCFTs, with entirely rational conformal weights.

We finally mention that  graph spectral theory \cite{sei} has been used
\cite{nsc} to identify
superconformal quadratic deformations at particular levels on almost all
edge-regular triangle-free graphs.

\subsection{Bosonic N=2 Superconformal Constructions \label{bks}}
\lam{bks}

Kazama and Suzuki \cite{ks} and Cohen and Gepner \cite{gep} have found
a graph theory unit of N=2 superconformal constructions
 in the interacting bosonic models (IBMs). It is suspected by the authors
that this set of IBMs live in the N=2 SME (see Section
\ref{sec8}), but this has not yet been shown.

In this construction, the ansatz for the supercurrents is \cite{ks},
\bs
\be
G_+(z) = \sum_{i=1}^n  g(\g^{(i)}) {\rm e}^{i \g^{(i)} \cdot \phi (z)}
\sp G_-(z) = \sum_{i=1}^n g(\g^{(i)})^{\dagger} {\rm e}^{-i \g^{(i)} \cdot
\phi (z)}
\ee
\be
\g^{(i)} \cdot \g^{(j)}
=\left\{  \matrix{3 & i=j \cr 0 \;\,\mbox{or}\;\, 1 & i\neq j \cr } \right.
\le{gac}
\es
where $\phi_A $ and $\g^{(i)}_A$, $A =1\ldots D$ are respectively a set of
bosons and any set of vectors which satisfy (\ref{gac}).
N=2 superconformal symmetry is obtained
when \cite{ks},
\bs
\be
\sum_{j=1}^n \Gamma_{ij} x_j =2 \sp x_i \equiv | g(\g^{(i)})|^2
\le{kse}
\be
c= \frac{3}{2} \sum_{i=1}^n x_i
\ee
\es
where $\Gamma_{ij} \equiv \g^{(i)} \cdot \g^{(j)} $.
Note that the off-diagonal part of $\Gamma_{ij}$ can be viewed as the
adjacency-matrix $\t_{ij}$ of a graph of order $n$, so this
system is also a graph theory unit of superconformal constructions \cite{gep}.

In particular, the central charge \cite{ks},
\be
c = { 3n \over 3 +d} \geq 3
\le{ksc}
is obtained on the regular graphs of order $n$ with degree $d$.
It was argued in \cite{gep} that there are an infinite number of CFTs at
each value (\ref{ksc}) of the central charge, and, among these, there are
an infinite number of new RCFTs, with entirely rational conformal weights.

Ref.\ \cite{gep} also discusses a method for obtaining the finite-level
fusion rules of new CFTs (see Section \ref{hc21} for the high-level fusion
rules of ICFT).

\newpage

\part{The Dynamics of ICFT}

\section{Dynamics on the Sphere \label{hc1}}

We turn now to the operator formulation of the dynamics of
ICFT, including the Ward
identities satisfied by the natural correlators and characters of
ICFT on the sphere and the torus.  Because the stress tensors come in
commuting K-conjugate pairs, the central notions here include
{\em biconformal field theory} and {\em factorization} of the
{\em bicorrelators} and {\em bicharacters} to obtain the correlators and
characters of the individual CFTs.

Biconformal fields were first
obtained for the coset constructions by Halpern in Ref.\  \cite{h3}.
Generalization of the biconformal fields to all ICFT was given by
Halpern and Obers in Ref.\  \cite{wi}, where these fields were used to
derive the {\em affine-Virasoro
Ward identities} for the bicorrelators on the sphere.
Factorization and solutions of the Ward identities were discussed
in Ref.\  \cite{wi2}, including the general solution which shows a universal
braiding for
all ICFT.  The flat-connection form of the Ward identities was obtained
in Ref.\  \cite{fc}.  In this form, one sees that the Ward identities
of ICFT are {\em generalized Knizhnik-Zamolodchikov equations},
which include the
conventional Knizhnik-Zamolodchikov equation \cite{kz}
as a special case.  The parallel development
on the torus is reviewed in Section \ref{hc31}.

Short reviews of these developments are found in Refs.\ \cite{rp} and
\cite{paris}.

\subsection{Background:  Biconformal Field Theory\label{hc2}}

We recall from Part I that the general
affine-Virasoro construction gives two commuting
conformal stress tensors $T$ and $\tT$, called a K-conjugate pair \cite{bh,hk},
\begin{subequations}
\begin{equation}
T=L^{ab}\xx J_a J_b\xx =\sum_m L(m)z^{-m-2}
\end{equation}
\begin{equation}
\tT=\tL^{ab}\xx J_aJ_b\xx =\sum_m \tL(m)z^{-m-2}
\end{equation}
\begin{equation}
T_g=T+\tT=L_g^{ab}\xx J_aJ_b\xx =\sum_mL_g(m)z^{-m-2}
\end{equation}
\begin{equation}
[L(m),\tL(n)]=0
\end{equation}
\end{subequations}
where each pair sums to the affine-Sugawara construction $T_g$.  It
is clear that, as constructed on the affine Hilbert space, the
natural structure of ICFT is a large
set of {\em biconformal field theories} \cite{h3,wi},
where each biconformal field theory has two commuting stress tensors.

The decomposition $T_g=T+\tT$ strongly suggests that, for each
K-conjugate pair, the
affine-Sugawara construction is a tensor product CFT, formed
by tensoring the conformal field theories of $T$ and $\tT$.
In practice, one faces the inverse problem, namely the definition
of the $T$ theory by modding out the $\tT$ theory and vice versa.
In the operator approach to the dynamics of ICFT
\cite{wi,wi2,fc}, the biconformal structure is
central and one uses
null states of the Knizhnik-Zamolodchikov (KZ)
type \cite{kz} to derive Ward identities for the biconformal
correlators \cite{wi}.
Then one must learn to factorize \cite{wi,wi2}
the biconformal correlators into the
conformal correlators of $T$ and $\tT$.

This discussion cannot be complete without mention of the
chiral null-state approach \cite{bpz,df,df2},
to which we owe a deep understanding
of conventional RCFT.  In this approach, one uses
null states in
modules of extended Virasoro algebras \cite{ns,r,z1,fz,fl} to
bypass the biconformal structure $T_g=T_h+T_{g/h}$ of $h$ and the
$g/h$ coset constructions, obtaining
directly the BPZ equations for the coset correlators.
Although beautifully crafted for the coset constructions, this
technique apparently has little to say about the affine-Sugawara
constructions and the general ICFT, for which one is led to
consider the more general (biconformal) dynamics of the KZ type.
Moreover, the approach through biconformal field theory gives an
alternate description of the coset constructions, which has led to new
results for the coset constructions on the sphere and the torus
(see Sections \ref{hc14} and \ref{hc38}).

\subsection{Biprimary States \label{hc3}}

Given a biconformal field theory, it is natural to decompose the
affine Hilbert space into Virasoro
biprimary and bisecondary states \cite{h3,wi},
where {\em Virasoro biprimary states} $|\D,\tD\>$ are Virasoro primary
under both commuting K-conjugate Virasoro operators,
\begin{subequations}
\begin{equation}
L(m)|\D,\tD\> = \d_{m,0}\D|\D,\tD\>\quad , \quad m\ge 0
\end{equation}
\begin{equation}
\tL(m)|\D,\tD\> = \d_{m,0}\tD|\D,\tD\>\quad, \quad m\ge 0
\end{equation}
\begin{equation}
\D+\tD=\D_g.
\end{equation}
\label{biprimary}
\end{subequations}
In (\ref{biprimary}), $\D,\tD, $ and $\D_g$ are the conformal weights under
$T, \tT,$ and $T_g$ respectively.  Virasoro bisecondary states are then
formed as usual, by applying the negative modes of $T$ and $\tT$ to the
biprimary states.

A useful characterization of biprimary states is as follows.

\ni \un{Proposition}.  Necessary and sufficient conditions for a state
$|\f\>$ to be Virasoro biprimary are
\begin{subequations}
\begin{equation}
L_g(m)|\f\>=\d_{m,0}\D_g|\f\>\quad, \quad m\ge0
\label{primary}
\end{equation}
\begin{equation}
L(0)|\f\>=\D|\f\>
\end{equation}
\label{conditions}
\end{subequations}
where any state which satisfies the condition (\ref{primary}) is called
an {\em affine-Sugawara primary state}. \newline
\ni \un{Proof}.  According to (\ref{biprimary}) and $T_g=T+\tT$,
the conditions (\ref{conditions}) are necessary.  To prove
sufficiency, one checks that $|\f\>$ is Virasoro primary under $T$,
\begin{equation}
L(m)|\f\>={1\over m}[L(m),L(0)]|\f\>=
{1\over m}[L_g(m),L(0)]|\f\>=0\quad, \quad m>0
\end{equation}
using the Virasoro algebra of $T$, $T_g=T+\tT$ and
$[L(m),\tL(n)]=0$.  Using $T_g=T+\tT$ once more, one easily checks
that $|\f\>$ is also Virasoro primary under $\tT$.
\ep

An explicit construction of all biprimary states has not yet
been found, but many examples are known.  The canonical examples
are the affine primary states
$|\f\>=|R_\T\>^\a, \a=1\ldots\dim \T$, which transform in matrix
irrep $\T$ of $g$.  In an {\em $L$- basis}
of $\T$ \cite{h3,nuc,wi}, these biprimary
states are called the {\em $L^{ab}$-broken affine primary states},
which also satisfy
\begin{subequations}
\begin{equation}
J_a(m)|R_\T\>^\a=\d_{m,0}|R_\T\>^\b(\T_a)_\b{}^\a\quad, \quad m\ge0
\label{Jbrok}
\end{equation}
\begin{equation}
\tL^{ab}(\T_a\T_b)_\a{}^\b=\tD_\a(\T)\d_\a{}^\b\quad, \quad
L^{ab}(\T_a\T_b)_\a{}^\b=\D_\a(\T)\d_\a{}^\b
\label{eigenbases}
\end{equation}
\begin{equation}
\tL(0)|R_\T\>^\a=\tD_\a(\T)|R_\T\>^\a\quad, \quad
L(0) |R_\T\>^\a=\D_\a(\T)|R_\T\>^\a
\end{equation}
\begin{equation}
\tD_\a(\T)+\D_\a(\T)=\D_g(\T)
\end{equation}
\label{broken}
\end{subequations}
where $\D_g(\T)$ is the $\a$-independent conformal weight under $T_g$ and
$\tD_\a(\T)$ and $\D_\a(\T)$ are the $L^{ab}$-broken
conformal weights under $\tT$ and $T$ respectively.
More generally, $L$-bases are the eigenbases of the conformal weight
matrices, such as (\ref{eigenbases}), which occur at each level of
the affine irrep.
Other examples of biprimary states include the one-current states
$J_A(-1)|0\>, A=1\ldots\dim g$, whose conformal weights
satisfy $\tD_A+\D_A=\D_g=1$.

\subsection{Biprimary Fields \label{hc4}}

In biconformal field theory, the natural generalization of
Virasoro primary fields are the Virasoro biprimary fields, which
are simultaneously Virasoro primary under both commuting stress
tensors $\tT$ and $T$.  These fields were first constructed
for $h$ and the $g/h$ coset constructions in Ref.\  \cite{h3}, where
they were called bitensor fields.  Generalization to all
ICFT was given in Ref.\  \cite{wi}.

{}To construct the biprimary fields, one begins with an
{\em affine-Sugawara primary field} $\f_g$, which satisfies
\begin{subequations}
\begin{equation}
T_g(z)\f_g(w)=\left({\D_g\over(z-w)^2}+{\pa_w\over z-w}\right)
\f_g(w) + {\rm reg.}
\end{equation}
\begin{equation}
\f_g(0)|0\>=|\f\>
\end{equation}
\end{subequations}
where an $L$-basis for $\f_g$ is assumed so that $|\f\>$ is
Virasoro biprimary under $\tT$ and $T$.  Examples of affine-Sugawara
primary fields include the affine primary fields and the currents,
\begin{subequations}
\begin{equation}
T_g(z)R_g^\a(\T,w)=\left( {\D_g(\T)\over (z-w)^2} + {\pa_w \over z-w} \right)
R_g^\a(w)+{\rm reg.}
\label{affprim}
\end{equation}
\begin{equation}
T_g(z)J_A(w)=\left( {1 \over (z-w)^2} + {\pa_w \over z-w} \right)
J_A(w) + {\rm reg.}
\end{equation}
\end{subequations}
in their respective $L$-bases.  It should be noted that, although
(\ref{affprim}) is usually assumed \cite{kz} for the affine primary fields,
the form is strictly correct only for integer levels of affine compact
$g$.  This subtlety is discussed in Ref.\  \cite{wi}, which finds an extra
zero-norm operator contribution for non-unitary affine-Sugawara constructions.

Although they create biprimary states,
the affine-Sugawara primary
fields are not Virasoro primary under $\tT$ and $T$.
Instead, they
satisfy the relations \cite{sta},
\begin{subequations}
\begin{equation}
\tT(z)\f_g(w)={\tD\f_g(w) \over (z-w)^2} + {\pa_w \f_g(w)+\tdf_g(w) \over z-w}
+ {\rm reg.}
\end{equation}
\begin{equation}
T(z)\f_g(w)={\D\f_g(w) \over (z-w)^2} + {\pa_w \f_g(w)+\df_g(w) \over z-w}
+ {\rm reg.}
\end{equation}
\label{affnotvir}
\end{subequations}
where $\tD$ and $\D$ are the conformal weights of $|\f\>$ under $\tT$ and $T$.
The extra terms $\tdf_g$ and $\df_g$ in (\ref{affnotvir}),
\begin{equation}
\tdf_g=-[L(-1),\f_g]\quad, \quad
\df_g=-[\tL(-1),\f_g]
\end{equation}
are generated by the existence of the non-trivial K-conjugate theories.

We turn now to the {\em Virasoro biprimary fields}
$\f(\tz,z)$, whose job it is
 to compensate for these extra terms.  In particular, the biprimary
fields are Virasoro primary under both $\tT$ and $T$,
\begin{subequations}
\begin{equation}
\tT(\tz)\f(\tw,w)=\left( {\tD \over (\tz-\tw)^2} + {\pa_\tw \over \tz-\tw}
\right) \f(\tw,w) + \mbox{reg. in $(\tz-\tw)$}
\end{equation}
\begin{equation}
T(z)\f(\tw,w)=\left( {\D \over (z-w)^2} + {\pa_w \over z-w}
\right) \f(\tw,w) + \mbox{reg. in $(z-w)$}
\end{equation}
\begin{equation}
\f(0,0)|0\>=\f_g(0)|0\>=|\f\>
\end{equation}
\label{bfields}
\end{subequations}
where $\tT$ and $T$ operate on $\tw$ and $w$
respectively.

The explicit construction of these bilocal fields is remarkably simple
\cite{h3,wi},
involving only $SL(2,\R)$ boosts of the affine-Sugawara primary
fields.  Because the biprimary fields are of central interest, we give a
number of equivalent forms,
\begin{subequations}
\label{biprim}
\begin{eqnarray}
\label{firstbiprim}
\f(\tz,z)&=& z^{L(0)}\tz^{\tL(0)}\f_g(1)z^{-L(0)-\D}\tz^{-\tL(0)-\tD} \\
&=& \left({\tz\over z}\right)^{\tL(0)}\f_g(z)
    \left(z\over\tz\right)^{\tL(0)+\tD} \\
&=& \left({z\over \tz}\right)^{L(0)}\f_g(\tz)
    \left(\tz\over z\right)^{L(0)+\D} \\
&=& e^{(\tz-z)\tL(-1)}\f_g(z)e^{(z-\tz)\tL(-1)} \\
&=& e^{(z-\tz)L(-1)}\f_g(\tz)e^{(\tz-z)L(-1)}.
\end{eqnarray}
\end{subequations}
The first line (\ref{firstbiprim}) is the original form of the
biprimary fields \cite{h3}, and the alternate forms of (\ref{biprim}d,e),
\begin{eqnarray}
\f(\tz,z)&=&\exp\left[(\tz-z)\oint_z{dw\over2\p i}\tT(w)\right]\f_g(z)
  \nonumber \\
&=&\exp\left[(z-\tz)\oint_\tz{dw\over2\p i}T(w)\right]\f_g(\tz)
\label{opforms}
\end{eqnarray}
have also been employed \cite{fc}.

Other useful properties of the biprimary fields include
\begin{subequations}
\begin{equation}
\< \f(\tz,z)\>=0
\label{average}
\end{equation}
\begin{equation}
\f(z,z)=\f_g(z)
\label{zz}
\end{equation}
\end{subequations}
where the average (\ref{average}) is in the affine vacuum $|0\>$ and
(\ref{zz}) says that, on the affine-Sugawara line $\tz=z$,
 the biprimary fields are equal to the
affine-Sugawara primary fields.

Examples of biprimary fields include the biprimary fields of the
$L^{ab}$-broken affine primary fields and the $L^{ab}$-broken
currents
\begin{subequations}
\begin{equation}
R^\a(\T,\tz,z)=\exp\left[(\tz-z)\oint_z{dw\over2\p i}\tT(w)\right]R_g^\a(\T,z)
\end{equation}
\begin{equation}
\tD_\a(\T)+\D_\a(\T)=\D_g(\T)
\end{equation}
\begin{equation}
\J_A(\tz,z)=\exp\left[(\tz-z)\oint_z{dw\over2\p i}\tT(w)\right]J_A(z)
\end{equation}
\begin{equation}
\tD_A+\D_A=1
\end{equation}
\end{subequations}
where $\a=1\ldots\dim\T$ and $A=1\ldots\dim g$.

\subsection{Bicorrelators \label{hc5}}

Biconformal correlators, or {\em bicorrelators}, are averages of biconformal
fields in the affine vacuum.  For the biprimary fields, one has the $n$-point
bicorrelators,
\begin{subequations}
\begin{equation}
\Fi(\tz,z)\equiv\<\f_1(\tz_1,z_1)\cdots\f_n(\tz_n,z_n)\>
\end{equation}
\begin{equation}
\Fi(z,z)=\Fi_g(z)=\<\f_{g,1}(z_1)\cdots\f_{g,n}(z_n)\>
\end{equation}
\label{bcorr}
\end{subequations}
which reduce, on the affine-Sugawara line $\tz=z$, to
the affine-Sugawara correlators of the affine-Sugawara
primary fields.  In general, the affine-Sugawara correlators also
satisfy a $g$-global Ward identity. For example, one has
\begin{subequations}
\begin{equation}
A^\a(\tz,z)=\<R^{\a_1}(\T^1,\tz_1,z_1)\cdots R^{\a_n}(\T^n,\tz_n,z_n)\>
\end{equation}
\begin{equation}
A^\a(z,z)=A_g^\a(z)=\<R^{\a_1}_g(\T^1,z_1)\cdots R^{\a_n}_g(\T^n,z_n)\>
\label{ascorr}
\end{equation}
\begin{equation}
A_g^\b(z)\left(\sum_{i=1}^n \T_a^i\right)_\b^{~\a}=0\quad, \quad
a=1\ldots\dim g
\label{gglobal}
\end{equation}
\end{subequations}
when broken affine primary fields $\f^\a=R^\a$ are chosen for the
bicorrelator.

The two- and three-point functions and leading term OPE's of
biprimary fields can be determined \cite{wi} from $SL(2,\R)\times SL(2,\R)$
covariance and knowledge of the corresponding quantities for the
affine-Sugawara primary fields.

As examples, we give the results for three broken affine primary fields,
\begin{subequations}
\begin{equation}
\< R^{\a_1}(\T^1,\tz_1,z_1)R^{\a_2}(\T^2,\tz_2,z_2)R^{\a_3}(\T^3,\tz_3,z_3)\>
= {Y_g^{\a_1\a_2\a_3}\over
\tz_{12}^{\tg_{12}} \tz_{13}^{\tg_{13}} \tz_{23}^{\tg_{23}}
z_{12}^{\g_{12}} z_{13}^{\g_{13}} z_{23}^{\g_{23}}  }
\end{equation}
\begin{equation}
R^{\a_1}(\T^1,\tz_1,z_1) R^{\a_2}(\T^2,\tz_2,z_2)
\limit{\sim}{\tz_{12},z_{12}\rightarrow 0}
\sum_{k,\b_k} {Y_g^{\a_1\a_2\a_k}\et_{\a_k\b_k}(\T^k)R^{\b_k}(\T^k,\tz_2,z_2)
\over \tz_{12}^{\tD_{12k}}z_{12}^{\D_{12k}}}\;.
\label{ROPE}
\end{equation}
\end{subequations}
Here, $\tz_{ij}=\tz_i-\tz_j, z_{ij}=z_i-z_j$, $\et_{\a\b}(\T)$ is the
carrier space metric of irrep $\T$ and
\begin{subequations}
\begin{equation}
\tg_{ij}=\tD_{\a_i}(\T^i)+\tD_{\a_j}(\T^j)-\tD_{\a_k}(\T^k)
\end{equation}
\begin{equation}
\g_{ij}=\D_{\a_i}(\T^i)+\D_{\a_j}(\T^j)-\D_{\a_k}(\T^k)
\end{equation}
\label{gamma}
\end{subequations}
while $\tD_{12k}$ and $\D_{12k}$ are given by (\ref{gamma}) with
$\a_k\rightarrow\b_k$.
The coefficient $Y_g$ is the invariant affine-Sugawara three-point
correlator, which satisfies the $g$-global Ward identity
$Y_g^\b(\sum_{i=1}^3 \T_a^i)_\b{}^\a=0$.  It follows that
$Y_g^{\a_1\a_2\a_k}$ are
the Clebsch-Gordon coefficients for $\T^1 \oplus \T^2$ into $\bar\T^k$, taken
in a simultaneous $L$-basis of the three irreps $\T$.

We also mention the bilocal current algebra \cite{wi},
{\samepage
\begin{eqnarray}
\J_A(\tz,z)\J_B(\tw,w)
& \limitt{\sim}{\tz\rightarrow\tw}{z\rightarrow w} &
{G_{AB} \over (\tz-\tw)^{2\tD_A}(z-w)^{2\D_A}} \hspace*{2in} \nonumber \\
& &+  \sum_C {if_{AB}{}^C \J_C(\tw,w) \over
  (\tz-\tw)^{\tD_A+\tD_B-\tD_C}(z-w)^{\D_A+\D_B-\D_C}}
\label{JOPE}
\end{eqnarray} }
satisfied by the biprimary fields of the $L^{ab}$-broken currents.  Here,
$G_{AB}$ and $f_{AB}{}^C$ are the generalized Killing metric and
structure constants of $g$, both in the $L$-basis of the one-current
states.

See Ref.\  \cite{wi} for further details on two- and three-point
bicorrelators.

More generally, one has the $SL(2,\R)\times SL(2,\R)$ decomposition
of the bicorrelators
\begin{subequations}
\begin{equation}
\<\f_1(\tz_1,z_1)\cdots\f_n(\tz_n,z_n)\> =
{Y(\tilde{u},u) \over \prod_{i<j} \tz_{ij}^{\tg_{ij}(\a)}z_{ij}^{\g_{ij}(\a)}}
\end{equation}
\begin{equation}
\sum_{j\ne i}\tg_{ij}(\a)=2\tD_{\a_i}\quad, \quad
\sum_{j\ne i}\g_{ij}(\a)=2\D_{\a_i}
\end{equation}
\begin{equation}
\tg_{ij}(\a)+\g_{ij}(\a)=\g_{ij}^g\quad, \quad
\sum_{j\ne i} \g_{ij}^g=2\D_i^g
\end{equation}
\begin{equation}
Y(u,u)=Y_g(u)
\end{equation}
\label{decomp}
\end{subequations}
where $\{\tilde{u}\}$ and $\{u\}$ are the sets of independent cross-ratios
constructed from $\{\tz_i\}$ and $\{z_i\}$ respectively, and
$Y(\tilde{u},u)$ are the invariant bicorrelators.  When $\phi^\a=R^\a$,
the invariant
affine-Sugawara correlators $Y_g$ also satisfy the $g$-global Ward identity
$Y_g\sum_{i=1}^n \T_a^i=0$.

The biconformal OPE's, such as (\ref{ROPE}) and (\ref{JOPE}), also
determine the most singular terms of the bicorrelators.  As an
example, we give the results for the invariant bicorrelators
of four broken affine primaries, using the KZ gauge \cite{kz}
\begin{subequations}
\begin{equation}
\tilde{u}={\tz_{12}\tz_{34} \over \tz_{14}\tz_{32}}\quad, \quad
 u={z_{12}z_{34} \over z_{14}z_{32}}
\end{equation}
\begin{equation}
\tg_{12}=\tg_{13}=\g_{12}=\g_{13}=0
\end{equation}
\end{subequations}
for the invariant decomposition (\ref{decomp}).  The result for the four-point
bicorrelators is \cite{hun},
\begin{equation}
 Y(\tilde{u},u)
\limit{\sim}{\tilde{u}, u\rightarrow0}
v_g^{(4)}u^{C(L_g)}\left({\tilde{u}\over u}\right)^{A(\tL)}
\left({\tilde{u}\over u}\right)^{-B(\tL)}
\label{sing1}
\end{equation}
where $v_g^{(4)}$ is the $g$-invariant tensor
\begin{subequations}
\begin{equation}
v_g^{(4)\a}\equiv (-1)^{C(L_g)}\sum_k Y_g^{\a_1\a_2\a_k} \et_{\a_k\b_k}(\T^k)
  Y_g^{\b_k\a_3\a_4}
\end{equation}
\begin{equation}
v_g^{(4)}\sum_{i=1}^4 \T_a^i = 0\quad, \quad a=1\ldots\dim g
\end{equation}
\label{ginvtens}
\end{subequations}
and $A, B, C$ are the matrices
\begin{subequations}
\begin{equation}
A(L)\equiv L^{ab}(\T_a^1+\T_a^2)(\T_b^1+\T_b^2)
\end{equation}
\begin{equation}
B(L)\equiv L^{ab}(\T_a^1\T_b^1+\T_a^2\T_b^2)
\end{equation}
\begin{equation}
C(L)\equiv A(L)-B(L)=2L^{ab}\T_a^1\T_b^2.
\end{equation}
\label{auxil}
\end{subequations}
In (\ref{auxil}) and below, the tensor product
$\T_a^i\T_b^j\equiv\r_a(\T^i)\otimes\r_b(\T^j)$ is understood when
matrix irreps $\T^i$ are in different spaces $i$.

\subsection{The Ward Identities of ICFT\label{hc6}}

{}To go beyond the simple considerations of the previous section,
one needs the
affine-Virasoro Ward identities \cite{wi}, which provide the central
dynamics of ICFT.

The form (\ref{opforms}) of the Virasoro biprimary fields indicates that the
biconformal correlators (\ref{bcorr}) can be evaluated as power series
about the affine-Sugawara line $\f(z,z)=\f_g(z)$.  Indeed, with
$\tpa_j=\pa/\pa\tz_j$ and $\pa_i=\pa/\pa z_i$, one obtains the formula
for the moments of the bicorrelators \cite{wi},
$$
\left.\tpa_{j_1}\cdots\tpa_{j_q} \pa_{i_1}\cdots\pa_{i_p}
   \<\f_1(\tz_1,z_1)\cdots\f_n(\tz_n,z_n)\>\right|_{\tz=z}
 \hspace*{2in}
$$
\begin{eqnarray}
&=& \oint_{\tz_{j_1}} {d\tw_1 \over 2\p i} \cdots
   \oint_{\tz_{j_q}} {d\tw_q \over 2\p i}
   \oint_{z_{i_1}} {dw_1 \over 2\p i} \cdots
   \oint_{z_{i_p}} {dw_p \over 2\p i} \cdot
\nonumber \\
& & \quad \cdot \<\tT(\tw_1)\cdots\tT(\tw_q)T(w_1)\cdots T(w_p)
  \f_{g,1}(z_1)\cdots\f_{g,n}(z_n) \>
\label{yuck}
\end{eqnarray}
because $\f(z,z)=\f_g(z)$ and each $\f(\tz,z)$ is Virasoro biprimary.
The stress tensors on the right of (\ref{yuck}) can be expressed in terms of
the currents, and the moments can be evaluated in principle by
computation of the averages $\<$ $\>$ in the affine-Sugawara theory on $g$.
The relations (\ref{yuck}) are called the {\em Ward identities} of ICFT.

We focus here on the simplest case of (\ref{yuck}),
that is the Ward identities for the
biprimary fields $\f^\a(\tz,z)=R^\a(\T,\tz,z)$ of the $L^{ab}$-broken
affine primary fields.  In this case, the Ward identities take the form
\cite{wi},
\begin{subequations}
\begin{equation}
\left.\tpa_{j_1}\cdots\tpa_{j_q}\pa_{i_1}\cdots\pa_{i_p} A^\a\right|_{\tz=z} =
A_g^\b(W_{j_1 \cdots j_q, i_1 \cdots i_p})_\b{}^\a
\label{avward}
\end{equation}
\begin{equation}
A^\a(\tz,z)\equiv \<R^{\a_1}(\T^1,\tz_1,z_1)\cdots R^{\a_n}(\T^n,\tz_n,z_n)\>
\end{equation}
\begin{equation}
A_g^\a(z)=A^\a(z,z)= \<R^{\a_1}_g(\T^1,z_1)\cdots R^{\a_n}_g(\T^n,z_n)\>
\end{equation}
\label{avcorr}
\end{subequations}
where the coefficients $W_{j_1 \cdots j_q i_1 \cdots i_p}$ are called the
affine-Virasoro {\em connection moments}.  The simple form of
(\ref{avward}), proportional to the affine-Sugawara correlators $A_g$,
 is due to the simple OPE of the currents with the affine
primary fields,
\begin{equation}
J_a(z)R_g^\a(\T,w)={R_g^\b(\T,w)\over z-w} (\T_a)_\b{}^\a + \mbox{reg.}
\end{equation}
whereas extra inhomogeneous
terms are generally obtained for the biprimary fields of broken
affine secondaries.  See Ref.\  \cite{wi} for discussion of the
corresponding Ward identities associated to the $L^{ab}$-broken currents.

All the connection moments
$W_{j_1\ldots j_q,i_1\ldots i_p}$
may be computed in principle by standard
dispersive techniques from the formula \cite{wi,wi2},
$$
A_g^{\b} (z) {W_{j_1 \ldots j_q, i_1 \ldots i_p } (z)_{\b}}^{\a} =
\;\; \;\;\;\;\;\;\;\;\;\;\; \;\;\;\;\;\;\;\;\;\;
$$
$$
 \left[ \prod_{r=1}^{q} \tL^{a_r b_r} \oint_{z_{j_r}}
{\mbox{d}\o_r \over 2\p i} \oint_{\o_r}  {\mbox{d}\et_r \over 2\p i} \;
{1 \over \et_r-\o_r} \right] \!
\left[ \prod_{s=1}^{p} L^{c_s d_s} \oint_{z_{i_s}} \!
{\mbox{d}\o_{q+s} \over 2\p i} \oint_{\o_{q+s}} \!
{\mbox{d}\et_{q+s} \over 2\p i} \; {1 \over \et_{q+s}-\o_{q+s}} \right]
$$
$$
 \times  \langle
J_{a_1}(\et_1)  J_{b_1}(\o_1)  \ldots J_{a_q}(\et_q)  J_{b_q}(\o_q)
J_{c_1}(\et_{q+1})  J_{d_1}(\o_{q+1})  \ldots
$$
\begin{equation}
 \;\;\;\; \;\;\;\;\;\; \; J_{c_p}(\et_{q+p})  J_{d_p}(\o_{q+p})
R_g^{\a_1} (\T^1,z_1) \ldots  R_g^{\a_n} (\T^n,z_n) \rangle
\label{huh}
\end{equation}
since the required averages are in the affine-Sugawara theory on $g$.
The computations of
the moments increase in complexity with their order $q+p$.

The Ward identities (\ref{avcorr}) prove the
existence of the biconformal correlators \cite{wi},
{\samepage
\begin{subequations}
\begin{equation}
A^\a(\tz,z)
=A_g^\b(z)\tF(\tz,z)_\b{}^\a
=A_g^\b(\tz)F(\tz,z)_\b{}^\a
\label{AFeq}
\end{equation}
\begin{equation}
\tF(\tz,z)=\sum_{q=0}^\infty {1\over q!} \sum_{j_1\ldots j_q}
 (\tz_{j_1}-z_{j_1})\cdots (\tz_{j_q}-z_{j_q})W_{j_1\ldots j_q,0}(z)
\end{equation}
\begin{equation}
F(\tz,z)=\sum_{p=0}^\infty {1\over p!} \sum_{i_1\ldots i_p}
 (z_{i_1}-\tz_{i_1})\cdots (z_{i_p}-\tz_{i_p})W_{0,i_1\ldots i_p}(\tz)
\end{equation}
\label{bicorex}
\end{subequations}}
as the indicated Taylor series expansions around the affine-Sugawara line.

\subsection{Properties of the Connection Moments \label{hc7}}

The connection moments have been computed explicitly through order
$q+p=2$.

The first-order connection moments are \cite{wi},
\begin{equation}
  W_{0,i} = 2 L^{ab} \sum_{j\neq i}^n
{\T_a^i \T_b^j \over z_{ij} }\quad, \quad
W_{i,0} = 2 \tL^{ab} \sum_{j\neq i}^n  { \T_a^i \T_b^j \over z_{ij} }
\label{conn}
\end{equation}
where $\T^i$ is the matrix irrep of the $i$th broken
affine primary field in the bicorrelator.
Note that, because of K-conjugation covariance,
 the sum of the first-order moments is
the affine-Sugawara connection $W_i^g$, which appears in the
Knizhnik-Zamolodchikov (KZ) equation \cite{kz} for $A_g$,
\begin{subequations}
\begin{equation}
W_{i,0}+W_{0,i}=W_i^g=2L_g^{ab}\sum_{j\ne i}^n {\T_a^i\T_b^j \over z_{ij}}
\label{asconn}
\end{equation}
\begin{equation}
\pa_i A_g^\a = A_g^\b(W_i^g)_\b{}^\a
\end{equation}
\begin{equation}
A_g\sum_{i=1}^n\T_a^i =0\quad, \quad a=1\ldots\dim g.
\end{equation}
\end{subequations}
Indeed, the KZ equation is implied by the sum of the first-order Ward
identities,
\begin{equation}
\pa_iA_g(z)=\left.(\tpa_i+\pa_i)A_g(\tz,z)\right|_{\tz=z}
=A_g(z)W_i^g(z)
\label{sum}
\end{equation}
where the first step in (\ref{sum}) is a chain rule.

The second-order connection moments are \cite{wi},
{\samepage
\begin{subequations}
\begin{equation}
W_{0,ij} = \pa_i W_{0,j}+\frac{1}{2} [W_{0,i},W_{0,j}]_+ +E_{0,ij}\quad,\quad
 W_{ij,0} = \pa_i W_{j,0} + \frac{1}{2} [W_{i,0},W_{j,0}]_+ +
E_{ij,0}
\end{equation}
\begin{equation}
W_{i,j} = W_{i,0} W_{0,j} + E_{i, j}
\end{equation}
\begin{equation}
E_{i, j} = - 2i L^{da}L^{e(b} {f_{de}}^{c)} \left\{
{ \T_c^j \T_b^j \T_a^i + \T_c^i \T_b^i \T_a^j\over z_{ij}^2}
- 2 \sum_{k \neq i,j}^n  {\T_c^k \T_b^i \T_a^j \over z_{ij} z_{ik} } \right\}
\quad,\quad i \neq j
\end{equation}
\begin{equation}
E_{0,ij} = -\frac{1}{2} ( E_{i, j} + E_{ j, i} )\quad, \quad
 E_{ij,0} = E_{0,ij}|_{L\ra\tL}\quad, \quad
 E_{ i, i} =-\sum_{j \neq i}^n E_{i ,j}
\end{equation}
\label{order2}
\end{subequations}}
where the Virasoro master equation (\ref{kME2}) was used to obtain the
terms $\pa_iW_{0,j}$ and $\pa_iW_{j,0}$, which are first order
in $L$ and $\tL$ respectively.

The connection moments have also been computed to all orders for the
$g/h$ coset constructions (see Section \ref{hc16}), the higher
affine-Sugawara nests
\cite{wi2,fc}, and the general ICFT at high level on simple $g$
(see Section \ref{hc23}).  Moreover, the leading singularities as
$u\rightarrow 0$ have been obtained for the all-order moments
of the general four-point invariant bicorrelators (See Section \ref{hc13}).

We turn now to some more general properties of the connection moments.

\noindent {\bf A.} Symmetry.
The connection moments $W_{j_1\cdots j_q, i_1\cdots i_p}$ are symmetric
under exchange of any two $j$ labels or any two $i$ labels \cite{wi}.

\noindent {\bf B.} K-conjugation covariance.
The connection moments satisfy \cite{fc}
\begin{equation}
W_{j_1\cdots j_q, i_1\cdots i_p}(\tL,L) =
W_{i_1\cdots i_p, j_1\cdots j_q}(L,\tL)
\label{Kconj}
\end{equation}
under exchange of the K-conjugate theories.

The {\em one-sided} connection moments $W_{j_1\cdots j_q,0}(\tL)$ and
$W_{0,i_1\cdots i_p}(L)$ are functions
of $\tL$ and $L$ respectively, and
satisfy the K-conjugation covariance \cite{fc}
\begin{subequations}
\begin{equation}
W_{j_1\cdots j_q,0}(\tL)=\left.W_{0,j_1\cdots j_q}(L)\right|_{L=\tL}
\end{equation}
\begin{equation}
W_{0,i_1\cdots i_p}(L)=\left.W_{i_1\cdots i_p,0}(\tL)\right|_{\tL=L}
\end{equation}
\end{subequations}
according to (\ref{Kconj}).

\noindent {\bf C.} Consistency relations.
The connection moments satisfy the consistency relations
\begin{equation}
(\pa_i+W_i^g)W_{j_1\cdots j_q, i_1\cdots i_p} =
W_{j_1\cdots j_q i,i_1\cdots i_p} +
W_{j_1\cdots j_q ,i_1\cdots i_p i}
\label{consrel}
\end{equation}
where $W_{00}=\one$ and $W_i^g$ is the affine-Sugawara connection
in (\ref{asconn}).
These relations were originally derived \cite{wi} from simple properties
of the biprimary fields, but they are also the integrability conditions
for the existence of the biconformal correlators.  To understand this,
the reader may wish to consider the simple example
$f_{qp}=\tpa^q\pa^p f(\tu,u)|_{\tu=u}$, which satisfies
$\pa f_{qp}=f_{q+1,p}+f_{q,p+1}$ for all $f(\tu,u)$.

The consistency relations relate the moments
at a fixed order to the moments at one
higher order.  For example,
\begin{subequations}
\begin{equation}
W_i^g=W_{i,0}+W_{0,i}
\label{repeat}
\end{equation}
\begin{equation}
(\pa_i+W_i^g)W_{j,0}=W_{ji,0}+W_{j,i}\quad, \quad
(\pa_i+W_i^g)W_{0,j}=W_{0,ij}+W_{i,j}
\label{consist}
\end{equation}
\label{cons}
\end{subequations}
are obtained through order $q+p\leq2$.  The first consistency
relation (\ref{repeat}) was
encountered in eq.(\ref{asconn}).

More generally, the consistency relations are deeply connected to the
Vira\-soro master equation.  To understand this, note that the moments
of order $q+p$ are naively of order $\tL^qL^p$, whereas the
derivative term in the
consistency relations mixes these orders.
In particular, the consistency relations
(\ref{consist}) are satisfied by the explicit forms (\ref{conn})
and (\ref{order2}) of the first- and second-order moments, but only
because the master equation, whose form is $L\sim L^2$,
 was used in the results for the
second-order moments.

The consistency relations also tell us about a large redundancy in
the connection moments.  In particular, one may solve the relations
to write the general moment in terms of $W_i^g$ and either of the
two sets of one-sided connection moments, for example
\begin{equation}
W_{j_1\cdots j_q,i_1\cdots i_p}=W_{j_1\cdots j_q, i_1\cdots i_p}
(W_i^g,W_{0,i_1,\cdots i_p}).
\end{equation}
As an illustration, one has from eq.(\ref{cons}) that
\begin{subequations}
\begin{equation}
W_{i,0}=W_i^g-W_{0,i}
\end{equation}
\begin{equation}
W_{i,j}=(\pa_i+W_i^g)W_{0,j}-W_{0,ij}
\end{equation}
\begin{equation}
W_{ij,0}=(\pa_i+W_i^g)(W_j^g-W_{0,j})-(\pa_j+W_j^g)W_{0,i}+W_{0,ij}
\end{equation}
\end{subequations}
through order $q+p=2$.  The all-order solution of the consistency
relations is given in Ref.\  \cite{wi2}.

\noindent {\bf D.} Crossing symmetry.
The connection moments satisfy the crossing symmetry \cite{wi2}
\begin{subequations}
\begin{equation}
\qquad
\left.W_{j_1\cdots j_q,i_1\cdots i_p}(z)\right|_{k\leftrightarrow l} =
W_{j_1\cdots j_q,i_1\cdots i_p}(z) \qquad
\end{equation}
\begin{equation}
k\lra l: \qquad
\T^k\leftrightarrow\T^l\quad, \quad
z_k\leftrightarrow z_l\quad, \quad
\left\{ \begin{array}{ll}
  i\ra l & \mbox{when } i=k \\
  i\ra k & \mbox{when } i=l
 \end{array} \right.
\label{exchange}
\end{equation}
\label{crossym}
\end{subequations}
where $i\in (j_1\cdots j_q,i_1\cdots i_p)$, and
$k\leftrightarrow l$ means exchange of $\T$'s, $z$'s, and
indices, as shown in (\ref{exchange}).

\noindent {\bf E.} $SL(2,\R)\times SL(2,\R)$ covariance.
$SL(2,\R)\times SL(2,\R)$ covariance in biconformal field theory was
discussed in Section \ref{hc5}.
For the connection moments, this covariance gives the
$L$-relations \cite{fc}
\begin{subequations}
\begin{equation}
A_g \sum_{i=1}^n W_{j_1  \ldots j_q, i_1 \ldots i_p i } = 0
\label{L0}
\end{equation}
\begin{equation}
A_g \left[ \sum_{i=1}^n z_i W_{j_1 \ldots j_q, i_1 \ldots i_p i} +
(  \sum_{i=1}^n \D_i + p ) W_{j_1 \ldots j_q ,i_1 \ldots i_p} \right] = 0
\end{equation}
\begin{eqnarray}
\;\;\;A_g \Bigl[ \sum_{i=1}^n z_i^2 W_{j_1 \ldots j_q,i_1 \ldots i_p i}
& + & 2( \sum_{i=1}^n z_i \D_i + \sum_{\n =1}^p z_{i_{\n}} )
W_{j_1 \ldots j_q,i_1 \ldots i_p }
\\
& + & 2 \sum_{\n =1}^ p (\D_{i_{\n}} + \sum_{\m = \n +1}^p \d_{i_{\n},i_{\m}} )
W_{j_1 \ldots j_q, i_1 .. \hat{i}_{\n} .. i_p } \Bigr] = 0
\nonumber
\label{omission}
\end{eqnarray}
\label{lrelations}
\end{subequations}
and analogous $\tL$ relations are obtained from (\ref{lrelations}) with
$\D\rightarrow\tD$.  In (\ref{omission}), the omission of an
index is denoted by a hat.  It is believed that the $L(0)$ relation
in (\ref{L0}) is also true without the factor $A_g$, but this
has not yet been demonstrated beyond order $q+p=2$.

\noindent {\bf F.} Translation sum rule.
{}To state this relation, one first
needs the (invertible) {\em evolution operator}
$A_g(z,z_0)$ of the affine-Sugawara construction on $g$, which satisfies
the KZ boundary value problem
\begin{equation}
\pa_i A_g(z,z_0)_\a{}^\b=A_g(z,z_0)_\a{}^\g W_i^g(z)_\g{}^\b\quad, \quad
A_g(z_0,z_0)_\a{}^\b=\d_\a{}^\b
\label{evolution}
\end{equation}
where $z_0$ is a regular reference point.  It follows that
{\samepage
\begin{subequations}
\begin{equation}
A_g(z,z_0)=P^*e^{\int_{z_0}^z dz_i' W_i^g(z')}
\end{equation}
\begin{equation}
A_g^{-1}(z,z_0)=A_g(z_0,z)
\end{equation}
\begin{equation}
A_g(z_3,z_1)A_g(z_2,z_3)=A_g(z_2,z_1)
\end{equation}
\begin{equation}
A_g^\a(z)=A_g^\b(z_0) A_g(z,z_0)_\b{}^\a
\end{equation}
\end{subequations} }
where $P^*$ is anti-ordering in $z$ and
$A_g^\a(z)$ is the affine-Sugawara correlator in (\ref{ascorr}).  Then
the translation sum rule \cite{wi2}
$$
\sum_{r,s=0}^{\infty} { 1 \over r ! \, s !}
\sum_{l_1 \ldots l_r}^n
\sum_{k_1 \ldots k_s}^n  \left[ \prod_{\m=1}^r (z_{l_{\m}} -z^0_{l_{\m}})
\right]  \left[\, \prod_{\n=1}^s (z_{k_{\n}} -z^0_{k_{\n}} )  \right]
W_{l_1 \ldots l_r j_1 \ldots j_q, k_{1} \ldots k_s i_1 \ldots i_p} (z_0)
$$
\begin{equation}
 = A_g(z,z_0) W_{j_1 \ldots j_q ,i_1 \ldots i_p} (z)
\label{transum}
\end{equation}
relates the connection moments at different points.

\subsection{Knizhnik-Zamolodchikov Null States \label{hc8}}

The Ward identities (\ref{avcorr}) and the result (\ref{huh}) for the
connection
moments were derived by standard OPE methods, using the
biprimary fields and current algebra, but many of these identities
can also be derived \cite{wi}
from null states of the Knizhnik-Zamolodchikov (KZ)
type, which live in the enveloping algebra of the affine algebra.
This is the method used by Knizhnik and Zamolodchikov in the original
derivation of the KZ equations \cite{kz}.

For example, the first KZ null state is
\begin{equation}
|\chi(\T)\>_1^\a=L(-1)|R_g(\T)\>^\a-2L^{ab}J_a(-1)|R_g(\T)\>^\b(\T_b)_\b{}^\a=0
\label{null}
\end{equation}
where $L^{ab}$ is any solution of the Virasoro master equation and
$|R_g(\T)\>^\a, \a=1\ldots\dim\T$ is the broken affine primary state
corresponding to matrix irrep $\T$.  Using the identities,
\begin{subequations}
\begin{equation}
\<0|R_g^{\a_1}(\T^1,z_1)\cdots R_g^{\a_n}(\T^n,z_n)|\chi(\T)\>_1^\a=0
\end{equation}
\begin{equation}
[J_a(-1),R_g^\a(\T,z)]=z^{-1}R_g^\b(\T,z)(\T_a)_\b{}^\a
\end{equation}
\begin{equation}
[\tL(-1),R_g^\a(\T,z)]=\left.\tpa R^\a(\T,\tz,z)\right|_{\tz=z}
\end{equation}
\label{identities}
\end{subequations}
this null state implies the first-order Ward identity
$\pa_iA|_{\tz=z}=A_gW_{0,i}$ and the form of the first moment $W_{0,i}$.
Similarly the Ward identity $\tpa_iA|_{\tz=z}=A_gW_{i,0}$ and the form of
$W_{i,0}$ is implied by the K-conjugate null state obtained
from (\ref{null}) by $L\rightarrow\tL$.
The special case of (\ref{null}--2)
with $L=L_g$ was used by Knizhnik and
Zamolodchikov in their derivation of the KZ equations for the
affine-Sugawara correlators.

More generally, one may define the $m$th null state of the KZ type as
\begin{subequations}
\begin{equation}
|\chi(\T)\>_m^\a = |m,L\>^\a-|m,J\>^\a=0\quad, \quad \a=1\ldots\dim\T
\end{equation}
\begin{equation}
|m,L\>^\a=\left(L(-1)\right)^m|R_g(\T)\>^\a
\end{equation}
\begin{equation}
L(-1)=2L^{ab}\sum_{n=0}^\infty J_a(-n-1)J_b(n)
\end{equation}
\end{subequations}
where $|m,J\>^\a$ is obtained by rewriting $|m,L\>^\a$ in terms of
negatively moded currents on $|R_g(\T)\>$.  This is done by moving the
non-negatively moded currents to the right, using eq.(\ref{Jbrok}) and
the current algebra (\ref{kaffine}).  As examples, one has
\begin{subequations}
\begin{equation}
|1,J\>^\a=2L^{ab}J_a(-1)|R_g(\T)\>^\b(\T_b)_\b{}^\a
\end{equation}
\begin{eqnarray}
|2,J\>^\a = 4L^{ab}L^{cd} &\Bigl\{ &
J_c(-1)J_e(-1)|R_g(\T)\>^\b if_{da}{}^e(\T_b)_\b{}^\a
\nonumber \\  &+ &
J_c(-1)J_a(-1)|R_g(\T)\>^\b(\T_d\T_b)_\b{}^\a \\ &+& \left.
J_c(-2)|R_g(\T)\>^\b\left(if_{da}{}^e(\T_e\T_b)_\b{}^\a+G_{da}(\T_b)_\b{}^\a
\right)\right\}. \nonumber
\end{eqnarray}
\end{subequations}
Then the null state identities
\begin{equation}
\<0|R_g^{\a_1}(\T^1,z_1)\cdots R_g^{\a_n}(\T^n,z_n)|\chi(\T)\>_m^\a=0
\end{equation}
imply the subset of Ward identities
\begin{equation}
\left.\pa_i^mA\right|_{\tz=z} = A_gW_{0,\underbrace{i\cdots i}_m}
\end{equation}
and the form of the associated connection moments.
Similarly, the null states with $L\ra\tL$ give the connection moments
$W_{i\cdots i,0}$, but it
is not yet clear whether the rest of the connection moments can be computed
from other null states of the KZ type.

\subsection{Invariant Ward Identities \label{hc9}}

Using the Ward identities (\ref{avcorr}) and the
$SL(2,\R)\times SL(2,\R)$ decomposition (\ref{decomp}) for the
four-point bicorrelators $A^\a$ in the KZ gauge, one obtains the
{\em invariant Ward identities} \cite{wi}
\begin{equation}
\left.\tpa^q\pa^p Y^\a(\tu,u)\right|_{\tu=u} =
Y_g^\b(u)W_{qp}(u)_\b{}^\a\quad, \quad
\a=(\a_1,\a_2,\a_3,\a_4)
\end{equation}
for the invariant four-point bicorrelators $Y(\tu,u)$.  Here $Y_g$ is the
invariant four-point affine-Sugawara correlator and
$W_{qp}$ are the {\em invariant connection moments}.  This simpler
system inherits many of the properties of the full Ward identities,
some of which are listed below.

\noindent {\bf A.} First-order moments and invariant KZ equation \cite{wi}.
\begin{subequations}
\begin{equation}
W_{10}=2\tL^{ab}\left({\T_a^1\T_b^2 \over u}
 +{\T_a^1\T_b^3 \over u-1}\right)\quad, \quad
W_{01}=2L^{ab}\left({\T_a^1\T_b^2 \over u}
 +{\T_a^1\T_b^3 \over u-1}\right)
\end{equation}
\begin{equation}
W_{10}+W_{01}=W^g=2L_g^{ab}\left({\T_a^1\T_b^2 \over u}
 +{\T_a^1\T_b^3 \over u-1}\right)
\label{invascorr}
\end{equation}
\begin{equation}
\pa Y_g(u)=\left.(\tpa+\pa)Y_g(\tu,u)\right|_{\tu=u}=Y_g(u)W^g(u)
\end{equation}
\begin{equation}
Y_g\sum_{i=1}^4 \T_a^i=0\quad, \quad a=1\ldots\dim g
\end{equation}
\label{1andKZ}
\end{subequations}
\noindent {\bf B.} Second-order moments \cite{wi}.
\begin{subequations}
\begin{equation}
W_{02}=\pa W_{01}+W_{01}^2+E_{02}\quad, \quad
W_{20}=\pa W_{10}+W_{10}^2+E_{20}
\end{equation}
\begin{equation}
W_{11}=W_{10}W_{01}-E_{02}=W_{01}W_{10}-E_{20}
\end{equation}
\begin{equation}
E_{02}=-2iL^{da}L^{e(b}f_{de}{}^{c)}V_{abc}\quad, \quad
E_{20}=\left.E_{02}\right|_{L\rightarrow\tL}
\end{equation}
\begin{eqnarray}
V_{abc}&=&{1\over u^2} \left[\T_a^1\T_b^2\T_c^2+\T_a^2\T_b^1\T_c^1\right]
 +{2\over u(u-1)}\T_a^1\T_b^2\T_c^3
\nonumber \\
& &+{1\over(u-1)^2}\left[\T_a^1\T_b^3\T_c^3+\T_a^3\T_b^1\T_c^1\right]
\end{eqnarray}
\label{2mom}
\end{subequations}
\noindent {\bf C.} Invariant bicorrelators in terms of invariant moments
\cite{fc}.
\begin{subequations}
\begin{equation}
Y(\tu,u)
=Y_g(u)\tilde F(\tu,u)
=Y_g(\tu)F(\tu,u)
\label{YFeq}
\end{equation}
\begin{equation}
\tF(\tu,u)=\sum_{q=0}^\infty {(\tu-u)^q \over q!} W_{q0}(u)
\end{equation}
\begin{equation}
F(\tu,u)=\sum_{p=0}^\infty {(u-\tu)^p \over p!} W_{0p}(\tu)
\end{equation}
\end{subequations}
\noindent {\bf D.} K-conjugation covariance \cite{wi2}.
\begin{subequations}
\begin{equation}
W_{qp}(\tL,L)=W_{pq}(L,\tL)
\end{equation}
\begin{equation}
W_{q,0}(\tL)=\left.W_{0,q}(L)\right|_{L=\tL}\quad, \quad
W_{0,p}(L)=\left.W_{p,0}(\tL)\right|_{\tL=L}
\end{equation}
\end{subequations}
\noindent {\bf E.} Consistency relations \cite{wi}.
\begin{equation}
(\pa+W^g)W_{qp}=W_{q+1,p}+W_{q,p+1}\quad, \quad W_{00}=\one
\label{invcons}
\end{equation}
The solution of these relations in terms of the one-sided
invariant moments $W_{q0}$ or $W_{0p}$ is given in Ref.\  \cite{wi2}.

\noindent {\bf F.} Crossing symmetry \cite{wi2}.
\begin{subequations}
\begin{equation}
W_{qp}(1-u)=(-1)^{q+p}P_{23}W_{qp}(u)P_{23}
\end{equation}
\begin{equation}
P_{23}\T^2 P_{23}=\T^3\quad, \quad
P_{23}^2=1
\end{equation}
\end{subequations}
\noindent {\bf G.} Translation sum rule \cite{wi2}.
\begin{equation}
\sum_{r,s=0}^\infty {(u-u_0)^{r+s} \over r!s!}W_{r+q,s+p}(u_0) =
 Y_g(u,u_0)W_{qp}(u)
\label{invevo}
\end{equation}
The invariant evolution operator $Y_g(u,u_0)$ in (\ref{invevo}) satisfies
\begin{subequations}
\begin{equation}
\pa Y_g(u,u_0)_\a{}^\b=Y_g(u,u_0)_\a{}^\g W^g(u)_\g{}^\b\quad, \quad
Y_g(u_0,u_0)_\a{}^\b=\d_\a{}^\b
\end{equation}
\begin{equation}
Y_g(u,u_0)=U^* e^{\int_{u_0}^u du' W^g(u')}
\end{equation}
\begin{equation}
Y_g^{-1}(u,u_0)_\a{}^\b=Y_g(u_0,u)_\a{}^\b
\end{equation}
\begin{equation}
Y_g^\a(u)=Y_g^\b(u_0) Y_g(u,u_0)_\b{}^\a
\end{equation}
\label{invevop}
\end{subequations}
where $U^*$ denotes anti-ordering in $u$ and $W^g$ is the invariant
affine-Sugawara connection in (\ref{invascorr}).

See also the all-order invariant coset connection moments
in Section \ref{hc19}, the invariant high-level connection moments in
Section \ref{hc26}, and the all-order $u\rightarrow 0$ singularities of
the invariant connections in Section \ref{hc13}.

\subsection{The Generalized KZ Equations of ICFT \label{hc10}}

The Ward identities (\ref{avcorr}) may be reorganized as linear
differential systems with flat connections \cite{fc}.  In this form,
it is clear that the Ward identities are {\em generalized KZ equations} which
include the
usual KZ equations as a special case.

The generalized KZ equations for the bicorrelators are
\begin{subequations}
\begin{equation}
\tpa_i A(\tz,z)=A(\tz,z)\tilde W_i(\tz,z)
\end{equation}
\begin{equation}
\pa_i A(\tz,z)=A(\tz,z) W_i(\tz,z)
\end{equation}
\begin{equation}
A(z,z)=A_g(z)
\label{notagain}
\end{equation}
\label{bdiffeq}
\end{subequations}
where $\tilde W_i$ and $W_i$ are the {\em flat connections of ICFT}.  For full
equivalence of this system
with the Ward identities, one must include the
affine-Sugawara boundary condition in (\ref{notagain}).

The equations
(\ref{bdiffeq}a,b) can be derived by differentiating
the relations (\ref{AFeq}), which also gives the explicit form of the
connections
\begin{subequations}
\begin{equation}
\qquad\qquad
\label{WFeq}
\tilde W_i=\tilde F^{-1}\tpa_i \tilde F\quad, \quad
W_i=F^{-1}\pa_i F
\qquad\qquad
\end{equation}
\begin{equation}
\qquad
\tilde F(\tz,z) = \sum_{q=0}^{\infty} {1 \over q !} \sum_{j_1 \ldots j_q}
\prod_{\n=1}^q (\tz_{j_\n} - z_{j_\n } ) W_{j_1 \ldots j_q,0} (z)
\qquad
\end{equation}
\begin{equation}
\qquad
 F(\tz,z) = \sum_{p=0}^{\infty} {1 \over p !} \sum_{i_i \ldots i_p}
\prod_{\m=1}^p (z_{i_\m} - \tz_{i_\m } ) W_{0,i_1 \ldots i_p} (\tz)
\qquad
\end{equation}
\label{explconn}
\end{subequations}
in terms of the connection moments.  The flatness conditions
\begin{subequations}
\begin{equation}
\tpa_i \tilde W_j - \tpa_j \tilde W_i + [\tilde W_i,\tilde W_j] =0
\end{equation}
\begin{equation}
\pa_i W_j - \pa_j W_i + [W_i,W_j] =0
\end{equation}
\begin{equation}
(\pa_i + W_i) \tilde W_j = (\tpa_j + \tilde W_j) W_i
\end{equation}
\label{expflat}
\end{subequations}
follow from the generalized KZ equations
 or the explicit forms of the connections in (\ref{explconn}).

According to eq.(\ref{WFeq}), the quantities $\tilde F$ and $F$ satisfy
\begin{subequations}
\begin{equation}
\tpa_i \tilde F=\tilde F\tilde W_i\quad, \quad \pa_i F=F W_i
\end{equation}
\begin{equation}
\tilde F(z,z)=F(z,z)=\one
\end{equation}
\label{Fdiff}
\end{subequations}
which identifies these quantities as the (invertible) {\em evolution operators
of the flat connections},
\begin{equation}
\tilde F(\tz,z) = P^* e^{\int_z^\tz d\tz_i' \tilde W_i(\tz',z) }\quad, \quad
F(\tz,z) = P^* e^{\int_\tz^z dz_i' W_i(\tz,z')}
\end{equation}
where $P^*$ is anti-path ordering.

\vskip .4cm
\noindent \un{Properties of the generalized KZ equations}
\vskip .3cm

\noindent {\bf A.} Formulae for the connections.
The formulae (\ref{huh}) for the connection moments can be
reexpressed as formulae for the flat connections,
\begin{subequations}
$$
 A^{\b}  (\tz,z) \tW_i (\tz,z)_{\b}{}^{\a}
\hspace*{4in}
$$
\begin{equation}
 = \oint_{\tz_i} { d w \over 2 \pi i} \oint_w { d \eta \over 2\pi i}
 {\tL^{ab} \over \eta -w} \langle J_a(\eta) J_b (w)
R^{\a_1} (\T^1,\tz_1,z_1) \cdots R^{\a_n} (\T^n ,\tz_n,z_n) \rangle
\end{equation}
$$
 A^{\b}  (\tz,z) W_i (\tz,z)_{\b}{}^{\a}
\hspace*{4in}
$$
\begin{equation}
 = \oint_{z_i} {d w \over 2 \pi i} \oint_w { d \eta \over 2\pi i} {
L^{ab} \over \eta -w} \langle J_a(\eta) J_b (w)
R^{\a_1} (\T^1,\tz_1,z_1) \cdots R^{\a_n} (\T^n ,\tz_n,z_n) \rangle.
\end{equation}
\label{genfform}
\end{subequations}
The explicit forms of the flat connections for the coset constructions, the
higher
affine-Sugawara nests, and the general ICFT at high level on compact $g$ are
discussed in Sections \ref{hc16}  and \ref{hc23}.

\noindent {\bf B.} Inversion formula.
The commuting differential operators formed from the flat connections,
\begin{subequations}
\begin{equation}
  \ctD_i (\tz,z) = \tpa_i + \tW_i(\tz,z)\quad, \quad
\cD_i (\tz,z) = \pa_i + W_i(\tz,z)
\end{equation}
\begin{equation}
[\ctD_i,\ctD_j ] = [\cD_i ,\cD_j] = [\cD_i, \ctD_j] =0
\end{equation}
\end{subequations}
are called the covariant derivatives.  The connection moments may
be recovered from the flat connections by the inversion formula
\begin{equation}
 W_{j_1\ldots j_q,i_1\ldots i_p} = \left.\ctD_{j_1} \cdots \ctD_{j_q}
\cD_{i_1} \cdots \cD_{i_p} \one \right|_{\tz = z}
\label{inversion}
\end{equation}
which is the inverse of eq.(\ref{explconn}).  As examples of the inversion
formula, one has
\begin{equation}
W_{0,0}=\one\quad, \quad W_{i,0}(z)=\tW_i(z,z)\quad, \quad W_{0,i}(z)=W_i(z,z)
\end{equation}
where $\tW_i(z,z)$ and $W_i(z,z)$ are called the {\em pinched connections}.
{}From eq.(\ref{conn}), one obtains the explicit form of the
pinched connections,
\begin{subequations}
\begin{equation}
\tW_i(z,z)=2\tL^{ab}\sum_{j\ne i}{\T_a^i \T_b^j \over z_{ij} }\quad, \quad
W_i(z,z)=2L^{ab}\sum_{j\ne i}{\T_a^i \T_b^j \over z_{ij} }
\end{equation}
\begin{equation}
\tW_i(z,z)+W_i(z,z)=W_i^g(z)=2L^{ab}_g \sum_{j\ne i}
      {\T_a^i \T_b^j \over z_{ij}}
\label{pinchedaff}
\end{equation}
\end{subequations}
where $W_i^g$ are the affine-Sugawara connections.  The
inversion formula is worked out to higher order in Ref.\  \cite{fc}.

\noindent {\bf C.} $SL(2,\R)\times SL(2,\R)$ covariance.
The bicorrelators satisfy the  $SL(2,\R)\times SL(2,\R)$ relations,
\begin{subequations}
\begin{equation}
A \sum_{i=1}^n \tW_i = A \sum_{i=1}^n (\tz_i \tW_i + \tD_i )  =
A \sum_{i=1}^n (\tz_i^2 \tW_i + 2 \tz_i \tD_i ) = 0
\end{equation}
\begin{equation}
 A \sum_{i=1}^n W_i = A \sum_{i=1}^n (z_i W_i + \D_i )  =
A \sum_{i=1}^n (z_i^2 W_i + 2 z_i \D_i ) = 0 .
\end{equation}
\label{brelations}
\end{subequations}
The  $SL(2,\R)\times SL(2,\R)$ moment relations (\ref{lrelations})
follow by multiple differentiation from (\ref{brelations}),
using the inversion
formula (\ref{inversion}).

\noindent {\bf D.} $\tL$ and $L$ dependence.
The one-sided connection moments $W_{j_1\cdots j_q,0}(\tL,z)$ and
$W_{0,i_1\cdots i_p}(L,z)$ are functions of $\tL$ and $L$ respectively,
so the evolution operators and flat connections are also functions only
of $\tL$ or $L$,
\begin{equation}
\tF(\tL,\tz,z)\quad, \quad F(L,\tz,z)\quad, \quad
\tW_i(\tL,\tz,z)\quad, \quad W_i(L,\tz,z).
\end{equation}

\noindent {\bf E.} K-conjugation covariance.
Under K-conjugation the evolution operators, flat connections, and
bicorrelators satisfy
\begin{subequations}
\begin{equation}
\tF (\tL,\tz,z) = F(L ,z,\tz)|_{L=\tL}\quad, \quad
F (L ,\tz,z) = \tF (\tL ,z,\tz)|_{\tL=L}
\end{equation}
\begin{equation}
\tW_i (\tL ,\tz,z) = W_i (L,z,\tz)|_{L=\tL}\quad, \quad
W_i (L,\tz,z) = \tW_i (\tL ,z,\tz)|_{\tL=L}
\end{equation}
\begin{equation}
\left.A(\tz,z) \right|_{\tL \lra L}  = A(z,\tz).
\end{equation}
\end{subequations}
\noindent {\bf F.} Crossing symmetry.
The evolution operators and flat connections satisfy the crossing relations
\begin{subequations}
\begin{equation}
\left.\tF (\tz,z) \right|_{k\lra l} = \tF(\tz,z)\quad, \quad
\left.F (\tz,z) \right|_{k\lra l} = F(\tz,z)
\end{equation}
\begin{equation}
\left.\tW_i (\tz,z) \right|_{k\lra l} = \tW_i (\tz,z)\quad, \quad
\left.W_i  (\tz,z) \right|_{k\lra l} = W_i (\tz,z)
\end{equation}
\begin{equation}
k\lra l: \qquad
\T^ k \lra \T^l\quad, \quad
\begin{array}{l}
\tz_k \lra \tz_l \cr
z_k \lra z_l
\end{array}\quad, \quad
\left\{ \begin{array}{ll}
i \ra l &\mbox{when }i=k \cr
i \ra k &\mbox{when }i=l
\end{array}\right.
\label{exch2}
\end{equation}
\end{subequations}
which follows from (\ref{crossym}).  Note that $k\lra l$ now includes the
exchange of $\tz$'s, as shown in (\ref{exch2}).

\noindent {\bf G.} More on the evolution operators.
The evolution operators $\tF$ and $F$ are related by the evolution
operator $A_g$ of the affine-Sugawara construction,
\begin{equation}
\tF (\tz,z) = A_g (\tz,z) F(\tz,z)\quad, \quad
F(\tz,z) = A_g (z,\tz) \tF (\tz ,z )
\end{equation}
and hence the evolution operator of the affine-Sugawara construction
is composed of the evolution operators of the flat connections,
\begin{equation}
A_g(\tz,z)=\tF(\tz,z)F^{-1}(\tz,z)=F(z,\tz)\tF^{-1}(z,\tz).
\label{composition}
\end{equation}
The relations (\ref{composition}) give a decomposition of the
affine-Sugawara operator for each pair of K-conjugate theories, mirroring
the basic composition law $T_g=\tT+T$ of ICFT.

The evolution operators also satisfy the differential relations
\begin{subequations}
\begin{equation}
(\pa_i + W_i^g (z) ) \tF (\tz,z)
= \tF(\tz,z) W_i (z)
\end{equation}
\begin{equation}
(\tpa_i + W_i^g (\tz) ) F(\tz,z)  = F(\tz,z) \tW_i (\tz)
\end{equation}
\end{subequations}
which supplement the differential relations in (\ref{Fdiff}).

\noindent {\bf H.} Relation to the conventional KZ equations.
The generalized KZ equations imply the conventional KZ equations \cite{kz}
by chain rule,
\begin{equation}
\pa_i A_g(z)=(\tpa_i+\pa_i)A(\tz,z)|_{\tz=z}=A_g(z)W_i^g(z)
\label{chainrule}
\end{equation}
using (\ref{bdiffeq}) and (\ref{pinchedaff}).

Moreover, the conventional KZ equation is included as the simplest case
of the generalized KZ equations, which read
\begin{subequations}
\begin{equation}
\tpa_i A=AW_i^g(\tz)\quad, \quad \pa_i A=0
\end{equation}
\begin{equation}
A(z,z)=A_g(z)
\end{equation}
\label{ordkz}
\end{subequations}
when $\tL=L_g$ and $L=0$.  It follows that the bicorrelator is the
affine-Sugawara correlator $A(\tz,z)=A_g(\tz)$ in this case.

\subsection{Non-Local Conserved Quantities\label{hc11}}

A remarkable set of new non-local conserved quantities \cite{fc}
is associated to the generalized KZ equations.

{}To understand these quantities, we first review the $g$-global Ward
identity \cite{kz} of the affine-Sugawara construction,
\begin{subequations}
\begin{equation}
A_g Q_a^g=0\quad, \quad Q_a^g=\sum_{i=1}^n \T_a^i\quad, \quad a=1\ldots\dim g
\label{conserved}
\end{equation}
\begin{equation}
[Q_a^g,Q_b^g]=if_{ab}{}^c Q_c^g
\end{equation}
\end{subequations}
where $Q_a^g$ are the conserved global generators of Lie $g$.  The
global generators are conserved by the KZ equation
in the sense that $A_gQ_a^g=0$
follows for
all $z$ if it is imposed at an initial point $z_0$,
\begin{equation}
A_g(z_0)Q_a^g=0 \rightarrow A_g(z)Q_a^g=0.
\end{equation}
It follows that the complete KZ system may be written as an initial
value problem,
\begin{subequations}
\begin{equation}
\pa_i A_g(z)=A_g(z)W_i^g(z)
\label{diffsys}
\end{equation}
\begin{equation}
A_g(z_0)Q_a^g=0
\label{asbound}
\end{equation}
\label{KZsys}
\end{subequations}
by including the $g$-global Ward identity only at some initial point $z_0$.

Through the affine-Sugawara boundary condition (\ref{notagain}), the
generalized KZ equations (\ref{bdiffeq}a,b) inherit this structure as
a set of conserved non-local generators $Q_a(\tz,z)$ of $g$,
\begin{subequations}
\begin{equation}
A(\tz,z)Q_a(\tz,z)=0
\label{stillcons}
\end{equation}
\begin{eqnarray}
Q_a(\tz,z)=\tF^{-1}(\tz,z)Q_a^g \tF(\tz,z)
  =  F^{-1}(\tz,z)Q_a^g F(\tz,z)
\label{Qadef}
\end{eqnarray}
\begin{equation}
[Q_a(\tz,z),Q_b(\tz,z)]=if_{ab}{}^c Q_c(\tz,z)
\end{equation}
\begin{equation}
Q_a(z,z)=Q_a^g(z)
\end{equation}
\end{subequations}
where $\tF$ and $F$ are the evolution operators of the flat
connections.
This result, which is easily verified from (\ref{AFeq}) and
(\ref{conserved}), is the lift
of the $g$-global Ward identity into the generalized KZ equations.

The non-local generators are again conserved in the sense that
(\ref{stillcons})
follows for all ($\tz,z$) if the condition is imposed at a reference
point $(\tz_0,z_0)$,
\begin{equation}
A(\tz_0,z_0)Q_a(\tz_0,z_0)=0\rightarrow A(\tz,z)Q_a(\tz,z)=0.
\end{equation}
This relation follows from the covariant constancy of the non-local
generators,
\begin{subequations}
\begin{equation}
\tilde D_i Q_a = \tpa_iQ_a+[\tW_i,Q_a]=0\quad, \quad
D_iQ_a=\pa_iQ_a+[W_i,Q_a]=0
\end{equation}
\end{subequations}
which in turn implies the covariant constancy of $AQ_a$.

It follows that the complete generalized KZ system
 (\ref{bdiffeq}) may
be cast as the initial value problem,
\begin{subequations}
\begin{equation}
\tpa_iA=A\tW_i\quad, \quad
\pa_iA=AW_i
\label{usethis}
\end{equation}
\begin{equation}
A(\tz_0,z_0)Q_a(\tz_0,z_0)=0
\label{eventhis}
\end{equation}
\label{flatsys}
\end{subequations}
in analogy to the complete KZ system (\ref{KZsys}).

In the initial value
formulation (\ref{flatsys}), the
affine-Sugawara correlator $A_g$ is recovered by the definition
$A_g(z)\equiv A(z,z)$,
and the complete KZ system (\ref{KZsys}) is implied by the complete
generalized KZ system
(\ref{flatsys})
as follows:  The KZ equation (\ref{diffsys}) follows by the
chain rule (\ref{chainrule}), and
the $g$-global Ward identity (\ref{asbound}) follows from (\ref{eventhis})
at $\tz_0=z_0$.

Another property of the non-local generators is that they reduce to the
expected global generators of $h$ when $h\subset g$ is an ordinary (spectral)
symmetry of the construction, that is, for the Lie $h$-invariant CFTs
discussed in Section \ref{ca}.  We will review this for the $g/h$
coset constructions in Sections \ref{hc17} and \ref{hc18}.

A challenging next step in this
direction is to understand these non-local conserved
quantities
at the level of the generic world-sheet action of ICFT
(see Section \ref{hc41}), or
at the operator level,
where they may be related to the parafermionic currents
\cite{par1,par2,par3,par4,par5}
of the coset constructions (see also Section \ref{hc18}).

\subsection{Invariant Flat Connections \label{hc12}}
\label{invflatsec}

Following Sections \ref{hc9} and \ref{hc10}, one arrives at the generalized KZ
equations for the invariant four-point bicorrelators \cite{fc},
\begin{subequations}
\begin{equation}
\tpa Y (\tu,u) = Y (\tu,u)  \tW (\tu,u)\quad, \quad
\pa Y (\tu,u)  = Y (\tu,u) W (\tu,u)
\end{equation}
\begin{equation}
 Y(\tu,u) Q_a (\tu,u) =0\quad,\quad a=1  \ldots \dim g
\end{equation}
\end{subequations}
where $\tW, W$ are the invariant flat connections and $Q_g(\tu,u)$ are
the invariant non-local conserved generators of $g$.

We give a partial list of results for the invariant systems, analogous
to those of Sections \ref{hc10} and \ref{hc11}.

\noindent {\bf A.} Flatness condition.
\begin{equation}
(\pa + W) \tW = (\tpa + \tW) W
\end{equation}

\noindent {\bf B.} Inversion formula.
\begin{equation}
W_{qp} = \left.(\tpa + \tW )^q (\pa + W)^p \one \right|_{\tu = u}
\label{flatinv}
\end{equation}

\noindent {\bf C.} Pinched connections and KZ equation.
\begin{subequations}
\begin{equation}
\tW (u,u) = W_{10}(u) = 2\tL^{ab} \left( { \T_a^1 \T_b^2  \over u}
+ { \T_a^1 \T_b^3  \over u-1} \right)
\end{equation}
\begin{equation}
 W (u,u) = W_{01}(u) = 2L^{ab} \left( { \T_a^1 \T_b^2  \over u}
+ { \T_a^1 \T_b^3  \over u-1} \right)
\end{equation}
\begin{equation}
\tW (u,u) + W(u,u) =  W^g(u) = 2L_g^{ab} \left( { \T_a^1 \T_b^2  \over u}
+ { \T_a^1 \T_b^3  \over u-1} \right)
\end{equation}
\begin{equation}
Y^\a(u,u)=Y_g^\a(u)\quad, \quad
\pa Y_g^\a(u)=Y_g^\b(u)W^g(u)_\b{}^\a
\end{equation}
\label{pinchconn}
\end{subequations}
\noindent {\bf D.} Invariant evolution operators of the flat connections.
\begin{subequations}
\begin{equation}
 Y(\tu,u) = Y_g (u) \tF (\tu,u) = Y_g(\tu) F (\tu,u)
\label{YFeq2}
\end{equation}
\begin{equation}
 \tF(\tu,u) \equiv U^* e^{\int_{u}^{\tu} {\rm d} \tu ' \tW(\tu ',u) }
= \sum_{q=0}^{\infty}
 { (\tu-u)^q \over q!} W_{q0}(u)
\end{equation}
\begin{equation}
 F(\tu,u) \equiv U^* e^{\int_{\tu}^u {\rm d} u ' W(\tu,u') }
= \sum_{p=0}^{\infty}
  { (u-\tu)^p \over p!}  W_{0p} (\tu)
\end{equation}
\label{invevof}
\end{subequations}
\begin{subequations}
\begin{equation}
\tF (\tu,u)  = Y_g (\tu,u)  F (\tu,u)
\end{equation}
\begin{equation}
Y_g(\tu,u)=\tF(\tu,u)F^{-1}(\tu,u)
\end{equation}
\begin{equation}
\tW = \tF^{-1} \tpa \tF = F^{-1} (\tpa + W^g(\tu) ) F
\label{useless1}
\end{equation}
\begin{equation}
 W = F^{-1} \pa F = \tF^{-1} (\pa + W^g(u) ) \tF
\label{useless2}
\end{equation}
\label{invevof2}
\end{subequations}
where $U^*$ is anti-ordering in $u$ and $Y_g(u,u_0)$ is the invariant
evolution operator of $g$ in (\ref{invevop}).

\noindent {\bf E.} Invariant non-local conserved generators.
\begin{subequations}
\begin{equation}
Q_a (\tu,u)
= \tF^{-1}(\tu,u)Q^g_a \tF(\tu,u)
= F^{-1}(\tu,u) Q^g_a F(\tu,u)
\end{equation}
\begin{equation}
[Q_a (\tu,u) , Q_b (\tu,u) ] = i f_{ab}{}^c Q_c(\tu,u)
\end{equation}
\begin{equation}
Y_g (u) Q^g_a =0 \lra Y(\tu_0,u_0) Q_a(\tu_0,u_0) =0 \lra
Y(\tu,u) Q_a(\tu,u) =0
\end{equation}
\end{subequations}
\noindent {\bf F.} K-conjugation.
\begin{subequations}
\begin{equation}
\tF (\tL ,\tu,u) = F(L,u,\tu)|_{L=\tL} \quad, \quad
F (L,\tu,u) = \tF (\tL ,u,\tu)|_{\tL=L}
\end{equation}
\begin{equation}
\tW (\tL ,\tu,u) = W (L ,u,\tu)|_{L=\tL} \quad, \quad
W (L,\tu,u) = \tW (\tL,u,\tu)|_{\tL=L}
\end{equation}
\begin{equation}
\left.Y(\tu,u) \right|_{\tL \lra L} = Y(u,\tu)
\end{equation}
\begin{equation}
\left.Q_a(\tu,u) \right|_{\tL \lra L} = Q_a(u,\tu)
\end{equation}
\end{subequations}
\noindent {\bf G.} Crossing symmetry.
\begin{subequations}
\begin{equation}
 \tF (1-\tu,1-u) =  P_{23} \tF (\tu,u) P_{23}\quad, \quad
 F (1-\tu,1-u) = P_{23} F(\tu,u) P_{23}
\end{equation}
\begin{equation}
\tW (1-\tu,1-u) =- P_{23} \tW (\tu,u) P_{23}\quad, \quad
 W (1-\tu,1-u) = - P_{23} W (\tu,u) P_{23}
\end{equation}
\begin{equation}
 Q_a(1-\tu,1-u) = P_{23} Q_a (\tu,u) P_{23}
\end{equation}
\end{subequations}
where $P_{23}$ is the exchange operator which satisfies $P_{23}T^2 P_{23}=T^3$,
$P_{23}^2=1$.

\subsection{Singularities of the Flat Connections \label{hc13}}

The leading singularities \cite{hun} of the bicorrelators follow
from the OPE's of the biprimary fields, and the leading singularities of the
invariant bicorrelators of four broken affine primary fields were given
in (\ref{sing1}).  This result may be translated into the singularities
of the evolution operators and the flat connections as follows.

One begins by discussing the singularities of the invariant evolution
operator $Y_g(u,u_0)$ of $g$ and the corresponding
invariant affine-Sugawara correlator $Y_g(u)$.
It follows by standard arguments that
\begin{subequations}
\begin{equation}
Y_g(u,u_0)=f_g^{-1}(u_0)f_g(u)
\end{equation}
\begin{equation}
\pa f_g(u)=f_g(u)W^g(u)
\end{equation}
\begin{equation}
f_g(u) \limit{=}{u\rightarrow0}  u^{C(L_g)}\quad, \quad
C(L)=2L^{ab}\T_a^1 \T_b^2
\end{equation}
\begin{equation}
Y_g(u,u_0) \limit{=}{u,u_0 \ra0} \left({u\over u_0}\right)^{C(L_g)}
\label{whyme}
\end{equation}
\begin{equation}
Y_g^\a(u) \limit{=}{u\ra0} v_g^{(4)\b}(u^{C(L_g)})_\b{}^\a
\label{comparethis}
\end{equation}
\end{subequations}
where $v_g^{(4)}$
is the $g$-invariant tensor in (\ref{ginvtens}).

Comparing (\ref{comparethis}) and (\ref{YFeq2}) with the result
(\ref{sing1}) for the leading
singularities of the invariant bicorrelators, one finds the leading
singularities of the invariant evolution operators \cite{hun},
\begin{subequations}
\begin{equation}
\tF(\tu,u) \limit{=}{\tu,u\ra0} \left({\tu\over u}\right)^{A(\tL)}
 \left({\tu\over u}\right)^{-B(\tL)}
\end{equation}
\begin{equation}
F(\tu,u) \limit{=}{\tu,u\ra0}  \left({u\over\tu}\right)^{A(L)}
  \left({u\over\tu}\right)^{-B(L)}
\end{equation}
\begin{equation}
A(L)=L^{ab}(\T_a^1+\T_a^2)(\T_b^1+\T_b^2)\quad, \quad
B(L)=L^{ab}(\T_a^1 \T_b^1 + \T_a^2 \T_b^2)
\end{equation}
\begin{equation}
A(L)-B(L)=C(L).
\end{equation}
\label{invsing}
\end{subequations}
Conversely, these results are equivalent to eq.(\ref{ROPE}),
which guarantees the
correct conformal weights at the singularities of the invariant
bicorrelators.

{}From the form (\ref{invsing})
of the invariant evolution operators and the relations (\ref{useless1},d),
one also obtains the leading singularities of the invariant flat connections
\cite{hun},
\begin{subequations}
\begin{equation}
\tW(\tu,u)_\a{}^\b \limit{=}{\tu,u\ra0}
   \left({\tu\over u}\right)^{\tD_{\a_1}(\T^1) + \tD_{\a_2}(\T^2) -
    \tD_{\b_1}(\T^1) - \tD_{\b_2}(\T^2) }
  { (2\tL^{ab}\T_a^1\T_b^2)_\a{}^\b \over \tu }
\end{equation}
\begin{equation}
W(\tu,u)_\a{}^\b \limit{=}{\tu,u\ra0}
   \left({u\over \tu}\right)^{\D_{\a_1}(\T^1) + \D_{\a_2}(\T^2) -
    \D_{\b_1}(\T^1) - \D_{\b_2}(\T^2) }
  { (2L^{ab}\T_a^1\T_b^2)_\a{}^\b \over u }
\end{equation}
\label{leadsing}
\end{subequations}
where we have used the fact that each irrep $\T$ is in its
appropriate $L$-basis.
The $L^{ab}$-broken conformal weights in (\ref{leadsing}) can be exchanged
for those of the $K$-conjugate partner
\begin{equation}
\tD_\a(\T)\lra -\D_\a(\T)
\end{equation}
because the affine-Sugawara conformal weights $\D_g(\T)$ are
independent of $\a$.

The results (\ref{leadsing})
show what appears to be a non-Fuchsian shielding (by the
$L^{ab}$-broken conformal weights) of the pole terms of the generic ICFT.
At least for the
 coset constructions, it is expected that these singularities are
equivalent to Fuchsian singularities (see Section \ref{hc20}).

Finally, the singularities (\ref{invsing}) of the evolution operators give the
singularities of the all-order one-sided connection moments,
\begin{subequations}
\begin{equation}
W_{q0}(u)=\left.\tpa^q \tF(\tu,u)\right|_{\tu=u} \,
  \limit{=}{u\ra0} { {h_{q0}(\tL) \over u^q} }
\end{equation}
\begin{equation}
W_{0p}(u)=\left.\pa^p F(\tu,u)\right|_{\tu=u} \,
  \limit{=}{u\ra0} { h_{0p}(L) \over u^p }
\end{equation}
\begin{equation}
h_{q+1,0}=[B(\tL),h_{q0}]+\left(C(\tL)-q\right)h_{q0}
\end{equation}
\begin{equation}
h_{0,p+1}=[B(L),h_{0p}]+\left(C(L)-p)\right)h_{0p}
\end{equation}
\label{allsing}
\end{subequations}
where $[A,B]$ is commutator and $h_{00}=\one$.  These results agree with
the known form (\ref{1andKZ}--3) of the invariant connection moments
through order $q+p=2$.  The singularities of the other connection
moments can be computed from the formula
\begin{equation}
W_{qp}(u) \limit{=}{u\ra0} {h_{qp}\over u^{q+p} }\quad, \quad
h_{q+1,p} + h_{q,p+1} = \left(C(L_g)-(q+p)\right)h_{qp}
\end{equation}
which follows from the invariant consistency relations (\ref{invcons}).

\section{Coset Correlators \label{hc14}}

\subsection{Outline \label{hc15}}

The $g/h$ coset constructions \cite{bh,h1,gko} are among the
simplest conformal field
theories in ICFT.  In this case, exact expressions have been obtained
[103-105] 
for the flat connections, the general coset correlators, and the
general coset blocks.  The coset blocks, now derived from the
Ward identities of ICFT, were originally conjectured by Douglas \cite{do}
and further discussed by Gaw\c edzki and Kupiainen \cite{gk,gk2}.

\subsection{Coset and Nest Connections \label{hc16}}

The flat connections of the $g/h$ coset constructions
 can be computed in an iterative scheme \cite{fc}
which generalizes to the flat connections of the higher affine-Sugawara nests
$g/h_1/\ldots /h_n$.

The scheme begins with the
trivial connection
\begin{equation}
W_i(L=0,\tz,z)=0
\end{equation}
of the trivial theory $L=0$ and uses only the relations,
\begin{subequations}
\begin{equation}
W_i (L,\tz,z) = \tW_i (\tL,z,\tz)|_{\tL\ra L}
\label{kcovflat}
\end{equation}
\begin{equation}
\pa_i F = F W_i\quad, \quad  F(z,z) = \one
\label{wow1}
\end{equation}
\begin{equation}
\tW_i (\tz,z) = F^{-1} (\tz,z) (\tpa_i + W_i^g(\tz))F(\tz,z)
\label{wow2}
\end{equation}
\end{subequations}
collected from above.  The relation (\ref{kcovflat}) is the K-conjugation
covariance of the flat connections.
The first few steps of the iteration procedure are as follows.

\noindent {\bf 1)}  Choose $\tL=L_g, L=0$.
Then one may compute
\begin{subequations}
\begin{equation}
F(L=0,\tz,z)=\one
\end{equation}
\begin{equation}
\tW_i(\tL=L_g,\tz,z)=W_i^g(\tz)=2L_g^{ab}\sum_{j\ne i}
   {\T_a^i\T_b^j \over \tz_{ij} }
\label{firststep}
\end{equation}
\end{subequations}
from (\ref{wow1},c)
where $W_i^g$ is the affine-Sugawara connection on $g$.

This first step also verifies the conventional KZ equations (\ref{ordkz})
when $\tL=L_g$ and $L=0$.

\noindent {\bf 2)} Choose $\tL=L_{g/h}, L=L_h$.
When $h\subset g$, one may rename the groups to obtain
\begin{equation}
W_i ( L=L_h,\tz,z) = W_i^h (z)
= 2 L^{ab}_h \sum_{j \neq i} { \T_a^i \T_b^j \over z_{ij} }
\label{nextstep}
\end{equation}
from (\ref{firststep}) and (\ref{kcovflat}).
Then one computes from (\ref{wow1},c) that
\begin{subequations}
\begin{equation}
F ( L=L_h,\tz,z) =  A_h (z,\tz)
= A_h^{-1} (\tz,z)
\label{HiMom}
\end{equation}
\begin{eqnarray}
\tW_i  [\tL = L_{g/h},\tz,z] & = & A_h(\tz,z) (\tpa_i + W_i^g(\tz) ) A_h(z,\tz)
\nonumber \\
& = &
  A_h (\tz,z) W_{i}^{g/h}  (\tz) A_h^{-1} (\tz,z)
\label{newevo}
\end{eqnarray}
\begin{equation}
W_i^{g/h}\equiv W_i^g-W_i^h =
2L_{g/h}^{ab}\sum_{j\ne i}{\T_a^i\T_b^j\over z_{ij}}\;\,.
\end{equation}
\end{subequations}
The quantity $A_h$ is the (invertible) evolution operator of $h$, which
satisfies
\begin{equation}
\pa_i A_h(z,z_0) = A_h(z,z_0) W_i^h (z)\quad, \quad A_h(z_0,z_0) = \one
\label{Aevo}
\end{equation}
in analogy to the evolution operator of $g$ in (\ref{evolution}).

The connections in (\ref{nextstep}) and (\ref{newevo}) are
the {\em flat connections of $h$ and $g/h$},
which explicitly satisfy the flatness condition (\ref{expflat}).
Note in particular
that the {\em coset connection}
 (\ref{newevo}) is the first coset connection moment
$W_{i,0}=W_i^{g/h}$, dressed by the evolution operator of the $h$ theory.

\noindent {\bf 3)} Choose $\tL=L_{g/h_1/h_2}, L=L_{h_1/h_2}$.
Here $g\supset h_1\supset h_2$ and $L_{g/h_1/h_2}=L_g-L_{h_1/h_2}$ is
the first of the higher affine-Sugawara nests.  One first obtains
\begin{equation}
 W_i ( L=L_{h_1/h_2},\tz,z)
  = A_{h_2} (z,\tz) W_{i}^{h_1/h_2}  (z) A_{h_2}^{-1} (z,\tz)
\end{equation}
by renaming the groups in the coset connection (\ref{newevo})
 and using the K-conjugation relation (\ref{kcovflat}).
Then one computes the evolution operator and the flat connection of
the nests,
\begin{subequations}
\begin{equation}
F ( L=L_{h_1/h_2},\tz,z) = A_{h_1}(z,\tz) A_{h_2} (\tz,z)
\end{equation}
\begin{equation}
  \tW_i  ( \tL=L_{g/h_1/h_2},\tz,z)
  = A_{h_2} (z,\tz) A_{h_1} (\tz,z) W_{i}^{g/h_1}  (\tz) A_{h_1} (z,\tz)
A_{h_2}(\tz,z) + W_i^{h_2} (\tz)
\end{equation}
\end{subequations}
from (\ref{wow1}) and (\ref{wow2}) respectively.

These examples are the first few steps in the general iterative procedure
$$
\tW_i ( \tL = L_{g/h_1/\ldots /h_n} ) \ra
\tW_i (\tL = L_{h_1 /\ldots /h_{n+1}} )
$$
$$
\ra W_i ( L= L_{h_1 / \ldots / h_{n+1}} ) \ra F ( L =L_{h_1 / \ldots /h_{n+1}})
$$
\begin{equation}
\ra \tW_i ( \tL = L_{g /h_1 /\ldots /h_{n+1}} )
\end{equation}
where the first step
 is a renaming of the groups, followed by the application of
eqs.(\ref{kcovflat},b,c)
in that order.  Continuing the iteration, one finds that
\begin{subequations}
\begin{equation}
F(L=L_{h_1/\ldots/h_{2n+1}},\tz,z)
 = A_{h_1}(z,\tz) A_{h_2} (\tz,z) \cdots A_{h_{2n+1}} (z,\tz)
\label{iterat}
\end{equation}
\begin{equation}
F(L=L_{h_1/\ldots/h_{2n}},\tz,z)
  =  A_{h_1}(z,\tz) A_{h_2} (\tz,z) \cdots A_{h_{2n}} (\tz,z)
\end{equation}
\begin{equation}
\tF (\tL=L_{g/h_1/\ldots /h_n } ,\tz,z) = A_g(\tz,z)
F( L=L_{h_1/\ldots /h_n} ,\tz,z)
\end{equation}
\end{subequations}
where $A_{h_i}(z,z_0)$ is the evolution operator of the subgroup $h_i$,
defined in analogy to eq.(\ref{Aevo}).
{}From these relations and (\ref{WFeq}), the flat connections of all the
affine-Sugawara nests are easily computed \cite{fc}.

With these results and the inversion relation (\ref{inversion}),
one may also compute
the all-order connection moments of the nests.  For simplicity, we give
the results only for $h$ and the $g/h$ coset constructions,
\begin{subequations}
\begin{equation}
 W_{ j_1 \ldots j_q, i_1 \ldots i_p  }  = W^{g/h}_{j_1 \ldots j_q}
 W^{h}_{i_1 \ldots i_p} \equiv W_{\{ q\}}^{g/h} W_{\{p\}}^h
\label{factprop}
\end{equation}
\begin{equation}
\qquad
W_{\{p\}i_{p+1}}^h  = (\pa_{i_{p+1}} + W_{i_{p+1}}^h ) W_{\{p\}}^h
\qquad
\end{equation}
\begin{equation}
 W_{ \{q\}j_{q+1}}^{g/h} = \pa_{j_{q+1}}  W_{\{q\}}^{g/h}+
W_{j_{q+1}}^g W_{\{q\}}^{g/h} - W_{\{q\}}^{g/h} W_{j_{q+1}}^h
\end{equation}
\label{factprop2}
\end{subequations}
which show the simple {\em factorization property} in (\ref{factprop}).
This result was first given in Ref.\  \cite{wi}.  The connection moments
for the higher nests are given in \cite{wi2}.

\boldmath
\subsection{The Non-Local Conserved Quantities of $h$ and $g/h$ \label{hc17}}
\unboldmath

The non-local conserved quantities of ICFT were discussed in Section
\ref{hc11}.
We turn now to the special case of the
 non-local conserved quantities of $h$ and $g/h$ \cite{fc},
\begin{subequations}
\begin{equation}
A(\tz,z)Q_a(\tz,z)=0\quad, \quad a=1\ldots\dim g
\end{equation}
\begin{equation}
[Q_a(\tz,z),Q_b(\tz,z)]=if_{ab}{}^c Q_c(\tz,z)
\end{equation}
\begin{equation}
Q_a(\tz,z)=A_h(\tz,z)Q_a^g A_h^{-1}(\tz,z)\quad, \quad
\label{qAqg}
Q_a^g=\sum_{i=1}^n \T_a^i
\end{equation}
\begin{equation}
Q_a(\tz,z)=Q_a^g\quad, \quad
a=1\ldots\dim h
\label{qaqg}
\end {equation}
\end{subequations}
which follow from (\ref{Qadef}) and (\ref{HiMom}).
Here, $Q_a^g$ are the global
generators of $g$, $A_h$ is the evolution operator of $h$ in eq.(\ref{Aevo}),
and $Q_a(\tz,z)$ are the non-local conserved generators of $g$.
The result in (\ref{qaqg}) follows from
(\ref{qAqg}) because $W_i^h$ and $A_h$, being $h$-invariant,
commute with the global generators of $h$.

The result (\ref{qaqg}) illustrates the general phenomenon that the
non-local conserved generators of $h\subset g$
simplify to the global generators
of $h$ for all the Lie $h$-invariant CFTs, while the $g/h$
generators remain generically non-local.
This aspect of the affine-Sugawara nests is discussed in Ref.\  \cite{fc}.

The implication of these conserved quantities for the coset constructions alone
is discussed, after factorization, in the following subsection.

\subsection{Factorization and the Coset Correlators \label{hc18}}

The bicorrelators of $h$ and $g/h$ may be obtained by solving
the generalized KZ equations,
\begin{subequations}
\begin{equation}
\tpa_i A(\tz,z)=A(\tz,z) \tW_i(L_{g/h},\tz,z)\quad, \quad
\pa_i A(\tz,z) = A(\tz,z) W_i(L_h,\tz,z)
\end{equation}
\begin{equation}
\tW_i(L_{g/h},\tz,z)=A_h(\tz,z)W_i^{g/h}(\tz)A_h^{-1}(\tz,z)\quad, \quad
W_i(L_h,\tz,z)=W_i^h(z)
\end{equation}
\begin{equation}
A(z,z)=A_g(z)
\end{equation}
\end{subequations}
where $A_h(\tz,z)$ is the evolution operator of $h$ and
 $\tW_i$ and $W_i$ are the flat connections of $g/h$ and $h$, collected
from Section \ref{hc16}.

The solution for the bicorrelators is \cite{wi,fc},
\begin{subequations}
\begin{equation}
A^\a(\tz,z)=A_{g/h}^\b(\tz,z_0)A_h(z,z_0)_\b{}^\a
\end{equation}
\begin{equation}
A^\a_{g/h}(\tz,z_0)=A_g^\b(\tz)A_h^{-1}(\tz,z_0)_\b{}^\a
\end{equation}
\label{bisolution}
\end{subequations}
where $A_g(\tz)$ is the affine-Sugawara correlator, $A_{g/h}(\tz,z_0)$ is
the {\em coset correlator}, and $z_0$ is a regular reference point,
on which the bicorrelators do not depend.

Note that the bicorrelators (\ref{bisolution}) show a simple
{\em factorization} into the $\tz$-dependent coset correlator and a
$z$-dependent $h$-factor.  For more general discussion of factorization in
ICFT, see Sections \ref{hc24}, \ref{hc25}, \ref{hc26}  and \ref{hc39}.

Other properties of the coset correlators $A_{g/h}$ include the following.

\noindent {\bf A.} $SL(2,\R)$ covariance \cite{wi}.

\noindent {\bf B.} The coset equations \cite{wi}.  The coset correlators
satisfy the linear differential equations,
\begin{subequations}
\begin{equation}
\qquad
\tpa_i A_{g/h}^{\a}(\tz,z_0)  = A_{g/h}^{\b} (\tz,z_0)
\tW_i (L_{g/h}, \tz,z_0 )_{\b}{}^{\a}
\qquad
\end{equation}
\begin{equation}
\tW_i(L_{g/h},\tz,z_0) = A_h (\tz,z_0) W_i^{g/h} (\tz) A_h^{-1} (\tz,z_0)
\label{indccon}
\end{equation}
\label{indccon2}
\end{subequations}
where the {\em induced} coset connections $\tW_i(L_{g/h},\tz,z_0)$
are also flat connections, induced from the flat connections
of the bicorrelators by choosing a reference point in $\tW_i$.

\noindent {\bf C.} Induced non-local conserved generators.
In parallel with the induced coset connections (\ref{indccon}), the coset
correlators enjoy their own set of non-local conserved generators,
\begin{subequations}
\begin{equation}
A_{g/h}^{\b} (\tz,z_0) Q_a (\tz,z_0)_{\b}{}^{\a} = 0\quad, \quad
a=1\ldots \dim g
\end{equation}
\begin{equation}
[ Q_a(\tz,z_0) , Q_b(\tz,z_0) ] = i f_{ab} {}^c Q_c(\tz,z_0)
\end{equation}
\begin{equation}
Q_a(\tz,z_0) = \left\{
\matrix{ Q^g_a & ,\;\;a \in h \cr
A_h(\tz,z_0) Q^g_a A_h^{-1}(\tz,z_0) & ,\;\; a \in g/h \cr}
\right.
\label{nlcgen}
\end{equation}
\begin{equation}
A_h(\tz,z_0)=P^* e^{\int_{z_0}^\tz dz_i' W_i^h(z')} \quad , \quad
Q_a^g=\sum_{i=1}^n \T_a^i
\end{equation}
\label{coscongen}
\end{subequations}
which are induced in the same way from the non-local generators
of $h$ and $g/h$.

Since these induced conserved quantities are directly associated to the
coset correlators, the non-local coset generators in (\ref{nlcgen}) may
be related to the parafermionic currents in
Refs.\ \cite{par1,par2,par3,par4,par5}.

\noindent {\bf D.} Affine-Sugawara nests.
The bicorrelators of the higher affine-Sugawara
 nests are discussed in Refs.\ \cite{wi2,fc}, and
Ref.\  \cite{wi2} discusses the factorization of the nest bicorrelators and the
conformal blocks of the nests.
The conclusion, as anticipated in Section \ref{sec444},
is that the higher affine-Sugawara
nests are tensor-product theories formed by tensoring the conformal blocks
of appropriate subgroups and cosets.

\subsection{Four-Point Coset Correlators \label{hc19}}

Following Section \ref{invflatsec},
the development above has been completed
for the invariant four-point correlators of the coset constructions
\cite{wi,fc} and the higher affine-Sugawara nests \cite{wi2,fc}.

The invariant generalized KZ equations of $h$ and $g/h$ are \cite{fc},
\begin{subequations}
\begin{equation}
\tpa Y = Y\tW\quad, \quad
\pa Y=YW
\end{equation}
\begin{equation}
\tW(L_{g/h},\tu,u)=Y_h(\tu,u)W^{g/h}(\tu) Y_h^{-1}(\tu,u)\quad,
\quad W(L_h,\tu,u)=W^h(u)
\label{invcc}
\end{equation}
\begin{equation}
W^{g/h}(\tu)=2L^{ab}_{g/h} \left( {\T_a^1 \T_b^2 \over \tu} +
  {\T_a^1 \T_b^3 \over \tu-1} \right)\quad, \quad
W^h(u)=2L^{ab}_h \left( {\T_a^1 \T_b^2 \over u} + {\T_a^1 \T_b^3 \over u-1}
\right)
\label{ccm}
\end{equation}
\begin{equation}
Y_h(u,u_0)=U^* e^{\int_{u_0}^u du' W^h(u')}
\end{equation}
\label{invkz}
\end{subequations}
where $\tW$ and $W$ are the invariant flat connections of $g/h$ and $h$
respectively and
$Y_h$ is the invariant evolution operator of $h$.

{}From the invariant flat connections (\ref{invcc}), one obtains the
evolution operators of the invariant connections,
\begin{equation}
\tF(\tu,u)=Y_g(\tu,u)Y_h^{-1}(\tu,u)\quad, \quad
F(\tu,u)=Y_h^{-1}(\tu,u)
\end{equation}
from their definitions in (\ref{invevof2}).  Using the inversion
formula (\ref{flatinv}), the invariant flat connections also give the invariant
connection moments of $h$ and $g/h$ \cite{wi},
\begin{subequations}
\begin{equation}
W_{qp}=W_{q0}^{g/h} W_{0p}^h
\label{invfact}
\end{equation}
\begin{equation}
W_{q+1,0}^{g/h}=\pa W_q^{g/h} + W^g W_{q,0}^{g/h}-W_{q,0}^{g/h}W^h
\end{equation}
\begin{equation}
W_{0,p+1}^h=(\pa+W^h)W_{0,p}^h
\end{equation}
\end{subequations}
where $W^g=W^{g/h}+W^h$ is the invariant affine-Sugawara connection.
Note that the invariant connection moments inherit the simple
factorization property (\ref{invfact})
already seen for the $n$-point connection moments in (\ref{factprop}).

The factorized form of the connection moments anticipates the factorized
form of the invariant four-point bicorrelators \cite{wi},
\begin{subequations}
\begin{equation}
Y(\tu,u)=Y_{g/h}(\tu,u_0) Y_h(u,u_0)
\end{equation}
\begin{equation}
Y_{g/h}(\tu,u_0)=Y_g(\tu)Y_h^{-1}(\tu,u_0)
\label{invcc4}
\end{equation}
\end{subequations}
which are the solution of the generalized KZ equations (\ref{invkz}).
The quantity
$Y_{g/h}$ in (\ref{invcc4}) is the invariant coset correlator.

The invariant
coset correlators inherit the induced invariant coset equations,
\begin{equation}
\tpa Y_{g/h}(\tu,u_0)=Y_{g/h}(\tu,u_0) W(L_{g/h},\tu,u_0)
\label{invceq}
\end{equation}
where $W(L_{g/h},\tu,u_0)$ is the induced coset connection.  The invariant
coset
correlators also inherit the induced non-local conserved
quantities,
\begin{subequations}
\begin{equation}
Y_{g/h}(\tu,u_0)Q(\tu,u_0)=0\quad, \quad a=1\ldots \dim g
\end{equation}
\begin{equation}
[Q_a(\tu,u_0),Q_b(\tu,u_0)]=if_{ab}{}^c Q_c(\tu,u_0)
\end{equation}
\begin{equation}
Q_a(\tu,u_0) = \left\{
\matrix{ Y_h(\tu,u_0) Q_a^g Y_h^{-1}(\tu,u_0) & , \;\; a \in g/h \cr
Q_a^g & , \;\; a \in h \cr}
\right.
\end{equation}
\end{subequations}
in correspondence with the $n$-point coset correlators in (\ref{coscongen}).

Finally, it is instructive to check the singularities in the coset structures
against the general results in Section \ref{hc13}.  For this, one needs
\cite{hun},
\begin{subequations}
\begin{equation}
Y_h(u,u_0) \limit{\sim}{u,u_0\ra 0} \left( {u \over u_0} \right)^{C(L_h)}
\label{xyzzy}
\end{equation}
\begin{equation}
\tF(\tu,u) \limit{\sim}{\tu,u\ra 0} \left( {\tu \over u} \right)^{C(L_g)}
\left( {\tu \over u} \right)^{-C(L_h)}
\end{equation}
\begin{equation}
\tW(L_{g/h},\tu,u)
\limit{\sim}{\tu,u\ra 0} \left( {\tu \over u} \right)^{C(L_h)}
{C(L_{g/h}) \over \tu} \left( {\tu \over u} \right)^{-C(L_h)}
\label{xyzzz}
\end{equation}
\begin{equation}
Y_{g/h}(\tu,u_0) \limit{\sim}{u,u_0\ra0} v_g^{(4)} \tu^{C(L_g)}
\left(\tu \over u_0\right)^{-C(L_h)}
\end{equation}
\begin{equation}
C(L)=2 L^{ab} \T_a^1 \T_b^2
\end{equation}
\label{singchk}
\end{subequations}
where (\ref{xyzzy}) follows in analogy to (\ref{whyme}) and
$v_g^{(4)}$ is defined in (\ref{ginvtens}).  Using
special properties of $h$ and $g/h$, for example,
\begin{subequations}
\begin{equation}
u^{C(L_g)} u^{-C(L_h)} = u^{A(L_{g/h})}u^{-B(Lg/h)}
\end{equation}
\begin{equation}
A(L)=L^{ab}(\T_a^1+\T_a^2)(\T_b^1+\T_b^2)\quad, \quad
B(L)=L^{ab}(\T_a^1\T_b^1 + \T_a^2\T_b^2)
\end{equation}
\end{subequations}
it is not difficult to see that the results in (\ref{singchk}) are in
agreement with the general forms (\ref{invsing}--3).

\subsection{Coset Blocks \label{hc20}}

So far, the dynamics of ICFT has been discussed in the Lie algebra
basis $\a=\a_1\ldots\a_n$ of the biprimary fields,
where $\a_i=1\ldots \dim \T_i$.
For $g$, $h,$ and $g/h$, the {\em conformal blocks} of the correlators
are obtained by
changing to a basis of $g$- and $h$-invariant tensors.  We
review this procedure here for the four-point invariant coset
correlators, whose coset blocks \cite{wi} turn out to be the blocks
conjectured for coset constructions by Douglas \cite{do}.

Because $Y_g$ and $Y_{g/h}$ satisfy global $g$ and $h$ invariance
respectively, one may expand the various quantities of the problem
in terms of the complete sets $\{v_m\}$ and $\{v_M\}$ of $g$-invariant
and $h$-invariant tensors,
\begin{subequations}
\begin{equation}
v_m \sum_{i=1}^4 \T_a^i =0\quad, \quad a=1\ldots \dim g
\end{equation}
\begin{equation}
v_M \sum_{i=1}^4 \T_a^i =0\quad, \quad a=1\ldots \dim h
\end{equation}
\end{subequations}
whose embedding matrix is $e_m{}^M=(v_M,v_m)$.
Using these expansions, it is shown in \cite{wi} that the coset correlators
have the form
\begin{equation}
Y^\a_{g/h}(u,u_0)=d^r \C(u)_r{}^M w_M^\a(u_0,h)
\end{equation}
where $d^r$ is a set of arbitrary constants, $\{w_M(u_0,h)\}$ is
proportional to $\{v_M\}$, and $\C(u)_r{}^M$ are the {\em coset blocks},
\begin{subequations}
\begin{equation}
\C(u)_r{}^M=\F_g(u)_r{}^m \F_h^{-1}(u)_m{}^M
\label{cblok}
\end{equation}
\begin{equation}
\left(\F_h^{-1}(u)\right)_m{}^M = e_m{}^N\left(\F_h^{-1}(u)\right)_N{}^M\;\;.
\end{equation}
\end{subequations}
Here, $(\F_g)_m{}^n$ and $(\F_h)_M{}^N$ are matrices of blocks of
the affine-Sugawara constructions on $g$ and $h$ respectively,
labelled so that the lower left index of $\F_g$ and $\F_h$  corresponds to
irreps $\T$ of $g$ and irreps of $h$ in $\T$ respectively.
$\F_h^{-1}$ is the inverse of the matrix of $h$-blocks, so the
coset blocks $\C_r{}^M$ are labelled by irreps of $g$ and $h$.
The coset blocks, now derived from the generalized KZ equations of ICFT,
 were originally conjectured by Douglas \cite{do}, who
argued that they define physical non-chiral correlators for the coset
constructions.

The matrix convention for the blocks is illustrated in a detailed
example
\begin{equation}
\<(n,1)(\bar n,1) (\bar n,1) (n,1)\>\, \mbox{ in }\,
{SU(n)_{x_1} \times SU(n)_{x_2} \over SU(n)_{x_1+x_2} }
\label{detex}
\end{equation}
in Ref.\  \cite{wi}.  In this case, the matrix $\F_g$ is $2\times 2$,
containing
the four blocks (hypergeometric functions) obtained by
Knizhnik and Zamolodchikov \cite{kz} for $SU(n)_{x_1}$, and $\F_h$ is the same
matrix with $x_1 \ra x_1+x_2$.  It follows that the coset block matrix
(\ref{cblok})
is also $2\times 2$, so the example in (\ref{detex})
is a four-block problem, each of which is a
sum of squares of hypergeometric functions. The explicit form of the
four coset blocks is given in Ref.\  \cite{wi}.
This result (and its truncation to
a two-block problem when $x_1=1$) includes a large number of sets of
conformal blocks \cite{bpz,df,df2,z1,z2,fz,fl,miz}  computed by other methods
in RCFT.  See also Ref.\  \cite{la} for a recent application of coset
blocks.

More generally, it seems that the coset block approach is the
ultimately practical solution for coset correlators.  The form
(\ref{cblok})
shows that the general coset blocks are sums of products of generalized
hypergeometric functions (that is, solutions to KZ equations), a
conclusion which has not yet been reproduced by the earlier chiral
null state \cite{bpz} and free field methods \cite{df,df2}.

An open direction here is as follows.  It is expected that the
coset blocks satisfy Fuchsian differential equations, and this has
been explicitly checked for the example (\ref{detex}) of Ref.\  \cite{wi}.
It would be interesting to understand
 how the Fuchsian structure emerges from the
general coset equations (\ref{invceq}), with their apparently non-Fuchsian
singularities in (\ref{xyzzz}).

\section{The High-Level Correlators of ICFT\label{hc21}}

High-level expansion \cite{hl} of the master equation was reviewed in
Section \ref{hla}.  Using this development, the high-level correlators of
ICFT were obtained in \cite{wi2,fc}. After some preliminary
remarks, we will review these
results in parallel for the correlators and the invariant correlators.

\subsection{Preliminaries \label{hc22}}
At high level on simple $g$, the high-level smooth ICFTs have the
form \cite{hl,gva},
\begin{subequations}
\begin{equation}
\tL^{ab}={\tP^{ab} \over 2k} + \cO(k^{-2})\quad, \quad
L^{ab}={P^{ab} \over 2k} + \cO(k^{-2})
\end{equation}
\begin{equation}
\tL^{ab}+L^{ab}=L_g^{ab}={\et^{ab} \over 2k} + \cO(k^{-2})
\end{equation}
\begin{equation}
\tilde c = \mbox{rank } \tP + \cO(k^{-1})\quad, \quad
c= \mbox{rank } P + \cO(k^{-1})\quad, \quad
c_g=\dim g + \cO(k^{-1})
\end{equation}
\begin{equation}
\tP^2=\tP \quad , \quad P^2=P \quad , \quad
\tP P =0 \quad , \quad \tP+P=1
\end{equation}
\end{subequations}
where $\et^{ab}$ is the inverse Killing metric of $g$ and $\tP, P$
are the high-level projectors of the $\tL$ and the $L$ theory
respectively.
In the partial classification of ICFT by generalized
graph theory (See Section \ref{nhl}),
the projectors are the edge-functions of the graphs, each of which labels
a level-family of ICFTs.
We remind the reader that the class of high-level smooth ICFTs (all ICFTs
for which $L=\cO(k^{-1})$ at high level)
is believed to include the generic level-family and
all unitary level-families on simple compact $g$.

High-level expansion of the low-spin bicorrelators is also
straightforward in principle by expansion of the basic formulae (\ref{huh})
or (\ref{genfform}) for the connections.
Low spin means that one allows only external irreps $\T$ of $g$ whose
``spin'' (or square root of the quadratic Casimir) is much less than the
level.  In this case, one may expand the connections, correlators,
etc. in a power series in $k^{-1}$, counting all representation
matrices $\T$ as $\cO(k^0)$.  These results are summarized in the
following subsections.

\newpage
\noindent \un{High-level abelianization}
\vskip .3cm

It is clear that the high-level limit is a contraction of the affine
algebra, but this contraction is {\em not} the naive contraction of
the algebra,
\begin{equation}
[\J_a(m),\J_b(n)]=\et_{ab}\d_{m+n,0}\quad, \quad
\J_a(m)\equiv {J_a(m) \over \sqrt{k} }
\end{equation}
in which all modes of the currents are abelian.  Instead, the fact that the
representation matrices $\T$ are treated as $\cO(k^0)$
in the expansion tells us
that the leading term corresponds to the contraction,
\begin{subequations}
\begin{equation}
[J_a(0),J_b(0)]=i f_{ab}{}^c J_c(0)\quad, \quad
[J_a(0),\J_b(m)]=i f_{ab}{}^c \J_c(m)
\end{equation}
\begin{equation}
[\J_a(m),\J_b(n)] = \et_{ab} \d_{m+n,0}\quad, \quad
\J_a(m)\equiv {J_a(m) \over \sqrt{k} } \quad, \quad m\ne 0
\end{equation}
\end{subequations}
in which only the higher modes are abelian.

\subsection{High-Level Connections \label{hc23}}

The high-level connection moments of ICFT
show the factorized form \cite{wi2}
\begin{subequations}
\begin{equation}
W_{j_1 \ldots j_q , i_1 \ldots i_p} = W_{j_1 \ldots j_q,0}
W_{0,i_1 \ldots i_p} + \cO(k^{-2})
\end{equation}
\begin{equation}
W_{qp}=W_{q0}W_{0p} + \cO(k^{-2})
\end{equation}
\label{highmom}
\end{subequations}
which was seen to all orders in $k^{-1}$
for $h$ and $g/h$ in Sections \ref{hc16}
and \ref{hc19}.
The explicit form of the one-sided connection moments is \cite{wi2},
\begin{subequations}
\begin{equation}
W_{j_1 \ldots j_q , 0} = \left(\prod_{r=1}^{q-1} \pa_{j_r}\right)
 W_{j_q,0} + \cO(k^{-2})\quad, \quad q\ge 1
\end{equation}
\begin{equation}
W_{0, i_1 \ldots i_p } = \left(\prod_{r=1}^{p-1} \pa_{i_r}\right)
 W_{0, i_p} + \cO(k^{-2})\quad, \quad p\ge 1
\end{equation}
\begin{equation}
W_{q,0}=\pa^{q-1}W_{1,0}+\cO(k^{-2})\quad, \quad q\ge 1
\end{equation}
\begin{equation}
W_{0,p}=\pa^{p-1}W_{0,1}+\cO(k^{-2})\quad, \quad p\ge 1
\end{equation}
\label{highmom2}
\end{subequations}
where the first-order connection moments $W_{j_q,0}$ and $W_{0, i_p}$
are given in eqs.(\ref{conn}) and
(\ref{pinchconn}).

Using the development of Sections \ref{hc10} and \ref{hc11},
the connection moments
(\ref{highmom2})
give the following high-level results \cite{wi2,fc}.

\noindent {\bf A.} Flat connections.
\begin{subequations}
\begin{equation}
\tW_i(\tz,z) = W_{i,0} (\tz) + \cO (k^{-2})
= {\tilde{P}^{ab} \over k} \sum_{j \neq i} {\T_a^i \T_b^j \over \tz_{ij}} +
\cO (k^{-2} )
\end{equation}
\begin{equation}
W_i(\bz,z) = W_{0,i} (z) + \cO (k^{-2})
= {P^{ab} \over k} \sum_{j \neq i} {\T_a^i \T_b^j \over z_{ij}} + \cO (k^{-2})
\end{equation}
\begin{equation}
\tW(\tu,u)=W_{10}(\tu)+\cO(k^{-2}) =
{\tP^{ab}\over k} \left( {\T_a^1\T_b^2 \over \tu}+{\T_a^1\T_b^3 \over \tu-1}
 \right) +\cO(k^{-2})
\end{equation}
\begin{equation}
W(\tu,u)=W_{01}(u)+\cO(k^{-2}) =
{P^{ab}\over k} \left( {\T_a^1\T_b^2 \over u}+{\T_a^1\T_b^3 \over u-1}
  \right) +\cO(k^{-2})
\end{equation}
\label{highflat}
\end{subequations}
\noindent {\bf B.} Evolution operators.
\begin{subequations}
\begin{equation}
 \tF(\tz,z) = \one + { \tilde{P}^{ab} \over k} \sum_{i <j}
  \T_a^i \T_b^j
\ln \left( {\tz_{ij} \over z_{ij}} \right)
+ \cO (k^{-2} )
\end{equation}
\begin{equation}
 F(\tz,z) = \one + {P^{ab} \over k}
 \sum_{i < j}
  \T_a^i \T_b^j
\ln \left( {z_{ij} \over \tz_{ij}} \right)
+ \cO (k^{-2} )
\end{equation}
\begin{equation}
 \tF(\tu,u) = \one + { \tilde{P}^{ab} \over k} \left(
\T_a^1 \T_b^2 \ln \left( { \tu \over u} \right)
+ \T_a^1 \T_b^3 \ln \left( {1-\tu \over 1-u} \right)  \right)
   + \cO (k^{-2} )
\end{equation}
\begin{equation}
 F(\tu,u) = \one + { P^{ab} \over k} \left( \T_a^1 \T_b^2
\ln \left( { u \over \tu} \right)
+ \T_a^1 \T_b^3 \ln \left( {1-u \over 1-\tu} \right)  \right)
   + \cO (k^{-2} )
\end{equation}
\end{subequations}
\noindent {\bf C.} Non-local conserved generators.
\begin{subequations}
\begin{eqnarray}
Q_a(\tz,z) &=& Q^g_a + \left[Q^g_a, {P^{ab} \over k} \sum_{i <j}
\ln \left( { z_{ij} \over \tz_{ij} } \right) \T_a^i \T_b^j \right]
+ \cO (k^{-2} ) \nonumber \\
 & =& Q^g_a + \left[Q^g_a, {\tilde{P}^{ab} \over k} \sum_{i <j}
\ln \left( { \tz_{ij} \over z_{ij} } \right) \T_a^i \T_b^j \right]
+ \cO (k^{-2} )
\label{hlnlcg}
\end{eqnarray}
\begin{eqnarray}
Q_a (\tu,u)& =& Q_a^g + \left[Q_a^g, {P^{ab} \over k} \left(
\T_a^1 \T_a^2 \ln \left( {u \over \tu} \right) +
\T_a^1 \T_b^3 \ln \left( {1-u \over 1-\tu} \right)\right) \right] +
\cO (k^{-2} )
\nonumber \\
 & = &Q_a^g + \left[Q_a^g, {\tilde{P}^{ab} \over k} \left(
\T_a^1 \T_a^2 \ln \left( {\tu \over u} \right) +
\T_a^1 \T_b^3 \ln \left( {1-\tu \over 1-u} \right)\right) \right] +
\cO (k^{-2} )
\nonumber \\
\end{eqnarray}
\end{subequations}
We remark in particular that the high-level flat connections in
(\ref{highflat})
satisfy the flatness condition (\ref{expflat}) in the form,
\begin{subequations}
\begin{equation}
\tpa_i\tW_j-\tpa_j\tW_i=\cO(k^{-2})\quad, \quad
\pa_i W_j-\pa_j W_i=\cO(k^{-2})
\end{equation}
\begin{equation}
\pa_i\tW_j-\tpa_j W_i=\cO(k^{-2})
\end{equation}
\begin{equation}
[\tW_i,\tW_j]=\cO(k^{-2})\quad, \quad
[W_i, W_j]=\cO(k^{-2})
\end{equation}
\begin{equation}
[\tW_i, W_j]=\cO(k^{-2})
\end{equation}
\end{subequations}
so the flat connections are abelian flat at high level.

\subsection{High-Level Bicorrelators \label{hc24}}

The high-level form of the generalized KZ equations is
\begin{subequations}
\begin{equation}
\tpa_i  A (\tz,z) = A(\tz,z)  \left( {\tilde{P}^{ab} \over k} \sum_{j \neq i}
{ \T_a^i \T_b^j \over \tz_{ij} }
+ \cO (k^{-2} ) \right )
\end{equation}
\begin{equation}
\pa_i  A (\tz,z) = A (\tz,z)\left( {P^{ab} \over k} \sum_{j \neq i}
{ \T_a^i \T_b^j \over z_{ij} }
+ \cO (k^{-2} ) \right )
\end{equation}
\begin{equation}
A (\tz,z) \, Q_a (\tz,z) = \cO (k^{-2})
\label{highcons}
\end{equation}
\label{highsys}
\end{subequations}
where the high-level non-local conserved generators $Q_a$ are given in
(\ref{hlnlcg}).
The solutions of this system are the high-level $n$-point bicorrelators
\cite{fc},
\begin{equation}
A(\tz,z) = A_g(z_0)
\left( \one + \sum_{i<j} \T_a^i \T_b^j \left[
{ P^{ab} \over k}  \ln \left( {z_{ij} \over z_{ij}^0} \right)
+ {\tilde{P}^{ab} \over k} \ln \left( {\bz_{ij} \over z_{ij}^0
} \right)  \right]   + \cO (k^{-2}) \right)
\label{hnbic}
\end{equation}
where $A_g$ (with $A_gQ^g=0$) is the affine-Sugawara correlator and $z_0$ is
a regular reference point.  In further detail, the solution to the
generalized KZ equations
 (\ref{highsys}a,b) gives (\ref{hnbic}) with the left factor as an
undetermined row vector $A^\b(z_0)$ instead of $A_g^\b(z_0)$.  The
row vector is then fixed to be the affine-Sugawara correlator $A_g(z_0)$
by the non-local conservation law (\ref{highcons}).

Here is a partial list of known properties of the high-level bicorrelators
\cite{fc}.

\noindent {\bf A.} $z_0$-independence.
The high-level bicorrelators are independent of the reference point $z_0$,
\begin{equation}
{\pa \over \pa z_i^0} A(\tz,z) = \cO (k^{-2} )
\end{equation}
as they should be.  To see this, one may use the high-level form of the
KZ equation,
\begin{equation}
{\pa \over \pa z_i^0} A_g(z_0) = A_g(z_0) \left( {(\tilde{P} + P )^{ab}
\over k} \sum_{j \neq i} {\T_a^i \T_b^j \over z^0_{ij} } + \cO (k^{-2}) \right)
\end{equation}
or one may rearrange the result in the equivalent forms
\begin{eqnarray}
A(\tz,z) & = &A_g(z) \left[ \one + {\tilde{P}^{ab}  \over k} \sum_{i<j}
\T_a^i \T_b^j
\ln
\left( { \tz_{ij} \over z_{ij} } \right)
\right] + \cO(k^{-2} )
\nonumber \\
 & = &A_g(\tz) \left[ \one + {P^{ab}  \over k} \sum_{i<j}
\T_a^i \T_b^j
\ln
\left( { z_{ij} \over \tz_{ij} } \right)
\right] + \cO(k^{-2} ).
\end{eqnarray}

\noindent {\bf B.} $SL(2,\R)\times SL(2,\R)$ covariance.  The bicorrelators
satisfy the $SL(2,\R)\times SL(2,\R)$ relations
\begin{subequations}
\begin{equation}
\sum_{i=1}^n \tpa_i A^{\a}
= \sum_{i=1}^n (\tz_i \tpa_i + \tD_{\a_i} ) A^{\a}
 = \sum_{i=1}^n (\tz_i^2 \tpa_i + 2 \tz_i \tD_{\a_i} ) A^{\a}  =\cO (k^{-2})
\end{equation}
\begin{equation}
 \sum_{i=1}^n \pa_i A^{\a}
= \sum_{i=1}^n (z_i \pa_i + \D_{\a_i} ) A^{\a}
 = \sum_{i=1}^n (z_i^2 \pa_i + 2 z_i \D_{\a_i} ) A^{\a}  =\cO (k^{-2}).
\end{equation}
\end{subequations}
\noindent {\bf C.} Invariant four-point bicorrelators.
We also give the result for the
invariant high-level four-point bicorrelators \cite{wi2,fc},
\begin{subequations}
\begin{eqnarray}
Y(\tu,u)=Y_g(u_0) \Biggl[ \one &+& {\tP^{ab}\over k} \left(
  \T_a^1\T_b^2 \ln\left(\tu\over u_0\right) +
  \T_a^1\T_b^3 \ln\left(1-\tu \over 1-u_0 \right) \right) \nonumber \\
&+& {P^{ab}\over k} \left(
  \T_a^1\T_b^2 \ln\left(u\over u_0\right) +
  \T_a^1\T_b^3 \ln\left(1-u \over 1-u_0 \right) \right)
\Biggr] + \cO(k^{-2})
\nonumber \\
\end{eqnarray}
\begin{equation}
\pa_{u_0}Y(\tu,u)=\cO(k^{-2})
\end{equation}
\end{subequations}
where $Y_g$ is the invariant affine-Sugawara correlator at a
reference point $u_0$.

\noindent {\bf D.}  Factorized forms.
The high-level bicorrelators can be written in the factorized forms,
\begin{subequations}
\begin {eqnarray}
 A(\tz,z)& =& A_g(z_0)
  \left[ \one  +
{ \tilde{P}^{ab} \over k}  \sum_{i <j} \T_a^i \T_b^j
\ln \left( {\tz_{ij} \over z_{ij}^0} \right)
\right] \nonumber \\
& & \ti  \left[ \one +
{P^{ab} \over k} \sum_{i <j} \T_a^i \T_b^j
\ln \left(
{z_{ij} \over z_{ij}^0
} \right)  \right]  + \cO (k^{-2} )
\end{eqnarray}
\begin{eqnarray}
Y(\tu,u)& =& Y_g(u_0)  \left[ \one +
  { \tilde{P}^{ab} \over k} \left(
 \T_a^1 \T_b^2 \ln  \left( { \tu \over u_0} \right)
+ \T_a^1 \T_b^3 \ln \left( {1-\tu \over 1-u_0} \right)  \right)
  \right]  \nonumber \\
& & \ti  \left[ \one + {P^{ab} \over k} \left(
\T_a^1 \T_b^2 \ln \left( { u \over u_0} \right)
+ \T_a^1 \T_b^3 \ln \left( {1-u \over 1-u_0} \right)  \right) \right]
   + \cO (k^{-2} )
\nonumber \\
\end{eqnarray}
\label{hibifact}
\end{subequations}
since the correction terms are $\cO(k^{-2})$.  Using this result, one
may obtain the high-level conformal correlators
$\tA(\tz,z_0)$ and $\tY(\tu,u_0)$ of the $\tL$ theory from the factorization
\begin{subequations}
\begin{equation}
A^{\a} (\tz,z) = \tA^{\b} (\tz,z_0) A(z,z_0)_{\b}{}^{\a} + \cO (k^{-2} )
\end{equation}
\begin{equation}
Y^\a(\tu,u)=\tY^\b(\tu,u_0) Y(u,u_0)_\b{}^\a +\cO(k^{-2}).
\end{equation}
\label{hiconfact}
\end{subequations}
These correlators are discussed in the following subsection.

\subsection{High-Level Conformal Correlators\label{hc25}}

Comparing the factorized forms (\ref{hibifact}) and (\ref{hiconfact})
of the biconformal
correlators, one reads off the high-level conformal correlators of
ICFT \cite{fc},
\begin{subequations}
\begin{equation}
\tA (\tz,z_0) =
 A_g(z_0)
 \left[ \one +
{ \tilde{P}^{ab} \over k}
 \sum_{i <j} \T_a^i \T_b^j \ln \left( {\tz_{ij} \over z_{ij}^0} \right)
\right] + \cO (k^{-2} )
\label{hicor1}
\end{equation}
\begin{equation}
\tY(\tu,u_0)=Y_g(u_0) \left[ \one + {\tP^{ab}\over k}
  \left(\T_a^1\T_b^2 \ln\left({\tu\over u_0}\right) +
\T_a^1\T_b^3 \ln\left({1-\tu \over 1-u_0}\right) \right)
\right] + \cO(k^{-2})
\label{hicor2}
\end{equation}
\begin{equation}
\tL^{ab} = { \tilde{P}^{ab} \over 2k } + \cO (k^{-2} )
\end{equation}
\label{hicorr}
\end{subequations}
where $\tL^{ab}(k)$ is any level-family which is high-level smooth on
simple $g$.

The high-level $n$-point correlators in (\ref{hicorr}) have the following
properties.

\noindent {\bf A.} $SL(2,\R)$ covariance.  The high-level
correlators verify the
expected $SL(2,\R)$ covariance
\begin{equation}
\sum_{i=1}^n \tpa_i \tA (\tz,z_0) = \sum_{i=1}^n (\tz_i \tpa_i + \tD_{\a_i})
\tA (\tz,z_0) = \sum_{i=1}^n (\tz_i^2 \tpa_i + 2 \tz_i \tD_{\a_i} )
\tA (\tz,z_0) = \cO (k^{-2} )
\end{equation}
using the global Ward identity (\ref{gglobal}) and the known conformal weights
$\tD = {\rm diag}(\tL^{ab} \T_a \T_b )$ of the broken affine
primary fields.

\noindent {\bf B.} Two-point correlators.
Choosing $n=2$ in (\ref{hicor1}), one obtains the high-level two-point
correlators of ICFT \cite{fc}
\begin{equation}
\tA^{\a_1 \a_2} (\tz_1\,z_1^0\,\T^1,\tz_2\, z_1^0 \,\T^2) =
{\eta^{\a_1 \a_2} (\T^1 ) \, \d(\T^2 - \bar{\T}^1 )
\over (z_{12}^0)^{2 \D_1^g}}
\left( { z^0_{12} \over \tz_{12} } \right)^{2 \tD_{\a_1} } + \cO (k^{-2})
\end{equation}
where $\eta_{\a\b}(\T)$ is the carrier space metric of irrep $\T$.

\noindent {\bf C.} Three-point correlators and fusion rules.
Choosing $n=3$ in (\ref{hicor1}), one obtains the high-level three-point
correlators of ICFT \cite{fc},
$$
\tA^{\a_1\a_2\a_3} (\tz_1 \, z_1^0 \,\T^1,\tz_2 \,z_2^0 \, \T^2,
\tz_3 \,z_3^0 \,\T^3)
 = A_g^{\a_1 \a_2 \a_3}
(z_1^0 \, \T^1,z_2^0 \, \T^2 ,z_3^0\, \T^3)
\hspace*{1in}
$$
\begin{equation}
\ti
\left( { z^0_{12} \over \tz_{12} } \right)^{ \tD_{\a_1} +\tD_{\a_2}-\tD_{\a_3}}
\left( { z^0_{13} \over \tz_{13} } \right)^{ \tD_{\a_1} +\tD_{\a_3}-\tD_{\a_2}}
\left( { z^0_{23} \over \tz_{23} } \right)^{ \tD_{\a_2} +\tD_{\a_3}-\tD_{\a_1}}
 + \cO (k^{-2})
\label{hithree}
\end{equation}
where the three-point affine-Sugawara correlators $A_g^{\a_1\a_2\a_3}$
are proportional to the Clebsch-Gordan coefficients of the
decomposition $\T^1\otimes \T^2$ into $\bar \T^3$.

This result
shows that the high-level fusion rules of the low-spin broken affine
primary fields follow the Clebsch-Gordan coefficients of the
representations.  It should be emphasized that the Clebsch-Gordan
coefficients are taken in the simultaneous $L$-basis (see Section \ref{hc3})
of the three representations.  The high-level fusion rules of ICFT
were first obtained in Ref.\  \cite{wi2}, as described in the next paragraph.

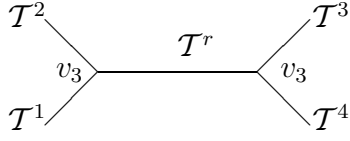
\begin{figure}
\begin{center}
\begin{picture}(120,50)(-50,-40)
\put(-30,0){\line(1,0){60}}
\put(0,6){$\T^r$}
\put(-46,-2){$v_3$}
\put(39,-2){$v_3$}
\put(-30,0){\line(-1,1){20}}
\put(-64,17){$\T^2$}
\put(-30,0){\line(-1,-1){20}}
\put(-64,-22){$\T^1$}
\put(30,0){\line(1,1){20}}
\put(51,17){$\T^3$}
\put(30,0){\line(1,-1){20}}
\put(51,-22){$\T^4$}
\end{picture}
\end{center}
\caption{Exchange of s-channel irreps.}
\label{kensfig}
\end{figure}

\noindent {\bf D.} Singularities of the four-point correlators.
The high-level invariant four-point
correlators (\ref{hicor2}) exhibit the correct physical singularities.
In the s-channel,

\noindent one finds \cite{wi2}
{\samepage
\begin{subequations}
$$
 \tilde Y^{\a} (u,u_0)  \smash{ \mathop{\simeq} \limits_{u \ra 0} }
   \sum_{{r,\x,\x' \atop \a_r \a_{\br} } } \F_g(r,\x,\x';u_0) \,
v_3^{\a_1 \a_2 \a_{\br}} (\x) \,
\left( { u \over u_0} \right)^{\tD_{\a_r} - \tD_{\a_1} - \tD_{\a_2}}
v_3^{\a_3 \a_4 \a_r}(\x') \,   \et_{\a_{\br}\a_r}
$$
\begin{equation}
\hspace*{2in}
+\cO(k^{-2})
\end{equation}
\vspace{5pt}
\begin{equation}
 v_3^{\b_i \b_j \b_r}(\x)  (\T_a^i + \T_a^j + \T_a^r)_{\b_i \b_j \b_r}
{}^{\a_i \a_j \a_r} =0
\end{equation}
\label{high4sing}
\end{subequations} }
where the conformal weight factor
$\left({u \over u_0}\right)^{\tD_{\a_r}-\tD_{\a_1}-\tD_{\a_2}}$
is correct for this channel.
The result (\ref{high4sing}) should be understood in terms of the
s-channel diagram
in Fig.\ref{kensfig},
which shows the external irreps $\T^1\ldots\T^4$ and the s-channel
irreps $\T^r$.  The conformal weights $\tD_{\a_1}$, $\tD_{\a_2}$, and
$\tD_{\a_r}$ are the $L^{ab}$-broken conformal weights of the
broken affine primary fields corresponding to irreps $\T^1$, $\T^2$,
and $\T^r$ respectively, while $v_3^{\a_i\a_j\a_r}$ is the
Clebsch-Gordan coefficient for the decomposition
$\T^i\otimes \T^j=\bigoplus_r \bar\T^r$.

The Clebsch factors in (\ref{high4sing}) reproduce the result seen in
(\ref{hithree}); the
high-level fusion rules of ICFT follow the Clebsch-Gordon
coefficients of the representations.
The singularities of the high-level $n$-point correlators are discussed
in Ref.\  \cite{fc}.

\noindent {\bf E.} Induced high-level connections and conserved quantities.
The high-level $n$-point correlators (\ref{hicorr})
satisfy a PDE with flat connection \cite{fc},
\begin{subequations}
\begin{equation}
\tpa_i\tA=\tA W_i[\tL] +\cO(k^{-2})
\end{equation}
\begin{equation}
W_i[\tL]=\tW_i(\tz,z_0)={\tP^{ab}\over k}
\sum_{j\ne i} {\T_a^i\T_b^j\over\tz_{ij}}
+ \cO(k^{-2})
\label{wPDE}
\end{equation}
\label{PDE}
\end{subequations}
which is induced from the flat connection
$\tW_i(\tz,z)$ by choosing the reference point $z_0$.

Similarly, the high-level correlators exhibit induced non-local
conserved generators of $g$,
\begin{subequations}
\begin{equation}
\tA(\tz,z_0) Q_a(\tz,z_0) = \cO (k^{-2} )\quad, \quad
a = 1\ldots \dim g
\end{equation}
\begin{equation}
[Q_a(\tz,z_0),Q_b(\tz,z_0)]=if_{ab}{}^c Q_c(\tz,z_0)+\cO(k^{-2})
\end{equation}
\begin{eqnarray}
Q_a(\tz,z_0) & = &Q^g_a + \left[Q^g_a, {\tilde{P}^{ab} \over k} \sum_{i <j}
\T_a^i \T_b^j \ln \left( {\tz_{ij} \over z_{ij}^0 } \right)\right] +
\cO (k^{-2})
\nonumber \\
& = & \sum_{i=1}^n \T_a^i + { i \tilde{P}^{b(c} f_{ab}{}^{d)} \over k}
\sum_{i < j} \T_c^i \T_d^j \ln \left( { \tz_{ij} \over z_{ij}^0 } \right)
+ \cO (k^{-2} )
\end{eqnarray}
\label{hiconsgen}
\end{subequations}
which are similarly obtained from the non-local conserved quantities
$Q_a(\tz,z)$.

\noindent {\bf F.} Exact induced flat connections for ICFT.

The $g/h$ coset constructions are included in the result (\ref{PDE}) when
$\tL=L_{g/h}$, and, indeed, eq.(\ref{wPDE}) is the high-level form
of the exact induced flat coset connection in eq.(\ref{indccon}).

On the basis of the high-level and coset results, it has been conjectured
\cite{fc} that
\begin{equation}
\tW_i(\tz,z_0)\quad, \quad W_i(z_0,z)
\end{equation}
are the finite-level induced flat connections of the $\tL$ and the
$L$ theory respectively.
This means, for example, that the conformal
correlators of the $\tL$ theory are correctly described by the
differential system,
\begin{subequations}
\begin{equation}
\tpa_i A(\tz,z_0)=A(\tz,z_0)\tW_i(\tz,z_0)
\end{equation}
\begin{equation}
\tpa_i \tW_j(\tz,z_0)-\tpa_j W_i(\tz,z_0) + [\tW_i(\tz,z_0),\tW_j(\tz,z_0)]=0
\end{equation}
\end{subequations}
which is obtained by picking a fixed reference point $z=z_0$ in
the generalized KZ
equations (\ref{bdiffeq}).  Similarly, the conformal correlators of the
$L$ theory would be described by $\pa_i A(z_0,z)=A(z_0,z)W_i(z_0,z)$.
This conjecture should be investigated vis-a-vis finite-level
factorization, as discussed in the following section.

Similarly, the induced non-local conserved quantities (\ref{hiconsgen})
are exact for the coset connections (see Section \ref{hc17}), which
suggests the conjecture \cite{fc} that
\begin{equation}
Q_a(\tz,z_0)\quad, \quad Q_a(z_0,z)
\end{equation}
are the finite-level induced non-local generators of the $\tL$ and
the $L$ theory respectively.

\noindent {\bf G.} Conformal blocks.
The outstanding open problem in high-level ICFT is the
analysis of the high-level invariant
four-point conformal correlators (\ref{hicor2}) at
the level of conformal blocks.  A first step in this direction would be to
use (\ref{hicor2}) to obtain the (as yet unknown) high-level
conformal blocks of the affine-Sugawara ($\tL=L_g$) and coset
constructions ($\tL=L_{g/h}$).

\section{Finite-Level Factorization in ICFT\label{hc26}}

\subsection{Orientation\label{hc27}}

The central problems in the operator formulation of ICFT are:

\noindent
{\bf a)} The exact computation of the flat connections $\tW_i, W_i$ and
the invariant flat connections $\tW, W$.

\noindent {\bf b)} The solution of the generalized KZ equations
\begin{subequations}
\begin{equation}
\tpa_i A=A\tW_i\quad, \quad \pa_i A=AW_i
\end{equation}
\begin{equation}
\tpa Y=Y\tW\quad, \quad \pa Y=YW
\end{equation}
\label{kzsol}
\end{subequations}
for the bicorrelators $A$ and the invariant bicorrelators $Y$.

\noindent {\bf c)} Analysis of the bicorrelators to identify the conformal
correlators, or conformal structures \cite{wi2}, of the $\tL$ and the $L$
theories.

\noindent
So far, these problems have been solved exactly only for the coset
constructions (see Section \ref{hc14}),
the higher affine-Sugawara nests \cite{wi2,fc},
and the general ICFT at high-level on simple $g$ (see
Section \ref{hc21}).  Present knowledge of the general flat connections
is reviewed in Section \ref{hc1}.

In this section, we review the development of Ref.\  \cite{wi2}, which
solves steps b) and c) above, assuming knowledge of the flat
connections as input data.
A short review of this development can also be found in Ref.\  \cite{rp}.

The central idea in this development is the {\em factorization} of the
bicorrelators
\begin{subequations}
\begin{equation}
A^\a(\tz,z)=\sum_\n \left( \tA_\n(\tz,z_0)A_\n(z,z_0)\right)^\a
\end{equation}
\begin{equation}
Y^\a(\tu,u)=\sum_\n \left( \tY_\n(\tu,u_0)Y_\n(u,u_0)\right)^\a
\end{equation}
\label{factcorr}
\end{subequations}
into the conformal structures (labelled by the
conformal-structure index $\n$) of the $\tL$ and the $L$ theories.
Factorization is then nothing but the search for a solution of the
generalized KZ equations (\ref{kzsol}) by {\em separation of variables}.
Operator
factorization at the level of the biprimary fields is discussed in \cite{wi}.

Early discussion of factorization was given in \cite{kp,gko2,gq,do,kw,h3}, and
we remind the reader that factorization has been
discussed above for the coset constructions \cite{wi} and the general ICFT
at high level on simple $g$ \cite{wi2,fc}.
The higher affine-Sugawara nests are factorized in Ref.\  \cite{fc}.
In all these cases, the conformal structure index $\n$ is closely related
to the Lie algebra label $\a=(\a_1\ldots \a_n)$, so that one finds a
finite number of distinct conformal structures.  In the generic case,
however, we will see that factorization requires an infinite number of
conformal structures, in accord with intuitive notions about ICFT.

\subsection{Partially-Factorized Solution of the Ward Identities\label{hc28}}

Given the flat connections in the generalized KZ equations,
\begin{subequations}
\begin{equation}
\tpa_i A=A\tW_i\quad, \quad \pa_i A=AW_i
\end{equation}
\begin{equation}
\tpa Y=Y\tW\quad, \quad \pa Y=YW
\end{equation}
\end{subequations}
there are a number of ways,
e.g. (\ref{bicorex}) and (\ref{invevof}),
to write the solutions for the bicorrelators $A$
or $Y$.

The present discussion is based on a related form of the
solution, called the {\em partially-factorized form} \cite{wi2},
\begin{subequations}
\begin{equation}
 A^{\a}  (\tz,z) =
  \sum_{q,p=0}^{\infty} \frac{1}{q!}
\sum_{j_1 \ldots j_q}^n \frac{1}{p!}
\sum_{i_1 \ldots i_p}^n \prod_{\m=1}^q (\tz_{j_{\m}} - z^0_{j_{\m}}) \,
[ A_g^{\b}(z_0) W_{j_1 \ldots j_q,i_1 \ldots i_p}(z_0 {)_{\b}}^{\a}]
 \prod_{\n=1}^p (z_{i_{\n}} -z^0_{i_{\n}} )
\end{equation}
\begin{equation}
 Y^{\a}(\tu,u) = \sum_{q,p=0}^{\infty} {(\tu -u_0)^q \over q!}\,
[Y_g^{\b}(u_0) W_{qp} (u_0 {)_{\b}}^{\a} ]\,{(u-u_0)^p \over p!}
\end{equation}
\begin{equation}
\pa_{z_i^0}A^\a=\pa_{u_0}Y^\a=0
\label{norefp}
\end{equation}
\label{partfact}
\end{subequations}
where $z_0, u_0$ are regular reference points and
$W_{j_1 \ldots j_q,i_1\ldots i_p}, W_{qp}$ are the connection moments
and invariant connection moments respectively.
Verification of these solutions  uses the
translation sum rules (\ref{transum}) and (\ref{invevo}),
while independence of the
reference point in (\ref{norefp}) follows from the consistency
relations (\ref{consrel}) and (\ref{invcons}).

The name ``partially-factorized'' derives from the fact that, in the
form (\ref{partfact}),
one has begun to separate the dependence on the twiddled variables from
the dependence on the untwiddled variables.  The complete
separation of variables is addressed below.

\subsection{Factorization of the Bicorrelators \label{hc29}}

Using the partially-factorized solutions in (\ref{partfact}),
factorization has been formulated as an algebraic problem \cite{wi2}.

If, at the reference point, one can factorize the connection moments into
sums of $q$ factors times $p$ factors,
\begin{subequations}
\begin{equation}
W_{j_1 \ldots j_q,i_1 \ldots i_p}(z_0) =
\sum_\n  \ta(z_0)_{j_1\ldots j_q}^\n a(z_0)_{i_1 \ldots i_p}^\n
\end{equation}
\begin{equation}
W_{qp}(u_0) = \sum_\n\ty(u_0)_q^\n y(u_0)_p^\n
\end{equation}
\label{Factmom}
\end{subequations}
then, with (\ref{partfact}), one also obtains
 the factorized forms (\ref{factcorr}) of the bicorrelators,
with conformal structures
\begin{subequations}
\begin{equation}
\tA_\n(\tz,z_0)=\sum_{q=0}^\infty {1\over q!} \sum_{j_1\ldots j_q}^n
  \prod_{\m=1}^q(\tz_{j_\m}-z_{j_\m}^0)
A_g(z_0) \ta(z_0)_{j_1\ldots j_q}^\n
\end{equation}
\begin{equation}
A_\n(z,z_0)=\sum_{p=0}^\infty {1\over p!} \sum_{i_1\ldots i_p}^n
  \prod_{\n=1}^p(z_{i_\n}-z_{i_\n}^0)a(z_0)_{i_1\ldots i_p}^\n
\end{equation}
\begin{equation}
\tY_\n(\tu,u_0)=\sum_{q=0}^\infty {1\over q!} (\tu-u_0)^q
Y_g(u_0) \ty_q^\n(u_0)
\end{equation}
\begin{equation}
Y_\n(u,u_0)=\sum_{p=0}^\infty {1\over p!} (u-u_0)^p y_p^\n(u_0).
\end{equation}
\end{subequations}
For simple examples of the factorization (\ref{Factmom}), see
eqs.(\ref{factprop}) and (\ref{invfact}) for
the coset constructions and eq.(\ref{highmom}) for the general high-level ICFT.

For generic connection moments, the factorization (\ref{Factmom}) is
intrinsically infinite-dimensional, and the factorization is not
unique \cite{wi2} in this case.
One source of ambiguity
is the assignment of Lie algebra indices in the products $\ta a$
or $\ty y$, which may vary over affine-Virasoro space.
Another source of ambiguity is possible infinite-dimensional basis
changes, and, indeed, this ambiguity has been used to find
unphysical factorizations \cite{wi2}.

In what follows, we focus on the eigenvector factorization of
Ref.\ \cite{wi2},
whose conformal structures have good physical properties so far as they
have been examined.

\subsection{Candidate Correlators for ICFT\label{hc30}}

We discuss the eigenvector factorization of Ref.\  \cite{wi2} for the
invariant four-point correlators of ICFT.

The invariant connection moments $W_{qp}(u_0)$ at the reference
point $u_0$ define a natural eigenvalue problem,
\begin{subequations}
\begin{equation}
\sum_p W_{qp} (u_0)_{\a}{}^{\b} \tps_{p \b}^{(\n)} (u_0)  =
E_{\n} (u_0) \tps_{ q \a}^{(\n)} (u_0)
\end{equation}
\begin{equation}
\sum_q \ps_{q (\n)}^{ \b }(u_0)  W_{qp} (u_0 {)_{\b}}^{\a}   =
E_{\n} (u_0) \ps_{p(\n) }^{ \a} (u_0)
\end{equation}
\end{subequations}
where the conformal-structure index $\n$ labels the eigenvectors.
Then the spectral resolution,
\begin{equation}
 W_{qp} (u_0 {)_{\a}}^{\b} = \sum_{\n =0}^{\infty}  \tps_{q \a}^{(\n)} (u_0)
E_{\n} (u_0) \, \ps_{p (\n)}^{ \b } (u_0)
\end{equation}
gives the desired algebraic factorization (\ref{Factmom}) of the
connection moments, and one obtains the conformal structures
\begin{subequations}
\begin{equation}
 Y^{\a} (\tu,u)
= \sum_{\n} \tY_{\n} (\tu,u_0) \, Y_{\n}^{\a} (u,u_0)
\end{equation}
\begin{equation}
\tY_{\n} (\tu,u_0) = \sqrt{E_{\n}(u_0)} \, Y_g^{\a} (u_0)  \tps_{\a}^{(\n)}
(\tu,u_0)\quad, \quad
Y_{\n}^{\a} (u, u_0) = \sqrt{E_{\n}(u_0)} \, \ps_{(\n)}^{\a} (u,u_0)
\end{equation}
\begin{equation}
\tps_{\a}^{(\n)} (\tu,u_0) \equiv\sum_{q=0}^{\infty}
 { (\tu -u_0)^q \over q!} \,\tps_{q \a}^{(\n)} (u_0)\quad,
\quad  \ps_{(\n)}^{\a} (u,u_0) \equiv \sum_{p=0}^{\infty}
{ (u-u_0)^p \over p!} \,\ps_{p(\n)}^{ \a } (u_0)
\end{equation}
\label{confstruct}
\end{subequations}
of the $\tL$ and the $L$ theories.  The objects $\tps^{(\n)}(\tu,u_0)$
and $\ps_{(\n)}(u,u_0)$ are
called the conformal eigenvectors of the $\tL$ and the $L$ theories
respectively.  Because the eigenvalue problem is
intrinsically infinite-dimensional, one finds an infinite
number of independent conformal structures for the generic
theory, in accord with
intuitive notions about ICFT.

The solution (\ref{confstruct}) verifies the following properties
\cite{wi2}.

\noindent {\bf A.}  Cosets and nests.
The solution reproduces the correct coset and nest correlators above.
The mechanism is a {\em degeneracy} of the conformal structures, in
which each $\tY_\n$ is proportional to the same known correlators.
In the case of the coset constructions, the degeneracy is easily
understood in terms of the factorized form,
\begin{equation}
W_{qp}=W_{q0}^{g/h}W_{0p}^h
\end{equation}
of the connection moments of $g/h$ and $h$.

\noindent {\bf B.} Good semi-classical behavior.  Because of the
factorized high-level form of the general connection moments,
\begin{equation}
W_{qp}=W_{q0}W_{0p}+\cO(k^{-2})
\end{equation}
one finds a similar degeneracy among the high-level conformal structures of
all ICFT.  In this case,
each $\tY_\n$ is proportional to the known high-level correlators,
\begin{subequations}
\begin{equation}
 \tY^{\a} (\tu,u_0)  =  Y_g^{\b} (u_0)  (
\one +  2 \tilde{L}_{ab}  \left[
\T_a^1  \T_b^2 \ln \left( {\tu \over u_0} \right)
+\T_a^1  \T_b^3 \ln \left( {1-\tu \over 1-u_0} \right) \right]
 )_{\b}{}^{\a} + \cO (k^{-2})
\end{equation}
\begin{equation}
\tL^{ab}={\tP^{ab}\over 2k} + \cO(k^{-2})
\end{equation}
\end{subequations}
 where $\tL^{ab}(k)$ is any level-family which is high-level smooth
on simple $g$.  The central outstanding
problem here is to compute the next order in $k^{-1}$, where the
high-level degeneracy of the irrational theories is expected to lift.

\noindent {\bf C.} Universal braiding.
The solution exhibits a braiding of the conformal
structures which is universal across all
ICFT.  The $u \lra 1-u, u_0\lra 1-u_0$ braiding
follows from the crossing symmetry of the connection
moments
\begin{subequations}
\begin{equation}
 W_{q p} (1-u) = (-1)^{q+p}  P_{23} W_{q p} (u) P_{23}
\end{equation}
\begin{equation}
P_{23} \T^2 P_{23} = \T^3\quad, \quad P_{23}^2 =1
\end{equation}
\end{subequations}
and the linearity of the eigenvalue problem.  The precise form of this
braiding is given in eq.(9.6) of Ref.\  \cite{wi2}.

Since the coset correlators are correctly included in the solution, this
universal
braiding includes and generalizes the braiding of RCFT.  In
particular, it will be interesting to see how the universal
eigenvector braiding reduces to Fuchsian braiding in those special
cases when the CFT is rational.

\section{ICFT on the Torus\label{hc31}}

\subsection{Background \label{hc32}}

ICFT on the torus was studied by Halpern and Sochen in Ref.\  \cite{tor}.
In some respects, this case is easier to understand
than ICFT on the sphere.  This is because non-trivial
zero-point correlators, the bicharacters of ICFT, may be studied first,
thereby postponing consideration of the $n$-point correlators --
for which the biprimary fields (see Section \ref{hc4}) will be necessary again.
The bicorrelators of ICFT on higher genus have not yet been studied.

The dynamics of ICFT on the torus follows the paradigm on the sphere.
One defines bicharacters using the Virasoro operators of
both members of the K-conjugate pair, and the
bicharacters satisfy a heat-like system of differential equations with
flat connections.  The heat-like system includes and generalizes the
heat equation of Bernard \cite{ber} for the affine-Sugawara characters.

The heat-like system has been solved for the coset constructions,
which gives a new integral representation for the general coset characters.
The system has also been solved for the general ICFT at high-level on
compact $g$, and a set of high-level candidate characters has been
proposed for the Lie $h$-invariant CFTs (see Section \ref{ca}).

We will also review the geometric formulation \cite{tor} of the system
on affine Lie groups, which
uses a new first-order differential representation of affine $g\times g$
to obtain closed-form expressions for the flat connections on the torus.
In this form, the flat connections are identified
as generalized Laplacians on the affine group.

A short review of ICFT on the torus is included in Ref.\  \cite{paris}.

\subsection{Bicharacters \label{hc33}}

For each K-conjugate pair $\tT, T$ of affine-Virasoro constructions on
integer level of affine
compact $g$, the {\em bicharacters} (or affine-Virasoro characters) are
defined as
\begin{equation}
\chi(\T,\~\tau,\tau,h)=\Tr_{\T}\left(\~q^{\~L(0)-\~c/24}q^{L(0)-c/24}h\right)
\label{bichar}
\end{equation}
where $\~q=e^{2\p i\~\tau}$ $(q=e^{2\pi i\tau}$) with Im$\~\tau>0$
(Im$\tau >0$), and
\begin{subequations}
\begin{equation}
\~L(0)=\~L^{ab}\big(J_a(0)J_b(0)+2\sum_{n>0}J_a(-n)J_b(n)\big)
\end{equation}
\begin{equation}
L(0)=L^{ab}\big(J_a(0)J_b(0)+2\sum_{n>0}J_a(-n)J_b(n)\big)
\end{equation}
\end{subequations}
are the zero modes of $\~T$ and $T$.  The source $h$ in (\ref{bichar}) is an
element of the compact Lie group $H\in G$, which may be
parametrized, for example, as
\begin{equation}
h=e^{i\beta^A(x)J_A(0)}\quad ,\quad A=1\ldots \dim h\quad
\end{equation}
where $x^i$, $i=1\ldots \dim h$ are coordinates on the $H$ manifold.
As special cases, one may choose, if desired, the standard sources on
$G$ or Cartan $G$ employed in Refs.\ \cite{eo,eo2} and \cite{ber} respectively.

In (\ref{bichar}), the trace is over the integrable affine irrep $V_\T$ whose
affine primary states $|R_\T\>$ correspond to matrix irrep $\T$ of $g$.
We choose the primary states in an $L$-basis of $\T$, where they are called
the $L^{ab}$-broken affine primary states (see Section \ref{hc3}),
\begin{subequations}
\begin{equation}
J_a(m)|R_\T\rangle^\alpha=\delta_{m,0}|R_\T\rangle^\beta(\T_a)_\beta^{\ \alpha}
\quad ,m\ge 0
\end{equation}
\begin{equation}
\~L^{ab}(\T_a\T_b)_\alpha^{\ \beta}=\~\Delta_\alpha(\T)\delta_\alpha^\beta\quad
,\quad L^{ab}(\T_a\T_b)_\alpha^{\ \beta}=
\Delta_\alpha(\T)\delta_\alpha^\beta
\label{confbasis}
\end{equation}
\begin{equation}
\~L(0)|R_\T\rangle^\alpha=\~\Delta_\alpha(\T)|R_\T\rangle^{\alpha}\quad ,\quad
L(0)|R_\T\rangle^\alpha=\Delta_\alpha(\T)|R_\T\rangle^{\alpha}
\end{equation}
\begin{equation}
\~\Delta_\alpha(\T)+ \Delta_\alpha(\T)= \Delta_g(\T).
\end{equation}
\label{brokaffp}
\end{subequations}
Here, $\~\Delta_\alpha(\T)$, $\Delta_\alpha(\T)$ and $\Delta_g(\T)$ are the
conformal weights of the broken affine primaries under $\~T$, $T$ and
$T_g=\~T+T$ respectively.  More generally, $L$-bases are the
eigenbases of the conformal weight matrices, such as (\ref{confbasis}), which
occur at each level of the affine irrep.

Here are some simple properties of the bicharacters.

\noindent {\bf A.} K-conjugation covariance.
The bicharacters satisfy
\begin{equation}
\chi(\T,\~\tau,\tau,h)|_{L\lra\~L\atop\tau\lra\~\tau}=
\chi(\T,\~\tau,\tau,h)
\end{equation}
under exchange of the K-conjugate theories.

\noindent {\bf B.} Affine-Sugawara boundary condition.  Since $\~T+T=T_g$ and
$\~c+c=c_g$, the affine-Virasoro characters reduce to the standard
affine-Sugawara
characters
\begin{equation}
\chi_g(\T,\tau,h)=
\chi(\T,\tau,\tau,h)=\Tr_{\T}\left(q^{L_g(0)-c_g/24}h\right)
\label{standchar}
\end{equation}
on the affine-Sugawara line $\~\tau=\tau$.
The affine-Sugawara characters are reviewed in Section \ref{hc35}.

\noindent {\bf C.} Small $\~q$ and $q$.
When $\~q$ and $q$ go to zero with $q/\~q$ fixed, one can use the
identity $\~q^{\tL(0)}q^{L(0)}=\~q^{L_g(0)}(q/\~q)^{L(0)}$ to see that
the bicharacters are dominated by the broken affine primary states.
It follows that the leading terms of the bicharacters in this limit are
\begin{subequations}
\begin{equation}
\chi(\T,\~\tau,\tau,h)=\sum_{\alpha=1}^{\dim \T}
\~q^{\~\Delta_\alpha(\T)-\~c/24}q^{\Delta_\alpha(\T)-c/24}
h(\T)_\alpha^{\ \alpha}+\cdots
\label{leadchar}
\end{equation}
\begin{equation}
\chi(\T=0,\~\tau,\tau,h)=1+\sum_{A=1}^{\dim g}
\~q^{\~\Delta_A-\~c/24}q^{\Delta_A-c/24}
h(\T^{adj})_A^{\ A}+\cdots
\label{vacbi}
\end{equation}
\end{subequations}
where $\D_\a,\~\D_\a$ are the $L^{ab}$-broken conformal weights in
(\ref{brokaffp})
and $h(\T)$ is the corresponding element of $H\subset G$ in matrix
irrep $\T$ of $g$.
For the vacuum bicharacter in (\ref{vacbi}), the computation of the
non-leading terms is performed in the $L$-basis
$J_A(-1)|0\rangle$ of the one-current states, so that $\~\Delta_A$ and
$\Delta_A$ (with $\~\Delta_A+\Delta_A=\Delta_g=1$) are the conformal weights
of these states under $\~T$ and $T$.

\subsection{The Affine-Virasoro Ward Identities \label{hc34}}

In analogy to the bicorrelators on the sphere (see Section \ref{hc5}), the
bicharacters satisfy the affine-Virasoro Ward identities
\begin{subequations}
\begin{equation}
\~\del^q\del^p\chi(\T,\~\tau,\tau,h)\vert_{\~\tau=\tau}=
D_{qp}(\tau,h)\chi_g(\T,\tau,h)
\end{equation}
\begin{equation}
\~\del\equiv\del_{\~\tau}=2\pi i\~q\del_{\~q }\quad, \quad
\del\equiv\del_{\tau}=2\pi iq\del_q
\end{equation}
\label{charward}
\end{subequations}
where $\chi_g$ is the affine-Sugawara character in eq.(\ref{standchar}).  In
this case, the affine-Virasoro connection moments $D_{qp}(\tau,h)$ are
differential operators on the $H$ manifold, which may be computed
in principle from the moment formula,
\begin{equation}
D_{qp}(\tau,h)\chi_g(\T,\tau,h)=
(2\pi i)^{q+p}\Tr_{\T}\left(q^{L_g(0)-c_g/24}
(\~L(0)-\~c/24)^q(L(0)-c/24)^ph\right).
\label{hmoments}
\end{equation}
 Note that the quantities on the right side of
(\ref{hmoments}) are averages in the affine-Sugawara theory,
so the connection moments may be computed by the methods of
Refs.\ \cite{eo,ber,eo2,tor}, reviewed below.

{}To set up the computational scheme, one considers the basic quantity,
\begin{equation}
\Tr_{\T}\left(q^{L_g(0)}J_a(-n){\cal O}h\right)\quad ,\quad n\in\Z
\label{BI}
\end{equation}
where $\cO$ is any vector in the enveloping algebra of the
affine algebra.  For simplicity, the source $h$ is restricted to those
subgroups $H$ for which $G/H$ is a reductive coset space.

The basic quantity (\ref{BI}) may be computed by iteration, using the
identities
\begin{subequations}
\begin{equation}
\Tr_{\T}\left(q^{L_g(0)}J_A(-n){\cal O}h\right)=
\big({q^n\rho(h)\over 1-q^n\rho(h)}\big)_A^{\ \ B}
\Tr_{\T}\left(q^{L_g(0)}\big[{\cal O},J_B(-n)\big]h\right)
\  ,n\ne 0
\end{equation}
\begin{equation}
\Tr_{\T}\left(q^{L_g(0)}J_I(-n){\cal O}h\right)=
\big({q^n\sigma(h)\over 1-q^n\sigma(h)}\big)_I^{\ \ J}
\Tr_{\T}\left(q^{L_g(0)}\big[{\cal O},J_J(-n)\big]h\right)
\  ,n\in \Z
\end{equation}
\begin{equation}
\Tr_{\T}\left(q^{L_g(0)}J_A(0){\cal O}h\right)=
E_A(h)\Tr_{\T}\left(q^{L_g(0)}{\cal O}h\right)
\end{equation}
\label{itident}
\end{subequations}
and the affine algebra (\ref{kaffine})
to reduce the number of currents by one.  In (\ref{itident}),
$E_A(h)$ is the
left-invariant Lie derivative on the $H$ manifold, and
the matrices $\r$ and $\s$  comprise the adjoint action of $h$,
\begin{subequations}
\begin{equation}
\Omega(h)_a^{\ b}=\pmatrix{
                       \rho(h)_A^{\ B} & 0 \cr
                          0  &\sigma(h)_I^{\ J} \cr}
\end{equation}
\begin{equation}
hJ_A(-n)=\rho(h)_A^{\ B}J_B(-n)h\quad ,\quad
hJ_I(-n)=\sigma(h)_I^{\ J}J_J(-n)h
\end{equation}
\end{subequations}
where $A=1\ldots\dim h$, $I=1\ldots \dim g/h$.

Iterating this step, the averages on the right side of (\ref{hmoments}) may be
reduced to differential operators on the one-current averages,
\begin{subequations}
\begin{equation}
\Tr_{\T}\left(q^{L_g(0)-c_g/24}J_A(0)h\right)=E_A(h)\chi_g(\T,\tau,h)
\end{equation}
\begin{equation}
\Tr_{\T}\left(q^{L_g(0)-c_g/24}J_I(0)h\right)=0
\end{equation}
\end{subequations}
which are proportional to the affine-Sugawara characters.

As an example, the first connection moment
\begin{eqnarray}
D_{01}(L,\tau,h)&=&2\pi i\Biggl\{ -c/24 + L^{AB}E_A(h)E_B(h)+
L^{IJ}({\sigma(h)\over 1-\sigma(h)})_I^{\ K}(if_{JK}^{\ \ \ A}E_A(h))
\nonumber \\
& & \qquad +2L^{AB}\sum_{n>0}
({q^n\rho(h)\over 1-q^n\rho(h)})_A^{\ \ C}(if_{BC}^{\ \ \ D}E_D(h)+nG_{BC})
\nonumber \\
& & \qquad +2L^{IJ}\sum_{n>0}
({q^n\sigma(h)\over 1-q^n\sigma(h)})_I^{\ K}(if_{JK}^{\ \ \ A}E_A(h)+nG_{JK})
\Biggr\}
\label{DOlII}
\end{eqnarray}
was computed from eq.(\ref{hmoments}) and the result for $D_{10}$ is
obtained from (\ref{DOlII}) by the substitution
$L\ra\~L$ and $c\ra \~c$.

\subsection{The Affine-Sugawara Characters \label{hc35}}

Following the development on
 the sphere, the affine-Virasoro Ward identities for the bicharacters
imply the dynamics of the
underlying affine-Sugawara characters.

Adding the (1,0) and (0,1) Ward identities, one finds the heat equation
for the affine-Sugawara characters,
\begin{subequations}
\begin{equation}
\del\chi_g(\T,\tau,h)=D_g(\tau,h)\chi_g(\T,\tau,h)
\end{equation}
\begin{eqnarray}
D_g(\tau,h)&=&D_{01}(\tau,h)+D_{10}(\tau,h)  \nonumber \\
&=&2\pi i\Bigg\{ -c_g/24 + L_g^{AB}E_A(h)E_B(h)+
L_g^{IJ}({\sigma(h)\over 1-\sigma(h)})_I^{\ K}(if_{JK}^{\ \ \ A}E_A(h))
\nonumber \\
& & \qquad +2L_g^{AB}\sum_{n>0}
({q^n\rho(h)\over 1-q^n\rho(h)})_A^{\ \ C}(if_{BC}^{\ \ \ D}E_D(h)+nG_{BC})
\nonumber \\
& & \qquad + 2L_g^{IJ}\sum_{n>0}
({q^n\sigma(h)\over 1-q^n\sigma(h)})_I^{\ K}(if_{JK}^{\ \ \ A}E_A(h)+nG_{JK})
\Bigg\} .
\end{eqnarray}
\label{charheat}
\end{subequations}
Bernard's heat equation \cite{ber} on a $G$-source
\begin{equation}
\del\chi_g(\T,\tau,g)=D_g(\tau,g)\chi_g(\T,\tau,g)
\end{equation}
is obtained from (\ref{charheat})
when the subgroup $H\subset G$ is taken to be $G$ itself.

In what follows, we will need the following properties of the
affine-Sugawara characters.

\noindent {\bf A.} Evolution operator of $g$.  The (invertible) evolution
operator of $g$,
\begin{equation}
\Omega_g(\tau,\tau_0,h)=Te^{\int_{\tau_0}^\tau d\tau^\prime
D_g(\tau^\prime,h)}
\end{equation}
determines the evolution of the affine-Sugawara
characters,
\begin{equation}
\chi_g(\T,\tau,h)=\Omega_g(\tau,\tau_0,h)\chi_g(\T,\tau_0,h)\ .
\end{equation}

\noindent {\bf B.} Explicit form.  For integrable irrep $\T$ of
simple affine $g$, the explicit form of the affine-Sugawara
characters is \cite{ber},
\begin{subequations}
\begin{equation}
\chi_g(\T,\tau,h)={1\over \Pi(\tau,\rho(h))\Pi(\tau,\sigma(h))}
\sum_{\T'}N_{\T'}^{\T}\Tr(h(\T'))q^{\Delta_g(\T')-{c_g\over 24}}
\label{HINO}
\end{equation}
\begin{equation}
\Pi(\tau,M)\equiv\prod_{n=1}^\infty\det(1-q^nM)
\end{equation}
\end{subequations}
where the sum in (\ref{HINO}) is over all matrix irreps $\T$ of $g$.
The coefficients in the sum satisfy
\begin{equation}
N_{\T'}^\T = \left\{
\matrix { \det\o & ,\;\; \l(\T')=\o(\l(\T)+\r)-\r+(x+\~h_g)\s \cr
          0 & ,\;\; \mbox{otherwise} \cr}
\right.
\label{coefsat}
\end{equation}
where $\l(\T)$ is the highest weight of irrep $\T$, $\o$ is some
element in the Weyl group of $g$, $\s$ is some element of the coroot
lattice, $\r$ is the Weyl vector, $x$ is the invariant level, and
$\~h_g$ is the dual Coxeter number.  For $g=\oplus_Ig_I$ and
$\T=\oplus_I \T^I$, the affine-Sugawara characters are $\chi_g(\T)
=\prod_I \chi_{g_I}(\T^I)$.

\subsection{General Properties of the Connection Moments\label{hc36}}

The following properties of the connection moments $D_{qp}$ are
easily established from their definition in (\ref{hmoments}).

\noindent {\bf A.} Representation independence.
The connection moments $D_{qp}(\tau,h)$ are independent of
irrep $\T$ of $g$, so that the representation dependence of the
bicharacters is determined entirely from their affine-Sugawara
boundary condition $\chi(\T,\tau,\tau,h)=\chi_g(\T,\tau,h)$.

\noindent {\bf B.} $\tL$ and $L$ dependence.
The one-sided connection moments $D_{q0}(\tL)$ and $D_{0p}(L)$ are
functions only of $\tL$ and $L$, while the mixed connection moments
$D_{qp}(\tL,L)$ with $q,p\ge 1$ are functions of both $\tL$ and $L$.

\noindent {\bf C.} K-conjugation covariance.
The connection moments satisfy
\begin{subequations}
\begin{equation}
D_{qp}(\~L,L)=D_{pq}(L,\~L)
\end{equation}
\begin{equation}
D_{q0}(\~L)=D_{0q}(L)|_{L\to\~L}
\end{equation}
\end{subequations}
under exchange of the K-conjugate CFTs.

\noindent {\bf D.} Consistency relations.
The connection moments satisfy the consistency relations
\begin{subequations}
\begin{equation}
\qquad d_gD_{qp}=D_{q+1,p}+D_{q,p+1}\quad ,\quad D_{00}=1 \qquad
\end{equation}
\begin{equation}
\~d_gf\equiv\~\del f+fD_g(\~\tau) \quad,\quad
d_gf\equiv\del f+fD_g(\tau)
\end{equation}
\end{subequations}
in analogy with the consistency relations on the sphere
(see Section \ref{hc7}).
When $q=p=0$, these relations reduce to the identity $D_g=D_{10}+D_{01}$
in eq.(\ref{charheat}).
Following the development on the sphere, the consistency
relations can be solved to express all $D_{qp}$ in terms of the
canonical sets $\{D_g,D_{0p}\}$ or $\{D_g,D_{q0}\}$.  See \cite{tor} for
other relations among the connection moments, in analogy to those on the
sphere.

\subsection{Flat Connections on the Torus\label{hc37}}

The Ward identities (\ref{charward}) can be reexpressed as heat-like
differential equations for the bicharacters,
\begin{subequations}
\begin{equation}
\~\del\chi(\T,\~\tau,\tau,h)=\~D(\~\tau,\tau,h)\chi(\T,\~\tau,\tau,h)
\end{equation}
\begin{equation}
\del\chi(\T,\~\tau,\tau,h)=D(\~\tau,\tau,h)\chi(\T,\~\tau,\tau,h)
\end{equation}
\begin{equation}
\chi(\T,\tau,\tau,h)=\chi_g(\T,\tau,h)
\label{ASbound}
\end{equation}
\label{chardiffeq}
\end{subequations}
whose solutions are unique given the affine-Sugawara boundary
condition (\ref{ASbound}).
The $h$-differential operators $\~D$ and $D$ are the flat
connections of ICFT on the torus,
\begin{subequations}
\begin{equation}
\~d D=d \~D
\label{dtdtdd}
\end{equation}
\begin{equation}
\~df \equiv\~\del f+f\~D\quad ,\quad df\equiv\del f+fD\quad ,\quad\forall f
\label{commcov}
\end{equation}
\end{subequations}
where $\~d$ and $d$ are commuting covariant derivatives.

The flat
connections can be computed in principle from the connection formulae,
\begin{subequations}
\begin{equation}
\~D(\~\tau,\tau,h)\chi(\T,\~\tau,\tau,h)=
2\pi i\Tr_{\T}\left(\~q^{\~L(0)-\~c/24}q^{L(0)-c/24}
\big(\~L(0)-\~c/24\big)h\right)
\end{equation}
\begin{equation}
D(\~\tau,\tau,h)\chi(\T,\~\tau,\tau,h)=
2\pi i\Tr_{\T}\left(\~q^{\~L(0)-\~c/24}q^{L(0)-c/24}
\big(L(0)-c/24\big)h\right)
\end{equation}
\label{connform}
\end{subequations}
or from the connection moments via the relations
\begin{subequations}
\begin{equation}
\~D=\~\del\~B\~B^{-1} \quad ,\quad
D=\del BB^{-1}
\end{equation}
\begin{equation}
\~B(\~\tau,\tau,h)=
\sum_{q=0}^\infty{(\~\tau-\tau)^q\over q!}D_{q0}(\tau,h)\quad ,\quad
B(\~\tau,\tau,h)=
\sum_{p=0}^\infty{(\tau-\~\tau)^p\over p!}D_{0p}(\~\tau,h)
\end{equation}
\label{HEc}
\end{subequations}
which follow in analogy to (\ref{explconn}) on the sphere.
According to (\ref{HEc}), the $h$-differential operators
$\~B$ and $B$ are
the (invertible) evolution operators of the flat connections,
\begin{subequations}
\begin{equation}
\~B(\~\tau,\tau,h)=\~T e^{\int_{\tau}^{\~\tau}d\~\tau'
\~D(\~\tau',\tau,h)}\quad, \quad
B(\~\tau,\tau,h)=T e^{\int_{\~\tau}^{\tau}d\tau'
D(\~\tau,\tau',h)}
\end{equation}
\begin{equation}
\tpa\~B=\~D\~B\quad, \quad \pa B=DB
\label{charevob}
\end{equation}
\label{charevo}
\end{subequations}
where $\~T$ and $T$ are ordering in $\~\tau$ and $\tau$ respectively.

As a result, one finds the formulae for the
bicharacters,
\begin{equation}
\chi(\T,\~\tau,\tau,h)=\~B(\~\tau,\tau,h)\chi_g(\T,\tau,h)=
B(\~\tau,\tau,h)\chi_g(\T,\~\tau,h)
\label{unchar}
\end{equation}
as the unique solution, given the affine-Sugawara characters,
 to the heat-like system (\ref{chardiffeq}).

We emphasize that the heat-like system (\ref{chardiffeq})
includes the special case of
Bernard's heat equation \cite{ber} for the affine-Sugawara characters.
Choosing $\tL=0$ and $L=L_g$ in eq.(\ref{chardiffeq}), one obtains the
heat-like system
\begin{subequations}
\begin{equation}
\pa\chi(\T,\~\tau,\tau,h)=D_g(\tau,h)\chi(\T,\~\tau,\tau,h)\quad, \quad
\tpa\chi(\T,\~\tau,\tau,h)=0
\end{equation}
\begin{equation}
\chi(\T,\tau,\tau,h)=\chi_g(\T,\tau,h)
\end{equation}
\end{subequations}
which is equivalent to the heat equation (\ref{charheat}).

Other properties of the flat connections include:

\noindent {\bf A.} Representation independence.  Like the connection moments,
the flat connections and their evolution operators are independent of
the representation $\T$.

\noindent {\bf B.} $\tL$ and $L$ dependence.  The flat connections
$\~D(\tL)$ and $D(L)$ and their evolution operators
$\~B(\tL)$ and $B(L)$ are functions only of $\tL$ and $L$
as shown.

\noindent {\bf C.} K-conjugation covariance.  The evolution operators and
connections satisfy the K-conjugation covariance
\begin{equation}
B(L,\~\tau,\tau,h)|_{\tau\lra\~\tau\atop L\to\~L}=
\~B(\~L,\~\tau,\tau,h)\quad ,\quad
D(L,\~\tau,\tau,h)|_{\tau\lra\~\tau\atop L\to\~L}=\~D(\~L,\~\tau,\tau,h).
\end{equation}

\noindent {\bf D.} Inversion formula.  The connection moments can be
computed from the flat connections by the inversion formula,
\begin{equation}
D_{qp}(\tau,h)={\~d}^qd^p1|_{\~\tau=\tau}
\label{charinv}
\end{equation}
where $\~d$ and $d$ are the covariant derivatives in (\ref{commcov}).
This relation is the inverse of (\ref{HEc}).
As examples, we list the first few moments,
\begin{subequations}
\begin{equation}
D_{00}(\tau)=1
\end{equation}
\begin{equation}
D_{10}(\tau)=\~D(\tau,\tau)\quad ,\quad D_{01}(\tau)=D(\tau,\tau)
\end{equation}
\begin{equation}
D_{20}(\tau)=(\~\del\~D+{\~D}^2)|_{\~\tau=\tau} \quad ,\quad
D_{02}(\tau)=(\del D+D^2)|_{\~\tau=\tau}
\end{equation}
\begin{equation}
D_{11}(\tau)=(\~\del D+D\~D)|_{\~\tau=\tau}=(\del\~D+\~D D)|_{\~\tau=\tau}
\end{equation}
\end{subequations}
noting that, as on the sphere, the pinched connections (at $\~\tau=\tau$) are
the first connection moments.

\noindent {\bf E.} Relations among the evolution operators.  It follows from
(\ref{charevo})
that the evolution operators of the flat connections
are related by the evolution operator of
$g$,
\begin{equation}
\~B(\~\tau,\tau,h)=
B(\~\tau,\tau,h)
\Omega_g(\~\tau,\tau,h)\quad ,\quad
B(\~\tau,\tau,h)=
\~B(\~\tau,\tau,h)
\Omega_g(\tau,\~\tau,h)
\end{equation}
and hence the evolution operator of $g$ is composed of the evolution
operators of the flat connections,
\begin{equation}
\Omega_g(\~\tau,\tau,h)=
B^{-1}(\~\tau,\tau,h)\~B(\~\tau,\tau,h).
\label{evog}
\end{equation}
We emphasize that, as on the sphere, the relation (\ref{evog}) is a
distinct decomposition of $\O_g$ for each K-conjugate pair of ICFTs.
The differential relations
\begin{equation}
(\~d_g-\~D)B=(d_g-D)\~B=0
\end{equation}
also hold, supplementing the differential relations in (\ref{charevo}).

\noindent {\bf F.} Behavior for small $\~q$ and $q$ \cite{hun}.  Beginning
with the small $\~q,q$ behavior (\ref{leadchar}) of the bicharacters on
a $G$ source, one infers that the flat connections are analytic
around $\~q=q=0$ with leading terms,
\begin{subequations}
\begin{equation}
\~D(\~\tau,\tau,g)=2\p i\left(\tL^{ab}E_aE_b-{\~c \over 24} \right)
 + \cO(\~q\mbox{ or } q)
\label{pow1}
\end{equation}
\begin{equation}
D(\~\tau,\tau,g)=2\p i\left(L^{ab}E_aE_b-{c \over 24} \right)
 + \cO(\~q\mbox{ or }q)
\label{pow2}
\end{equation}
\begin{equation}
\qquad D_g(\tau,g)=2\p i\left(L^{ab}_g E_aE_b-{c_g \over 24} \right)
 + \cO(q) \qquad
\end{equation}
\end{subequations}
where $E_a,\, a=1\ldots\dim g$ is the left-invariant Lie derivative
on the $G$ manifold.  The leading terms (\ref{pow1},b)
 of the connections
are flat and abelian flat.

\noindent {\bf G.} Other relations.
A number of other identities are discussed for the torus in
Ref.\  \cite{tor}, following the development on the sphere.  We note
in particular that the non-local conserved generators of $g$ are
also found on the torus.

\subsection{Coset Characters\label{hc38}}

The flat connections and bicharacters of $h$ and $g/h$ have been
obtained in closed form when $G/H$ is a reductive coset space and the
source $h$ is chosen in the same subgroup $H \subset G$.  The final
result is a new integral representation of the general coset
characters.

The flat connections of $h$ and $g/h$,
\begin{subequations}
\begin{eqnarray}
D(L_h,\tau,h)&=&D_{01}(L_h,\tau,h)
\nonumber \\
&=&  2\pi i\Biggl\{ -c_h/24+
L_h^{AB}E_A(h)E_B(h)
\nonumber \\
& &+ 2L_h^{AB}
\sum_{n>0}\big({q^n\rho(h)\over 1-q^n\rho(h)}\big)_A^{\ C}
(if_{BC}^{\ \ D}E_D(h)+nG_{BC})\Biggr\}\hspace*{1in}
\label{charflat}
\end{eqnarray}
\begin{equation}
\~D(L_{g/h},\~\tau,\tau,h)=\Omega_h(\tau,\~\tau,h)D_{g/h}(\~\tau,h)
\Omega_h^{-1}(\tau,\~\tau,h)
\label{Cconn}
\end{equation}
\end{subequations}
follow from the connection formulae (\ref{connform}),
where $\O_h$ is the invertible evolution operator of $h$
\begin{equation}
\Omega_h(\tau,\~\tau,h)=T e^{\int_{\~\tau}^\tau d\tau' D_h(\tau',h)}.
\end{equation}
As on the sphere, the
{\em coset connection} in (\ref{Cconn})
is an $h$-dressing of the first coset connection moment
$D_{g/h}=D_{10}(L_{g/h})$.

Using these connections in the inversion formula (\ref{charinv}),
one finds that
the connection moments of $h$ and $g/h$ have the factorized
form,
\begin{subequations}
\begin{equation}
D_{qp}(\tau)=D_{0p}^h(\tau)D_{q0}^{g/h}(\tau)
\end{equation}
\begin{equation}
D_{0p}^h(\tau)\equiv d^p1|_{\~\tau=\tau}\quad ,\quad
D_{q0}^{g/h}(\tau)\equiv {\~d}^q1|_{\~\tau=\tau}
\end{equation}
\end{subequations}
in analogy to the corresponding result (\ref{factprop2})
on the sphere.

The solution (\ref{unchar}) of
 the heat-like system then gives the bicharacters of
$h$ and $g/h$ in two equivalent forms.  The first form is simply
\begin{equation}
\chi(\T,\~\tau,\tau,h)=
\Omega_h(\tau,\~\tau,h)\chi_g(\T,\~\tau,h)
\end{equation}
where $\O_h$ is the evolution operator of $h$.  To understand the
second form, one needs the $\hat h$-characters for integrable
representations $\T^h$ of affine $h$ on an $h$-source,
\begin{subequations}
\begin{eqnarray}
\chi_h(\T^h,\tau,h)&=&\Tr_{\T^h}\big(q^{L_h(0)-c_h/24}h\big)
\\
&=& {1 \over \Pi (\tau,\rho(h))}
\sum_{\T'^h} N_{\T'^h}^{\T^h}\Tr(h(\T'^h))
q^{\Delta_h(\T'^h)-c_h/24}
\end{eqnarray}
\end{subequations}
where $h$ is a simple subalgebra of $g$, the sum is over all the
unitary irreps of $h$, and $N_{\T'^h}^{\T^h}$ is the $h$-analogue of
$N_{\T'}^\T$ in (\ref{coefsat}).

The second form of the bicharacters is the factorized form,
\begin{equation}
\chi(\T,\~\tau,\tau,h)=\sum_{\T^h}\vphantom{\Biggl(}'
\chi_{g/h}(\T,\T^h,\~\tau)\chi_h(\T^h,\tau,h)
\label{hcfform}
\end{equation}
where the primed sum is over the integrable representations $\T^h$ of $h$
at the induced level of the subalgebra and
$\chi_{g/h}(\T,\T^h,\~\tau)$ are the
{\em coset characters},
 which are independent of the
source.  The factorized form (\ref{hcfform}) includes the known
factorization of the affine-Sugawara characters \cite{kp,gko2,gq,kw},
\begin{equation}
\chi_g(\T,\~\tau,h)=\sum_{\T^h}\vphantom{\Biggl(}'
\chi_{g/h}(\T,\T^h,\~\tau)\chi_h(\T^h,\~\tau,h)
\end{equation}
on the affine-Sugawara line $\tau=\~\tau$.

Using the orthonormality relation for $\hat h$-characters \cite{tor},
\begin{subequations}
\begin{equation}
\int dh \chi_h^\dagger(\T'^h,\tau,h)\chi_h(\T^h,\tau,h)=\d(\T'^h,\T^h)
\end{equation}
\begin{equation}
\chi_h^\dagger(\T^h,\tau,h)\equiv {\Pi(\tau,\r(h)) \over f(\T^h,q)}
 \sum_{\T'^h}N_{\T'^h}^{\T^h} \Tr(h^*(\T'^h))q^{\D_h(\T'^h)+c_h/24}
\end{equation}
\begin{equation}
f(\T^h,\tau)\equiv\sum_{\T'^h} |N_{\T'^h}^{\T^h}|
q^{2\Delta_h(\T'^h)}
\end{equation}
\end{subequations}
where $dh$ is Haar measure on $H$,
one obtains the integral representation for the general $g/h$ coset
character,
\begin{subequations}
\begin{equation}
\chi_{g/h}(\T,\T^h,\~\tau)=
\int dh\chi_h^\dagger(\T^h,\~\tau,h)\chi_g(\T,\~\tau,h) \hspace*{4in}
\label{ODM}
\end{equation}
\begin{equation}
={\~q^{-{c_{g/h}\over 24}}\over f(\T^h,\~\tau) }
\sum_{\T', \T'^h}
N_{\T'}^{\T}N_{\T'^h}^{\T^h}
\left(\int dh{\Tr(h^*(\T'^h))\Tr(h(\T'))\over\Pi(\~\tau,\sigma(h))}\right)
\~q^{\Delta_g(\T')+\Delta_h(\T'^h)}\raisebox{-2pt}{.}
\label{ODM2}
\end{equation}
\end{subequations}
The general result (\ref{ODM}) holds for semisimple $g$ and simple $h$.  In
this form, the result is the analogue of the formula $\C_{g/h}=\F_g\F_h^{-1}$
for the coset blocks on the sphere
(see Section \ref{hc20}).  The special case in
(\ref{ODM2}) is the explicit form of (\ref{ODM}) for simple $g$.  See
also Ref.\  \cite{hr1,hr2}
for an apparently similar form of the coset characters,
obtained in the gauged WZW model.

The coset characters also satisfy a set of induced linear differential
equations \cite{tor}, the coset equations on the torus, in analogy to the
coset equations (\ref{indccon2}) on the sphere \cite{wi}.

An open problem in this direction is to obtain the flat connections
and characters of the higher affine-Sugawara nests.

\subsection{High-Level Characters\label{hc39}}

\subsubsection{Background:  The symmetry hierarchy of ICFT}

The low-spin bicharacters have been computed for all ICFT at high level on
simple $g$, and high-level candidate characters have been proposed for the
Lie $h$-invariant CFTs, which were reviewed in Section \ref{ca}.

In our review of this development, it will be helpful to bear in mind
the symmetry hierarchy of ICFT,
\begin{equation}
\hbox{\rm ICFT}\supset\supset\hbox{\rm $H$-invariant CFTs}
\supset\supset\hbox{\rm Lie $h$-invariant CFTs}\supset\supset{\rm RCFT}
\end{equation}
where $H \subset \au$ is any symmetry group, which may be a finite group
or a Lie group (see Section \ref{ca}).
We will also need the high-level forms of the
inverse inertia tensors,
\begin{subequations}
\begin{equation}
\~L^{ab}={\~P^{ab}\over 2k}+O(k^{-2})\quad ,\quad
L^{ab}={P^{ab}\over 2k}+O(k^{-2})
\end{equation}
\begin{equation}
\~c={\rm rank\,}\~P+O(k^{-1})\quad ,\quad c={\rm rank\,} P+O(k^{-1})
\end{equation}
\label{khlsi}
\end{subequations}
where $\tP$ and $P$ are the high-level projectors of the $\tL$ and the $L$
theories respectively (see Section \ref{slc}).
Then, the high-level characterization of the symmetric theories,
\begin{subequations}
\begin{equation}
\mbox{$H$-invariant CFTs: }\;\;\;  [\Omega(h),\~P]= [\Omega(h),P]=0
\quad ,\quad\forall \, \O(h)\in H\subset \au
\label{charchar}
\end{equation}
\begin{equation}
\mbox{Lie $h$-invariant CFTs: } \;\;\;
[\T_A^{\rm adj},\~P]= [\T_A^{\rm adj},P]=0
\quad , \quad A=1\ldots \dim h
\hspace*{.5in}
\label{charchar2}
\end{equation}
\end{subequations}
follows from eqs.(\ref{hia}) and (\ref{lhs}).

\subsubsection{High-level bicharacters}

We begin with the main results for the general ICFT at high level on
simple $g$, which follow from the connection formulae (\ref{connform}).

The leading terms of the flat connections are
\begin{subequations}
\begin{equation}
\~D(\~L,\~\tau,\tau,g)=2\pi i\left(\sum_{n>0}n\Tr\big({X_n\over 1-X_n}\~P\big)
-{\rank\~P\over 24}\right)+O(k^{-1})
\end{equation}
\begin{equation}
D(L,\~\tau,\tau,g)=2\pi i\left(\sum_{n>0}n\Tr\big({X_n\over 1-X_n}P\big)
-{\rank P\over 24}\right)+O(k^{-1})
\end{equation}
\begin{equation}
X_n(\~\tau,\tau,g)\equiv (\~q^n\~P+q^nP)\Omega(g)
\end{equation}
\label{leadchar2}
\end{subequations}
where $\O(g)$ is the adjoint action of $g$.  The leading terms in
(\ref{leadchar2})
are complex-valued
functions, so that, as seen earlier on the sphere, the high-level flat
connections are also abelian flat.

Given the high-level connections in (\ref{leadchar2}),
one obtains the high-level
evolution operators $\~B$ and $B$ by integrating eq.(\ref{charevob}).
The high-level
bicharacters then follow from eq.(\ref{unchar}),
\begin{equation}
\vphantom{\Biggl(}
\chi(\T,\~\tau,\tau,g)=\~B(\~\tau,\tau,g)\chi_g(\T,\tau,g)=
B(\~\tau,\tau,g)\chi_g(\T,\~\tau,g)
\label{hibichar}
\end{equation}
given the high-level form of the affine-Sugawara characters
$\chi_g(\tau)$.  It bears emphasis that the high-level connections and
evolution operators are valid for all irreps $\T$, but, so far, only the
high-level form of the low-spin affine-Sugawara characters
\begin{equation}
\chi_g(\T,\tau,g)\,\limit{=}{k}\,
q^{-{\dim g\over 24}}
{\Tr(g(\T))\over \Pi(\tau,\Omega(g))}
\label{hiaschar}
\end{equation}
has been worked out, where low-spin means that the invariant
Casimir of irrep $\T$ is $\cO(k^0)$.

One obtains the high-level form of the low-spin
bicharacters,
\begin{eqnarray}
\chi(\T,\~\tau,\tau,g)
&\limit{=}{k}&{\~q}^{-{\rank\~P\over 24}}q^{-{\rank P\over 24}}
\prod_{n=1}^\infty e^{
2\pi in\int_{\tau}^{\~\tau} d{\~\tau}^\prime
\Tr\left\{\big({X_n({\~\tau}',\tau,g)
\over 1-X_n({\~\tau}',\tau,g)}\big)\~P\right\}
}
{\Tr(g(\T))\over \Pi(\tau,\Omega(g))}
\nonumber \\
&\limit{=}{k}&{\~q}^{-{\rank\~P\over 24}}q^{-{\rank P\over 24}}
\prod_{n=1}^\infty e^{
2\pi in\int_{\~\tau}^\tau d\tau^\prime
\Tr\left\{\big({X_n(\~\tau,\tau',g)
\over 1-X_n(\~\tau,\tau',g)}\big)P\right\}
}
{\Tr(g(\T))\over \Pi(\~\tau,\Omega(g))}
\nonumber \\
\label{hilochar}
\end{eqnarray}
where the results in (\ref{hiaschar}--7)
are the leading terms of the
asymptotic expansion of these quantities.
We remark
on the simple intuitive form,
\begin{equation}
\chi(\T,\~\tau,\tau,g=1)
\,\limit{=}{k}\,{\dim \T\over\eta(\~\tau)^{\rank\~P}\eta(\tau)^{\rank P} }
\end{equation}
exhibited by the result (\ref{hilochar}) at unit source, where $\eta$ is the
Dedekind $\eta$-function.
An open direction here is to obtain the high-level
high-spin affine-Sugawara characters, which then determine the
high-level high-spin bicharacters via eq.(\ref{hibichar}).

\boldmath
\subsubsection{Simplification for the $H$-invariant CFTs}
\unboldmath

The results quoted above are valid for the general theory on a general
source $g\in G$, but a simplification has been found for the $H$-invariant
CFTs, which are those theories with a symmetry group.

In this case, one may choose the
source to be any element $h\in H$ of the symmetry group $H$ and use the
identity (\ref{charchar}).  One obtains the flat connections
of the $H$-invariant CFTs,
\begin{subequations}
\begin{equation}
\~D(\~L,\~\tau,h)=
2\pi i\left(\sum_{n>0}n
\Tr\left({\~q^n\Omega(h)\over 1-{\~q}^n\Omega(h)}\~P\right)
-{\rank\~P\over 24}\right) +O(k^{-1})
\end{equation}
\begin{equation}
D(L,\tau,h)=
2\pi i\left(\sum_{n>0}n
\Tr\left({q^n\Omega(h)\over 1-{q}^n\Omega(h)}P\right)
-{\rank P\over 24}\right) +O(k^{-1})
\end{equation}
\label{Hflat}
\end{subequations}
which are functions only of $\~\tau$ and $\tau$ respectively.  This
property is lost when the source is chosen on a manifold larger than
the symmetry group of the theories, a complication which should be
studied first for $h$ and the $g/h$ coset constructions.

The connections (\ref{Hflat}) can be further simplified
\begin{subequations}
\begin{equation}
\~D(\~L,\~\tau,h)=
-\left(2\pi i{\rank\~P\over 24}+\~\del\log\Pi(\~P,\~\tau,\Omega (h))\right)
+O(k^{-1})
\end{equation}
\begin{equation}
D(L,\tau,h)=
-\left(2\pi i{\rank P\over 24}+\del\log\Pi(P,\tau,\Omega (h))\right)
+O(k^{-1})
\end{equation}
\begin{equation}
\Pi(M,\tau,\Omega (h))\equiv\prod_{n=1}^{\infty}
e^{\Tr\left(M\log (1-q^n\Omega(h))\right)}
\label{genPI}
\end{equation}
\end{subequations}
by introducing the generalized $\Pi$-function in (\ref{genPI}).  Then
one obtains the evolution operators of the flat connections,
\begin{subequations}
\begin{equation}
\~B(\~\tau,\tau,h)=\left(q\over \~q\right)^{\rank\~P\over 24}
{\Pi(\~P,\tau,\Omega(h))\over\Pi(\~P,\~\tau,\Omega(h))}+O(k^{-1})
\end{equation}
\begin{equation}
B(\~\tau,\tau,h)=\left(\~q\over q\right)^{\rank P\over 24}
{\Pi(P,\~\tau,\Omega (h))\over\Pi(P,\tau,\Omega (h))}+O(k^{-1})
\end{equation}
\end{subequations}
and finally the high-level low-spin bicharacters of the $H$-invariant
CFTs,
\begin{equation}
\chi(\T,\~\tau,\tau,h)\,\limit{=}{k}\,
{1\over \~q^{\rank\~P\over 24}\Pi(\~P,\~\tau,\Omega(h))}
\Tr(h(\T))
{1\over q^{\rank P\over 24} \Pi(P,\tau,\Omega(h))}
\end{equation}
where eq.(\ref{hiaschar}) was used for the low-spin
affine-Sugawara characters in (\ref{hibichar}).

\boldmath
\subsubsection{Candidate characters for the Lie $h$-invariant CFTs}
\unboldmath

{}To obtain the characters of the individual ICFTs, it is necessary to
factorize the bicharacters
\begin{equation}
\chi(\T,\~\tau,\tau,h)=
\sum_{\nu}\chi_{\~L}^\nu(\T,\~\tau,h)\chi_{L}^{\vphantom{g}\nu}(\T,\tau,h)
\label{charfact}
\end{equation}
into the conformal characters $\chi_{\tL}$ and $\chi_L$ of the $\tL$ and
the $L$ theories respectively.  As on the sphere \cite{wi2}, the
factorization (\ref{charfact}) is not unique,
but one is
interested only in those factorizations for which the conformal
characters are modular covariant.

A modest beginning in this direction was given in \cite{tor}, using
intuition from the coset constructions to guess a set of high-level
candidate characters for the Lie $h$-invariant CFTs, which are those
theories with a Lie symmetry.

More precisely, this guesswork is limited to the Lie $h$-invariant
CFTs with simple $h\subset g$.
Then\footnote{The general phenomenon in (\ref{phen1}--15)
is discussed in Ref.\  \cite{lie}, and
Section \ref{lhg} illustrates this phenomenon on the Lie
$h$-invariant graphs in the graph theory unit of ICFTs on $SO(n)$.}
one of the theories,
say $\tL$, has a local Lie $h$ invariance (like $L_{g/h}$)
\begin{equation}
[J_A(m),\~L(n)]=0\quad,\quad m,n\in\Z\quad, \quad A=1\ldots\dim h
\label{phen1}
\end{equation}
while its K-conjugate partner $L$ (like $L_h$) carries only the
corresponding global invariance,
\begin{equation}
[J_A(0),L(n)]=0\quad,\quad n \in\Z\quad, \quad A=1\ldots\dim h.
\label{phen2}
\end{equation}
Then one may adopt as a working
hypothesis that, as in $h$ and $g/h$,
 all the source dependence of the bicharacters is
associated to the global theory.

One obtains the factorized high-level
bicharacters of the Lie $h$-invariant CFTs,
\begin{equation}
\chi(\T,\~\tau,\tau,h)\,\limit{=}{k}
\sum_{\T^h}\chi_{\~L}^{\vphantom\nu}(\T,\T^h,\~\tau)
\chi_{L}^{\vphantom g}(\T,\T^h,\tau,h)
\end{equation}
where the sum is over all unitary irreps $\T^h$ of $h$ and
\begin{subequations}
\begin{equation}
\chi_{\~L}^{\phantom g}(\T,\T^h,\~\tau) \,\limit{=}{k}
\int dh
{\Tr(h^*(\T^h))\Tr(h(\T))
\over
\~q^{\rank\~P\over 24}
\Pi(\~P,\~\tau,\Omega(h))}
\end{equation}
\begin{equation}
\qquad \chi_L^{\phantom g}(\T^h,\tau,h)\,\limit{=}{k}\,
{\Tr(h(\T^h))\over
q^{\rank P\over 24}
\Pi(P,\tau,\Omega(h))} \qquad
\end{equation}
\label{candidate}
\end{subequations}
are the high-level candidate characters for the Lie $h$-invariant
CFTs.

The candidate characters (\ref{candidate}) reduce to the correct high-level
characters of $h$ and $g/h$,
\begin{subequations}
\begin{equation}
\chi_{L_{g/h}}^{\phantom g}(\T,\T^h,\~\tau)\,\limit{=}{k}
\int dh
{\Tr(h^*(\T^h)) \Tr(h(\T))
\over
\~q^{{\rm dim}(g/h)\over 24}
\Pi(\~\tau,\sigma(h))}
\end{equation}
\begin{equation}
\qquad \chi_{L_h}^{\phantom g}(\T^h,\tau,h)\,\limit{=}{k}\,
{\Tr(h(\T^h))\over
q^{{\rm dim}h\over 24}
\Pi(\tau,\rho(h))} \qquad
\end{equation}
\end{subequations}
when $\tP=P_{g/h}$ and $P=P_h$.  The next step is to test the
candidate characters for modular covariance, or to further
decompose the candidates until modular covariance is obtained.
For this, it will be necessary to adjoin the corresponding set of
high-spin candidate characters, which may be obtained,  as described
above, from
the high-spin affine-Sugawara characters.

Although this proposal is technically involved, it is
expected that chiral modular covariant characters and non-chiral
modular invariants exist in ICFT, just as braid-covariant
correlators have been found on the sphere (see Section \ref{hc30}). This
expectation has further support in the case of the high-level
smooth ICFTs (\ref{khlsi}), because diffeomorphism-invariant
world-sheet actions are known for the generic theory of this type
(see Section \ref{hc41}).

\subsection{Formulation on Affine Lie Groups\label{hc40}}

The characters studied above were defined with a conventional
Lie source, but the problem takes a geometric form \cite{tor} on an affine
source $\hga$ in the affine Lie group $\hat L G$.  One finds a new
first-order differential representation of affine $g \times g$ and
closed form expressions for the flat connections, which are seen as
generalized Laplacians on the affine group.

We review first the new representation of affine $g \times g$.

It is convenient to write the algebra of simple affine $g$ as an
infinite-dimen\-sional Lie algebra
\begin{subequations}
\begin{equation}
[\J_L,\J_M]=if_{LM}{}^N\J_N
\label{summation}
\end{equation}
\begin{equation}
\J_L=(J_a(m),k)\quad, \quad L=(am,\ys)
\end{equation}
\begin{equation}
f_{am,bn}{}^{cp}=f_{ab}{}^c \d_{m+n,p}\quad, \quad
f_{am,bn}{}^\ys=-im\et_{ab} \d_{m+n,0}
\label{structure}
\end{equation}
\end{subequations}
where the central element $k$ is included among the generators and the
non-zero structure constants $f_{LM}{}^N$ are given in (\ref{structure}).
An arbitrary element $\hga$ in $\hat L G$ has the form,
\begin{equation}
\hga(x,y)=e^{iyk} \hg(J,x)
\label{source}
\end{equation}
where $y$ and $x^{i\m}, i=1\ldots\dim g, \m \in \Z$ are the
coordinates on the affine group manifold.

On the affine group manifold, one may define left and right invariant
vielbeins, inverse vielbeins, and affine Lie derivatives.  We
focus here on the reduced affine Lie derivatives \cite{tor},
\begin{subequations}
\begin{equation}
E_a(m)=-i e_{am}{}^{i\m}(\pa_{i\m}-ik e_{i\m}{}^\ys)\quad, \quad
\bE_a(m)=-i \bar e_{am}{}^{i\m}(\pa_{i\m}+ik\bar e_{i\m}{}^\ys)
\label{redaff}
\end{equation}
\label{quantE}
\begin{equation}
E_a(m)\hg=\hg J_a(m)\quad,\quad
\bE_a(m)\hg=-J_a(m)\hg
\label{indact}
\end{equation}
\label{raffld}
\end{subequations}
which describe the induced action of the affine Lie derivatives on the
reduced group element $\hg$.  The quantities
$e_\L{}^L, \bar e_\L{}^L$ and $e_L{}^\L, \bar e_L{}^\L$ with
$\L=(i\m,y), L=(am, \ys)$ are the vielbeins and inverse vielbeins on the
affine group manifold.  The reduced affine Lie derivatives satisfy
two commuting copies of the affine algebra
\begin{subequations}
\begin{equation}
[E_a(m),E_b(n)]=if_{ab}{}^c E_c(m+n)+mk\et_{ab}\d_{m+n,0}
\end{equation}
\begin{equation}
[\bE_a(m),\bE_b(n)]=if_{ab}{}^c \bE_c(m+n)-mk\et_{ab}\d_{m+n,0}
\end{equation}
\begin{equation}
[E_a(m),\bE_b(n)]=0
\label{bothcommute}
\end{equation}
\label{Ealgebra}
\end{subequations}
at level $k$ and $-k$ respectively.  One may obtain two commuting
copies of the affine algebra at the same level $k$ by defining
$\bE'_a(m)\equiv\bE_a(-m)$.

The result (\ref{raffld})
is a class of new first-order differential representations
of affine $g \times g$, one for each basis choice of $\hg$.  An
example of this class, for a particular basis of affine $SU(2)$, was
studied in Ref.\  \cite{ans}.

Other first-order differential representations of affine Lie algebra are
known, such as the coadjoint orbit representations in Refs.\ \cite{bf} and
\cite{t}, but these provide only a single chiral copy of the algebra.

We turn now to the generalized bicharacters,
\begin{subequations}
\begin{equation}
\qquad
\chi(\T,\~\tau,\tau,\hg)=\Tr_\T(\~q^{\tL(0)-\~c/24} q^{L(0)-c/24} \hg)
\qquad
\end{equation}
\begin{equation}
\chi_g(\T,\tau,\hg)=\chi(\T,\tau,\tau,\hg)=\Tr_\T(\~q^{L_g(0)-c_g/24} \hg)
\label{genaschar}
\end{equation}
\end{subequations}
whose source is the reduced affine group element $\hg$.
Following the development above, one finds first that the generalized
affine-Sugawara characters $\chi_g(\hg)$ in (\ref{genaschar})
satisfy a heat equation
on the affine group manifold,
\begin{subequations}
\begin{equation}
\pa\chi_g(\T,\tau,\hg)=D_g(\hg)\chi_g(\T,\tau,\hg)
\end{equation}
\begin{equation}
D_g(\hg)=-2\p i \D(\hg)=2\p i L^{ab}_g\Big(E_a(0)E_b(0)
 + 2\sum_{m>0}E_a(-m)E_b(m)\Big)
\end{equation}
\end{subequations}
where $E_a(m)$ is the left-invariant reduced affine Lie derivative in
(\ref{raffld}).
Moreover, one finds that the generalized bicharacters solve the heat-like
differential system,
\begin{subequations}
\begin{equation}
\~\del\chi(\T,\~\tau,\tau,\hg)=\~D(\hg)\chi(\T,\~\tau,\tau,\hg)\quad ,\quad
\del\chi(\T,\~\tau,\tau,\hg) =D(\hg)\chi(\T,\~\tau,\tau,\hg)
\end{equation}
\begin{equation}
\~D(\hg) =-2\pi i\~\Delta(\hg)=2\pi i\~L^{ab}\Big(E_a(0)E_b(0)
+2\sum_{m>0}E_a(-m)E_b(m)\Big)
\end{equation}
\begin{equation}
D(\hg) =-2\pi i\Delta(\hg)=2\pi iL^{ab}\Big(E_a(0)E_b(0)
+2\sum_{m>0}E_a(-m)E_b(m)\Big)
\end{equation}
\begin{equation}
\~D(\hg)+D(\hg)=D_g(\hg)
\end{equation}
\label{genheat}
\end{subequations}
where the closed form connections $\~D$ and $D$ are flat and abelian flat.

The second-order differential operators $\~\D(\hg)$, $\D(\hg)$, and
$\D_g(\hg)$, which represent $-\tL(0)$, $-L(0)$, and $-L_g(0)$ respectively,
are three mutually commuting generalized
Laplacians\footnote{Using other first-order differential
 representations of the current algebra,
the affine-Sugawara Laplacian $\D_g(\hg)$ is known in
mathematics.  See e.g. Ref.\  \cite{efk}.}
 on the affine Lie
group.  The simultaneous eigenbasis of the three Laplacians is the
simultaneous $L$-basis (see Section \ref{hc3})
for all levels of the affine modules.

With the affine-Sugawara boundary condition $\chi(\tau,\tau)=\chi_g(\tau)$,
the heat-like system (\ref{genheat}) has the unique solution for the
generalized bicharacters
\begin{equation}
\chi(\T,\~\tau,\tau,\hg)=e^{-2\pi i(\~\tau-\tau)\~\Delta(\hg)}
\chi_g(\T,\tau,\hg)
=e^{-2\pi i(\tau-\~\tau)\Delta(\hg)}\chi_g(\T,\~\tau,\hg)
\end{equation}
in terms of the generalized affine-Sugawara characters.  To our knowledge, the
explicit form of the generalized
affine-Sugawara characters $\chi_g(\hg)$ has not
yet been given.  See Ref.\  \cite{tor} for further discussion of this
solution.

The bicharacters of the earlier subsections can be obtained from the
generalized
bicharacters by restricting the affine source to a Lie source. The
advantage of the formulation on the affine group is that the closed form
connections may be useful in the investigation of global properties
such as factorization and modular covariance.

\section{The Generic World-Sheet Action\label{hc41}}

\subsection{Background\label{hc42}}

The world-sheet action of the generic ICFT was given by Halpern and
Yamron \cite{gva}.  The linearized form of this action was found by
de Boer, Clubok, and Halpern \cite{bch} and an alternative, presumably
equivalent, form of the linearized action has been given by
Tseytlin \cite{ts3}.
The question has been raised, but not yet answered, whether the
affine-Virasoro constructions can also be described by a generalized
Thirring model [168, 66, 169-173, 177, 14]. 

\subsection{Non-Chiral ICFTs\label{hc43}}

Given a solution $L^{ab}$ of the Virasoro master equation, one may
construct a non-chiral conformal field theory as follows.  The basic
Hamiltonian of the $L$ theory is taken as
\begin{equation}
H_0=L(0)+\bar{L}(0) = L^{ab} (\xx J_aJ_b + \bar{J}_a\bar{J}_b \xx)_0
\end{equation}
which is the sum of the zero modes of the left- and right-mover stress tensors
\begin{equation}
T=L^{ab}\xx J_aJ_b \xx\quad, \quad \bT=L^{ab}\xx\bar J_a\bar J_b\xx
\end{equation}
where the barred currents $\bar J$ are right-mover copies of the left-mover
currents $J$.

In general, the basic Hamiltonian $H_0$ admits a local gauge invariance,
described by the (symmetry) algebra of the commutant of $H_0$, and the
physical (gauge-fixed) Hilbert space of the $L$ theory may be taken as
the primary states with respect to the symmetry algebra.

As a simple example, consider the stress tensor
$T_{g/h}=L^{ab}_{g/h}\xx J_aJ_b\xx$
of the $g/h$ coset constructions
\cite{bh,h1,gko}, whose symmetry algebra is the affinization of $h$.  Then the
physical Hilbert space of the non-chiral coset constructions is the
set of states
\begin{equation}
J_a{}(m>0)|\mbox{phys}\rangle = \bar{J}_a{}(m>0)|\mbox{phys}\rangle = 0,
\hskip 10pt a=1\ldots\dim h
\label{ref1}
\end{equation}
which are primary under affine $h \times h$.

In the space of all CFTs, the coset constructions are only special
points of higher symmetry.  The symmetry algebra of the generic stress
tensor is the Virasoro algebra of its commuting K-conjugate theory,
\begin{equation}
\tilde{T}=\tilde{L}^{ab}\xx J_a J_b\xx=T_g-T, \hskip 10pt
\tilde{L}^{ab}=L_g^{ab}-L^{ab}, \hskip 10pt
c(\tilde L)=c_g-c(L)
\end{equation}
where $T_g$ is the affine-Sugawara construction on $g$.  Then the
physical Hilbert space of the generic theory $L$ may be taken as the
states
\begin{subequations}
\begin{equation}
\tilde{L}(m>0)|\mbox{phys}\rangle=
\bar{\tilde{L}}(m>0)|\mbox{phys}\rangle=0
\end{equation}
\begin{equation}
\tilde{T}=\tilde{L}^{ab} \xx J_a J_b\xx , \hskip 15pt
\bar{\tilde{T}}=\tilde{L}^{ab}\xx \bar{J}_a \bar{J}_b \xx
\end{equation}
\label{ref2}
\end{subequations}
which are Virasoro primary under the K-conjugate stress tensors
$\tilde T$ and $\bar{\tilde T}$.

Transcribing (\ref{ref1}) and (\ref{ref2}) into the language of
world-sheet actions, it is clear that one will obtain a spin-1 gauge
theory \cite{brs,gk,gk2,kahs,kas}
for the special cases of the coset constructions and a
spin-2 gauge theory \cite{gva} for the generic theory, where the generic theory
$L$ is gauged by its K-conjugate theory $\tL$.

One other preliminary is necessary to understand the generic
affine-Virasoro action.  Since an action begins as a semi-classical
description of a quantum system, the affine-Virasoro action will involve
the semi-classical (high-level) form of the solutions of  the master
equation.   In order to  obtain a smooth semi-classical description,
the  discussion  of \cite{gva} is  limited to the high-level smooth
CFTs, whose high-level form
\begin {subequations}
\begin{equation}
L^{ab}_\infty={1 \over 2} G^{ac}P_c{}^b, \hskip 10pt
\tilde{L}^{ab}_\infty={1 \over 2} G^{ac}\tilde{P}_c{}^b, \hskip 10pt
L^{ab}_\infty+\tilde L^{ab}_\infty=L^{ab}_{g,\infty}=\half G^{ab}
\end{equation}
\begin{equation}
c(L_\infty)=\rank P, \hskip 10pt c(\tilde L_\infty)=\rank \tilde P,
\hskip 10pt
c(L_\infty)+c(\tilde L_\infty)=c(L_{g,\infty})=\dim g
\end{equation}
\begin{equation}
P^2=P, \hskip 10pt \tilde{P}^2=\tilde{P}, \hskip 10pt
P+\tilde{P}=1, \hskip 10pt P\tilde P = \tilde P P = 0
\end{equation}
\label{mastersol}
\end{subequations}
was reviewed in Section \ref{slc}.  Here $P$ and $\tP$ are the
high-level projectors of the $L$ and the $\tL$ theories respectively,
whose high-level central charges are the ranks of their projectors.
These results are valid for high levels $k_I$ of $g=\oplus_Ig_I$, but
the  reader may wish to think in terms of simple $g$, for which
$G^{ab}=k^{-1}\et^{ab}$, where $\et^{ab}$ is the inverse Killing metric
of $g$.
We  remind the reader that  the high-level smooth CFTs are believed to
include the generic  level-family and all unitary level-families  on
simple $g$.

\subsection{The WZW Model\label{hc44}}

In this section we review the WZW model \cite{nov,wit1}, following the lines
which are used to construct the generic affine-Virasoro action.

One begins on the group manifold $G$ with $g(x)\in G$ in
matrix irrep $\T$,
\begin{equation}
[\T_a,\T_b] = if_{ab}{}^c, \hskip 10pt \Tr (\T_a\T_b)=yG_{ab}, \hskip 10pt
a,b,c=1\ldots\dim g
\end{equation}
and a set of coordinates $x^i(\tau,\sigma), i=1\ldots \dim g$
with associated canonical momenta $p_i(\tau,\s)$.
Introduce the left- and right-invariant vielbeins $e_i{}^a,\bar{e}_i{}^a$
and the antisymmetric tensor field $B_{ij}$ by
\begin{subequations}
\begin{equation}
e_i \equiv -ig^{-1}\partial_i g = e_i{}^a \T_a \hskip 15pt
\bar{e}_i \equiv -ig\partial_i g^{-1} = \bar{e}_i{}^a \T_a
\end{equation}
\begin{equation}
\hskip 15pt i \Tr(e_i[e_j,e_k])=
     -i \Tr(\bar e_i[\bar e_j,\bar e_k])
    =\partial_i B_{jk} + \partial_j B_{ki}
                         +\partial_k B_{ij}.
\end{equation}
\end{subequations}
The inverse vielbeins are $e_a{}^i, \bar e_a{}^i$.

In this notation, one has the canonical representation of the
currents \cite{bow}
\begin{subequations}
\begin{equation}
J_a = 2\pi e_a^{~i}\hat{p}_i + {1\over 2} G_{ab} e_i^{~b}x^{\prime i},
 \hskip 15pt
\bar{J}_a = 2\pi \bar{e}_a^{~i}\hat{p}_i
        - {1\over2} G_{ab} \bar{e}_i^{~b}{x'}^i
\end{equation}
\begin{equation}
\hat{p}_i \equiv p_i-{1 \over 4\pi y} B_{ij} x'^j
\label{pref}
\end{equation}
\label{currents}
\end{subequations}
which satisfy the bracket algebra of affine $g\times g$.  The classical
WZW Hamiltonian is then
\begin{subequations}
\begin{equation}
H_{WZW}=\int_0^{2\pi}d\sigma {\cal H}_{WZW}
\end{equation}
\begin{equation}
{\cal H}_{WZW} = {1 \over 2\pi} L^{ab}_{g,\infty}(J_aJ_b+\bar{J}_a\bar{J}_b),
\hskip 15pt  L^{ab}_{g,\infty} ={1 \over 2} G^{ab}
\end{equation}
\label{wzwham}
\end{subequations}
where $L^{ab}_{g,\infty}$ is the high-level form of the inverse inertia
tensor $L_g^{ab}$ of
the affine-Sugawara construction on $g$.

Using the Hamiltonian equations of motion, one eliminates $p$ in favor
of $\dot x$ to obtain the
component form of the WZW action,
\begin{subequations}
\begin{equation}
S_{WZW}  =  \int d\tau d\sigma ({\cal L}_{WZW} + \Gamma)
\end{equation}
\begin{equation}
{\cal L}_{WZW}  =  \frac{1}{8\pi}G_{ab} e_i{}^a e_j{}^b
   (\dot{x}^i\dot{x}^j - x'^ix'^j), \hskip 15pt
\Gamma  =  \frac{1}{4\pi y} B_{ij} \dot{x}^i x'^j
\end{equation}
\label{wzw1}
\end{subequations}
where $\Gamma$ is the WZW term.
The action can also be written in terms of the group variable
\begin{subequations}
\begin{equation}
S_{WZW}=-{1 \over 2\pi y}\int d^2z \, \Tr(g^{-1}\partial g g^{-1}
                   \bar\partial g)
       -{1 \over 12\pi y}\int_M \Tr(g^{-1}dg)^3
\end{equation}
\begin{equation}
\pa\equiv{1\over2}(\pa_\tau+\pa_\s)\quad, \quad
\bar\pa\equiv{1\over2}(\pa_\tau-\pa_\s)
\end{equation}
\end{subequations}
where we have defined $d^2z=d\tau d\s$.

\subsection{The Generic Affine-Virasoro Hamiltonian \label{hc45}}

The classical basic Hamiltonian of any high-level smooth
affine-Virasoro construction $L$
is
\begin{equation}
H_0=\int_0^{2\p} d\s {\cal H}_0\quad, \quad
{\cal H}_0 = {1 \over 2\pi}L^{ab}_\infty(J_aJ_b+\bar{J}_a\bar{J}_b)
\end{equation}
where $L^{ab}_\infty$ is the high-level form of $L^{ab}$ in (\ref{mastersol}),
and
\begin{equation}
 {1 \over 2\pi}L^{ab}_\infty J_aJ_b\quad, \quad
 {1 \over 2\pi}L^{ab}_\infty \bar{J}_a\bar{J}_b
\label{Stress1}
\end{equation}
are the classical analogues of the left- and right-mover stress tensors
of the $L$ theory.
For generic $L^{ab}\ra L^{ab}_\infty$,
the local symmetry algebra of $H_0$ is generated
by the stress tensors
\begin{equation}
{1\over2\pi}\tilde{L}^{ab}_\infty J_a J_b\quad, \quad
 {1\over2\pi}\tilde{L}^{ab}_\infty \bar{J}_a \bar{J}_b
\label{Stress2}
\end{equation}
of the commuting K-conjugate theory.  All four stress tensors
(\ref{Stress1}--3) satisfy commuting (bracket) Virasoro
algebras with no central terms, so the K-conjugate stress tensors in
(\ref{Stress2}) are first
class constraints of $H_0$.

Following Dirac \cite{dirac}, one obtains the full
Hamiltonian of the generic theory $L$ \cite{gva},
\begin{subequations}
\begin{equation}
H=\int_0^{2\pi} d\sigma {\cal H}\quad, \quad
{\cal H}  =  {\cal H}_0 + v \cdot K(\tilde{L}_\infty)
\end{equation}
\begin{equation}
v \cdot K(\tilde{L}_\infty) = {1\over2\pi}\tilde{L}^{ab}_\infty (v J_a J_b
    + \bar{v} \bar{J}_a \bar{J}_b)
\end{equation}
\label{fullham}
\end{subequations}
where the K-conjugate stress tensors play the role of Gauss' law, and
$v, \bar v$ are multipliers.  The multipliers form a spin-two gauge
field on the world-sheet, the so-called K-conjugate metric, which
couples only to the K-conjugate ``matter.''  It should be emphasized that
this Hamiltonian generalizes and includes the WZW Hamiltonian (\ref{wzwham}),
which is included when $L=L_g$, $\tL=0$, and the
 conventional world-sheet metric
formulation of the WZW model, which is included when $L=0$, $\tL=L_g$.

\subsection{The Generic Affine-Virasoro Action\label{hc46}}

{}To eliminate $p$ in favor of $\dot x$, one needs the adjoint action of $g$
\begin{equation}
g\T_a g^{-1}  = \omega_a{}^b \T_b, \hskip 15pt
\omega_a{}^c G_{cd} \omega_b{}^d = G_{ab}.
\end{equation}
Then one obtains the non-linear form of the generic affine-Virasoro
action \cite{gva}
\begin{subequations}
\begin{equation}
     S  =  \int d\tau d\sigma ({\cal L} + \Gamma)
\end{equation}
\begin{eqnarray}
     {\cal L} & = & {1\over 8\pi}e_i{}^aG_{bc}\e_j{}^c\biggl[
        \left[f(Z)
          +\alpha\alphab\omega\tilde{P}\omega^{-1}f(Z)\tilde{P}\right]_a{}^b
           \left(\dot{x}^i\dot{x}^j - x^{\prime i}x^{\prime j}\right)
\nonumber \\
       & &+\alpha\left[f(Z)\tilde{P}\right]_a{}^b
             \left(\dot{x}^i\dot{x}^j  + x^{\prime i}x^{\prime j}
                         +\dot{x}^{(i}x^{j)\prime}\right)
\nonumber \\
      &  &+\alphab
               \left[\omega\tilde{P}\omega^{-1}f(Z)\right]_a{}^b
             \left(\dot{x}^i\dot{x}^j  + x^{\prime i}x^{\prime j}
                         -\dot{x}^{(i}x^{j)\prime}\right)
\nonumber \\
      & &+\left[1-f(Z)
          +\alpha\alphab\omega\tilde{P}\omega^{-1}f(Z)\tilde{P}\right]_a{}^b
             \left(\dot{x}^{[i}x^{j]\prime}\right)\biggr]
\label{badl}
\end{eqnarray}
\label{eqafvirl}
\begin{equation}
f(Z)  \equiv  [1-\alpha \alphab Z]^{-1}, \hskip 15pt
Z\equiv \tilde{P}\omega\tilde{P}\omega^{-1}, \hskip 15pt
\alpha \equiv {1-v \over 1+v}, \hskip 15 pt
\bar{\alpha} \equiv {1-\bar{v} \over 1+ \bar{v}}
\end{equation}
\end{subequations}
which is the world-sheet description of the generic theory $L$.  This
action exhibits
 Lorentz, diffeomorphism, local Weyl and conformal
invariance, as discussed below.

\noindent {\bf A.}  Diff S$_2(K)$ invariance.
The affine-Virasoro action is invariant
under the Diff~S$_2(K)$ coordinate transformations,
\begin{subequations}
\begin{equation}
\delta x^i = \Lambda^a e_a{}^i + \bar\Lambda^a \bar e_a{}^i
\end{equation}
\begin{equation}
\delta g=gi\Lambda^a\T_a-i\bar\Lambda^a\T_ag
\label{gtrans}
\end{equation}
\begin{equation}
\delta J_a = f_{ab}{}^c \Lambda^b J_c + G_{ab}\partial_\sigma \Lambda^b,
     \hskip 10pt
\delta \bar{J}_a = f_{ab}{}^c \bar{\Lambda}^b \bar{J}_c
     -G_{ab} \partial_\sigma \bar{\Lambda}^b
\label{jtransf}
\end{equation}
\begin{equation}
\Lambda^a = 2\epsilon \tilde L^{ab}_\infty J_b, \hskip 15pt
\bar \Lambda^a = 2\bar\epsilon \tilde L^{ab}_\infty \bar J_b
\label{Lambda}
\end{equation}
\begin{equation}
\delta v  =  \dot{\epsilon} + \epsilon\lrp_\sigma v, \hskip 15pt
\delta \bar{v} = \dot{\bar{\epsilon}} + \bar{v}\lrp_\sigma \bar{\epsilon}
\label{diffv}
\end{equation}
\label{diffs2}
\end{subequations}
associated to the stress tensors of the $K$-conjugate theory.
$\epsilon(\tau,\s)$ and $\bar\epsilon(\tau,\s)$ are the infinitesimal
diffeomorphism parameters.

With Ref.\  \cite{gva}, we note the remarkable $\tilde L^{ab}$-dependent
embedding of Diff S$_2(K)$ in
local Lie $G$ $\times$ Lie $G$, which we believe will be of future
interest in mathematics.  The transformation of the
group element $g$ in eq.(\ref{gtrans}) shows that infinitesimal
Diff S$_2(K)$ transformations are particular transformations in
local Lie $G$ $\times$ Lie $G$, with the current-dependent local
Lie $g$ $\times$ Lie $g$ gauge parameters $\Lambda,\bar\Lambda$ in
eq.(\ref{Lambda}).  In this sense, Diff S$_2(K)$ is a large set of
distinct local embeddings of Diff S$_2$
in local Lie $G$ $\times$ Lie $G$, one for each $\tilde L^{ab}$.
Moreover, the result
(\ref{jtransf}) shows that the currents $J,\bar J$ transform under
Diff S$_2(K)$ as local Lie $g$ $\times$ Lie $g$ gauge fields, or
connections, with the same gauge parameters $\Lambda,\bar\Lambda$.
The embedding of Diff S$_2(K)$ in local Lie $G$ $\times$ Lie $G$
is the underlying geometry of the generic affine-Virasoro construction, and
this geometry continues to play
a central role in the linearized action below.

The transformation of $v,\bar{v}$ in (\ref{diffv}) allows the
identification of a second-rank tensor field,
the K-conjugate metric,
\begin{equation}
\tilde{h}_{mn}\equiv e^{-\phi}\pmatrix{ -v \bar{v} &
       {1 \over 2}(v- \bar{v}) \cr {1 \over 2}(v-\bar{v}) & 1 }, \hskip 10pt
\sqrt{-\tilde h} \tilde h^{mn} = {2 \over v+\bar v}
  \pmatrix{ -1 & {1\over2}(v-\bar v) \cr {1\over2}(v-\bar v) & v \bar v }
\label{tmetric}
\end{equation}
which couples only to the K-conjugate matter.  The Weyl mode $\phi$
does not appear in the action (\ref{eqafvirl}),
guaranteeing the classical invariance of the action
under local Weyl transformations.

\noindent {\bf B.}  WZW limits.
The affine-Virasoro action reduces to the WZW action when
$L^{ab}=L^{ab}_g, \tL=0$, and to the WZW action coupled to
gravity when $L=0, \tL=L_g$.
In the $L=0$ case, the K-conjugate metric (\ref{tmetric})
is the conventional world-sheet metric of the WZW model,
because all the matter belongs to the K-conjugate theory, and
the group element $g$ is a conventional scalar field under Diff S$_2$.
In this particular case, the embedding discussed above may be related
\cite{gva} to earlier work by Polyakov \cite{pol}.

For any $L^{ab}$,  the action also reduces to $S_{WZW}$
in the WZW gauge,
\begin{equation}
v=\bar v=1\quad, \quad
\alpha=\bar\alpha=0\quad, \quad
\sqrt{-\tilde h}\tilde h^{mn}=\pmatrix{-1 & 0 \cr 0 & 1}
\end{equation}
so that the WZW gauge is the conformal gauge of the generic action.

\noindent {\bf C.}  K-conjugate metric as a world-sheet metric \cite{bch}.
Following the usual prescription, one defines the symmetric, covariantly
conserved gravitational stress tensor of the theory as
\begin{equation}
\~\theta^{mn}={2\over \sqrt{-\tilde{h}}} {\delta S \over \delta
 \tilde{h}_{mn}}
\label{gravtens}
\end{equation}
where the differentiation precedes any choice of gauge.
In the WZW gauge, one finds that the stress tensor reduces to the form,
\begin{subequations}
\begin{equation}
\~\theta_{00}=\~\theta_{11}={1\over 2\pi} \tilde{L}^{ab}_\infty
     (J_a J_b + \bar{J}_a \bar{J}_b)
\end{equation}
\begin{equation}
\~\theta_{01}=\~\theta_{10}={1\over 2\pi} \tilde{L}^{ab}_\infty
     (J_a J_b - \bar{J}_a \bar{J}_b)
\end{equation}
\end{subequations}
which identifies the K-conjugate metric
as the world-sheet metric of the $\tilde{L}$ theory.

\noindent {\bf D.} Rigid conformal invariance.
The action is invariant under a rigid (ungauged) conformal invariance,
including a rigid world-sheet Lorentz invariance, of the usual form,
\begin{subequations}
\begin{equation}
\delta \xi^\pm = -\rho^\pm(\xi^\pm), \hskip 10pt
\delta x^i = (\rho^+\partial_+ + \rho^-\partial_-) x^i
\end{equation}
\begin{equation}
\delta\alpha=(\rho^+\partial_+ + \rho^-\partial_-)\alpha
    +(\partial_-\rho^- - \partial_+\rho^+) \alpha
\end{equation}
\begin{equation}
\delta\bar\alpha=(\rho^+\partial_+ + \rho^-\partial_-)\bar\alpha
    +(\partial_+\rho^+ - \partial_-\rho^-)\bar\alpha
\end{equation}
\label{contran}
\end{subequations}
where
$\xi^\pm\equiv(\tau\pm\sigma)/\sqrt{2}$.
The transformations in (\ref{contran})
identify $\alpha$ and $\bar\alpha$ as ($-1$,1) and (1,$-1$) conformal
tensors respectively, and $x$ as a conformal scalar.
With these identifications, each term in the action
density (\ref{badl}) is (1,1) on inspection.

The rigid conformal group is the conformal group of the $L$ theory,
generated by the stress tensors $L_\infty^{ab}J_a J_b/2\pi$ and
$L_\infty^{ab}\bar J_a \bar J_b/2\pi$.  The theory also has a gauged
conformal group in Diff S$_2(K)$, associated to the $K$-conjugate
stress tensors $\tilde L^{ab}_\infty J_a J_b /2\pi$ and
$\tilde L^{ab}_\infty \bar J_a \bar J_b /2\pi$.

\subsection{The Linearized Generic Affine-Virasoro Action\label{hc47}}

The original form (\ref{eqafvirl}) of the generic action is highly non-linear.
By introducing auxiliary fields $B_a, \bar B_a, a=1\ldots\dim g$,
it is possible to write the affine-Virasoro
action in a linearized form \cite{bch},
\begin{subequations}
\begin{equation}
      S'  = \int d\tau d\sigma ({\cal L}' + \Gamma)
\end{equation}
\begin{eqnarray}
       {\cal L}' & = & -{1\over 8\pi} G_{ab}(e_\tau{}^a e_\tau{}^b
               - e_\sigma{}^a e_\sigma{}^b)
              +{1\over\pi}G^{ab}B_a\omega_b^{~c}\bar{B}_c
\nonumber \\
        & &+{\alpha\over \pi}\tL^{ab}_\infty B_a B_b
              + {1\over 2\pi} (e_\tau{}^a - e_\s{}^a)B_a
\nonumber \\
       & & +{\bar{\alpha} \over \pi} \tL^{ab}_\infty\bar{B}_a \bar{B}_b
                + {1\over 2\pi} (\bar e_\tau{}^a + \bar e_\s{}^a)
                         \bar{B}_a
\label{aform}
\end{eqnarray}
\label{eqslamone}
\end{subequations}
where we have introduced $e_\tau{}^a\equiv e_i{}^a\dot{x}^i,
e_{\sigma}{}^{a}\equiv e_{i}{}^{a}x'^i$ and similarly for $\bar e$.
Note that the first term in ${\cal L}'$ and the WZW term $\Gamma$
almost comprise the action $S_{WZW}$, but the kinetic energy term has the
wrong sign.

Writing the action (\ref{eqslamone}) in terms of the group variable
$g$, one obtains
\begin{subequations}
\begin{equation}
S'=S_{WZW}+\int d^2z \Delta \sL_B
\end{equation}
\begin{eqnarray}
\D \sL_B &=& {\a\over\p y^2} \tL^{ab}_{\infty}\Tr(\T_aB)\Tr(\T_bB)
\nonumber \\
 & & + {\bar\a\over\p y^2} \tL^{ab}_\infty\Tr(\T_a\bB)\Tr(\T_b\bB)
\nonumber \\
 & & - {1\over\p y} \Tr(\bar D_BgD_Bg^{-1})
\end{eqnarray}
\begin{equation}
B\equiv B_aG^{ab}\T_b\quad, \quad \bB\equiv \bB_aG^{ab}\T_b\quad, \quad
D_B\equiv \pa+iB\quad, \quad \bar D_B=\bar\pa+i\bB
\end{equation}
\label{maction1}
\end{subequations}
where the covariant derivatives $D_B$ and $\bar D_B$ have been introduced.
In this form, the kinetic energy term in $S_{WZW}$
has the correct sign, but this sign is flipped by a similar term
in $\D \sL_B$.

The sign of the kinetic energy can be corrected by introducing
a second set of auxiliary connections,
\begin{equation}
A\equiv A_a G^{ab}\T_b = -i g D_B g^{-1}\quad, \quad
\bA\equiv \bA_a G^{ab}\T_b = -i g^{-1} \bar D_B g .
\label{secset}
\end{equation}
In terms of these auxiliary fields, one obtains the final form of the
linearized
affine-Virasoro action \cite{bch},
\begin{subequations}
\begin{equation}
S'=S_{WZW}+\int d^2z \Delta {\cal L}
\end{equation}
\begin{equation}
S_{WZW}=-{1 \over 2\pi y}\int d^2z \, \Tr(g^{-1}\partial g g^{-1}
                   \bar\partial g)
       -{1 \over 12\pi y}\int_M \Tr(g^{-1}dg)^3
\end{equation}
\begin{eqnarray}
\Delta{\cal L}& = &
     -{\alpha \over \pi y^2} \tilde L^{ab}_\infty
     \Tr \left(\T_ag^{-1}D_Ag\right)
     \Tr \left(\T_bg^{-1}D_Ag\right)
\nonumber \\
 & & -{\bar\alpha \over \pi y^2} \tilde L^{ab}_\infty
     \Tr \left(\T_ag\bar D_Ag^{-1}\right)
     \Tr \left(\T_bg\bar D_Ag^{-1}\right)
\nonumber \\
\nopagebreak
 & & +{1\over \pi y} \Tr(g\bar A g^{-1} A)
\label{mform}
\end{eqnarray}
\begin{equation}
D_A\equiv\partial+iA,\hskip 15pt \bar D_A\equiv\bar\partial+i\bar A
\label{hcovdir}
\end{equation}
\label{maction}
\end{subequations}
where the covariant derivatives
$D_A$ and $\bar D_A$ are defined in (\ref{hcovdir}).
The auxiliary fields can be integrated out of either
(\ref{maction1}) or (\ref{maction}) to recover the non-linear
form of the action (\ref{eqafvirl}).

In this form, the theory is clearly seen as a Diff S$_2$-gauged WZW model,
which bears an intriguing resemblance to the form of the usual
(Lie algebra) gauged WZW model \cite{brs,gk,gk2,kahs,kas}.
As we shall discuss in the following subsection, this resemblance
is due to the Diff S$_2(K)$
transformations of the auxiliary connections.

\subsection{Invariances of the Linearized Action\label{hc48}}

Because it is simpler, we discuss first the rigid conformal invariance,
under which the group element $g$ is a conformal scalar, $\a,\bar\a$
transform as in (\ref{contran}) and the auxiliary
fields $A, B$ and $\bar A, \bar B$ are (1,0) and (0,1) tensors
respectively.  With these assignments, each term in (\ref{maction1}) or
(\ref{maction}) is a
(1,1) tensor.  We note in particular the (2,0) and
(0,2) matter factors in (\ref{maction}),
which are proportional to $\a$ and $\bar\a$ respectively.
Such terms, which do not appear in the spin-one gauged WZW model,
are allowed when the gauge field is spin two.

The action (\ref{maction1}) is also invariant under the the
Diff S$_2(K)$ transformation
\begin{subequations}
\begin{equation}
\delta \alpha   =  -\bar\partial\xi+\xi\lrp\alpha, \hskip 15pt
\delta \bar\alpha = -\partial\bar\xi + \bar\xi\lrbp\bar\alpha
\label{alpha}
\end{equation}
\begin{equation}
\delta g=gi\lambda-i\bar\lambda g
\end{equation}
\begin{equation}
\delta B=\partial\lambda+i[B,\lambda], \hskip 10pt
\delta \bar B=\bar\partial\bar\lambda+i[\bar B,\bar\lambda]
\end{equation}
\begin{equation}
\lambda\ident\lambda^a\T_a,\hskip 15pt \bar\lambda\ident\bar\lambda^a\T_a
\end{equation}
\begin{equation}
\lambda^a  \ident  2\xi\tilde L^{ab}_\infty B_b, \hskip 10 pt
\bar\lambda^a \ident 2\bar\xi\tilde L^{ab}_\infty \bar B_b
\label{lambda}
\end{equation}
\label{trans1}
\end{subequations}
where $\xi,\bar\xi$ are the diffeomorphism parameters.
The final form (\ref{maction}) of the action keeps the transformations
(\ref{trans1}a,b) for $\a,\bar\a$ and the group element $g$, while
(\ref{trans1}c,e) are replaced by
\begin{subequations}
\begin{equation}
\delta A=\partial\bar\lambda+i[A,\bar\lambda], \hskip 10pt
\delta \bar A=\bar\partial\lambda+i[\bar A,\lambda]
\end{equation}
\begin{equation}
\lambda^a  =  2\xi\tilde L^{ab}_\infty B_b(A), \hskip 10 pt
\bar\lambda^a \ident 2\bar\xi\tilde L^{ab}_\infty \bar B_b(\bar A)
\end{equation}
\label{mtrans}
\end{subequations}
where $B(A)=-ig^{-1}D_Ag$ is the inverse of $A(B)$ in (\ref{secset}), and
similarly for $\bar B(\bar A)$.

We remark in particular that the auxiliary connections
in (\ref{trans1})
and (\ref{mtrans}) transform under Diff S$_2(K)$
as local
Lie $g$ $\times$ Lie $g$ connections,
with field-dependent local Lie $g$ $\times$ Lie $g$ parameters
$\lambda, \bar\lambda$, restricted as shown in (\ref{lambda}).
As emphasized above, this
is possible because Diff S$_2(K)$ is locally embedded in
local Lie $G$ $\times$ Lie $G$.
The transformation of the auxilary connections, as if they were gauge
fields of a local Lie algebra, accounts for the intriguing
resemblance of the linearized action (\ref{maction})
to the usual (Lie algebra) gauged WZW model.

Reference \cite{bch} also gives a semi-classical discussion of the
affine-Virasoro action following the analysis of Distler and
Kawai \cite{dk}.  The full quantum version of this argument may be
realized with standard BRST operators of the form
\begin{equation}
Q=\oint{dz\over2\p i} c (\tT+T_{FF}+{1\over2}T_G)\quad, \quad
c_{FF}=26-c(\tL)
\end{equation}
where $\tT$ is the quantum stress tensor of the $\tL$ theory,
$T_G$ is the stress tensor of the usual diffeomorphism ghosts, and
$T_{FF}$ is the stress tensor of the appropriate Feigen-Fuchs system.
Up to discrete states, these BRST operators reproduce the physical
state conditions (\ref{ref2}) for the generic theory $L$.

\subsection{Two World-Sheet Metrics\label{hc49}}

So far, we have reviewed the world-sheet action of the $L$ theory,
which is gauged by its commuting K-conjugate theory $\tL$, and we
have seen that the spin-2 gauge field $\tilde h_{mn}$ of this
theory is the world-sheet metric of the $\tL$ theory.  Halpern and
Yamron \cite{gva} have also indicated how to incorporate the world-sheet
metric of the $L$ theory,
\begin{equation}
h_{mn}\ident e^{-\chi}\pmatrix{ -u \bar{u} &
       {1 \over 2}(u- \bar{u}) \cr {1 \over 2}(u-\bar{u}) & 1 }, \hskip 10pt
\sqrt{-h} h^{mn} = {2 \over u+\bar u}
  \pmatrix{ -1 & \half(u-\bar u) \cr \half(u-\bar u) & u \bar u }
\end{equation}
which results in the {\em doubly-gauged} affine-Virasoro action, with a
K-conjugate pair of world-sheet metrics $\tilde h_{mn}$ and $h_{mn}$.
In what follows, we refer to $\tilde h_{mn}$ and $h_{mn}$ as the
$\tL$-metric and the $L$-metric, which couple to $\tL$ and $L$ matter
respectively.

One begins with the doubly-gauged Hamiltonian \cite{gva},
\begin{subequations}
\begin{equation}
H_D=\int d\sigma {\cal H}_D
\end{equation}
\begin{equation}
{\cal H}_D = {1\over 2\pi} \left[
  ( u L_\infty + v \tilde L_\infty )^{ab} J_a J_b +
      ( \bar{u} L_\infty + \bar{v} \tilde L_\infty )^{ab}
         \bar{J}_a \bar{J}_b \right ]
  \ident u\cdot T + v\cdot K
\end{equation}
\end{subequations}
which reduces to the Hamiltonian (\ref{fullham})
of the $L$ theory when $u=\bar u=1$
(the conformal gauge of the $L$-metric).

Following the development above, one obtains
 the linearized form of the doubly-gauged action \cite{bch},
\begin{subequations}
\begin{equation}
S_D=S_{WZW}+\int d^2z \Delta {\cal L_D}
\end{equation}
\begin{eqnarray}
\Delta{\cal L_D}& = &
     -\left({\alpha \over \pi y^2} \tilde L^{ab}_\infty
            + {\b \over \p y^2} L^{ab}_\infty \right)
     \Tr \left(\T_ag^{-1}D_Ag\right)
     \Tr \left(\T_bg^{-1}D_Ag\right)
\nonumber \\
 & & -\left({\bar\alpha \over \pi y^2} \tilde L^{ab}_\infty
         +{\bar\b\over\p y^2} L^{ab}_\infty \right)
     \Tr \left(\T_ag\bar D_Ag^{-1}\right)
     \Tr \left(\T_bg\bar D_Ag^{-1}\right)
\nonumber \\
 & & +{1\over \pi y} \Tr(g\bar A g^{-1} A)
\end{eqnarray}
\begin{equation}
D_A=\partial+iA\quad,\quad \bar D_A=\bar\partial+i\bar A\quad,\quad
\b={1-u\over1+u}\quad, \quad \bar\b={1-\bar u\over 1+\bar u}\;\,.
\end{equation}
\label{lindoub}
\end{subequations}
Note that this action may be obtained from (\ref{maction}) by the simple
substitution
\begin{equation}
\alpha\tilde L^{ab}_\infty \rightarrow \alpha\tilde L^{ab}_\infty
                                      +\beta L^{ab}_\infty, \hskip 15pt
\bar\alpha\tilde L^{ab}_\infty \rightarrow \bar\alpha\tilde L^{ab}_\infty
                                      +\bar\beta L^{ab}_\infty
\label{doubsub}
\end{equation}
and the same substitution may be used in the forms (\ref{eqslamone}) or
(\ref{maction1}) to obtain
the corresponding doubly-gauged forms.

The non-linear form of the doubly-gauged action \cite{bch},
\begin{subequations}
\begin{equation}
     S_D = \int d\tau d\sigma ({\cal L}_D+\Gamma)
\end{equation}
\begin{equation}
     {\cal L}_D  =
           -{1\over 8\pi}G_{ab}
                   \penclose{e_{\tau}{}^{a}e_{\tau}{}^{b}
                           - e_{\sigma}{}^{a}e_{\sigma}{}^{b}}
         -{1\over 8\pi}
               E^A  (C^{-1})_A{}^B E_B
\end{equation}
\begin{equation}
E^A= \pmatrix{(e_{\s}{}^{a}-e_{\tau}{}^{a}),
           & (e_{\s}{}^{a}+e_{\tau}{}^{a})}, \hskip 15pt
E_B= \pmatrix{G_{bc}(e_{\s}{}^{c}-e_{\tau}{}^{c})
           \cr G_{bc}(e_{\s}{}^{c}+e_{\tau}{}^{c})}
\end{equation}
\begin{equation}
  C^{-1}     = \pmatrix{-\omega(\alphab\tilde{P}+\bar{\beta} P)\omega^{-1} f
         &1+\omega(\alphab\tilde{P} + \bar\beta P)
                  \omega^{-1} f(\alpha\tilde{P}+\beta P)\cr
                  f &  -f(\alpha\tilde{P}+ \beta P)}
\end{equation}
\begin{equation}
f \ident \sbenclose{1-(\alpha\tilde{P}+\beta P) \omega
         (\alphab\tilde{P}+\bar{\beta} P) \omega^{-1}}^{-1}
\end{equation}
\label{nlindoub}
\end{subequations}
may be obtained from (\ref{lindoub})
by integrating out the auxiliary connections.
It is not difficult to check that this result agrees with the
non-linear form (\ref{eqafvirl}) when $u=\bar u=1$ so that $\b=\bar\b=0$.

The doubly-gauged actions (\ref{lindoub}) or (\ref{nlindoub})
are invariant under two commuting
diffeomorphism groups, called Diff S$_2(T)$ $\times$ Diff S$_2(K)$,
associated to  the two metrics
$h_{mn}$ and $\tilde h_{mn}$.  In particular, $h_{mn}$ is a
rank-two tensor under  Diff S$_2(T)$ but inert under Diff S$_2(K)$ and
vice versa for  $\tilde h_{mn}$.  The explicit forms of these
transformations are given in Refs.\ \cite{gva} and \cite{bch}.

In the non-linear form (\ref{nlindoub}), one defines the
K-conjugate pair of gravitational stress tensors,
\begin{equation}
\tilde\theta^{mn}=
{2\over\sqrt{-\tilde h}} {\d S_D\over\d \tilde h_{mn}}\quad, \quad
\theta^{mn}={2\over\sqrt{-h}} {\d S_D\over\d h_{mn}}
\label{doubgrav}
\end{equation}
and checks that $\~\theta$ ($\theta$) reduces
to the stress tensor of the $\tL$ ($L$) theory when $\a=\bar\a=0$
($\b=\bar\b=0$).
Thus $\tilde h_{mn}$ and $h_{mn}$ are the world-sheet metrics of the
$\tL$ and the $L$ theories respectively.
\subsection{Speculation on an Equivalent Sigma Model\label{hc50}}

An open problem in the action formulation is the possible equivalence of the
affine-Virasoro action to non-linear sigma models.  In this section, we
sketch a speculative, essentially classical derivation \cite{chun} of such a
sigma model, with surprising results.
It is our intuition that the ideas discussed here will be an important
direction in the future, but the details of the derivation cannot be
taken seriously until one-loop effects are properly included.

One begins with the doubly-gauged action
obtained by the substitution (\ref{doubsub}) in (\ref{eqslamone}),
and integrate out the gauge fields $\a, \bar \a$ of
the $\tL$ metric.  This gives the $\d$-function constraints
\begin{equation}
\tP^{ab}B_a B_b=\tP^{ab}\bar B_a \bar B_b=0
\end{equation}
which are solved by
\begin{equation}
B_a=P_a{}^b b_b\quad, \quad \bar B_a=P_a{}^b\bar b_b
\end{equation}
with unconstrained $b,\bar b$.
Using the inversion formula,
\begin{subequations}
\begin{equation}
\left( \matrix{\b P & P\o P \cr
               P\o^{-1}P & \bar\b P} \right)
\left( \matrix{-\bar\b PM(\o)P & P\o PM(\o^{-1})P \cr
             P\o^{-1}PM(\o)P & -\b PM(\o^{-1})P \cr } \right) =
\left( \matrix{ P & 0 \cr
                0 & P } \right)
\end{equation}
\begin{equation}
M(\o)\equiv \left(-\b\bar\b P +(1+P(\o-1)P)(1+P(\o^{-1}-1)P)\right)^{-1}
\end{equation}
\end{subequations}
one integrates $b,\bar b$ to obtain the
action,
\begin{subequations}
\begin{equation}
S_{\b,\bar\b}=\int d\tau d\s(\sL+\Gamma)
\end{equation}
\begin{equation}
\sL=-{1 \over 8\p}G_{ab}(e_\tau{}^a e_\tau{}^b-e_\s{}^a e_\s{}^b)
  -{1\over 8\p} E^A(C^{-1})_A{}^B E_B
\end{equation}
\begin{equation}
E^A=\left(\matrix{
(e_\tau{}^a-e_\s{}^a), & (\bar e_\tau{}^a+\bar e_\s{}^a)}\right)\quad,
\quad E_B= \left( \matrix{
G_{bc}(e_\tau{}^c-e_\s{}^c) \cr G_{bc}(\bar e_\tau{}^c + \bar e_\s{}^c)}\right)
\end{equation}
\begin{equation}
C^{-1}=\left( \matrix{-\bar\b PM(\o)P & P\o PM(\o^{-1})P \cr
             P\o^{-1}PM(\o)P & \b PM(\o^{-1})P \cr } \right)
\end{equation}
\label{Sbb}
\end{subequations}
which is now a function only of $\b,\bar\b$, that is,
the $L$ metric.  This is the conformal field theory of $L$ coupled to
its world-sheet metric.

One may compute the stress tensor of the
$L$ theory
\begin{subequations}
\begin{eqnarray}
\theta_{00}=\theta_{11}&=&{1\over16\pi}(e_\tau{}^a-e_\s{}^a)
(e_\tau{}^b-e_\s{}^b)
 G_{bc}(PF(\o)P)_a{}^c
\nonumber \\ & + &
 {1\over16\pi}(\bar e_\tau{}^a+\bar e_\s{}^a)(\bar e_\tau{}^b+\bar e_\s{}^b)
 G_{bc}(PF(\o^{-1})P)_a{}^c
\nonumber \\
\theta_{01}=\theta_{10}&=&{1\over16\pi}(e_\tau{}^a-e_\s{}^a)
(e_\tau{}^b-e_\s{}^b)
 G_{bc}(PF(\o)P)_a{}^c
\nonumber \\ & - &
 {1\over16\pi}(\bar e_\tau{}^a+\bar e_\s{}^a)(\bar e_\tau{}^b+\bar e_\s{}^b)
 G_{bc}(PF(\o^{-1})P)_a{}^c
\end{eqnarray}
\begin{equation}
F(\o)\equiv ((1+P(\o-1)P)(1+P(\o^{-1}-1)P))^{-1}
\end{equation}
\label{sigstress}
\end{subequations}
from eq.(\ref{doubgrav}) in the conformal gauge $\b=\bar\b=0$ of the $L$
theory.  Note that the degrees of freedom in $\theta_{mn}$ are
entirely projected onto the $P$ subspace, so
the semi-classical limit ($g=1+{i\over\sqrt{k}}\T\cdot\hat x+\cO(k^{-1})$)
of (\ref{sigstress}) gives the chiral stress tensors,
\begin{equation}
\theta_{\pm\pm}\;\limit{=}{k}\;{1\over4\p}((\pa_\tau \pm \pa_\s)P\hat x)^2
\quad, \quad
(P\hat x)^a=P^a{}_b \hat x^b.
\end{equation}
In this limit, the chiral stress tensors are conformal with high-level
central charge $c(L_\infty)=\rank P$, in agreement with the general
affine-Virasoro construction.

Finally, one obtains the conformal field theory of $L$ as the
effective sigma model,
\begin{subequations}
\begin{equation}
S_{\rm eff}=
\int d\tau d\s \left[{1\over8\p} G_{ij}^{\rm eff}(\dot x^i \dot x^j-x'^i x'^j)
+{1\over4\p y} B_{ij}^{\rm eff} \dot x^i x'^j \right]
\end{equation}
\begin{equation}
G_{ij}^{\rm eff}=e_i{}^a e_j{}^b G_{cb} (N+N^T-1)_a{}^c\quad, \quad
B_{ij}^{\rm eff}=B_{ij}-y e_i{}^a e_j{}^b G_{cb} (N-N^T)_a{}^c
\end{equation}
\begin{equation}
N=\o P(1+P(\o-1)P)^{-1}P\quad, \quad
N^T=P(1+P(\o^{-1}-1)P)^{-1}P\o^{-1}
\end{equation}
\label{Sigma}
\end{subequations}
by evaluating (\ref{Sbb})
in the conformal gauge of the $L$ metric.  In this sigma model,
$G_{ij}^{\rm eff}$ and $B_{ij}^{\rm eff}$ are the space-time metric and the
antisymmetric tensor field on the target space.  This action reduces to
$S_{WZW}$ when $P=1$, as it should.

Although the result (\ref{Sigma}) is an ordinary sigma model\footnote{Indeed,
Tseytlin \cite{ts3} and Bardak\c ci \cite{b}
 have studied a similar sigma model, which
is related to the bosonization of a generalized Thirring model
\cite{kps,kps2}.  When $Q$ in
Tseytlin's (3.1) is taken as twice the high-$k$ projector $P$, the two actions
are the same except for a dilaton term and
an overall minus sign for the kinetic term.},
the derivation above
highlights the fact that conformal invariance of a sigma model depends
on the choice of the world-sheet metric (the gravitational coupling)
and its associated stress tensor.
The correct stress tensor (\ref{sigstress}) of the $L$ theory
 followed from the unconventional
coupling of the $L$ metric to the $L$ matter in (\ref{Sbb}), but one may
also consider the same sigma model (\ref{Sigma}) with the
conventional choice \cite{cal},
\begin{equation}
\dot x^i \dot x^j - x'^i x'^j \ra g_C^{mn}\pa_m x^i\pa_n x^j \ra \theta_{mn}^C
\end{equation}
of the world-sheet metric $g_{mn}^C$
and its associated conventional stress tensor $\theta_{mn}^C$.
It is unlikely that $\theta_{mn}^C$ is conformal invariant in this case,
but the
question is not directly relevant because $g^C_{mn}$ and $\theta_{mn}^C$ are
 not the world-sheet
metric and stress tensor of the $L$ theory.

\newpage
\addcontentsline{toc}{section}{Acknowledgements}
\section*{Acknowledgements}

The authors wish to thank the theory groups of Ecole Polytechnique, CERN
and Berkeley for their gracious hospitality during the various stages of
this work.  For helpful discussions, we also thank K. Bardak\c ci,
D. Bisch, J. de Boer, P. Bouwknegt, S. Hwang, V. Jones,
N. Reshetikin, M. Shaposhnikov, N. Sochen, C. Thorn, A. Tseytlin and B. Zumino.

The work
of MBH and KC was supported in part by the Director, Office
of Energy Research, Office of High Energy and Nuclear Physics, Division of
High Energy Physics of the U.S. Department of Energy under Contract
DE-AC03-76SF00098 and in part by the National Science Foundation under
grant PHY90-21139.
The work of NO was supported in part by a
S\'ejour Scientifique de Longue Dur\'ee of the Minist\`ere des Affaires
Etrang\`eres.

\end{document}